\documentclass[%
 aip,
 amsmath,amssymb,
 reprint,]{revtex4-2}

\usepackage{graphicx}
\usepackage{dcolumn}
\usepackage{bm}
\usepackage{url}
\usepackage[utf8]{inputenc}
\usepackage[T1]{fontenc}
\usepackage{mathptmx}
\usepackage{etoolbox}

\usepackage{amsmath,amssymb}
\usepackage{amsfonts}
\usepackage{amsbsy}
\usepackage{mathtools}
\usepackage{color}
\usepackage{comment}
\usepackage{physics}
\usepackage{hyperref}

\newcommand{\sumal}{\sum_{\alpha=1}^{2}}

\newtheorem{statement}{Statement}
\makeatletter
\def\@email#1#2{%
 \endgroup
 \patchcmd{\titleblock@produce}
  {\frontmatter@RRAPformat}
  {\frontmatter@RRAPformat{\produce@RRAP{*#1\href{mailto:#2}{#2}}}\frontmatter@RRAPformat}
  {}{}
}%
\makeatother
\begin{document}

\preprint{AIP/123-QED}

\title[Classification of Sub-Shocks in the Shock Structure  of a  Binary Mixture of  Gases 
]
{A Complete Classification of Sub-Shocks in the Shock Structure  of a  Binary Mixture of Eulerian Gases with Different Degrees of Freedom
}
\author{Tommaso  Ruggeri}
\affiliation{ 
	Department of Mathematics $\&$ Alma Mater Research Center on Applied Mathematics, 
	University of Bologna, Bologna, Italy
}%
\author{Shigeru Taniguchi}
\email{taniguchi.shigeru@kct.ac.jp}
\affiliation{ 
Department of Creative Engineering, 
National Institute of Technology, Kitakyushu College, Kitakyushu, Japan
}%
\email{tommaso.ruggeri@unibo.it}

\date{\today}

\begin{abstract}
The shock structure in a binary mixture of polyatomic Eulerian gases with different degrees of freedom of a molecule is studied based on the multi-temperature model of rational extended thermodynamics.
Since the system of field equations is hyperbolic, the shock-structure solution is not always regular, and discontinuous parts (sub-shocks) can be formed.
For given values of the mass ratio and the specific heats of the constituents, we identify the possible sub-shocks as the Mach number $M_0$ of the shock wave and the concentration $c$ of the constituents change.
In the plane $(c,M_0)$, we identify the possible regions for the sub-shock formation. 
The analysis is obtained to verify when the velocity of the shock wave meets a characteristic velocity in the  unperturbed or perturbed equilibrium states that gives a necessary condition for the sub-shock formation. 
The condition becomes necessary and sufficient when the velocity of the shock becomes greater than the maximum characteristic velocity in the unperturbed state. 
Namely, the regions with no sub-shocks, a sub-shock for only one constituent, or sub-shocks for both constituents are comprehensively classified.
The most interesting case is that the lighter molecule has more degrees of freedom than that of the heavy one. 
In this situation, the topology of the various regions becomes different. 
We also solve the system of the field equations numerically using the parameters in the various regions and confirm whether the sub-shocks emerge or not.
Finally, the relationship between an acceleration wave in a constituent and the sub-shock in the other constituent is explicitly derived.  
\end{abstract}

\maketitle

\section{Introduction}
\label{sec:Intro}

The shock wave phenomenon has attracted many researchers in various fields because of its wide-range applications~\cite{VincentiKruger,Zeldovich}. 
An up-to-date survey of the shock waves based on the hyperbolic theories can be found in a review paper~\cite{ShRu} and from the mathematical point of view in the recent book of Liu \cite{Liubook}. 
One of the essential differences from the parabolic theory is that the hyperbolic theory does not always predict a continuous solution of the shock structure, and a discontinuous part (sub-shock) may appear depending on the velocity $s$ of the shock wave. 
Ruggeri showed that the shock-structure solution becomes, in principle, singular when $s$ meets any characteristic velocity of the hyperbolic system~\cite{Ruggeri1993}. 
Boillat and Ruggeri proved a theorem~\cite{Breakdown} stating that if the hyperbolic system of balance laws has a convex entropy, a continuous solution for the shock structure cannot exist when $s$ is greater than the maximum characteristic velocity evaluated in the equilibrium state in front of the shock wave and a sub-shock arises. 
The possibility of the sub-shock formation with smaller $s$ than the maximum characteristic velocity is still unclear. 
In the theory of Rational Extended Thermodynamics (RET) of a monatomic gas~\cite{RET}, 
Weiss proved numerically that when  $s$ meets the characteristic velocity, the solution becomes regular except for the maximum characteristic velocity evaluated in the unperturbed state in accordance with the Boillat-Ruggeri theorem~\cite{Weiss}. For this reason,  
it has been conjectured that the sub-shock can appear only when $s$ is greater than the maximum characteristic velocity.
This conjecture has also been confirmed for single polyatomic gases~\cite{IJNLM2017,RuggeriSugiyama,BookNew}.
Recently, by adopting several hyperbolic systems of field equations, counter-examples of this conjecture have also been found in the context of a mixture of rarefied monatomic gas~\cite{FMR,Bisi1,Bisi2} and also of simple toy models~\cite{IJNLM2017,subshock2}. 

In order to understand the nature of the sub-shock formation more deeply, we focus on the shock structure in a binary mixture of polyatomic Eulerian gases~\cite{RuggeriSimic2007}, which is modeled in the spirit of RET theory. 
For small Mach numbers (weak shocks), no sub-shock arises, as was proved in the case of a binary mixture of monatomic gases~\cite{SimicRuggeri2007,MRSimic}. 
A comparison with the experimental results was also made, and this analysis shows good agreement between the theoretical prediction and experimental data on the mass density~\cite{2013MS}. 
In the present analysis, we want to consider also strong shocks in a binary mixture of polyatomic gases with different degrees of freedom, and we want to give also a comparison with the results obtained in a binary mixture of monatomic gases~\cite{FMR,BookNew}.  


In the previous paper~\cite{shock_Lincei}, 
preliminary results of the sub-shock formation in a binary mixture of rarefied polyatomic gases are summarized, and the shock structure with multiple sub-shocks is shown when the degrees of freedom of a heavier molecule are larger or equal to the degrees of freedom of a lighter molecule.  
In the present paper, a complete classification of the sub-shock formation will be made with particular attention to the case with smaller degrees of freedom of the lighter molecule than the degrees of freedom of the heavy one. 
In these last circumstances, we show that a more complicated and interesting situation arises for the regions of possible sub-shocks formation varying the Mach number and the concentration of the two species for an assigned ratio of mass and prescribed degree of freedoms.

The organization of the present paper is summarized as follows.  
In section \ref{sec:Model}, the balance laws system and the constitutive equations are summarized. 
We adopt a multi-temperature Eulerian model of a binary mixture of rarefied gases in which the entropy inequality is satisfied, and the entropy density is a convex function (thermodynamical stability). 
The shock structure problem is explained in section \ref{sec:ShockStruct}, and we classify possible regions from the viewpoint of the necessary conditions for the sub-shock formation in section \ref{sec:classification_regions}. 
Section \ref{sec:Numerical} is devoted to the confirmation of all possible cases by solving numerical field equations. 
Moreover, we prove that an acceleration wave appears in a constituent when the sub-shock emerges in the other constituent.
Furthermore, the profiles of the average temperature are also shown. 
The summary and concluding remarks will be shown in section \ref{sec:summary}. 

\section{Basic equations} 
\label{sec:Model}

We consider a binary mixture of rarefied polytropic gases described by the following thermal and caloric equations of state: 
\begin{equation}\label{coneq}
	p_{\alpha} =  \frac{k_B}{m_{\alpha}} \rho_{\alpha} T_{\alpha}, \qquad 
	\varepsilon_{\alpha} = \frac{1}{\gamma_{\alpha} - 1}\frac{k_B}{m_{\alpha}} T_{\alpha}, 
\end{equation} 
where  $k_B$, $p_\alpha$, $m_\alpha$, $\varepsilon_\alpha$, $\rho_\alpha$ and $\gamma_\alpha$ are, respectively,  the Boltzmann constant, the pressure, the molecular mass, the specific internal energy, the mass density and the ratio of the specific heats of the species $\alpha=1,2$.  
The global pressure $p$ and the global specific intrinsic  internal energy $\varepsilon_{I}$ are: 
\begin{equation}\label{eq:state}
	p = p_{1} + p_{2} ,
	\qquad
	\rho \varepsilon_{I} = \rho_{1} \varepsilon_{1}
	+ \rho_{2} \varepsilon_{2}, 
\end{equation}
where $\rho$ is the total mass density given by 
\begin{equation}\label{eq:totalrho}
	\rho = \rho_1 + \rho_2. 
\end{equation}
In an equilibrium state where $T_1 = T_2 = T$ we can impose the same form of \eqref{coneq} on $p^E$ and $ \varepsilon_{I}^E$: 
\begin{equation}\label{eq:state_E}
	p^E  = \rho \frac{k_{B}}{m} T,
	\qquad
	 \varepsilon_{I}^E	= \frac{1}{\gamma - 1}  \frac{k_{B}}{m}T 
\end{equation}
provided that we introduce the average mass $m \equiv m(c)$ and the average ratio of the specific heats $\gamma \equiv \gamma(c)$ of the mixture
in the following form~\cite{RuggeriSimic2007}:
\begin{align}\label{equ:avgTemp}
\begin{split}
	&	\frac{1}{m(c)} = \frac{c}{m_{1}} + \frac{1 - c}{m_{2}}, \\
&	\frac{1}{\gamma - 1} = \frac{m(c)}{m_{1}} \frac{c}{\gamma_1 - 1} + \frac{m(c)}{m_{2}} \frac{1 - c}{\gamma_2 - 1},
\end{split}
\end{align}
where $c$ ($0 \leq c \leq 1$) is the concentration related to the mass densities:
\begin{equation*}
	\rho_{1} = \rho c, \quad \rho_{2} = \rho (1 - c). 
\end{equation*}

For the analysis of mixtures, it is convenient to introduce the diffusion velocity $\mathbf{u}_\alpha$: 
\begin{equation*}\label{def:u}
	\mathbf{u}_\alpha = \mathbf{v}_\alpha - \mathbf{v},  \qquad \left(
	\sumal \rho_\alpha \mathbf{u}_\alpha = \mathbf{0}
	\right), 
\end{equation*}
where the mixture velocity $\mathbf{v}$ is defined by
\begin{equation}\label{def:v}
	\mathbf{v} = \frac{1}{\rho} \sumal \rho_\alpha \mathbf{v}_\alpha
\end{equation}
with $\mathbf{v}_\alpha$ being the velocity of the constituent $\alpha$. 
We focus on the one-dimensional problem in the $x$-direction such that the velocities are expressed as 
\begin{equation*}\label{eq:va}
	\mathbf{v} = (v, 0,0), \quad 
	\mathbf{v}_{\alpha} = (v_{\alpha}, 0,0), \quad
	\mathbf{u}_{\alpha} = (u_{\alpha}, 0,0). 
\end{equation*}
Furthermore, it is assumed that chemical reactions are not present. 
Then the system  of a mixture of Eulerian gases is~\cite{RuggeriSimic2007,RuggeriSimic2009,BookNew}: 
\begin{widetext}
\begin{equation}\label{finale1}
	\begin{split}
		& \frac{\partial \rho_1 }{\partial t} + \frac{\partial \rho_1  v_1 }{\partial x} 
		= 0, \\
		&\frac{\partial \rho_1 v_1 }{\partial t}+ \frac{\partial}{\partial x} \left( \rho_1  v^2_1 + p_1 \right) = \hat{m}_1, \\
		&\frac{\partial}{\partial t} \left(\rho_1  v^2_1 + 2 \rho_1  \varepsilon_1 \right) +
		\frac{\partial}{\partial x}\left\{\left( \rho_1 v^2_1 + 2\rho_1  \varepsilon_1  + 2 p_1 \right) v_1  \right\} 
		= 2(\hat{e}_1 +  \hat{m}_1 v),
		\\
		& \frac{\partial \rho_2 }{\partial t} + \frac{\partial \rho_2  v_2 }{\partial x} 
		= 0, \\
		&\frac{\partial \rho_2 v_2 }{\partial t}+ \frac{\partial}{\partial x} \left( \rho_2  v^2_2 + p_2 \right) =  \hat{m}_2, \\
		&\frac{\partial}{\partial t} \left(\rho_2  v^2_2 + 2 \rho_2  \varepsilon_2 \right) +
		\frac{\partial}{\partial x}\left\{\left( \rho_2 v^2_2 + 2\rho_2  \varepsilon_2  + 2 p_2 \right) v_2  \right\} 
		=  2(\hat{e}_2 +  \hat{m}_2 v). 
	\end{split}
\end{equation}
The production terms $\hat{m}_\alpha$ and $\hat{e}_\alpha$ represent the interchange of the momentum and the energy, respectively, and we have to take into account the  following relations: 
\begin{equation*}
	\sumal \hat{m}_\alpha = \sumal \hat{e}_\alpha = 0.
\end{equation*}
The system \eqref{finale1} is equivalent to the following system~\cite{RuggeriSimic2007,RuggeriSimic2009,BookNew}: 
\begin{equation}\label{finale}
	\begin{split}
		&\frac{\partial \rho}{\partial t} + \frac{\partial \rho v}{\partial x} = 0,\\
		&\frac{\partial \rho v}{\partial t}+ \frac{\partial }{\partial x} \left(\rho v^2 + p + \Pi - \sigma
		\right) = 0,\\
		&\frac{\partial }{\partial t} \left(\frac{1}{2}\rho v^2 + \rho \varepsilon\right)
		+ \frac{\partial}{\partial x}\left\{\left(\frac{1}{2}\rho v^2 + \rho \varepsilon +  p + \Pi -\sigma
		\right)v + q \right\} = 0,\\
		& \frac{\partial \rho_1 }{\partial t} + \frac{\partial \rho_1  v_1 }{\partial x} 
		= 0, \\
		&\frac{\partial \rho_1 v_1 }{\partial t}+ \frac{\partial}{\partial x} \left( \rho_1  v^2_1 + p_1  \right) = \hat{m}_1, \\
		&\frac{\partial}{\partial t} \left(\rho_1  v^2_1 + 2 \rho_1  \varepsilon_1 \right) +
		\frac{\partial}{\partial x}\left\{\left( \rho v^2_1 + 2\rho_1  \varepsilon_1  + 2 p_1  \right) v_1  \right\} 
		=  2(\hat{e}_1 +  \hat{m}_1 v),
	\end{split}
\end{equation}
\end{widetext}
where in the one-dimensional case the deviatoric part of the global shear stress $\sigma$, the global dynamic pressure $\Pi$,
 the global specific internal energy $\varepsilon$ and 
   the global heat flux $q$ are given by 
\begin{equation} \label{campiG}
	\begin{split}
		&\sigma = { - \frac{2}{3} \sumal\rho_\alpha u_\alpha^2} =  - \frac{2}{3} \frac{\rho_{1} \rho}{\rho_{2}} u_1^2,\\
		&\Pi  =\sumal \frac{1}{3}\rho_\alpha u_\alpha^2 = -\frac{\sigma}{2},\\
		&\rho \varepsilon =  \rho \varepsilon_{I} +\frac{1}{2} \sumal\rho_\alpha u_\alpha^2 = \rho \varepsilon_{I} + \frac{1}{2}\left( \Pi - \sigma \right),  \\
		&q = \sumal\left(\frac{1}{2}\rho_\alpha u_\alpha^2 + \rho_\alpha \varepsilon_\alpha + p_\alpha \right) u_\alpha,
	\end{split}
\end{equation}
and the global $p$ and intrinsic energy $\varepsilon_{I}$ are given by \eqref{eq:state}.

The system \eqref{finale} is compatible with the supplementary entropy law~\cite{RuggeriSimic2007}: 
\begin{equation}
	\frac{\partial h}{\partial t} + 	\frac{\partial k}{\partial x} = \Sigma,
\end{equation}
where $h$ and $k$ are, respectively, the entropy density and the entropy flux, and 
the entropy production $\Sigma$ is given by
\begin{equation*}
	\Sigma = \hat{m}_1 \hat{\Lambda}^{v_1} +  \hat{e}_1 (2\hat{\Lambda}^{\varepsilon_1})
\end{equation*}
with $\hat{\Lambda}^{v_1}$ and $\hat{\Lambda}^{\varepsilon_1}$ being the  components of the Lagrange multipliers  (\emph{main field}) evaluated in the rest frame relative to the last two balance laws in \eqref{finale}\cite{RuggeriSimic2007}: 
\begin{equation}\label{eq:main_field}
	\hat{\Lambda}^{v_1} = - \frac{u_1}{T_1} + \frac{u_2}{T_2}, \quad
	\hat{\Lambda}^{\varepsilon_1} = \frac{1}{2 T_1} - \frac{1}{2 T_2}.  
\end{equation}
The expressions of $\hat{m}_1$ and $\hat{e}_1$ are obtained such that $\Sigma$ is positive and quadratic as usual in non-equilibrium thermodynamics: 
\begin{align}\label{eq:production_interaction}
	\begin{split}
	&	\hat{m}_1 = \psi  \hat{\Lambda}^{v_1} = \psi \left( \frac{u_2}{T_2} - \frac{u_1}{T_1} \right), \\
	&	\hat{e}_1 = 2 \theta  \hat{\Lambda}^{\varepsilon_1}=\theta \left(\frac{1}{  T_1} - \frac{1}{  T_2}\right), 
	\end{split}
\end{align}
where the phenomenological coefficients $\psi >0$ and $\theta>0 $ depend on the mass densities and temperatures. 
Their functional forms may be determined by kinetic theoretical consideration and/or experimental data. 

Inserting the constitutive equations \eqref{coneq} and the production terms \eqref{eq:production_interaction}  into the system \eqref{finale1} or the equivalent system \eqref{finale}, we obtain a closed system with the unknown $(\rho_1,v_1, T_1, \rho_2, v_2, T_2)$ or $(\rho, v, T, \rho_1, v_1, T_1)$.
It is noticeable that there appear in the system \eqref{finale} through \eqref{campiG} the global viscous stress $\sigma$, the dynamical pressure $\Pi$  and the global heat flux $q$ due to the diffusion of velocities and temperatures even if we neglect these dissipative quantities in the field equations for each constituent. 

From \eqref{eq:production_interaction} and \eqref{eq:main_field}, we have $v = v_1 = v_2$ and $T = T_1 = T_2$ in an equilibrium state from the condition that the production terms $\hat{m}_1$ and $\hat{e}_1$ vanish. 
The associate {\em equilibrium subsystem}~\cite{Boillat-1997} of \eqref{finale} is given by the four first equations of \eqref{finale} with fields evaluate in equilibrium:
\begin{align}
		&\frac{\partial \rho}{\partial t} + \frac{\partial \rho v}{\partial x} = 0, \nonumber\\
		& \frac{\partial \rho_1 }{\partial t} + \frac{\partial \rho_1 v}{\partial x} = 0, \nonumber\\
		&\frac{\partial \rho v}{\partial t}+ \frac{\partial }{\partial x} \left(\rho v^2 + p^E \right) = 0, \label{subsystem:finale}\\
		&\frac{\partial }{\partial t} \left(\frac{1}{2}\rho v^2 + \rho \varepsilon_I^E\right)
		+\frac{\partial}{\partial x}\left\{\left(\frac{1}{2}\rho v^2 + \rho \varepsilon_I^E +  p^E \right)v \right\} = 0,\nonumber
\end{align}
where $p^E$ and $\varepsilon_{I}^E$ are the equilibrium pressure and the specific internal energy \eqref{eq:state} with $m$ and $\gamma$ given in \eqref{equ:avgTemp}.

\section{Shock Structure}
\label{sec:ShockStruct}
The system of field equations \eqref{finale1}, or equivalently \eqref{finale}, belongs to a particular case of general first order hyperbolic quasi-linear system of balance laws: 
\begin{equation}\label{sistemagenerale}
	\frac{\partial \mathbf{u}}{\partial t} + \frac{\partial \mathbf{F}(\mathbf{u})}{\partial x} = \mathbf{P}(\mathbf{u}), 
\end{equation}
where $\mathbf{u}$, $\mathbf{F}$, and $\mathbf{P}$ are column vectors of $R^N$ representing the density, the flux, and the production, respectively. 

The shock structure is described as a traveling-wave solution depending on a single variable $\varphi = x - s t$,
which asymptotically connects equilibrium states in front and
behind the shock wave. 
Namely, 
\begin{equation}\label{eq:travaling}
	\mathbf{u}\equiv \mathbf{u}(\varphi), \qquad \varphi= x - s t
\end{equation}
with constant equilibrium  boundary conditions at  infinity:
\begin{equation}\label{contorno}
	\lim_{\varphi\rightarrow  + \infty  } \mathbf{u} =\mathbf{u}_0, \qquad \lim_{\varphi \rightarrow  - \infty } \mathbf{u} = \mathbf{u}_\mathrm{I}, 
\end{equation} 
where
\begin{equation}\label{eq:up_and_p}
	\mathbf{P}(\mathbf{u}_0) = \mathbf{P}(\mathbf{u}_\mathrm{I})= 0.
\end{equation}
We call the equilibrium states $\mathbf{u}_0$ and the $\mathbf{u}_\mathrm{I}$  {\it unperturbed state} and   {\it perturbed state}, respectively. 
Hereafter, the quantities with the subscript 0 represent the quantities evaluated in the unperturbed state, and the quantities with subscript I represent the ones evaluated in the perturbed state. 
By inserting \eqref{eq:travaling} into \eqref{sistemagenerale}, we have the following ODE system:
\begin{equation}\label{struttura}
	\left( \mathbf{A}( \mathbf{u}) - s \mathbf{I}
	\right)\frac{d\mathbf{u}}{d\varphi}= \mathbf{P}( \mathbf{u}), \qquad \mathbf{A}= \frac{\partial \mathbf{F}}{\partial \mathbf{u}}
\end{equation}
with boundary conditions given by \eqref{contorno}.

As pointed out by Ruggeri~\cite{Breakdown}, when $s$ meets a characteristic velocity $\lambda$, which is the eigenvalue of the matrix $\mathbf{A}$, the solution may have a breakdown, in other words, a singularity (sub-shock) may appear.
In order to discuss the necessary condition for the sub-shock formation, we need to analyze the meeting points between $s$ and $\lambda$. 

Following \cite{Breakdown}, by taking the typical features of rational extended thermodynamics into account, we may split the system \eqref{sistemagenerale} into the blocks of $M$  conservation laws and of $N-M$ balance equations as follows (\eqref{finale} belongs in this kind of system):   
\begin{equation}\label{blocks}
	\begin{split}
		&\frac{\partial \mathbf{V}(\mathbf{u})}{\partial t} + \frac{\partial \mathbf{R}(\mathbf{u})}{\partial x} = 0, \\
		&\frac{\partial \mathbf{W}(\mathbf{u})}{\partial t} + \frac{\partial \mathbf{Q}(\mathbf{u})}{\partial x} = \mathbf{g}(\mathbf{u}). 
	\end{split}
\end{equation}
Taking \eqref{blocks} and \eqref{eq:travaling} into account, we may rewrite \eqref{struttura} as 
\begin{align}\label{eq:shockstructure_general}
	\begin{split}
		& \frac{d}{d\varphi}\left\{- s \mathbf{V}(\mathbf{u}) +\mathbf{R}(\mathbf{u}) \right\} = 0, 
		\\
		& - s \frac{d \mathbf{W}(\mathbf{u})}{d \varphi} + \frac{d \mathbf{Q}(\mathbf{u})}{d \varphi} = \mathbf{g}(\mathbf{u}). 
	\end{split}
\end{align}
By integrating \eqref{eq:shockstructure_general}$_1$, we have
\begin{equation}
	- s \mathbf{V}(\mathbf{u}) +\mathbf{R}(\mathbf{u}) = \text{const.}
\end{equation}
and in particular taking into account \eqref{contorno} we have
\begin{equation}\label{eq:RH_eqsub2}
	- s \mathbf{V}(\mathbf{u}_0) +\mathbf{R}(\mathbf{u}_0) 
	= - s \mathbf{V}(\mathbf{u}_{\rm I}) +\mathbf{R}(\mathbf{u}_{\rm I}). 
\end{equation}
The conditions \eqref{eq:RH_eqsub2} is nothing else the  Rankine-Hugoniot conditions of the equilibrium subsystem:
\begin{equation}
\frac{\partial \mathbf{V}(\mathbf{u})}{\partial t} + \frac{\partial \mathbf{R}(\mathbf{u})}{\partial x} = 0, \qquad \mathbf{g}(\mathbf{u})=0.
\end{equation} 
For a given unperturbed state $\mathbf{u}_{0}$, and the shock velocity $s$, one may determine the perturbed state $\mathbf{u}_\mathrm{I}$ as function of $\mathbf{u}_{0}$, and the shock velocity $s$.  
Due to the Galilean invariance, we can choose the unperturbed velocity as $v_0 = 0$ without any loss of generality.
Taking into account the conservation laws, we have the following boundary conditions:
\begin{equation}
	\mathbf{u}_{0} =
	\left[%
	\begin{array}{l}
		\rho_{0} \\
		v_{0} =0\\
		T_{0} \\
		\left(\rho_1\right)_{0} = c_0 \rho_{0} \\
		\left(v_1\right)_{0}=v_0=0 \\
		\left(T_1\right)_{0} = T_0 \\
	\end{array}%
	\right], \quad
	\mathbf{u}_\mathrm{I} =
	\left[%
	\begin{array}{c}
		\rho_\mathrm{I} \\
		v_\mathrm{I} \\
		T_\mathrm{I} \\
		\left(\rho_1\right)_\mathrm{I} = c_\mathrm{I} \rho_\mathrm{I}\\
		\left(v_1\right)_\mathrm{I}=v_I \\
		\left(T_1\right)_\mathrm{I}=T_I \\
	\end{array}%
	\right] 
	\label{Model:EqStates}
\end{equation}
with $\sigma_0=\Pi_0=q_0= \sigma_I=\Pi_I=q_I =0$.
We notice that the system \eqref{subsystem:finale} except for the second equation is the Euler system of a single fluid and therefore the RH conditions gives:
\begin{equation}\label{eq:RH1}
	\begin{split}
		&\rho_\mathrm{I} = \frac{(\gamma_0 + 1) M_{0}^{2}}{2 + (\gamma_0 - 1) M_{0}^{2}}\rho_0, \quad
		v_\mathrm{I}  = \frac{2 (M_{0}^{2} - 1)}{(\gamma_0 + 1) M_{0}} a_0, \\
		&T_\mathrm{I} = \frac{\{2 + (\gamma_0 - 1) M_0^2\}\{1 + \gamma_0(2 M_0^2 - 1)\}}{(\gamma_0 + 1)^2 M_0^2} T_0, \\
	\end{split}
\end{equation}
with $\gamma_0$ being the function $\gamma$ given in \eqref{equ:avgTemp} evaluated in the unperturbed state $\gamma_0 = \gamma (c_0)$, 
\begin{equation}\label{MN}
	M_{0} =\frac{s - v_0}{a_{0}}= \frac{s}{a_0}, 
\end{equation}
$M_0$ is the unperturbed Mach number, 
 $a_0$ is the sound velocity in the unperturbed state 
\begin{equation*}
	a_{0} = \sqrt{ \gamma_0 \frac{k_{B}}{m_{0}} T_{0} }
\end{equation*}
and $m_0$ is   being the equilibrium average mass \eqref{equ:avgTemp} evaluated in  the unperturbed state: $m_{0} = m(c_{0})$.
The RH equation for the second equation    combined with the RH of the first equation of \eqref{subsystem:finale} give immediately  that the concentration is the same in both equilibrium states  $c_\mathrm{I} = c_{0}$, and in the sequel, it will be termed
equilibrium concentration, without special regard to unperturbed or
perturbed state.

\section{Regions classified by the possibility of sub-shock formation}
\label{sec:classification_regions}
To analyze the regularity of the shock-structure solution in a binary mixture, we recall the characteristic velocities of the full system \eqref{finale1} in equilibrium in which both constituents have common velocity $v$ and temperature $T$:  
\begin{equation}\label{EquCharSpeeds}
	v \pm
	\sqrt{\gamma_1  \frac{k_{\mathrm{B}}}{m_{1}} T}, \quad	
	v \,\, ({\rm multiplicity\,\,2}), \quad
	v \pm \sqrt{\gamma_2  \frac{k_{\mathrm{B}}}{m_{2}} T}. 
\end{equation}
The characteristic velocities of the equilibrium subsystem \eqref{subsystem:finale} are 
\begin{equation*}
	 v - \sqrt{\gamma  \frac{k_{\mathrm{B}}}{m} T},
	\quad
	v \,\, ({\rm multiplicity\,\,2}), 
	\quad
	v + \sqrt{\gamma  \frac{k_{\mathrm{B}}}{m} T}, 	
\end{equation*}
where $\gamma$ and $m$ are given by \eqref{equ:avgTemp}. 

We focus only on positive velocities with respect to the frame moving with the fluid as for symmetry  
the negative ones have the same properties.
Let define the characteristic velocities in equilibrium of species $1$ and $2$:
\begin{equation}\label{eq:char_2species}
\lambda_1=	v +
\sqrt{\gamma_1  \frac{k_{\mathrm{B}}}{m_{1}} T}, \quad  \lambda_2=	v + \sqrt{\gamma_2  \frac{k_{\mathrm{B}}}{m_{2}} T}
\end{equation}
and we call 
\begin{equation}\label{eq:char_eqsub}
\bar{	\lambda} =	v + \sqrt{\gamma  \frac{k_{\mathrm{B}}}{m} T}
\end{equation}
the characteristic velocity   of the equilibrium subsystem.
It is noticeable that the magnitude relation between the characteristic velocities for species 1 and 2 depends on the values of $\gamma_1/m_1$ and  $\gamma_2/m_2$. But
by the Theorem
\cite{Boillat-1997}, the subcharacteristic conditions hold and we have
\begin{equation*}
\bar{	\lambda} \leq \max{\left(\lambda_1,\lambda_2\right)}.
\end{equation*}
We consider the shock family that bifurcates from the trivial zero shock solution when $s=\bar{\lambda}_{0}$. 
In this case, by using the Lax condition~\cite{Lax}, we can determine the admissibility and obtain the following admissible equilibrium states (\ref{Model:EqStates}) $\bar{\lambda}_{0} < s <
\bar{\lambda}_\mathrm{I}$. 
This conditions is equivalently expressed as $M_{0} > 1$ as is expected. 
Hereafter we call species $1$ the one with smaller or equal mass between the two species: $m_1 \leq m_2$.
In order to classify the possible cases, we introduce the ratio of the masses of the constituents,
\begin{equation*}
	\mu = \frac{m_{1}}{m_{2}}, \qquad (0 < \mu \leq 1),
\end{equation*}
and the ratio of the specific heats $g$, 
\begin{equation*}
	g = \frac{\gamma_1}{\gamma_2}. 
\end{equation*} 
Moreover we recall that the degree of freedom of a species $D_\alpha$ is related to the ratio of specific heats $\gamma_\alpha$ by the relation
\begin{equation}\label{dd}
D_\alpha = \frac{2}{\gamma_\alpha - 1}, \qquad \alpha=1,2.
\end{equation}
In the following, we consider the following two cases separately
\begin{equation*}
	\begin{split}
		&{\rm Case \,\,\,\, A: } \quad \gamma_1\geq \gamma_2, \\
		&{\rm Case \,\,\,\, B: } \quad \gamma_1< \gamma_2,
	\end{split}
\end{equation*}
In Case A, the lighter species has fewer or equal degrees of freedom than the heavier, while in Case B, the species with smaller mass has more degrees of freedom than the species with larger mass.

To our knowledge, only the preliminary analysis of the case A was done; in particular, there are some results for mixtures of both monatomic gases ($\gamma_1=\gamma_2=5/3$)~\cite{MRSimic,FMR,Bisi1}, while the case B has never been studied at least for a complete classification of sub-shock formation.

\subsection{Case A: $\gamma_1\geq \gamma_2$ }

The preliminary results on the shock structure of Case A in a binary mixture of polyatomic gases were summarized in the previous paper~\cite{shock_Lincei}. A
binary mixture of monatomic gases in which  $\gamma_1 = \gamma_2 = 5/3$, ($ g  = 1$) belong in this case. 

\begin{figure}[]
	\centering
	\includegraphics[width=0.9\linewidth]{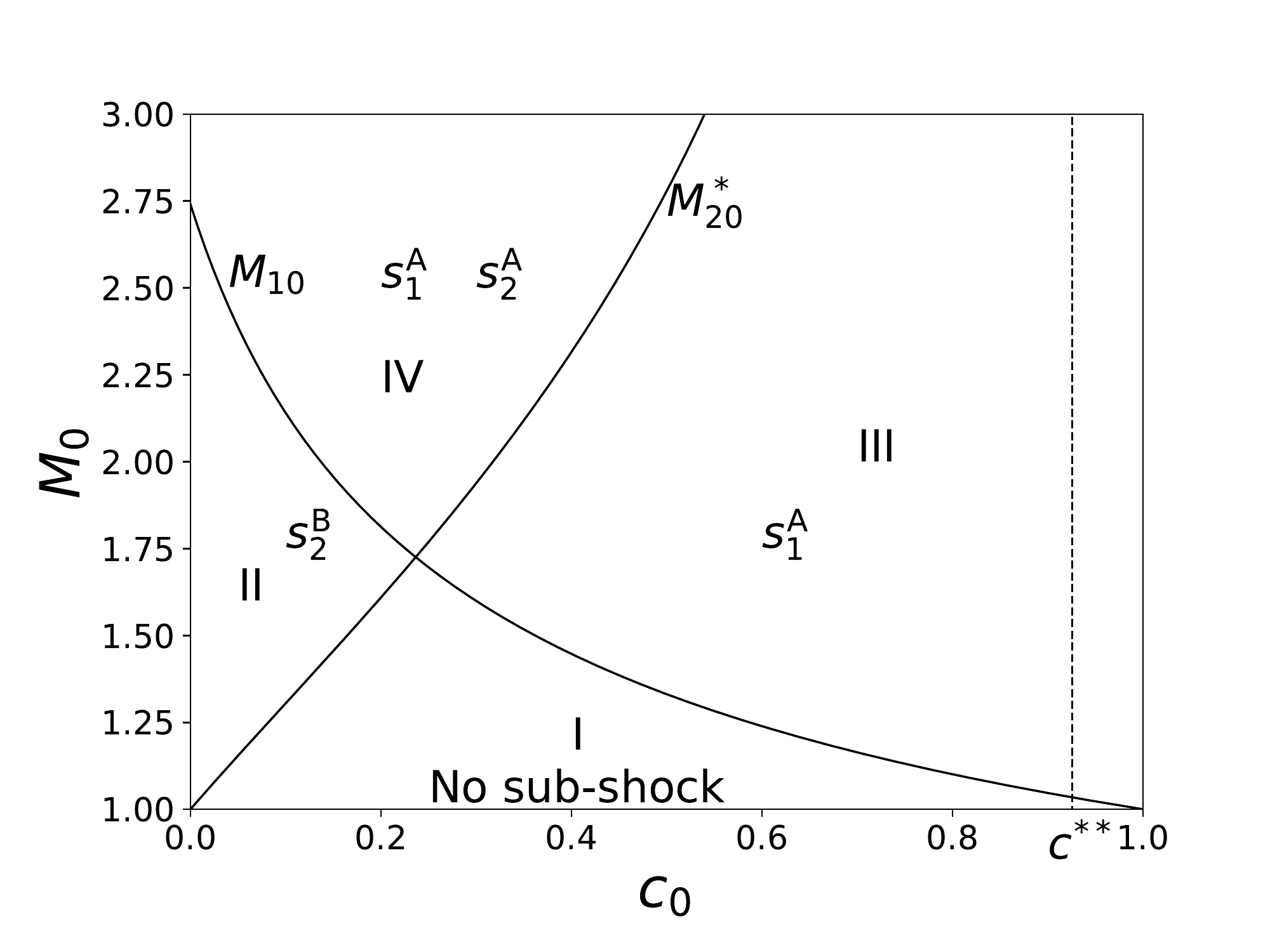}
	\caption{Case A: Four regions in the plane $(c_0,M_0$)  of possible sub-shocks in the case that $M_{20}^*$ have a vertical asymptote in $c_0=c_0^{*}<1$. Values are 
		$\gamma_1=7/5$, $\gamma_2=9/7$, and $\mu = 0.145 < \mu^*$. }
	\label{fig:subshockEuler_mu0145}
\end{figure}

\begin{figure}[]
	\centering
	\includegraphics[width=0.9\linewidth]{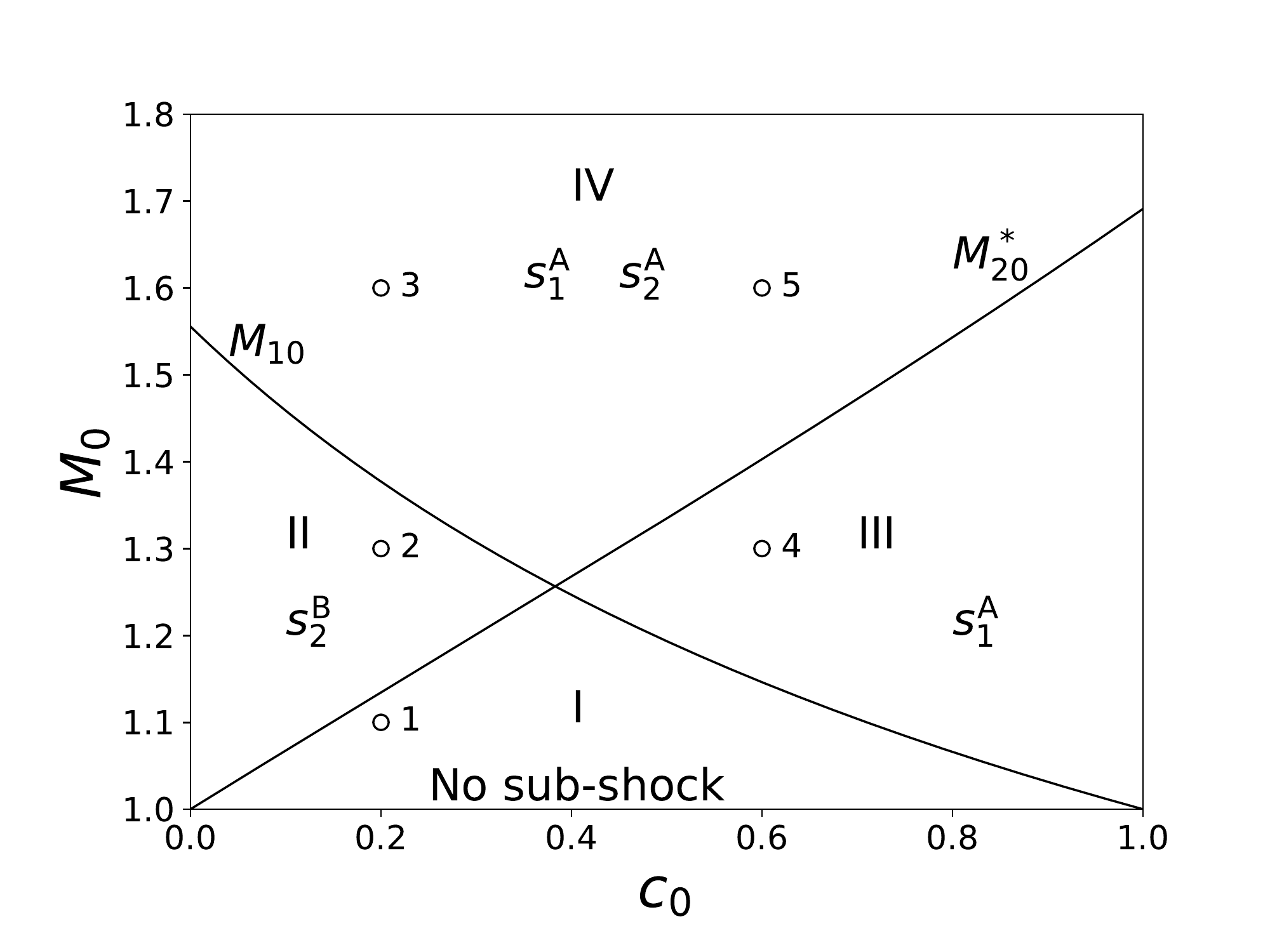}
	\caption{Case A: Four regions in the plane $(c_0,M_0$) with $c_0^*>1$ classified by the possibility of the sub-shock formation. The values are
		$\gamma_1 = 7/5$, $\gamma_2 = 9/7$, and $\mu = 0.45 > \mu^*$. }
	\label{fig:subshockEuler_mu045}
\end{figure}

In the Case A, the characteristic velocities in equilibrium in front the shock satisfies the inequalities for any $c_0\in ]0,1[ $ and any $\mu \in ]0,1[$:
\begin{equation}\label{diseg}
\lambda_{20}< \bar{	\lambda}_0<	\lambda_{10}, \qquad s> \bar{	\lambda}_0.
\end{equation}
According to the Theorem proved in \cite{Breakdown}, a sub-shock appears if the shock velocity is faster than the maximum characteristic velocity in the unperturbed state $s > \lambda_{10}$ and in the present case, we obtain the following condition:
\begin{equation*}
	\begin{split}
		& 
		M_{0} > M_{10} = \frac{\lambda_{10} }{a_0} = \sqrt{\frac{m_0 \gamma_1}{m_1 \gamma_0}}
		= \sqrt{\frac{\gamma_1}{\gamma_0\left\{c_{0} + (1 - c_{0}) \mu\right\}}},
	\end{split}
\end{equation*}
where $M_{10}$ is the dimensionless characteristic velocity for the constituent $1$ evaluated in the unperturbed state. 
Taking into account \eqref{equ:avgTemp}, it is easy to see  that 
\begin{equation*}
	\lim_{c_0 \rightarrow 1} M_{10} = 1.
\end{equation*} 
As typical examples, the dependence of $M_{10}$ on $c_0$ for $\mu = 0.45$, $\gamma_1 = 7/5$, and $\gamma_2 = 9/7$ is given in Figures \ref{fig:subshockEuler_mu0145} and  \ref{fig:subshockEuler_mu045}. 
We see that if the parameters lie in the Regions III and IV of the figures, there exists a sub-shock involving the variables of the first constituent that we call $S_1^A$ the index indicates the constituent $1$ and A means that the sub-shock emerges after the maximum eigenvalues in the unperturbed state.

To know if other sub-shocks can exists we recall   the  observation  that a sub-shock can arise when $s$ meets an eigenvalue $\lambda$ evaluated along with the shock structure (see \eqref{struttura}). 

For the genuine-non linearity along the solution  both $\lambda_2(\mathbf{u}(\varphi))$ and $\lambda_1(\mathbf{u}(\varphi))$ are increasing function of $s$ \cite{NuovoCimento}, 
and therefore taking into account \eqref{diseg},  only $\lambda_2({\bf u}^*)=s$ can be verified  for some value of $\varphi^*$. As $\lambda_{20}\leq \lambda_2({\bf u}^*) \leq \lambda_{2\mathrm{I}}$,   a necessary condition (but not sufficient) is that
$\lambda_{2\mathrm{I}} (s)> s$ that 
in term of Mach numbers is equivalent that $M_0> M_{20}^*$ where $M_{20}^*$ is solution of 
\begin{equation}\label{eq:M05}
	M_{2\mathrm{I}} (M_{20}^*,c,\mu)=  M_{20}^*,
\end{equation}
where we have indicates with $M_{2\mathrm{I}}=   \lambda_{2\mathrm{I}} /a_0$, the dimensionless characteristic velocity behind the shock.
We have the following expression: 
\begin{equation*}
	\begin{split}
		M_{2\mathrm{I}}  
		&= \frac{2 (M_{0}^{2} - 1)}{(\gamma_0 + 1) M_{0}} \\
		& \,\, + \sqrt{ \frac{\{2 + (\gamma_0 - 1) M_0^2\}\{1 + \gamma_0(2 M_0^2 - 1)\}}{(\gamma_0 + 1)^2 M_0^2} \frac{\gamma_2 m_0}{\gamma_0 m_{2}}},
	\end{split}
\end{equation*}
and we obtain the solution of \eqref{eq:M05} as follows:
\begin{equation}\label{Mast}
	M_{20}^*  
	= \sqrt{\frac{2 m_2 \gamma_0 + m_0 \gamma_2 (\gamma_0 - 1)}{\gamma_0 \{2 m_0 \gamma_2 - m_2(\gamma_0 - 1) \}}}.
\end{equation}
From the definitions of the average mass and the global ratio of the specific heats \eqref{equ:avgTemp}, we have
\begin{equation}\label{M0star-0}
	\lim_{c_0 \rightarrow 0} M_{20}^* = 1.
\end{equation}  
The existence of real value of $M_{20}^*$ require $2 m_0 \gamma_2 - m_2(\gamma_0 - 1) > 0$, which implies
\begin{equation*}
	\begin{split}
		c_0 < c_0^* 
		= &\frac{\mu}{(\gamma_1 - 1)(\gamma_2 - 1) (1-\mu)^2}\\
		&\times \bigg(
		- 1 + \mu + \gamma_1 (1 - \mu) - \gamma_2 (\gamma_1 - \gamma_2) + \sqrt{\Gamma}
		\bigg), 
	\end{split}
\end{equation*}
where $\Gamma = \gamma_2 \{(\gamma_2 - 1) [2 \gamma_1 (\gamma_1 - 1) + \gamma_2 - 2 \gamma_1 \gamma_2 + \gamma_2^2 ]  - 2 \mu \gamma_1 (\gamma_1 - 1)(\gamma_2 - 1) + \mu^2 \gamma_2 (\gamma_1 - 1)^2\}$. 
Let
\begin{equation}
\mu^* = \frac{\gamma_1-1}{2 \gamma_2}
\end{equation}
and we need to distinguish two cases: if $\mu < \mu^*$ then $c_0^* <1$ and we have the Figure \ref{fig:subshockEuler_mu0145}, while if $\mu > \mu^*$  then $c_0^* >1$ and we show a typical example of the case for $\mu = 0.45$, $\gamma_1 = 7/5$, and $\gamma_2 = 9/7$ in Figure \ref{fig:subshockEuler_mu045}.

In both  Figures  \ref{fig:subshockEuler_mu0145} and \ref{fig:subshockEuler_mu045},  we have $4$ regions in the plane $(c_0,M_0)$: 
in Region I, no sub-shock can exist, and the shock structure is always regular. 
In Region II, we may have a sub-shock for the variables of the second constituent $S_2^B$, where B represents before the maximum eigenvalue. 
In Region III, a sub-shock $S_1^A$ for constituent $1$ after the maximum characteristic velocity is predicted according to the theorem~\cite{Breakdown}. 
Finally, in Region IV, we have by sure the sub-shock $S_1^A$ but may also have a sub-shock  $ S_2^A$; in other words, in this region, we may have multiple sub-shocks.
We summarize  Case A as a statement:
\begin{statement}
- If the species with the smallest mass has fewer or equal degrees of freedom than the heavier one, we have for each value of the concentration
the inequality \eqref{diseg} and the possible regions in the plane $(c_0, M_0)$ where there may be sub-shocks are those described in Figure  \ref{fig:subshockEuler_mu0145}  and Figure  \ref{fig:subshockEuler_mu045}.  The curve $M_{20}^* (c_0)$ has a vertical  asymptote in $c_0=c _0^* \,$  for values of the ratio of the masses $\mu <\mu ^ *$  (Figure  \ref{fig:subshockEuler_mu0145}) and this implies that multiple sub-shocks can exists only for $c_0 <c _0^*$ while for $\mu > \mu ^ *$ there is no asymptote (Figure  \ref{fig:subshockEuler_mu045}) and the multiple sub-shocks may exist for any value of concentration.
\end{statement}

For completeness of the classification, we remark that there exists, at least mathematically, a very special case with $\mu =  g  = 1$. 
In this case, the characteristic velocities in equilibrium for both species \eqref{eq:char_2species} and also the characteristic velocity for the equilibrium subsystem \eqref{eq:char_eqsub} have the same value. 
Therefore $M_{10} = M^{*}_{20} = 1$ hold and there exists only Region IV. 
According to the theorem~\cite{Breakdown}, we conclude that no continuous solution exists and multiple sub-shocks connecting the unperturbed and perturbed states always appear for any  $M_0  > 1$.

\subsection{Case B: $\gamma_1 < \gamma_2$}

Qualitatively different behavior is predicted when the degrees of freedom of lighter species is greater than the one of the heavy species because the topology of the regions changes with respect to Case A. 
In fact, we need to distinguish three cases as we increase the value of $\mu$.

\subsubsection{Case B$_1$: $\mu <  g $}

In this case we have as in Case A  
\begin{equation}\label{disegb1}
	\lambda_{20} <	\lambda_{10},
\end{equation}
but the value of $\tilde{\lambda}_0$ depends on the concentration.
In fact, there exists a value of concentration $\hat{c}_0$ for which
\begin{align}
	&\text{If} \,\,\ 0<c_0<\hat{c}_0,  \quad 
	\bar{	\lambda}_0 < \lambda_{20}< 	\lambda_{10}, \quad s> \bar{	\lambda}_0, \label{b11}\\
	&\text{If} \,\,\ \hat{c}_0<c_0<1,  \quad 
	\lambda_{20}< 	\bar{	\lambda}_0 <\lambda_{10}, \quad s> \bar{	\lambda}_0. \label{b22}
\end{align}
\begin{figure}[h]
	\centering
	\includegraphics[width=0.9\linewidth]{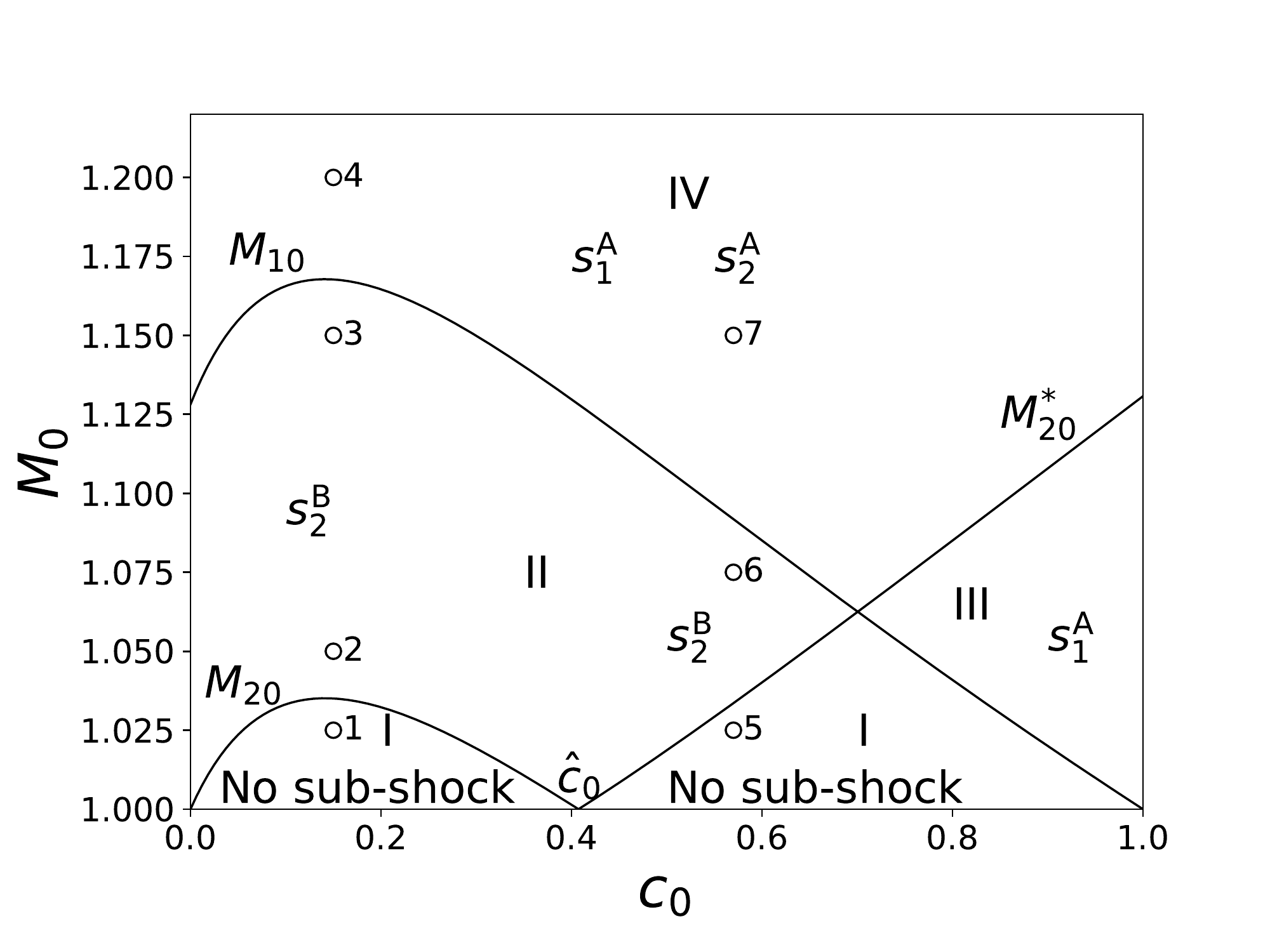}
	\caption{Case B$_1$: Four regions in the plane $(c_0,M_0$) of possible sub-shocks. $\gamma_1=7/6$, $\gamma_2=5/3$, and $\mu = 0.55< g  $. }
	\label{fig:subshockEuler_mu055}
\end{figure}
In the first case \eqref{b11} a new curve $M_{20} (c_0)>1$ may emerge, and a sub-shock $S_2^B$ can exist if $ M_{20} < M_0 <M_{10}  $ with $M_{20}$ being the dimensionless   characteristic velocity of species 2: 
\begin{equation*}
	M_{20} = \frac{\lambda_{20}}{a_0}
	= \sqrt{\frac{m_0 \gamma_2}{m_2 \gamma_0}}
	= \sqrt{\frac{\gamma_2 \mu}{\gamma_0 \left\{c_{0} + (1 - c_{0}) \mu\right\}}}.
\end{equation*}
By solving $M_{20} \geq 1$, we obtain  
\begin{equation}\label{sol:c0_zero_and}
	0<	c_0 < \hat{c}_0 =
	\frac{\mu \{ \gamma_1 \gamma_2 (2 - \mu) - \gamma_1 - \gamma_2 (\gamma_2 - \mu)\}}{(1 - \mu)\{\gamma_1 - \gamma_1 \gamma_2 (1 - \mu) - \gamma_2 \mu\}}. 
\end{equation}
It is simple to verify that $\hat{c}_0 <1$ as $\mu< g $.

For value of concentration $c_0 > \hat{c}_0$ we have the inequalities \eqref{b22} similar to the case A  and we have the curve $M_{20}^* (c_0)$ given by \eqref{Mast}. 
It is easy to verify   that $M_{20}^* (\hat{c}_0)=1$.
An example with $\gamma_1 = 7/6$ and $\gamma_2 = 5/3$ and $\mu = 0.55$ is shown in Figure \ref{fig:subshockEuler_mu055}. 

Although the topology of the regions is different from the one in Case A, we have the four regions I, II, III, and IV  with the same possible sub-shocks that the one described in Case A: in Region I, no sub-shock emerges, in Region II, possible sub-shock in the species 2 may be formed, in Region III, sub-shock of species 1 appears, and in Region IV, both species can have sub-shocks.

\subsubsection{Case B$_2$: $\mu =  g $}

\begin{figure}[]
	\centering
	\includegraphics[width=0.9\linewidth]{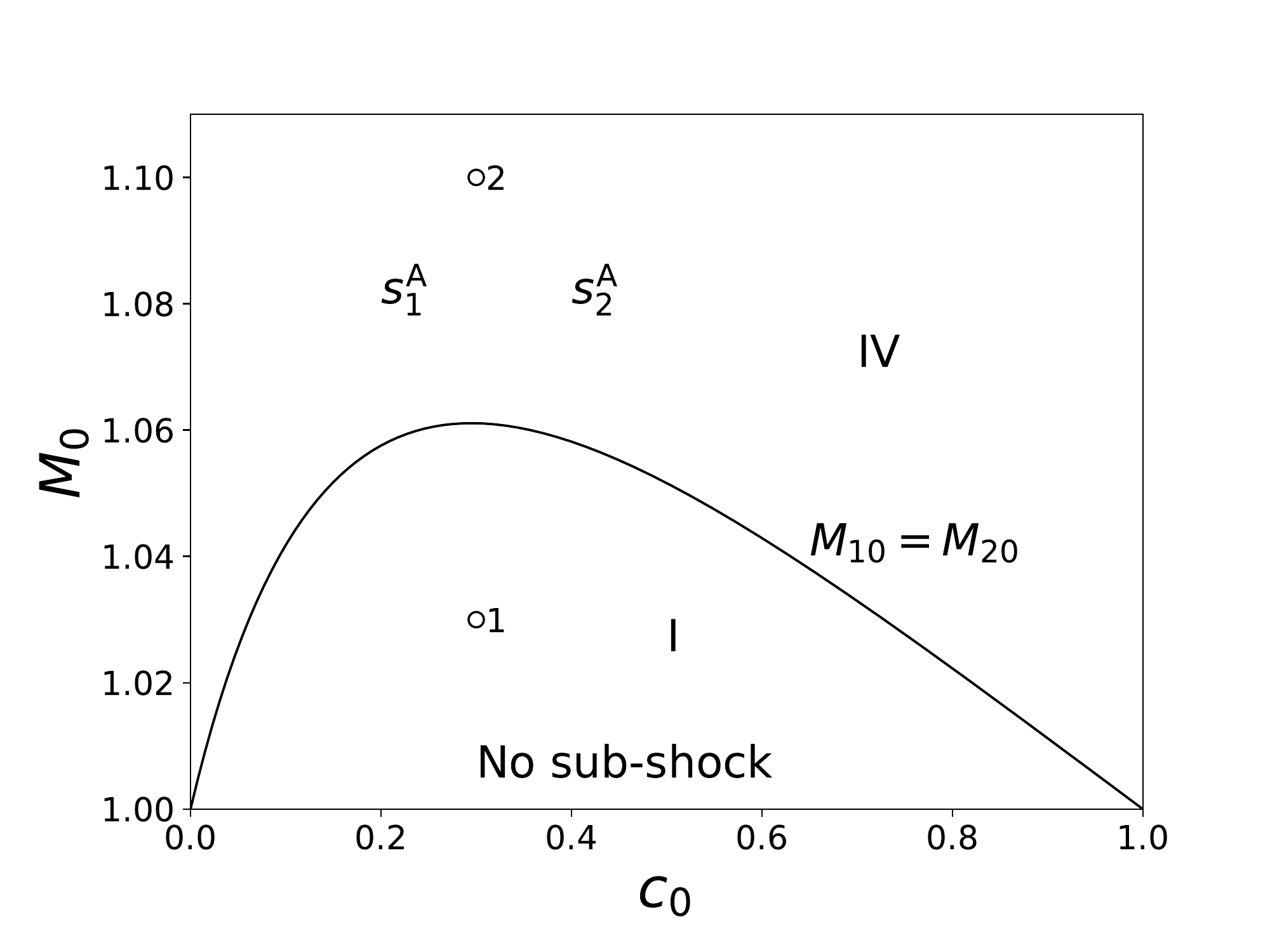}
	\caption{Case B$_2$: Two regions in the plane $(c_0,M_0$)  of possible sub-shocks. 
		$\gamma_1 = 7/6$, $\gamma_2 = 5/3$, and $\mu = g  = 0.7$. }
	\label{fig:subshockEuler_mu07}
\end{figure}

When $\mu =  g $, i.e. $m_1/m_2 = \gamma_1/\gamma_2$, we have the
special situation in which  the characteristic velocities of the two species in equilibrium are equal and $\hat{c}_0=1$ then \eqref{b11} becomes:
\begin{equation}
	\bar{	\lambda}_0 < \lambda_{20} =	\lambda_{10}, \quad s> \bar{	\lambda}_0, \label{bspec}
\end{equation}
and as consequence  $M_{20} = M_{10}$.
We show this special case in Figure \ref{fig:subshockEuler_mu07} by adopting $\mu = 0.7$, $\gamma_1 = 7/6$ and $\gamma_2 = 5/3$.  
This is an exceptional case where the regions become only two:   regions I have no sub-shocks and region IV in which multiple sub-shock arise.

\begin{figure}[]
	\centering
	\includegraphics[width=0.9\linewidth]{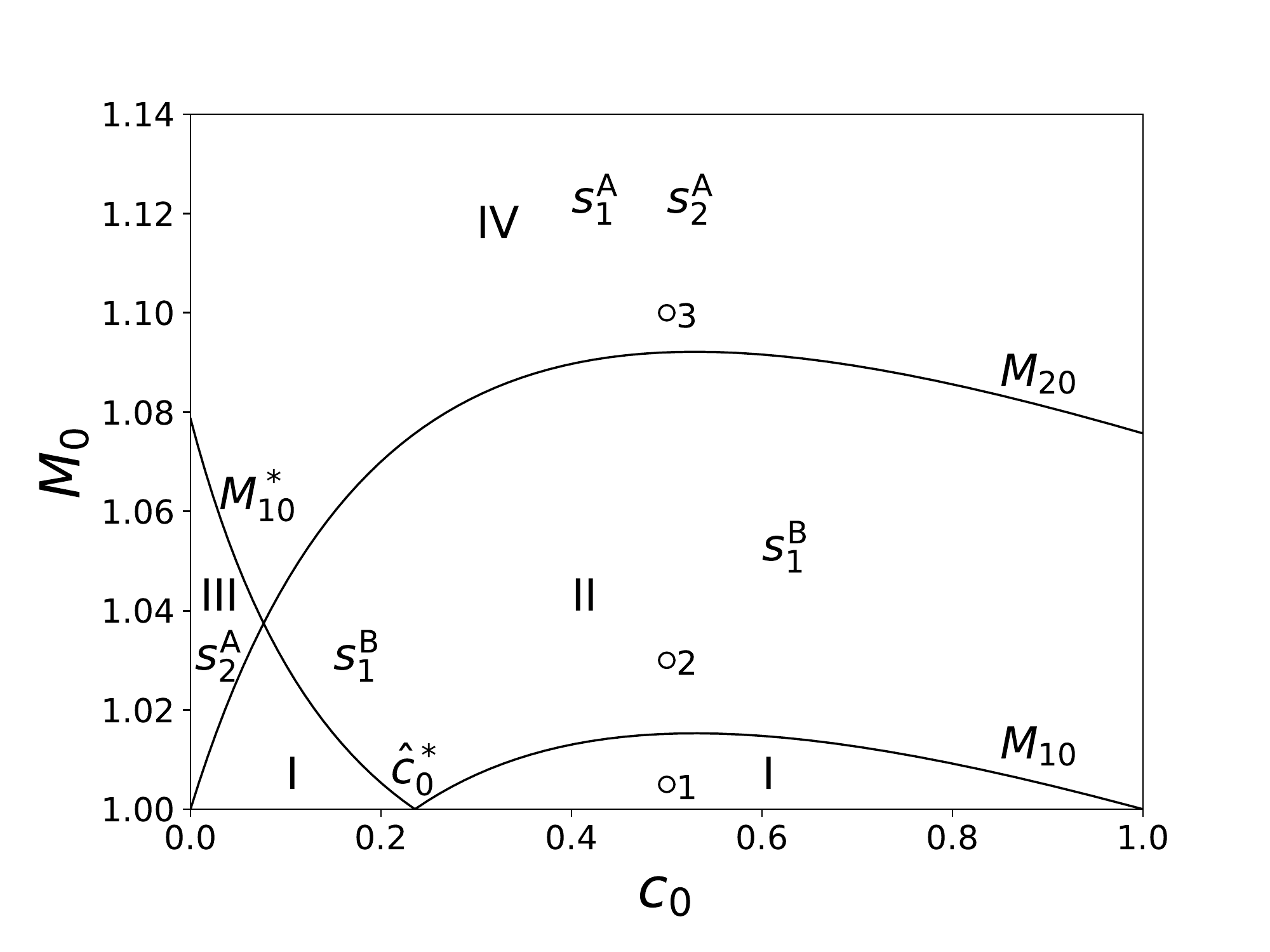}
	\caption{Case B$_3$ -  Four regions in the plane $(c_0,M_0$)  of possible sub-shocks. 
		$\gamma_1 = 7/6$, $\gamma_2 = 5/3$, and $\mu = 0.81 <\mu^{**}$. }
	\label{fig:subshockEuler_mu081}
\end{figure}
\subsubsection{Case B$_3$: $ \mu >  g $}
Now the inequality \eqref{disegb1} becomes opposite and the characteristic velocity of species 1 becomes greater than the one of the species 2:
\begin{equation}\label{diseg2}
	\lambda_{10} <	\lambda_{20},
\end{equation}
A sub-shock after the maximum characteristic velocity is observed when
\begin{equation*}
	\begin{split}
		&s > \lambda_{20}  \\
		&\Leftrightarrow \quad
		M_{0} > M_{20}  
		= \frac{\lambda_{20} }{a_0} 
		= \sqrt{\frac{m_0 \gamma_2}{m_2 \gamma_0}}
		= \sqrt{\frac{\gamma_2 \mu}{\gamma_0 \left\{c_{0} + (1 - c_{0}) \mu\right\}}}.
	\end{split}
\end{equation*}
It can be verified  that 
\begin{equation*}
	\lim_{c_0 \rightarrow 0} M_{20}  = 1, \quad \lim_{c_0 \rightarrow 1} M_{20}  = \sqrt{\frac{\mu}{ g }}.
\end{equation*}
Now we have a specular situation as the case B$_1$. There exists a value of the concentration $\hat{c}_0^*$ such that 
\begin{align}
	&\text{If} \,\,\ 0<c_0<\hat{c}_0^*,  \quad 
	\lambda_{10}< \bar{	\lambda}_0 <	\lambda_{20}, \quad s> \bar{	\lambda}_0, \label{b111}\\
	&\text{If} \,\,\ \hat{c}_0^* <c_0<1,  \quad 
	\bar{	\lambda}_0 <\lambda_{10} <\lambda_{20}, \quad s> \bar{	\lambda}_0. \label{b222}
\end{align}
In the case  \eqref{b111} we have the possibility that $\lambda_{1\mathrm{I}} =s $ and as consequence the curve $M_{1\mathrm{I}}(c_0)$ exist solution of 
\begin{equation}\label{eq:M055}
	M_{1\mathrm{I}} (M_{10}^*,c,\mu)=  M_{10}^*.
\end{equation}
where we have indicates with $M_{1\mathrm{I}}=   \lambda_{1\mathrm{I}} /a_0$, the dimensionless characteristic velocity in the equilibrium state behind the shock. 
We have the following expression of $M_{1\mathrm{I}}$: 
\begin{equation*}
	\begin{split}
		M_{1\mathrm{I}}
		= &\frac{2 (M_{0}^{2} - 1)}{(\gamma_0 + 1) M_{0}}\\ 
		&+ \sqrt{ \frac{\{2 + (\gamma_0 - 1) M_0^2\}\{1 + \gamma_0(2 M_0^2 - 1)\}}{(\gamma_0 + 1)^2 M_0^2} \frac{\gamma_1 m_0}{\gamma_0 m_{1}}},
	\end{split}
\end{equation*}
and in this case the solution of \eqref{eq:M055} is given by
\begin{equation*}
	M_{10}^*  
	= \sqrt{\frac{2 m_1 \gamma_0 + m_0 \gamma_1 (\gamma_0 - 1)}{\gamma_0 \{2 m_0 \gamma_1 - m_1(\gamma_0 - 1) \}}}.
\end{equation*}
It is easy to verify that $M_{10}^* $ is always a real number and is greater or equal to $1$ if $0 < c_0 \leq \hat{c}_0^*$ with
\begin{equation}
	\hat{c}_0^* = 	\frac{\mu (\gamma_1 - 1)(\gamma_1 - \gamma_2 \mu)}{(1 - \mu)\{\gamma_1 - \gamma_1 \gamma_2 (1 - \mu) - \gamma_2 \mu\}}.
\end{equation}
Moreover  
\begin{equation*}
	\lim_{c_0 \rightarrow 0} M_{10}^*=\sqrt{\frac{\gamma_1(\gamma_2 - 1) + 2 \gamma_2 \mu}{\gamma_2 \left\{ 2 \gamma_1 - \mu (\gamma_2 - 1)\right\}}}, \quad \lim_{c_0 \rightarrow \hat{c}_0^*} M_{10}^* = 1.
\end{equation*}
It is easy to verify that $\hat{c}_0^* <1$ if $\mu < \mu^{**}$ with 
\begin{equation}\label{mustarstar}
	\mu^{**} = \frac{\gamma_1 \left(1-\gamma_2\right)}{\gamma_1^2 + \gamma_2 - 2 \gamma_1 \gamma_2}. 
\end{equation}
Then if $\mu<\mu^{**}$ and $c_0 > \hat{c_0}^*$ we have the inequalities \eqref{b222} and in this case emerges the curve $M_{10}\geq 1$
\begin{equation*}
	M_{10} = \frac{\lambda_{10}}{a_0}
	= \sqrt{\frac{m_0 \gamma_1}{m_1 \gamma_0}}
	= \sqrt{\frac{\gamma_1}{\gamma_0 \left\{c_{0} + (1 - c_{0}) \mu\right\}}},
\end{equation*}
with the property that $M_{10}=1$ when $c_0=\hat{c}_0*$ and $c_0=1$.

As a typical example, we depict the curves with $\mu = 0.81$, $\gamma_1 = 7/6$, $\gamma_2 = 5/3$ in Figure \ref{fig:subshockEuler_mu081}. 

\begin{figure}[]
	\centering
	\includegraphics[width=0.9\linewidth]{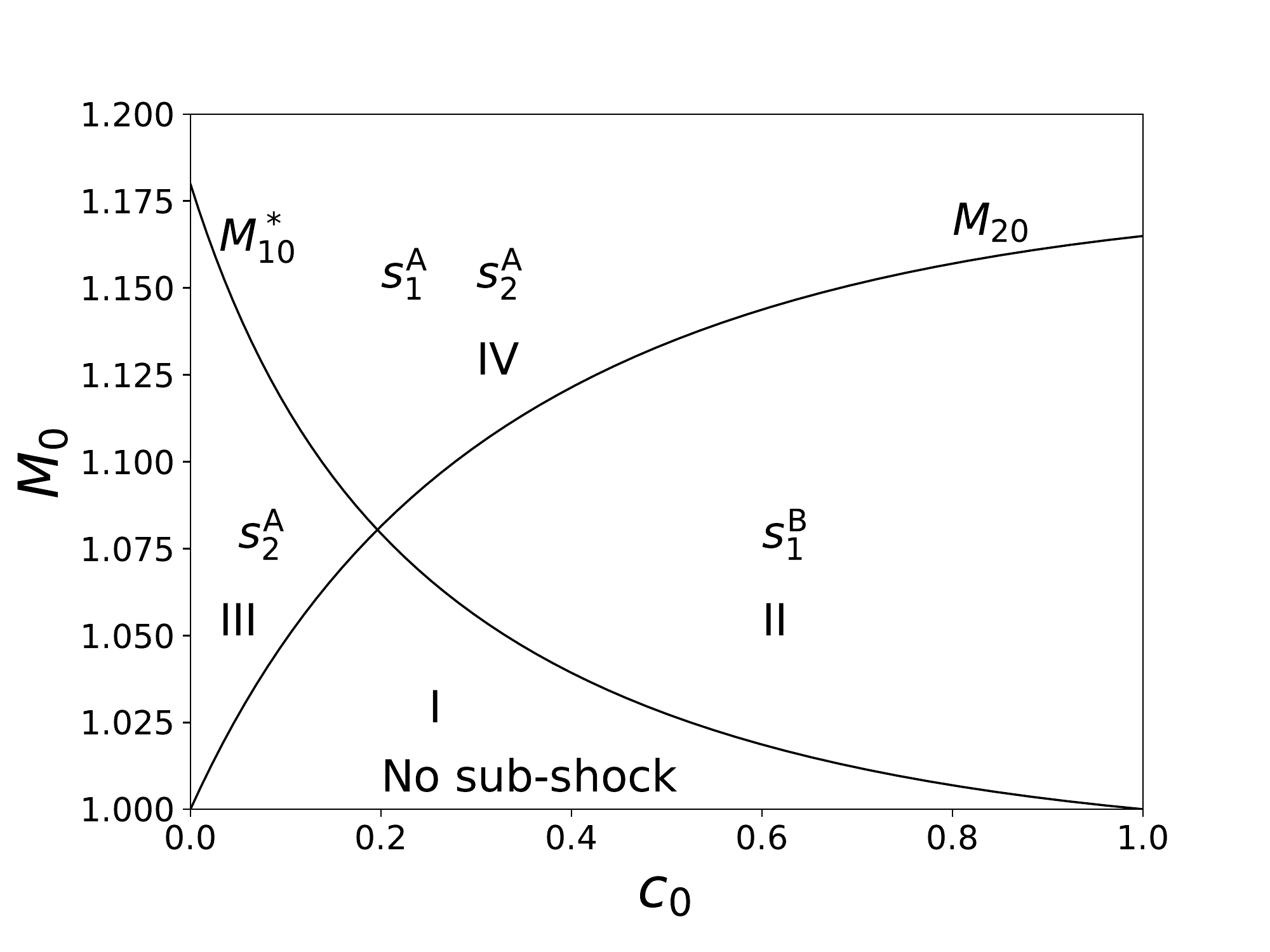}
	\caption{Case B$_3$ - Four regions in the plane $(c_0,M_0$)  of possible sub-shocks in a binary Eulerian mixture. 
		$\gamma_1 = 7/6$, $\gamma_2 = 5/3$, and $\mu = 0.95 > \mu^{**}$. }
	\label{fig:subshockEuler_mu095}
\end{figure}

While if $\mu >\mu^{**}$ we have $\hat{c}_0>1$ and only the case \eqref{b111} hold and we have only the curve $M_{10}^*$ in all the interval of concentration.
As a typical example, we depict the curves with $\mu = 0.95$, $\gamma_1 = 7/6$, and $\gamma_2 = 5/3$ in Figure \ref{fig:subshockEuler_mu095}. 

In Case B$_3$ we have: in Region I, shock structure is always continuous, 
in Region II, there may exist a sub-shock for the variables of first species $S_1^B$, 
in Region III, we  have a sub-shock $S_2^A$ for constituent $2$ and 
finally, in Region IV, we may have multiple sub-shocks for both constituent $S_1^A, S_2^A$ after the maximum eigenvalue in the unperturbed state. 
Summarizing the case B as a statement:
\begin{statement}
	- If the species with the smallest mass has more degrees of freedom than the heavier one, we have a  different topology of the regions for increasing values of the ratio of masses $\mu$. For $0<\mu < g $, the regions are the ones described in Figure \ref{fig:subshockEuler_mu055}, and the characteristic velocity of heavy species is less than the one with smallest mass \eqref{disegb1}. For $\mu = g $,  the characteristic velocity of both species are equal, and the possible regions reduce only to two as in Figure \ref{fig:subshockEuler_mu07}. Then for $\mu > g $ we have the inversion between the two characteristic velocities \eqref{diseg2} and we have the Figures \ref{fig:subshockEuler_mu081} and 
	\ref{fig:subshockEuler_mu095} respectively for $\mu <\mu^{**}$ and for $\mu >\mu^{**}$  with $\mu^{**}$ given by \eqref{mustarstar}.
\end{statement}

\section{Numerical results on the shock structure}
\label{sec:Numerical}

In the previous section, from the possibility for the sub-shock formation, we have identified regions for different values of the parameters in the plane $(c_0,M_0)$. 
However, we are sure that sub-shocks exist only when the shock velocity is greater than the maximum characteristic velocities evaluated in the equilibrium state in front of the shock by the Boillat and Ruggeri theorem \cite{Breakdown}, or we are sure in Region I that we have a regular structure without sub-shock formation. In the other case, we have only a necessary condition for the existence of sub-shocks, and we do not know theoretically if these sub-shocks arise. 
We understand the reason why the condition $s=\lambda$ is not in general sufficient in the following way.  
If we multiply the system \eqref{struttura} by the left eigenvector $\mathbf{l}$ of $\mathbf{A}$ corresponding to a given eigenvalue $\lambda$, we obtain
\begin{equation*}
	\mathbf{l} \cdot \frac{d \mathbf{u}}{d \varphi} = \frac{\mathbf{l}\cdot \mathbf{P}}{\lambda -s}. 
\end{equation*}
Therefore the shock structure can be regular if $\mathbf{l}\cdot \mathbf{P}$ should tend to zero as $s$  approaches  the eigenvalue   $\lambda$. 
This is not an exceptional case because, at least in cases of a single fluid in the framework of RET studied in the literature~ \cite{Weiss,RET,BookNew,ShRu}, this is always present except for the maximum eigenvalue evaluated in equilibrium. 
Therefore we need to solve the system of field equations numerically and make the possibility of the sub-shock formation clear in the different points of each region.

\subsection{Dimensionless field equations and phenomenological coefficients}
For convenience, we introduce the following dimensionless variables scaled by the values evaluated in the unperturbed state:
\begin{equation}\label{Model:DimLess}
	\begin{split}
		&\hat{\rho} = \frac{\rho}{\rho_{0}}, \quad
		\hat{v} = \frac{v}{a_{0}}, \quad
		\hat{T} = \frac{T}{T_{0}}, \quad
		\hat{x} = \frac{x}{t_c \, a_0}, \quad 
		\hat{t} = \frac{t}{t_c}, \\
		&\hat{\psi} = \frac{t_c}{\rho_0 T_0}\psi , \quad
		\hat{\theta} = \frac{t_c}{\rho_0 \frac{k_B}{m_0} T_0^2}\theta , \quad
	\end{split}
\end{equation}
where $t_c$ is arbitrary characteristic time for numerical computations. 
The system \eqref{finale} is rewritten as 
\begin{widetext}
\begin{equation}\label{dimless:finale1}
	\begin{split}
		& \frac{\partial \hat{\rho}_1 }{\partial \hat{t}} + \frac{\partial \hat{\rho}_1  \hat{v}_1 }{\partial \hat{x}} 
		= 0, \\
		&\frac{\partial \hat{\rho}_1 \hat{v}_1 }{\partial \hat{t}}+ \frac{\partial}{\partial \hat{x}} \left\{\hat{\rho}_1\left( \hat{v}^2_1 + \frac{\hat{T}_1}{\gamma_0 \left\{c + (1 - c) \mu \right\}} \right) \right\} 
		= - \hat{\psi} \frac{(\hat{v}_1 - \hat{v}_2)(\hat{\rho}_1 \hat{T}_1 + \hat{\rho}_2 \hat{T}_2)}{(\hat{\rho}_1 + \hat{\rho}_2) \hat{T}_1 \hat{T}_2}, \\
		&\frac{\partial}{\partial \hat{t}} \left\{\hat{\rho}_1\left( \hat{v}^2_1 + \frac{2 \hat{T}_1}{\gamma_0 (\gamma_1 - 1) \left\{c + (1 - c) \mu \right\}} \right) \right\} 
		+ \frac{\partial}{\partial \hat{x}}\left\{\hat{\rho}_1 \hat{v}_1\left( \hat{v}^2_1 + \frac{2 \gamma_1 \hat{T}_1}{\gamma_0 (\gamma_1 - 1) \left\{c + (1 - c) \mu \right\}} \right) \right\} \\
		&\quad = - 2 \frac{\hat{\theta}}{\gamma_0} \frac{(\hat{T}_1 - \hat{T}_2)}{\hat{T}_1 \hat{T}_2}
		-2 \hat{\psi} \frac{(\hat{v}_1 - \hat{v}_2)(\hat{\rho}_1 \hat{v}_1 + \hat{\rho}_2 \hat{v}_2)(\hat{\rho}_1 \hat{T}_1 + \hat{\rho}_2 \hat{T}_2)}{(\hat{\rho}_1 + \hat{\rho}_2)^2 \hat{T}_1 \hat{T}_2},
		\\
		& \frac{\partial \hat{\rho}_2 }{\partial \hat{t}} + \frac{\partial \hat{\rho}_2  \hat{v}_2 }{\partial \hat{x}} 
		= 0, \\
		&\frac{\partial \hat{\rho}_2 \hat{v}_2 }{\partial \hat{t}} 
		+ \frac{\partial}{\partial \hat{x}} \left\{\hat{\rho}_2\left( \hat{v}^2_2 + \frac{\mu \hat{T}_2}{\gamma_0 \left\{c + (1 - c) \mu \right\}} \right) \right\} 
		= \hat{\psi} \frac{(\hat{v}_1 - \hat{v}_2)(\hat{\rho}_1 \hat{T}_1 + \hat{\rho}_2 \hat{T}_2)}{(\hat{\rho}_1 + \hat{\rho}_2) \hat{T}_1 \hat{T}_2}, \\
		&\frac{\partial}{\partial \hat{t}} \left\{\hat{\rho}_2\left( \hat{v}^2_2 + \frac{2 \mu \hat{T}_2}{\gamma_0 (\gamma_2 - 1) \left\{c + (1 - c) \mu \right\}} \right) \right\} 
		+ \frac{\partial}{\partial \hat{x}}\left\{\hat{\rho}_2 \hat{v}_2\left( \hat{v}^2_2 + \frac{2 \gamma_2 \mu \hat{T}_2}{\gamma_0 (\gamma_2 - 1) \left\{c + (1 - c) \mu \right\}}  \right) \right\} \\
		&\quad = 2 \frac{\hat{\theta}}{\gamma_0} \frac{(\hat{T}_1 - \hat{T}_2)}{\hat{T}_1 \hat{T}_2}
		+ 2 \hat{\psi} \frac{(\hat{v}_1 - \hat{v}_2)(\hat{\rho}_1 \hat{v}_1 + \hat{\rho}_2 \hat{v}_2)(\hat{\rho}_1 \hat{T}_1 + \hat{\rho}_2 \hat{T}_2)}{(\hat{\rho}_1 + \hat{\rho}_2)^2 \hat{T}_1 \hat{T}_2},
	\end{split}
\end{equation}
\end{widetext}

Next we need to estimate the values of the dimensionless phenomenological coefficients $\hat{\psi}$ and $\hat{\theta}$ defined in \eqref{Model:DimLess}. 
Both $\hat{\psi}$ and $\hat{\theta}$ are proportional to the characteristic time $t_c$ and 
therefore only the ratio $\hat{\psi}/\hat{\theta}$ is important for the present analysis. 
There are at least two kinds of expressions from kinetic-theoretical considerations for rarefied monatomic/polyatomic gases~\cite{Bose,Milana,Symmetry}. 
While the detailed expressions proposed of these two models are different from each other,  the values of $\hat{\psi}$ and $\hat{\theta}$ seem the same order of magnitude. 
For discussing typical examples of the shock structure and the sub-shock formation, we adopt the following constant values of the phenomenological coefficients: $\hat{\psi} = \hat{\theta} = 0.1$.  

\subsection{Numerical methods}

In the present analysis, in order to obtain the shock-structure solution with or without sub-shocks, we use a different procedure solving ad hoc Riemann problem for the PDE system \eqref{finale1} instead of solving the ODE system \eqref{struttura}. 
This strategy is based on the conjecture about the large-time behavior of the Riemann problem and the Riemann problem with structure \cite{Liu_struct1,Liu_struct2} for a system of balance laws proposed by Ruggeri and coworkers \cite{Brini_Osaka,BriniRuggeri,MentrelliRuggeri} -- following an idea of Liu \cite{Liu_conjecture}.
According to this conjecture, the solutions of both Riemann problems with and without structure, for large time, instead of converging to the corresponding  Riemann problem of the equilibrium subsystem  (i.e combination of shock and rarefaction waves), converge to solutions that represent a combination of shock structures (with and without sub-shocks) of the full system and rarefaction waves of the equilibrium subsystem. 

In particular, if the Riemann initial data correspond to a shock family $\mathcal{S}$ of the equilibrium subsystem, for a large time, the solution of the Riemann problem of the full system converges to the corresponding shock structure.
This strategy allows us to use the Riemann solvers~\cite{toro} for numerical calculations of the shock structure and was adopted in several shock phenomena of RET \cite{RuggeriSugiyama}. 
In particular, the conjecture was tested numerically for a Grad 13-moment system and a mixture of fluids \cite{Brini_Osaka, Brini_Wascom} and was verified in a simple $ 2\times 2$ dissipative models \cite{MentrelliRuggeri,IJNLM2017,subshock2}. 

We perform numerical calculations on the shock structure obtained after a long time for the Riemann problem consisting of two equilibrium states $\mathbf{u}_0$ and $\mathbf{u}_{\rm I}$ satisfying \eqref{Model:EqStates} and \eqref{eq:RH1}. 
The initial discontinuity between $\mathbf{u}_0$ and $\mathbf{u}_{\rm I}$ is set to be located at $\hat{x} = 0$.
We have developed numerical code written in Python on the basis of the Uniformly accurate Central Scheme of order 2 (UCS2) proposed by Liotta, Romano, and Russo~\cite{UCS2} for analyzing the hyperbolic balance laws with production terms. 
As the UCS2 is an implicit scheme, we need to solve the derived nonlinear difference equations iteratively for every time step by using an appropriate library such as SciPy~\cite{SciPy}. 
We check the convergence of the asymptotic profile and the independence of the size of spatial mesh $\Delta \hat{x}$ and temporal mesh $\Delta \hat{t}$ carefully. 

\subsection{Shock structure in Case A}

\begin{figure}
	\begin{center}
		\includegraphics[width=0.45\linewidth]{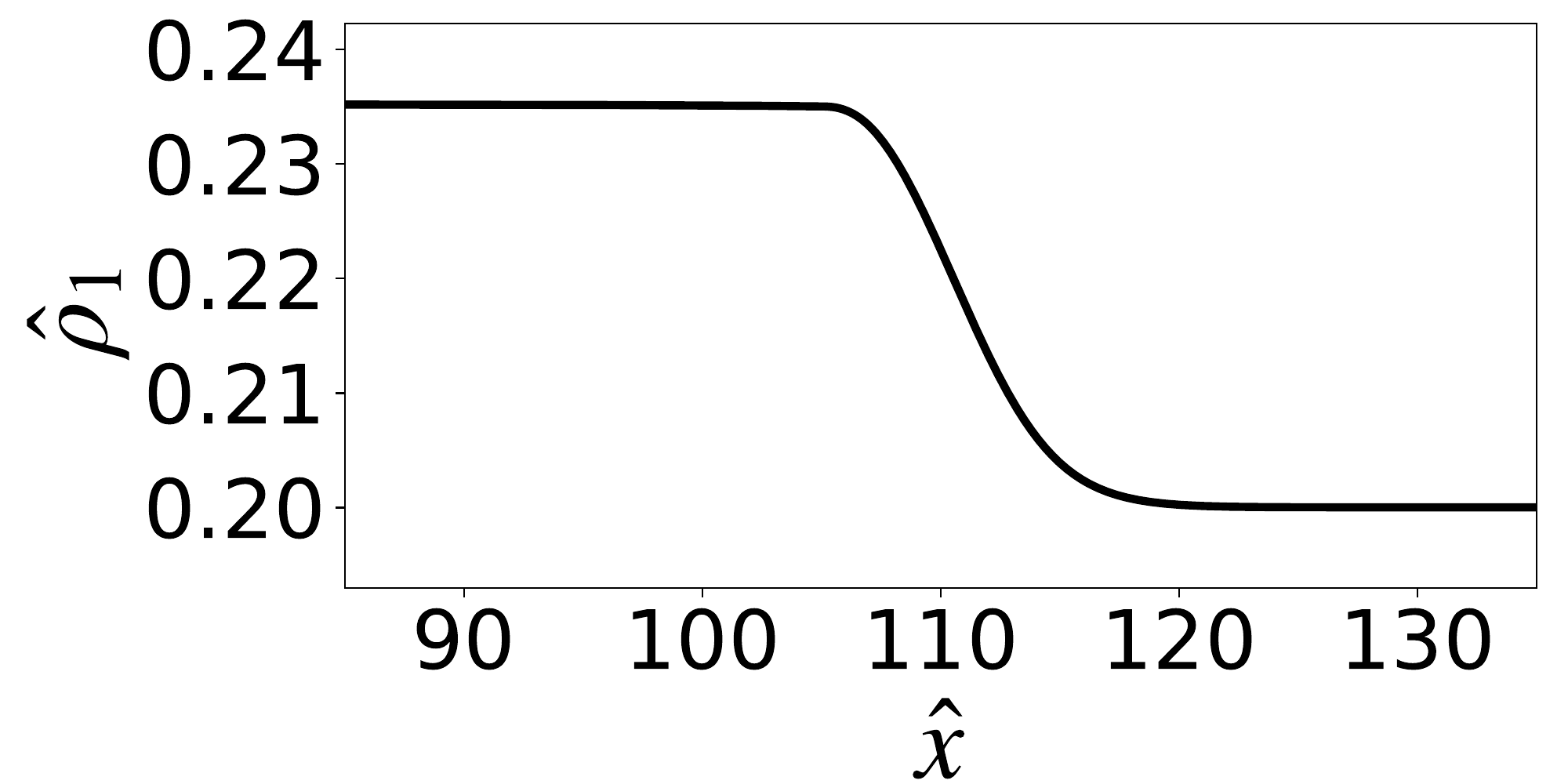} %
		\includegraphics[width=0.45\linewidth]{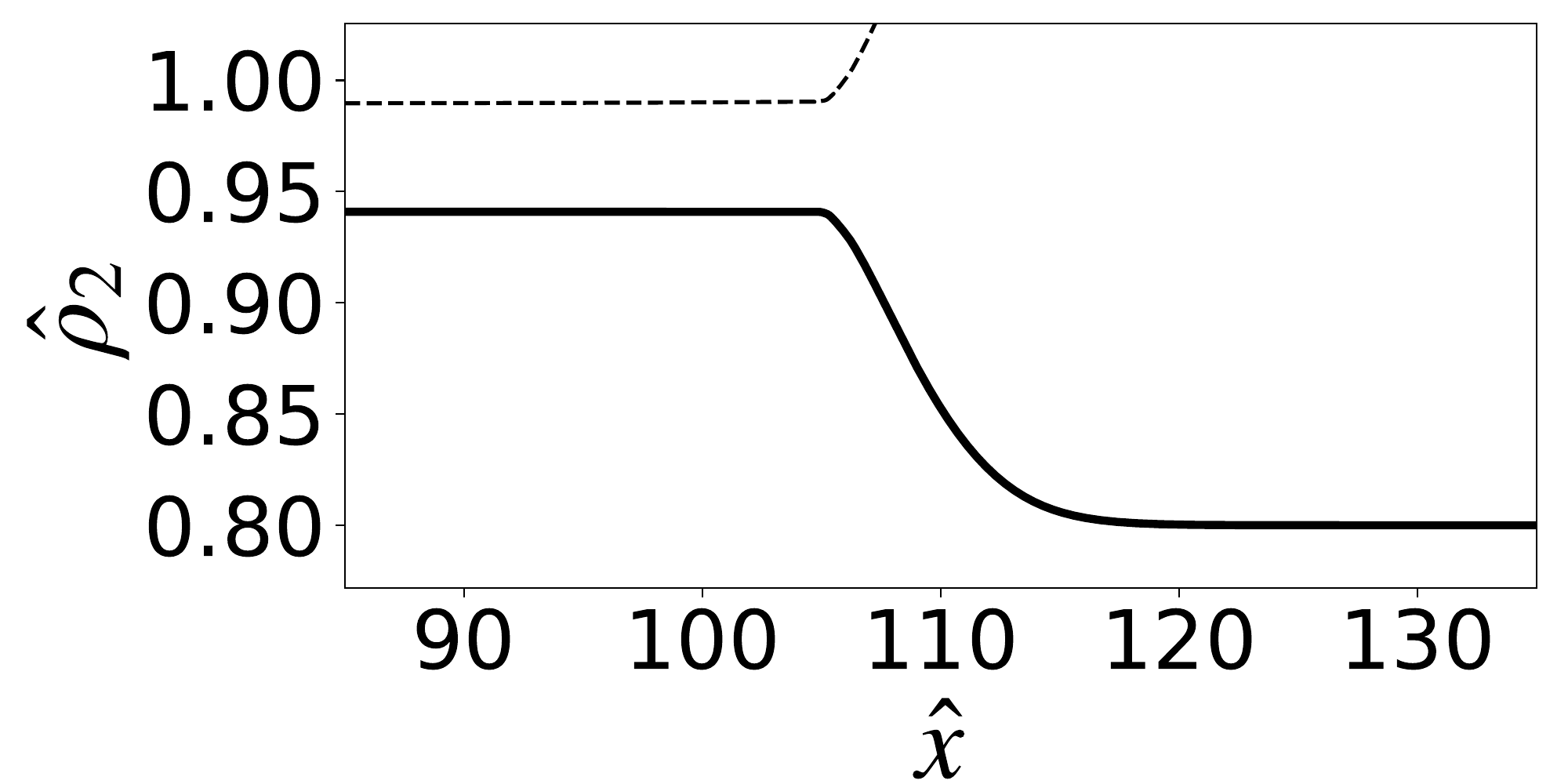}
		\includegraphics[width=0.45\linewidth]{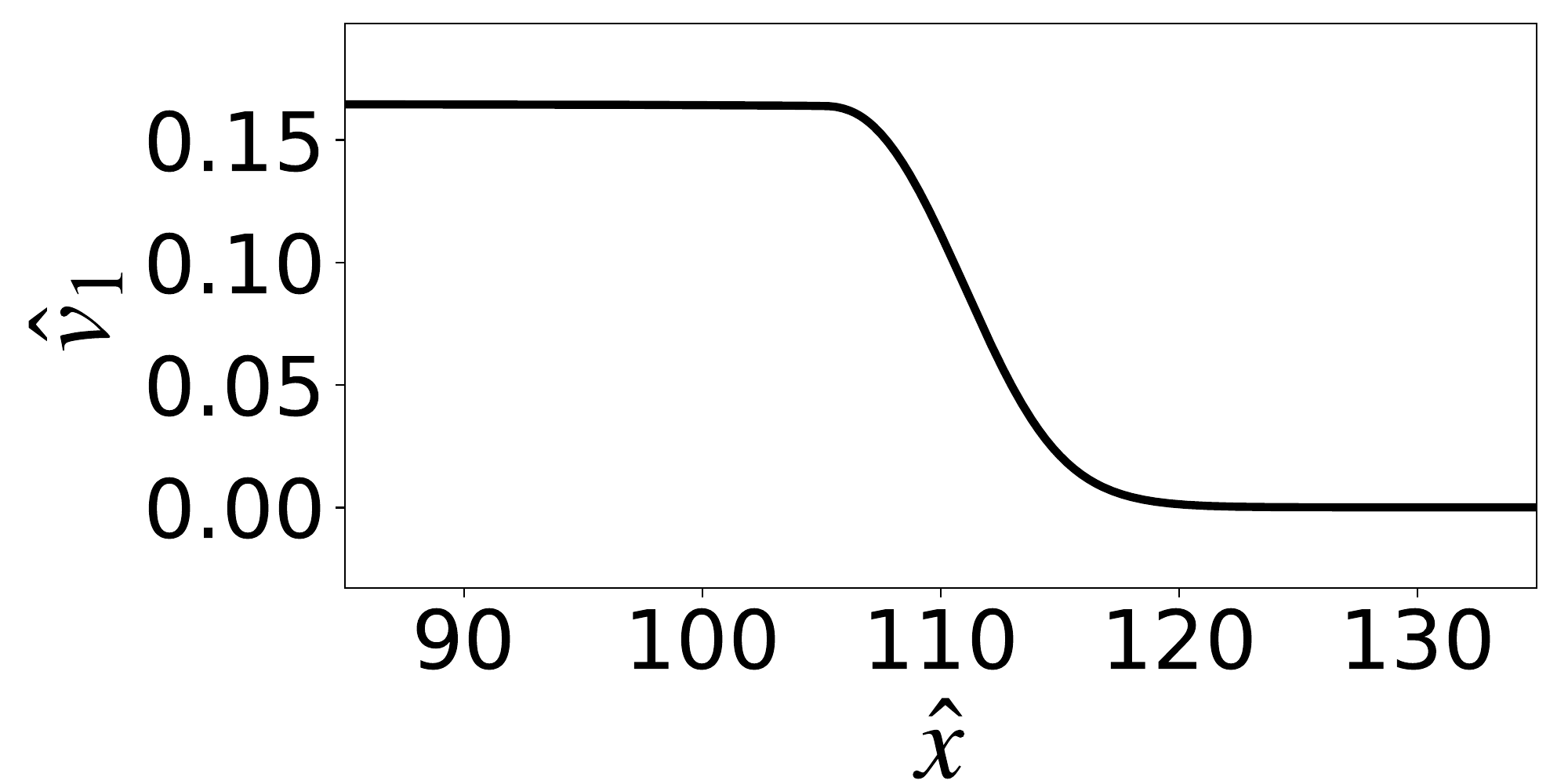} %
		\includegraphics[width=0.45\linewidth]{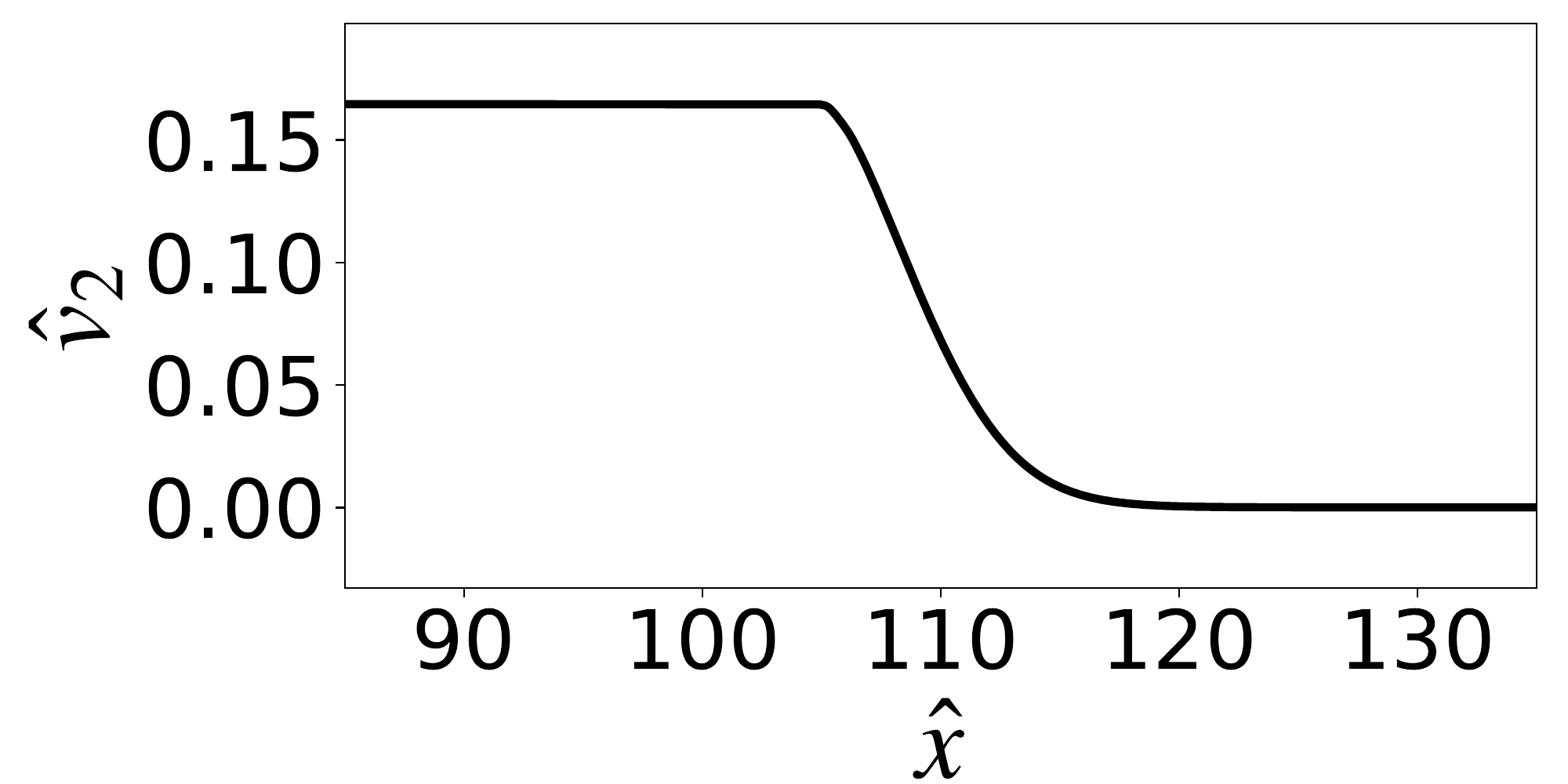}
		\includegraphics[width=0.45\linewidth]{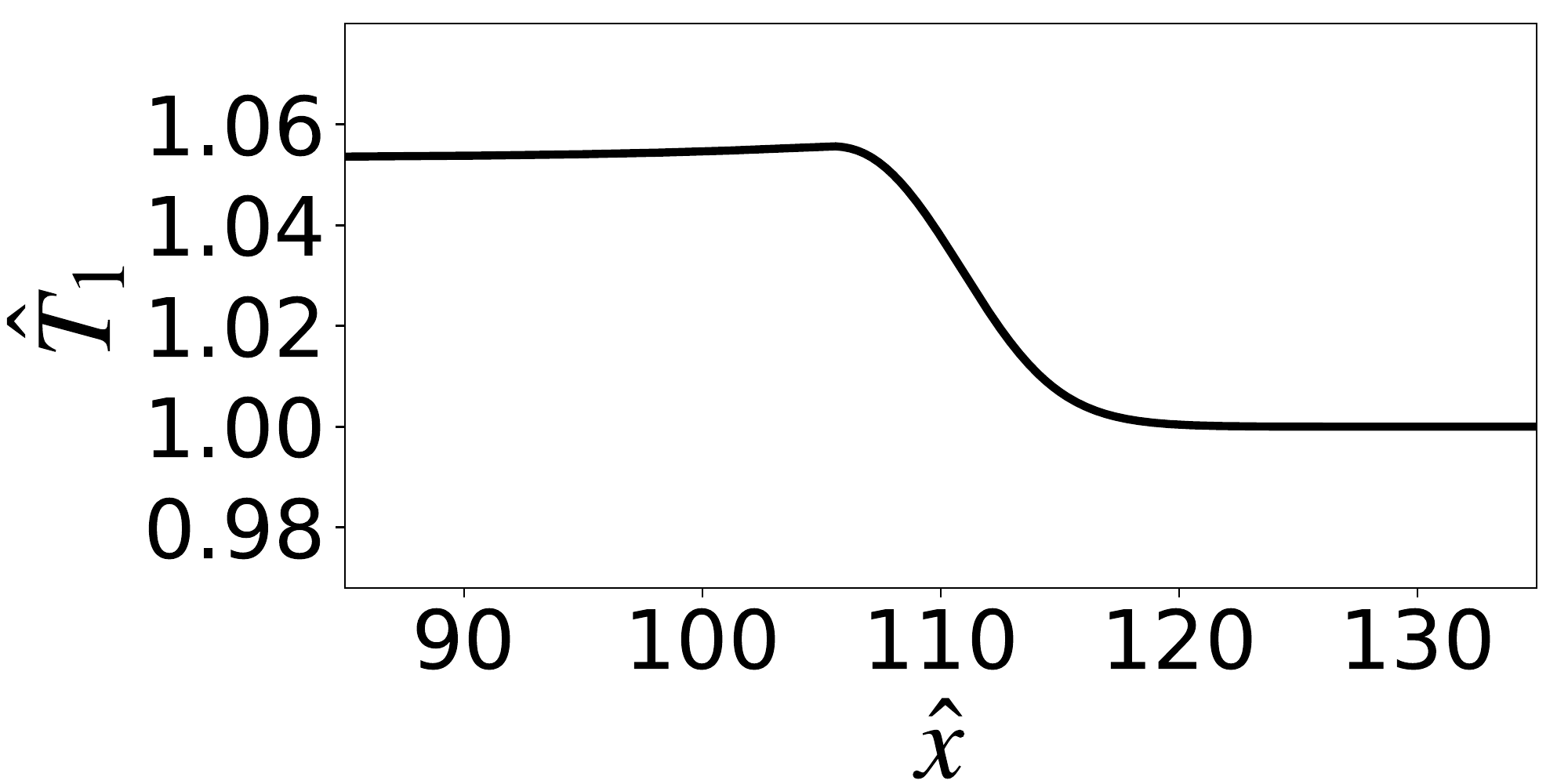} %
		\includegraphics[width=0.45\linewidth]{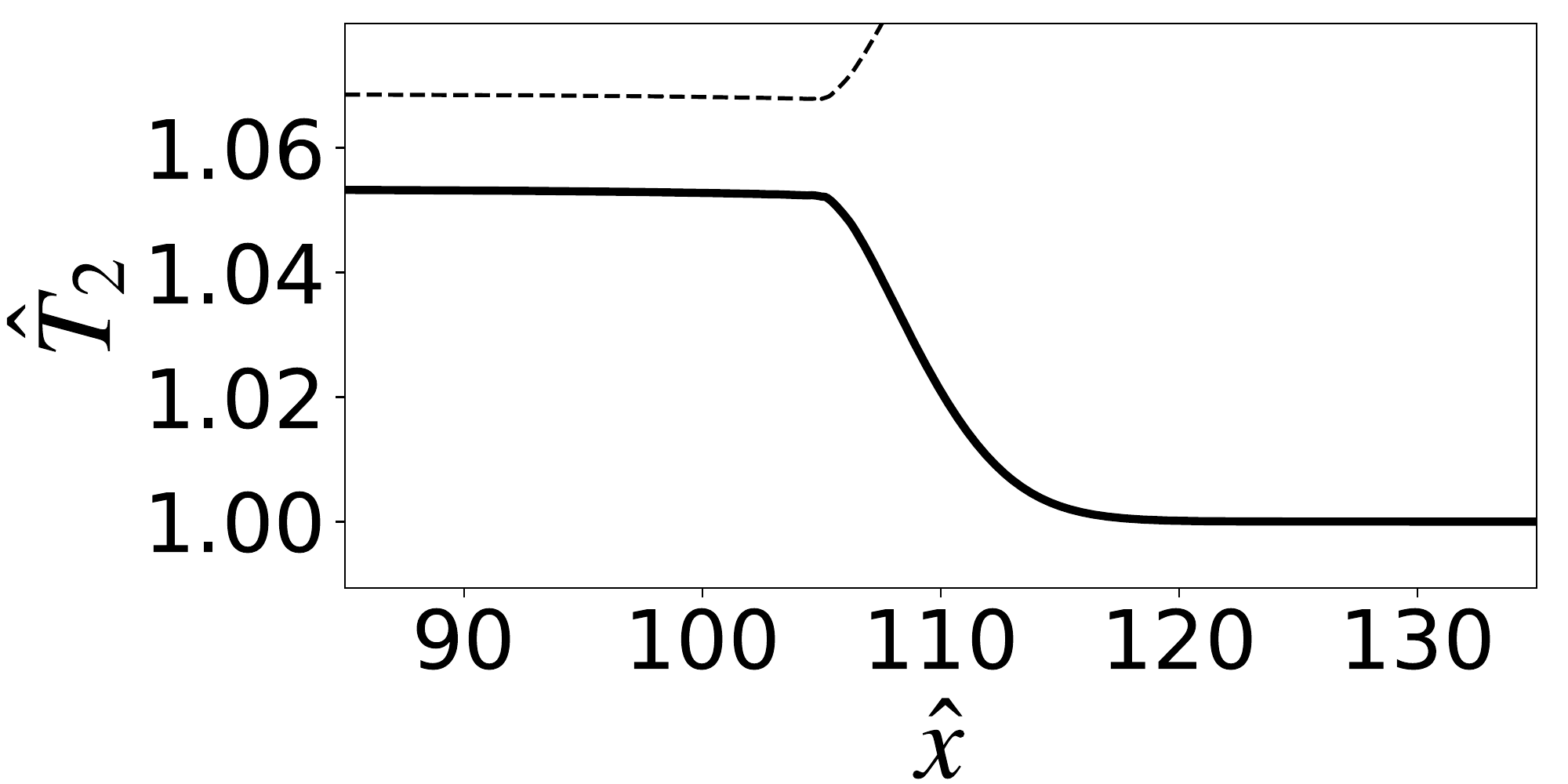}
	\end{center}
	\caption{Case A: Profiles of the dimensionless mass density, velocity and temperature for the constituent 1 of light molecules (left) and for the constituent 2 of heavy molecules (right) at $\hat{t} = 100$.  
	The parameters are set to be in Region I and correspond to No. 1 in Figure \ref{fig:subshockEuler_mu045} of Case A$_1$; $\gamma_1 = 7/5$, $\gamma_2 = 9/7$, $\mu = 0.45$, $c_0 = 0.2$, and $M_0 = 1.1$. 
	The temporal and spatial numerical meshes are $\Delta \hat{t} = 0.025$ and $\Delta \hat{x} = 0.1$. }
	\label{fig:c02_M0-1_1}
\end{figure}

\begin{figure}
	\begin{center}
		\includegraphics[width=0.45\linewidth]{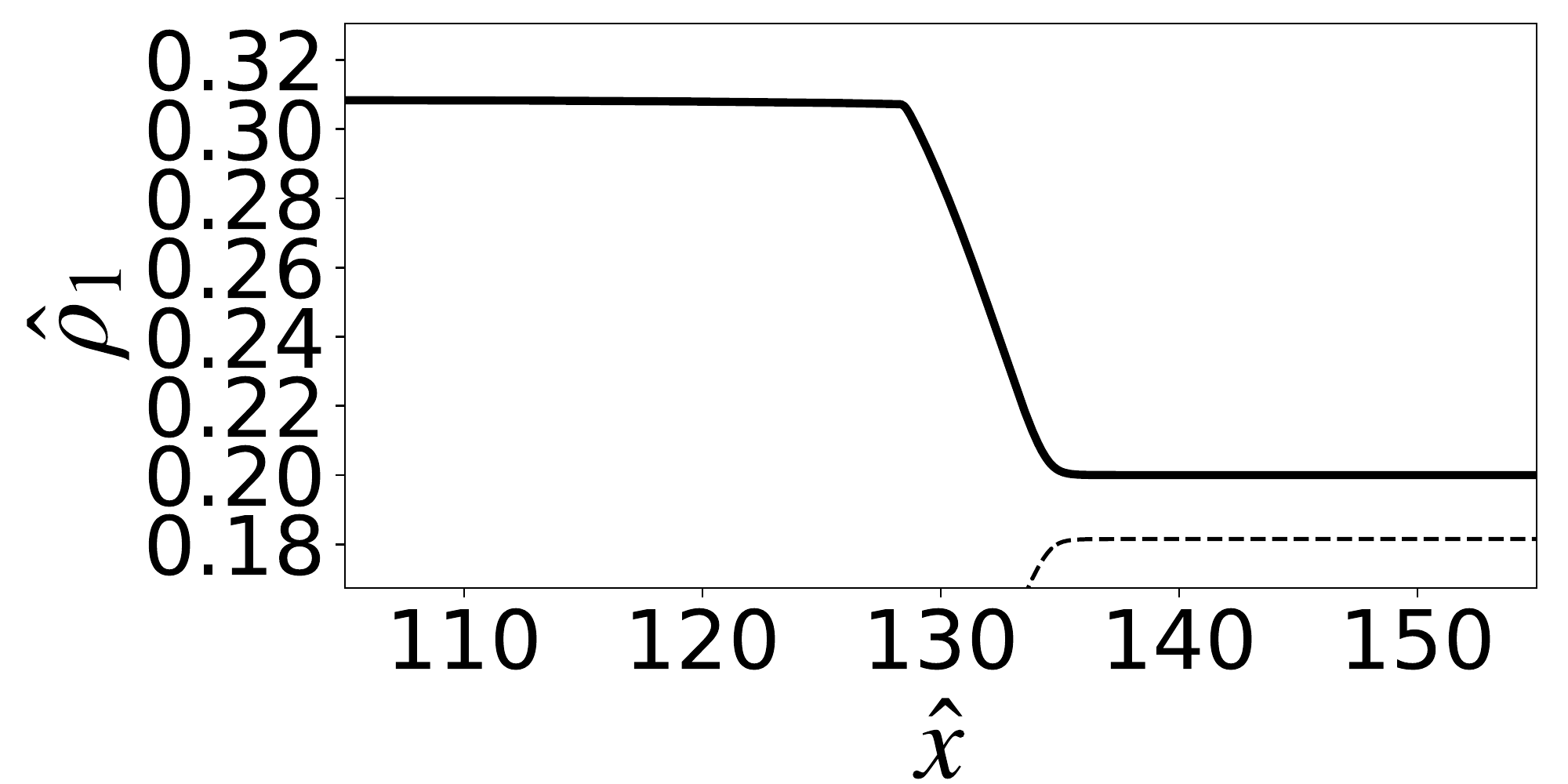} %
		\includegraphics[width=0.45\linewidth]{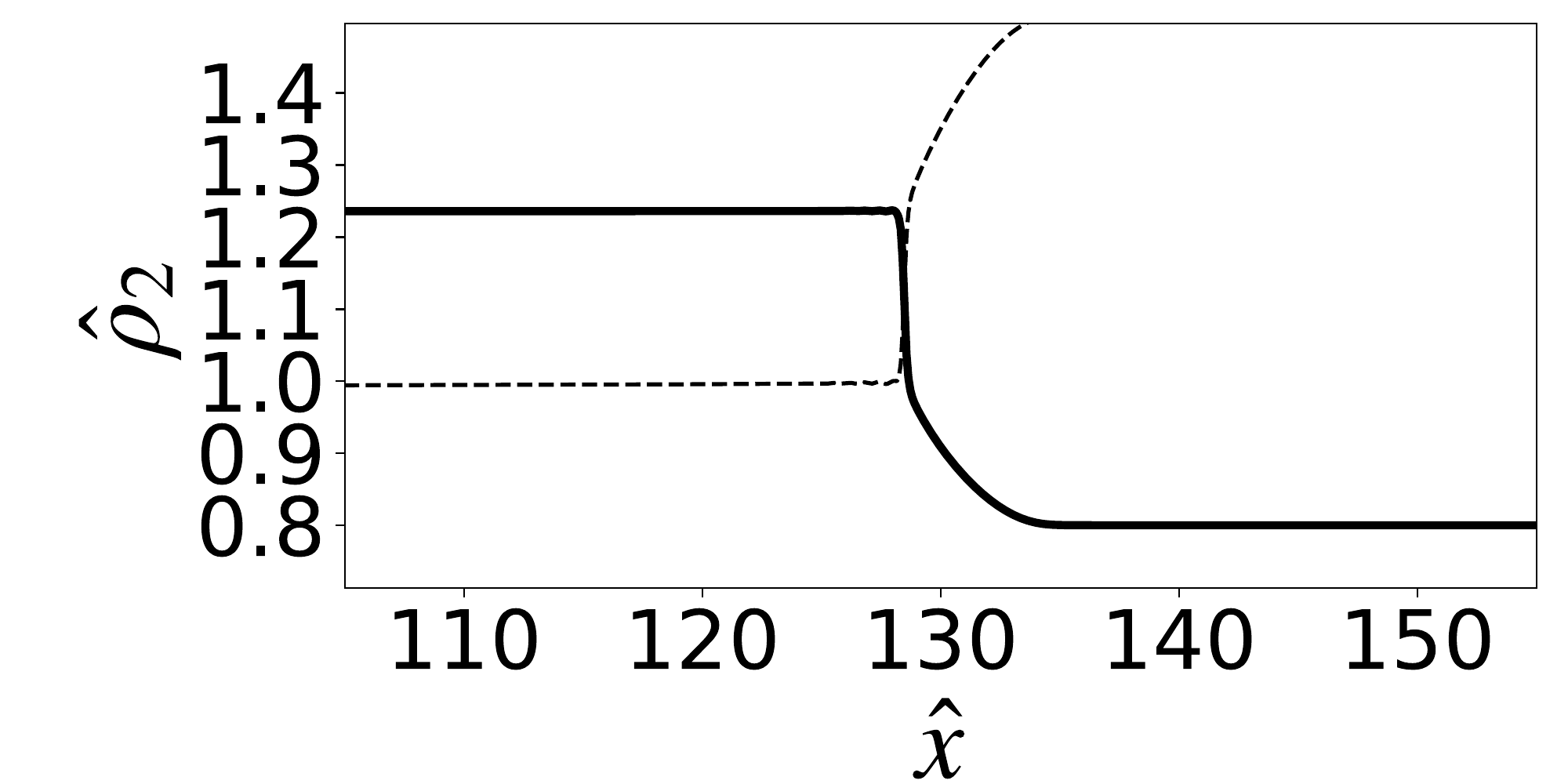}
		\includegraphics[width=0.45\linewidth]{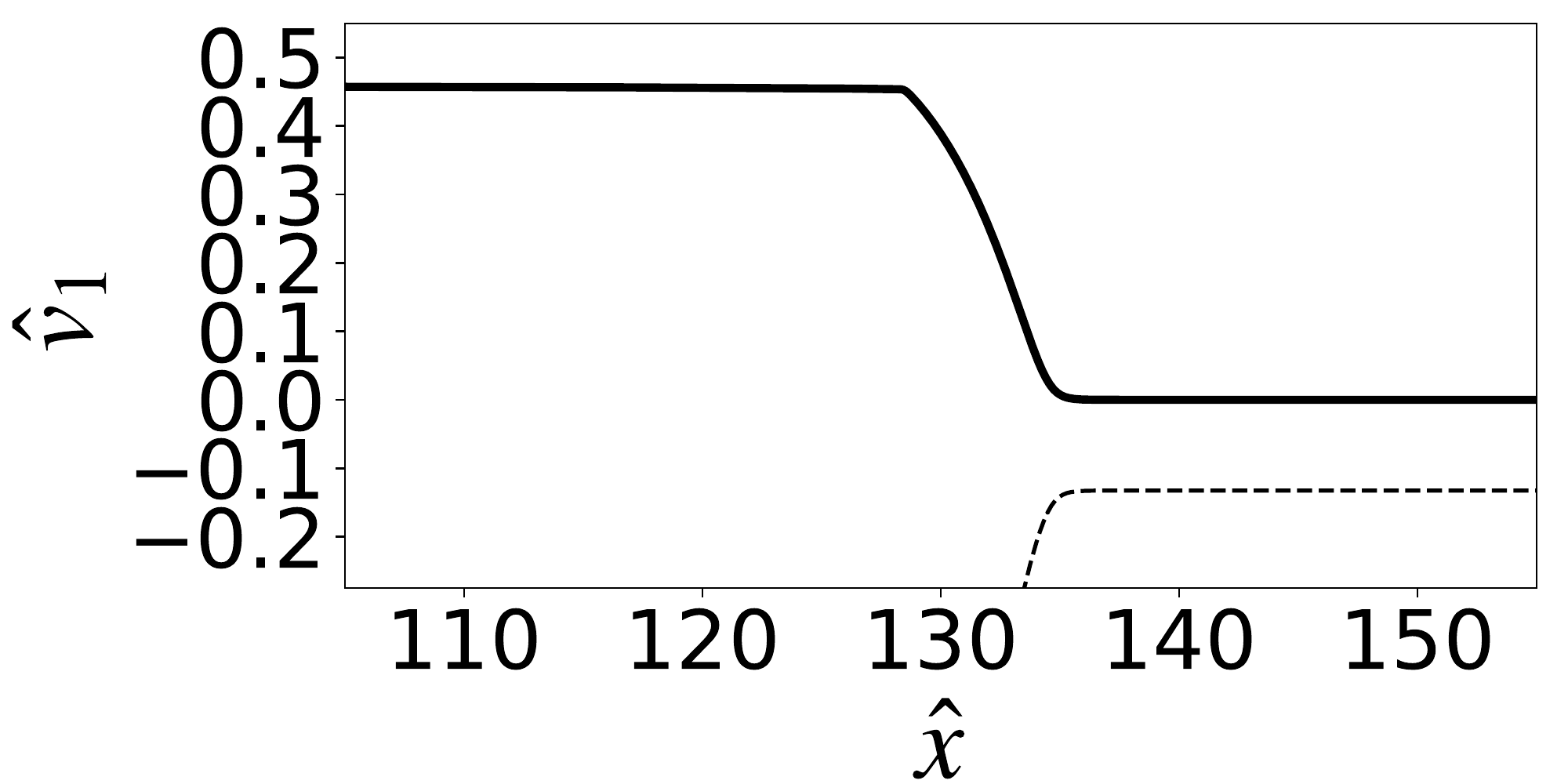} %
		\includegraphics[width=0.45\linewidth]{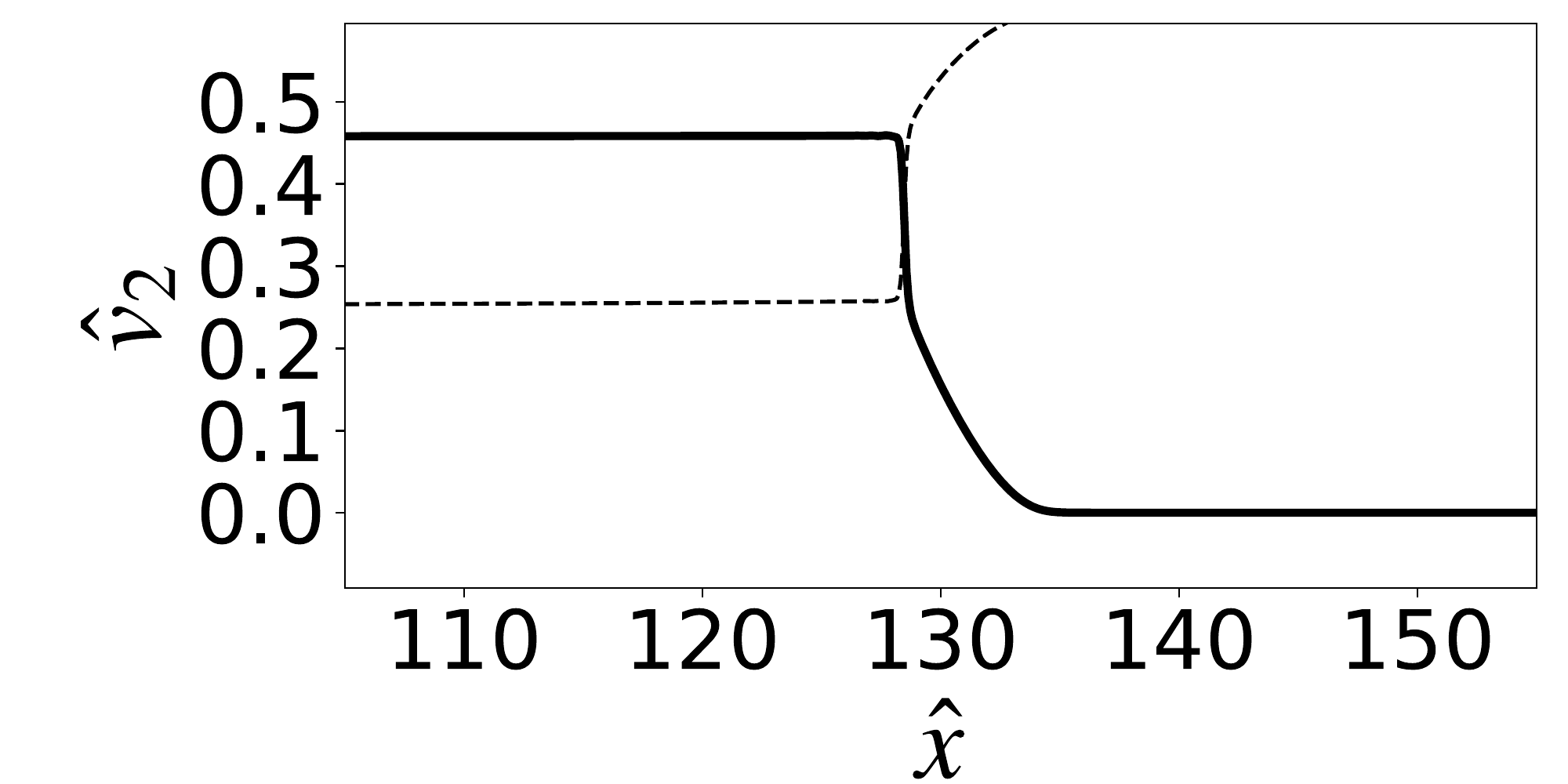}
		\includegraphics[width=0.45\linewidth]{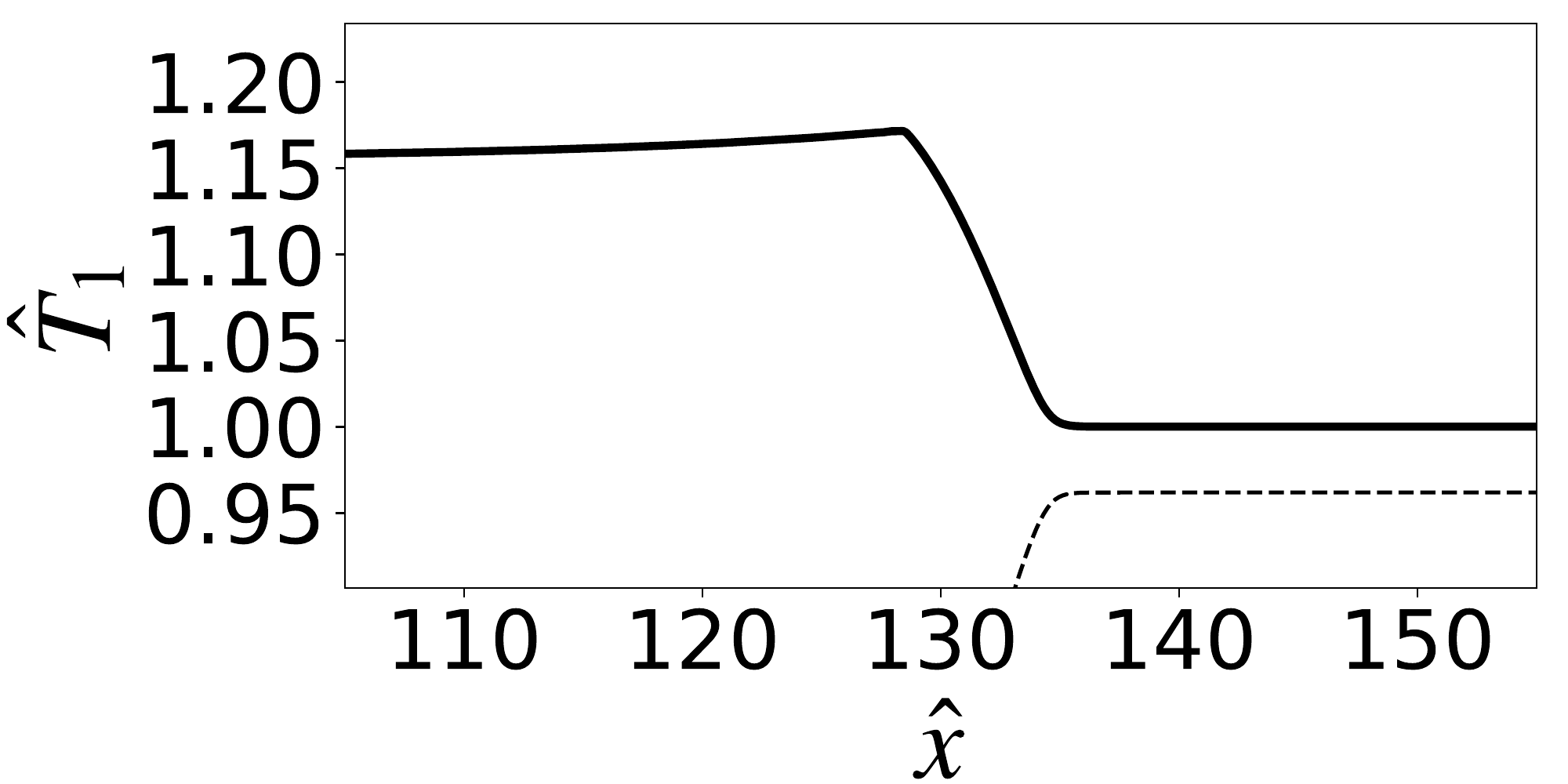} %
		\includegraphics[width=0.45\linewidth]{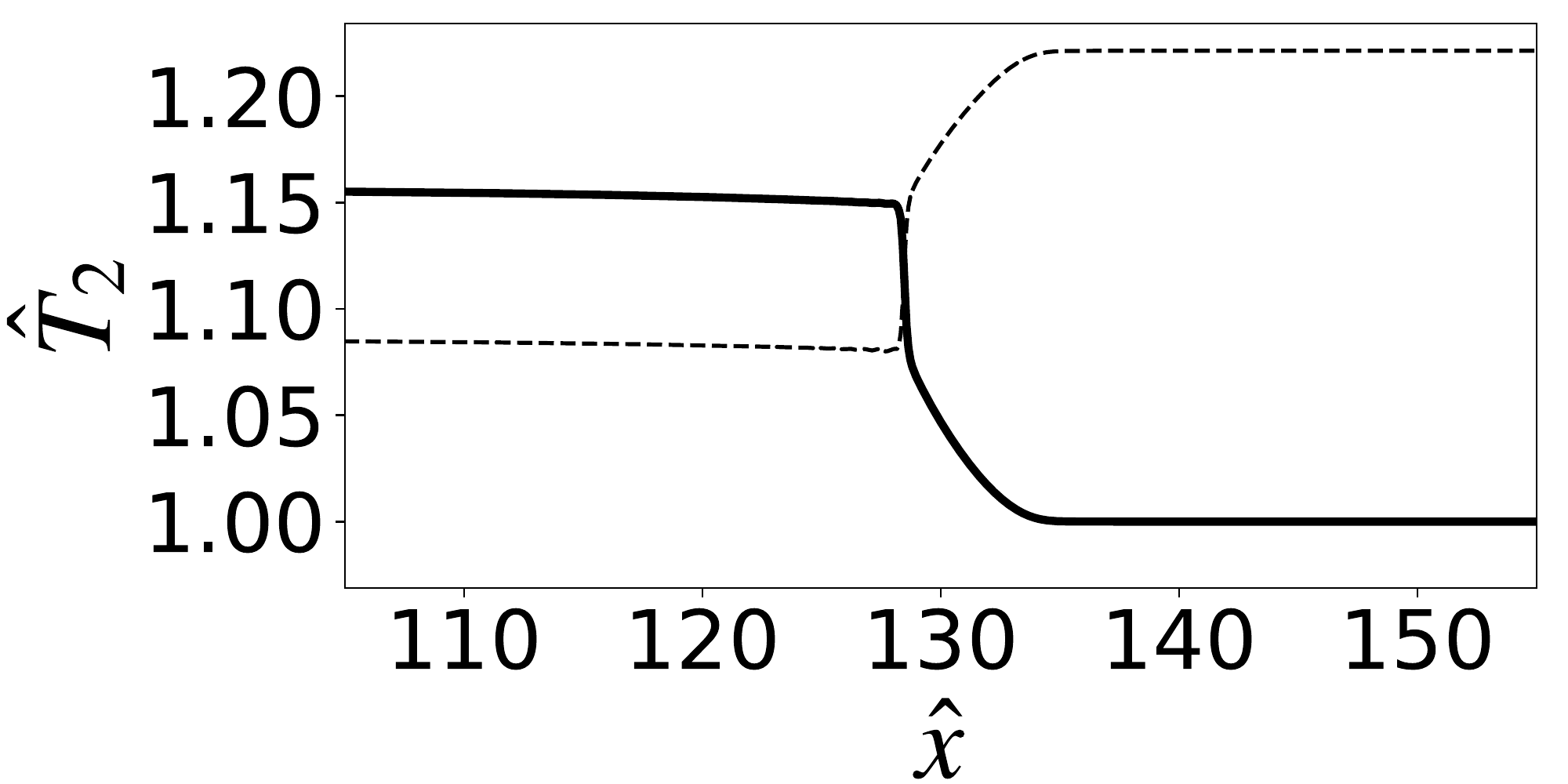}
	\end{center}
	\caption{Case A: Shock structure in a binary Eulerian mixture of polyatomic gases obtained at $\hat{t} = 100$ (Solid curves). 
	The parameters are in Region II and correspond to the mark of No. 2 shown in Figure \ref{fig:subshockEuler_mu045}; $\gamma_1 = 7/5$, $\gamma_2 = 9/7$, $\mu = 0.45$, $c_0 = 0.2$, and $M_0 = 1.3$. 
	The numerical conditions are $\Delta \hat{t} = 0.01$ and $\Delta \hat{x}=0.08$. 
	The curves of the potential sub-shock predicted any point of the shock structure are also shown (dotted curves). }
	\label{fig:c02_M0-1_3}
\end{figure}

\begin{figure}
	\begin{center}
		\includegraphics[width=0.45\linewidth]{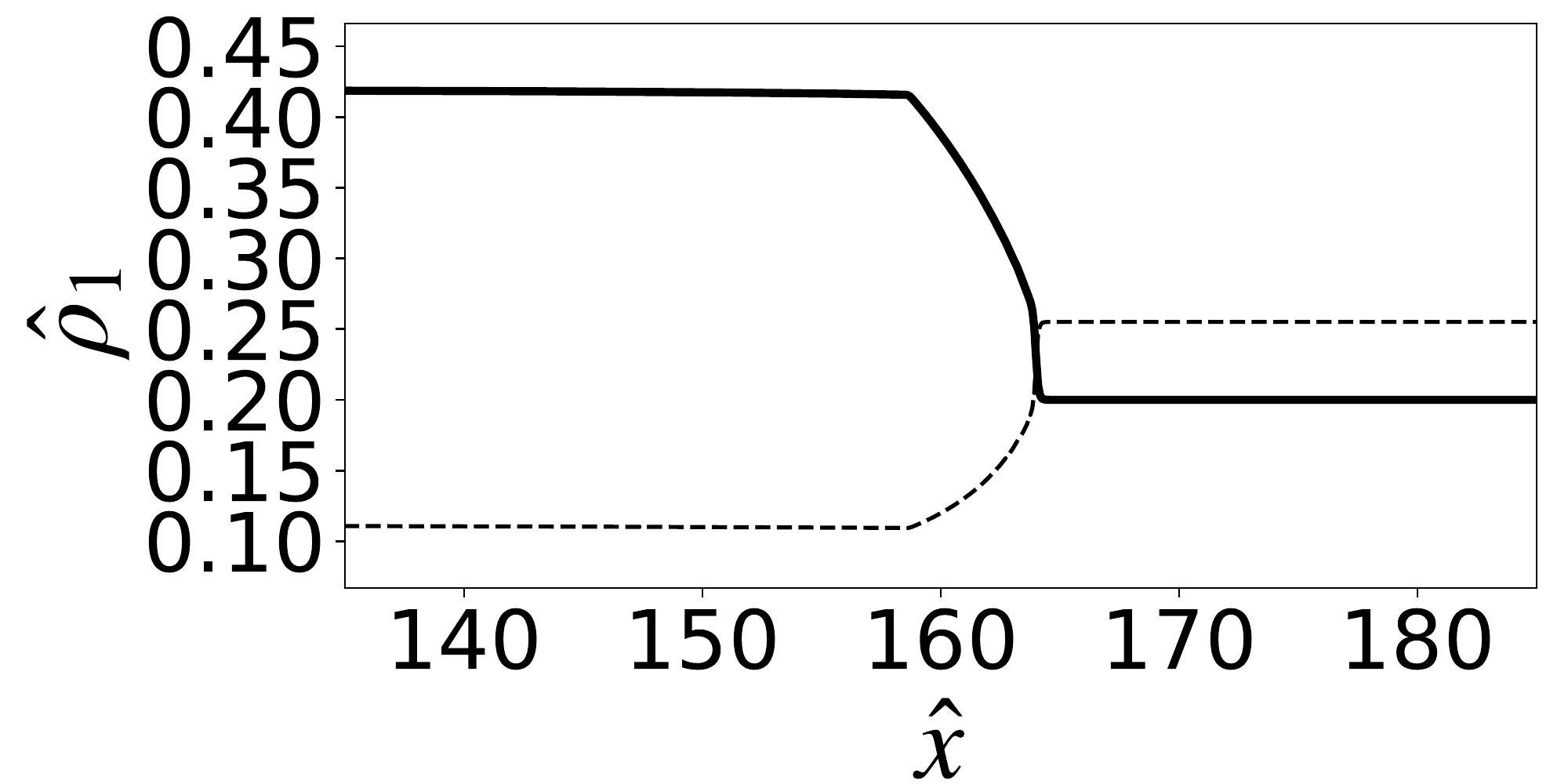} %
		\includegraphics[width=0.45\linewidth]{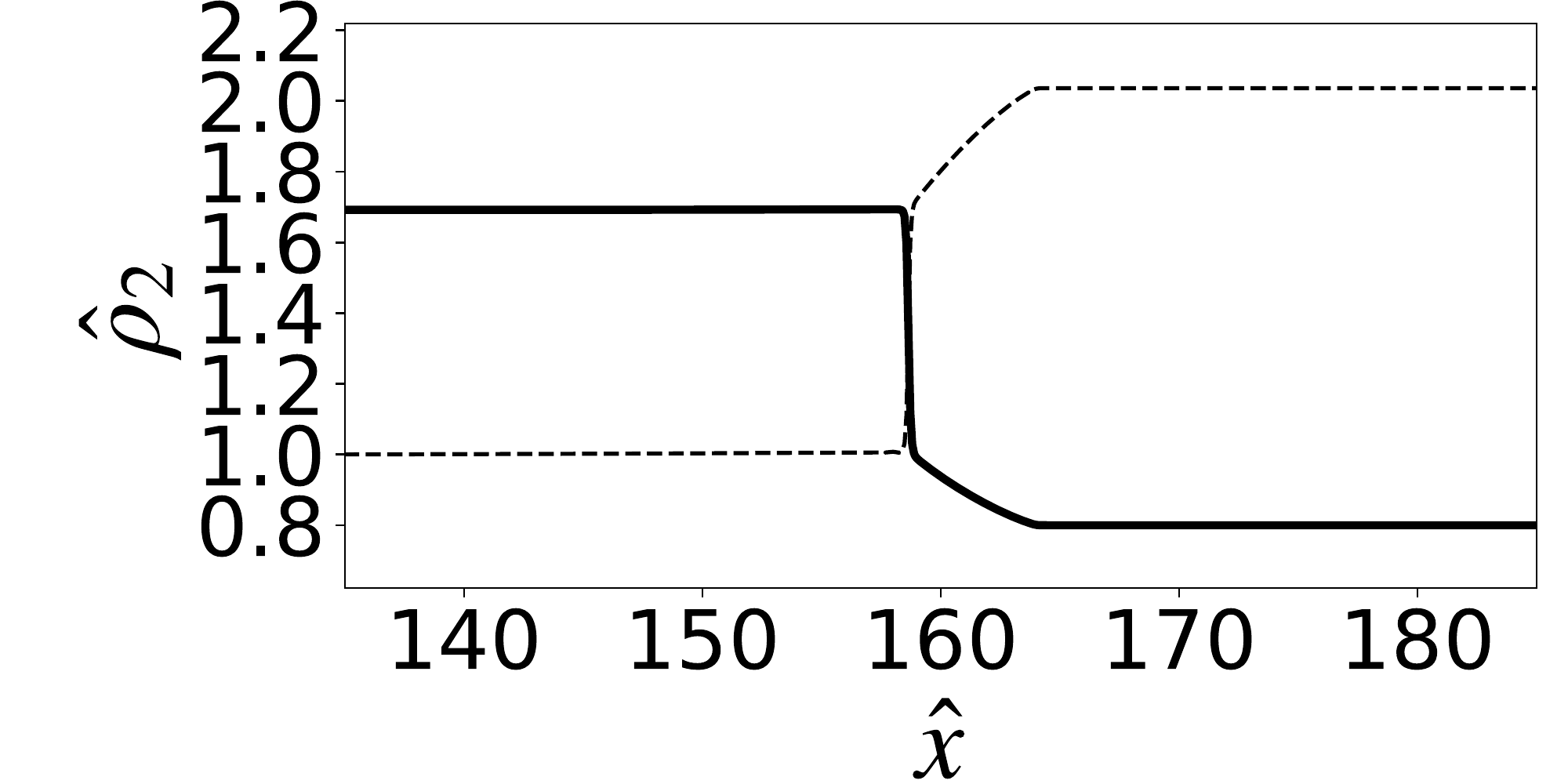}
		\includegraphics[width=0.45\linewidth]{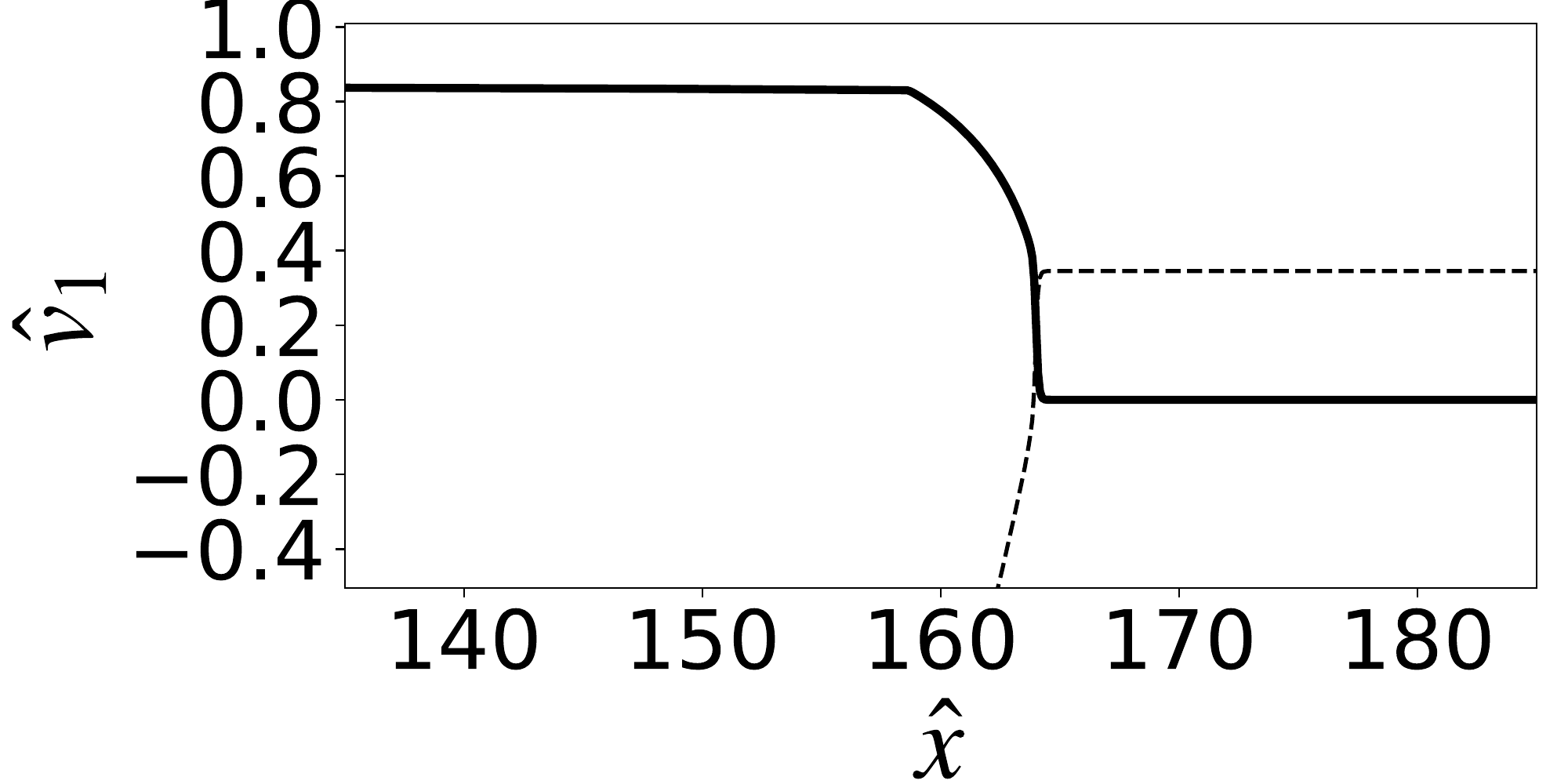} %
		\includegraphics[width=0.45\linewidth]{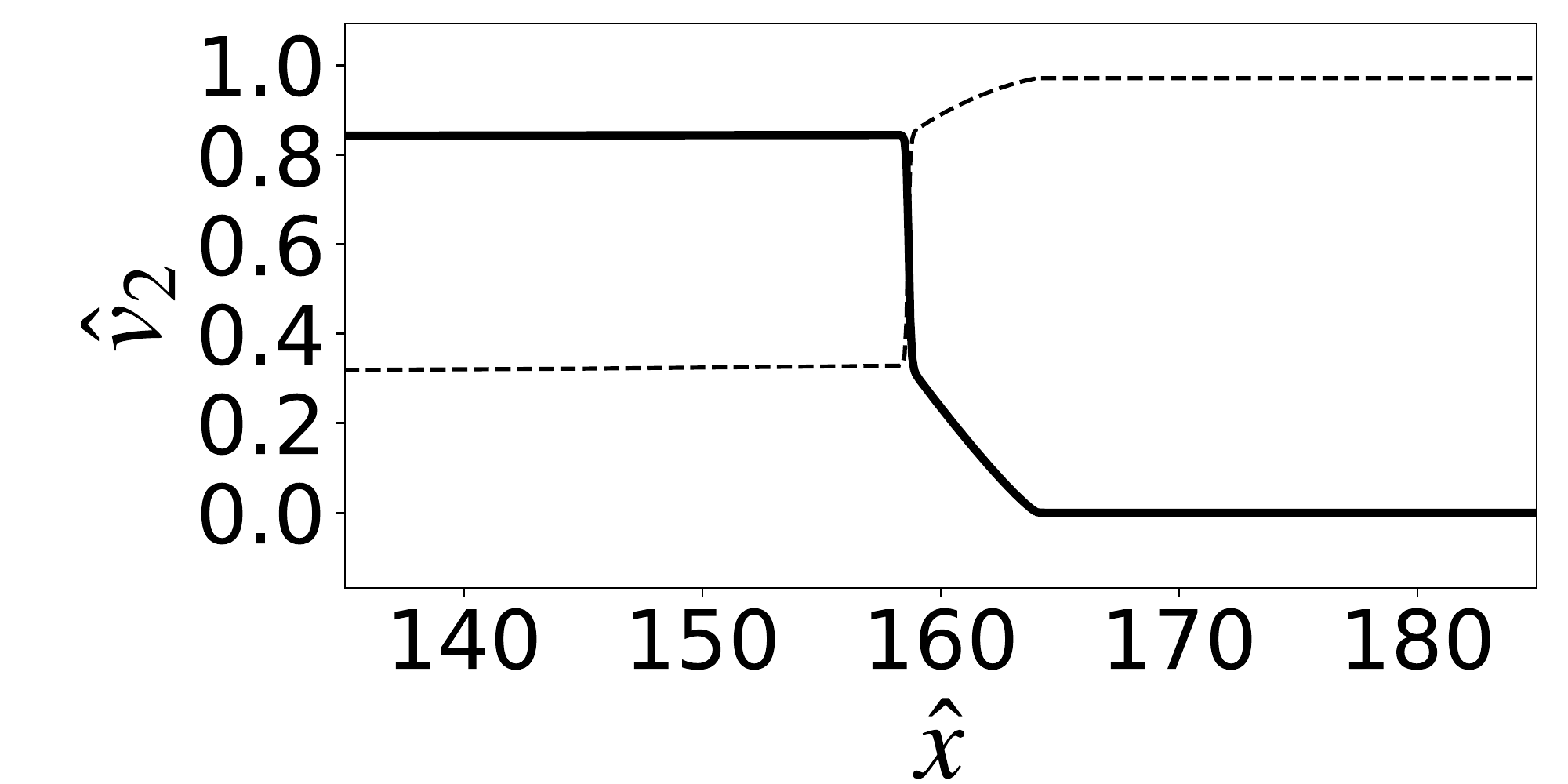}
		\includegraphics[width=0.45\linewidth]{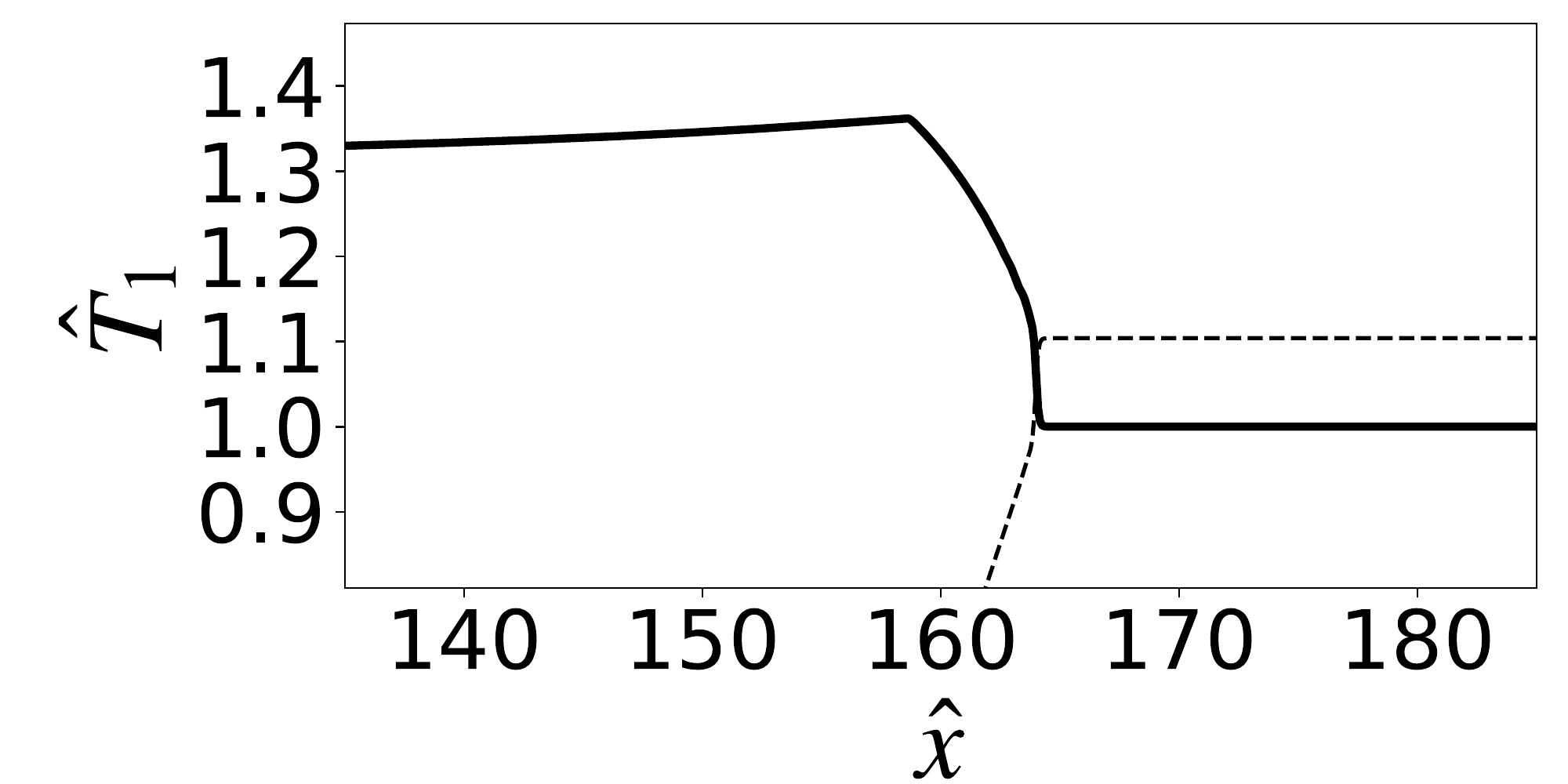} %
		\includegraphics[width=0.45\linewidth]{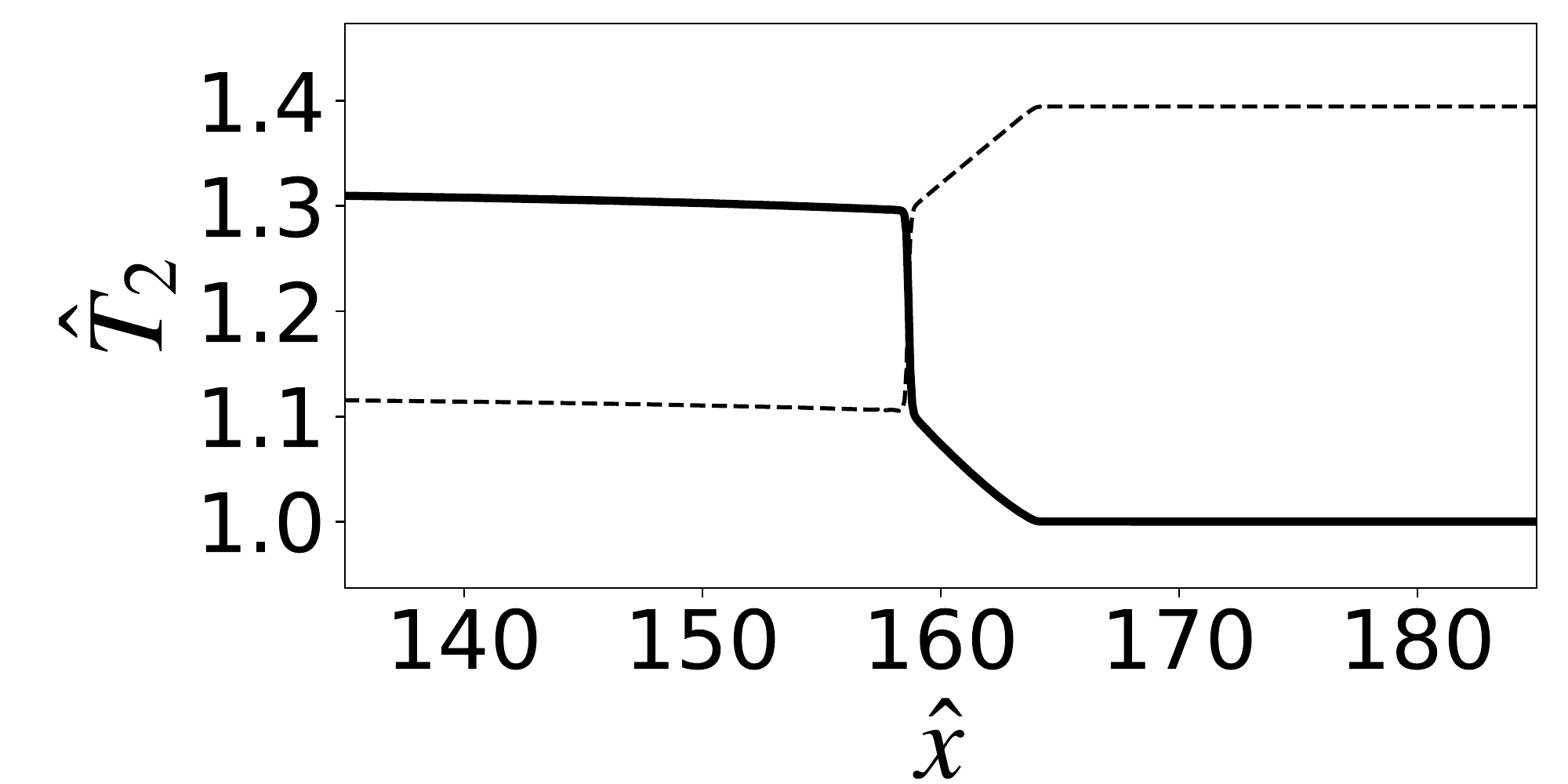}
	\end{center}
	\caption{Case A: Shock structure in a binary Eulerian mixture of polyatomic gases obtained at $\hat{t} = 100$. 
	The parameters are in Region IV and correspond to the mark of No. 3 shown in Figure \ref{fig:subshockEuler_mu045}; 
	$\gamma_1 = 7/5$, $\gamma_2 = 9/7$, $\mu = 0.45$, $c_0 = 0.2$, and $M_0 = 1.6$. 
	The numerical conditions are $\Delta \hat{t} = 0.01$ and $\Delta \hat{x}=0.08$. }
	\label{fig:c02_M0-1_6}
\end{figure}

In this subsection, for discussing the sub-shock formation in Case A, we adopt the following parameters; $\gamma_1 = 7/5$, $\gamma_2 = 9/7$, and $\mu = 0.4$. 
In Figures \ref{fig:c02_M0-1_1} -- \ref{fig:c02_M0-1_6}, we show the profiles of the dimensionless mass density, the dimensionless velocity and the dimensionless temperature for several Mach numbers in the case of $c_0 = 0.2$. 
Figure \ref{fig:c02_M0-1_1} shows the shock structure for $M_0 = 1.1$, which corresponds to the parameter in Region I and, as expected, we confirm that the shock structure is continuous and no sub-shock appears. 

Figure \ref{fig:c02_M0-1_3} shows the shock structure for $M_0 = 1.3$, which corresponds to a parameter in Region II in which the necessary condition of the sub-shock formation is satisfied, and we see that very steep change of the physical quantities of constituent $2$. 
In order to distinguish a real sub-shock from a continuous part with a very steep gradient, we adopt a method proposed in a paper on the shock structure in a binary mixture of monatomic gases~\cite{FMR}, which uses the fact that the jump of a sub-shock must satisfy the compatibility conditions, namely, the Rankine-Hugoniot (RH) conditions for the full system. 
For the full system~\eqref{dimless:finale1}, we have the following expression of the state just after the sub-shock: 
\begin{equation}\label{eq:subshockRH}
	\begin{split}
		&\rho_1^{*} = \frac{(\gamma_1 + 1) M_{1}^2}{2 + (\gamma_1 - 1) M_{1}^2}\hat{\rho}_1, \\
		&\hat{v}_1^{*} = \hat{v}_1 + \frac{2 \left( M_{1}^2 - 1 \right)}{(\gamma_1 + 1) M_{1}} \frac{a_1}{a_0}, \\
		&\hat{T}_1^{*} = \frac{\left\{2 + (\gamma_1 - 1) M_{1}^2\right\}\left\{1 + \gamma_1 (2 M_{1}^2 - 1)\right\}}{(\gamma_1 + 1)^2 M_{1}^2} \hat{T}_1, \\
		&\rho_2^{*} = \frac{(\gamma_2 + 1) M_{2}^2}{2 + (\gamma_2 - 1) M_{2}^2}\hat{\rho}_2, \\
		&\hat{v}_2^{*} = \hat{v}_2 + \frac{2 \left( M_{2}^2 - 1 \right)}{(\gamma_2 + 1) M_{2}} \frac{a_2}{a_0}, \\
		&\hat{T}_2^{*} = \frac{\left\{2 + (\gamma_2 - 1) M_{2}^2\right\}\left\{1 + \gamma_2 (2 M_{2}^2 - 1)\right\}}{(\gamma_2 + 1)^2 M_{2}^2} \hat{T}_2, \\
	\end{split}
\end{equation}
where the quantities with superscript $*$ represent the quantities in the state just after a sub-shock predicted by the RH condition of the full system. 
Here $M_{1}$ and $M_{2}$ are the Mach numbers for the constituent 1 and 2: 
\begin{equation*}
	M_{1} \equiv \frac{s - v_1}{a_1} = (M_{0} -\hat{v}_1)\frac{a_0}{a_1}, \quad 
	M_{2} \equiv \frac{s - v_2}{a_2}(M_{0} -\hat{v}_2)\frac{a_0}{a_2}, 
\end{equation*}
where $a_1$ and $a_2$ are the sound velocity for the constituent $1$ and $2$, respectively and the following relations hold: 
\begin{equation*}
	\begin{split}
		\frac{a_0}{a_1} = \sqrt{\frac{\gamma_0}{\gamma_1 \hat{T}_1}\left\{ c + (1-c) \mu \right\}}, \quad
		\frac{a_0}{a_2} =\sqrt{\frac{\gamma_0}{\gamma_2 \hat{T}_2}\left\{ \frac{c}{\mu} + (1-c) \right\}}. 
	\end{split}
\end{equation*}

In all figures for the shock structure, we depict the profile of the physical quantities with the curve representing potential state just after the sub-shock predicted by \eqref{eq:subshockRH} from any point of the shock structure.  
Although a sub-shock is captured with a very steep but continuous change in numerical calculations due to the numerical viscosity, it can be expected that the part of a sub-shock is approximately described as a part joining the curve of the potential sub-shock. 
Therefore the curve representing potential sub-shock can be regarded as a useful indicator of the behavior of the shock-structure solution. 

By drawing two curves together, from Figure \ref{fig:c02_M0-1_3}, we see that a sub-shock appears in the profile of the constituent 2 for $M_0 = 1.3$. 
As discussed in the previous section, in the present case in Region II, the shock velocity for the Mach number $M_0 = 1.3$ is slower than the maximum characteristic velocity evaluated in the unperturbed state. 
Therefore, we conclude that there exists a clear counter-example of the conjecture on the sub-shock formation also in the case of a mixture of rarefied polyatomic gases. 

We show the shock structure for $M_0 = 1.6$ in Figure \ref{fig:c02_M0-1_6} and understand that sub-shocks are observed in both the profiles for constituents $1$ and $2$. 
It is concluded that multiple sub-shocks are predicted with the parameters in Region IV. 

\begin{figure}
	\begin{center}
		\includegraphics[width=0.45\linewidth]{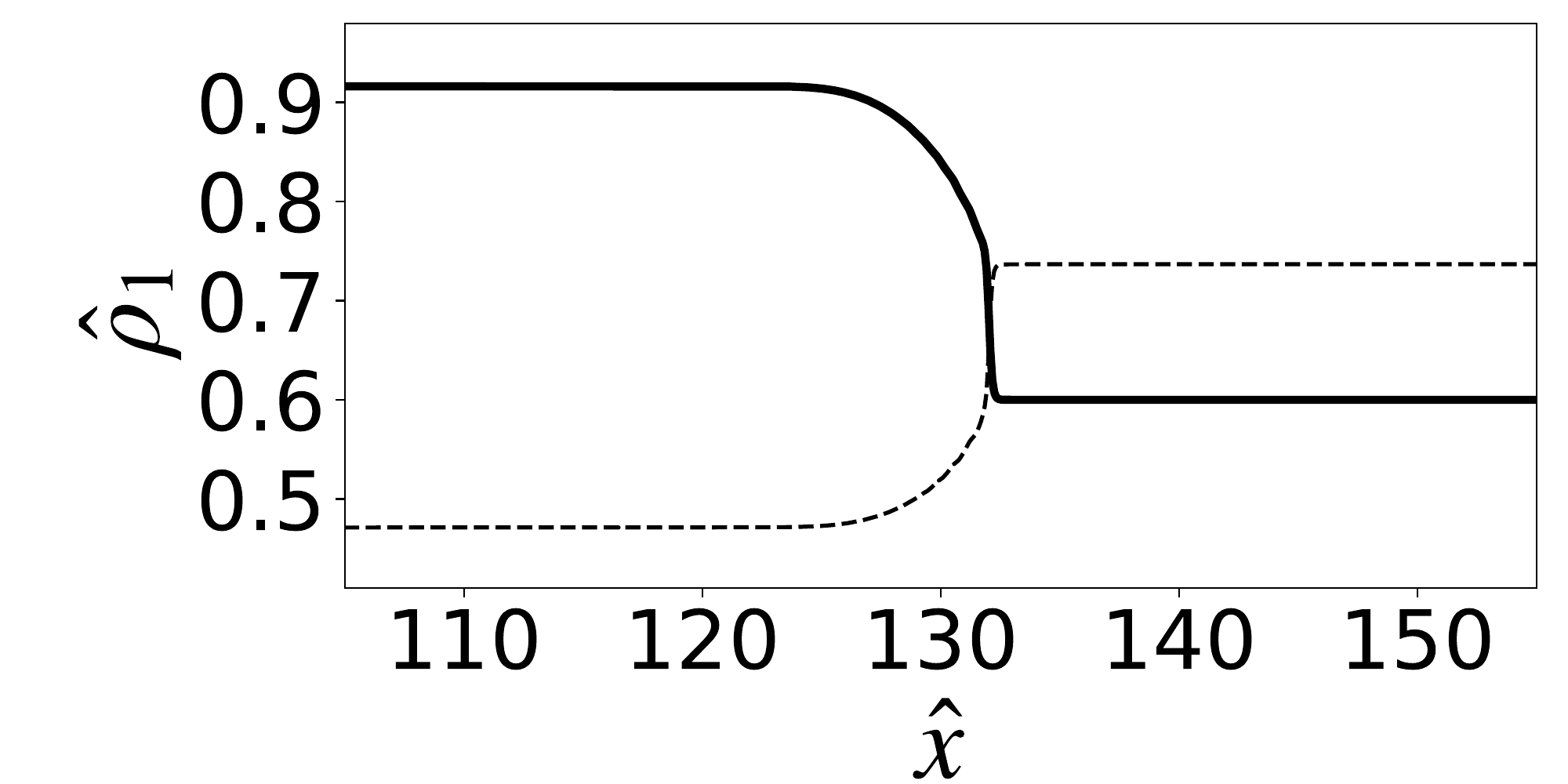} %
		\includegraphics[width=0.45\linewidth]{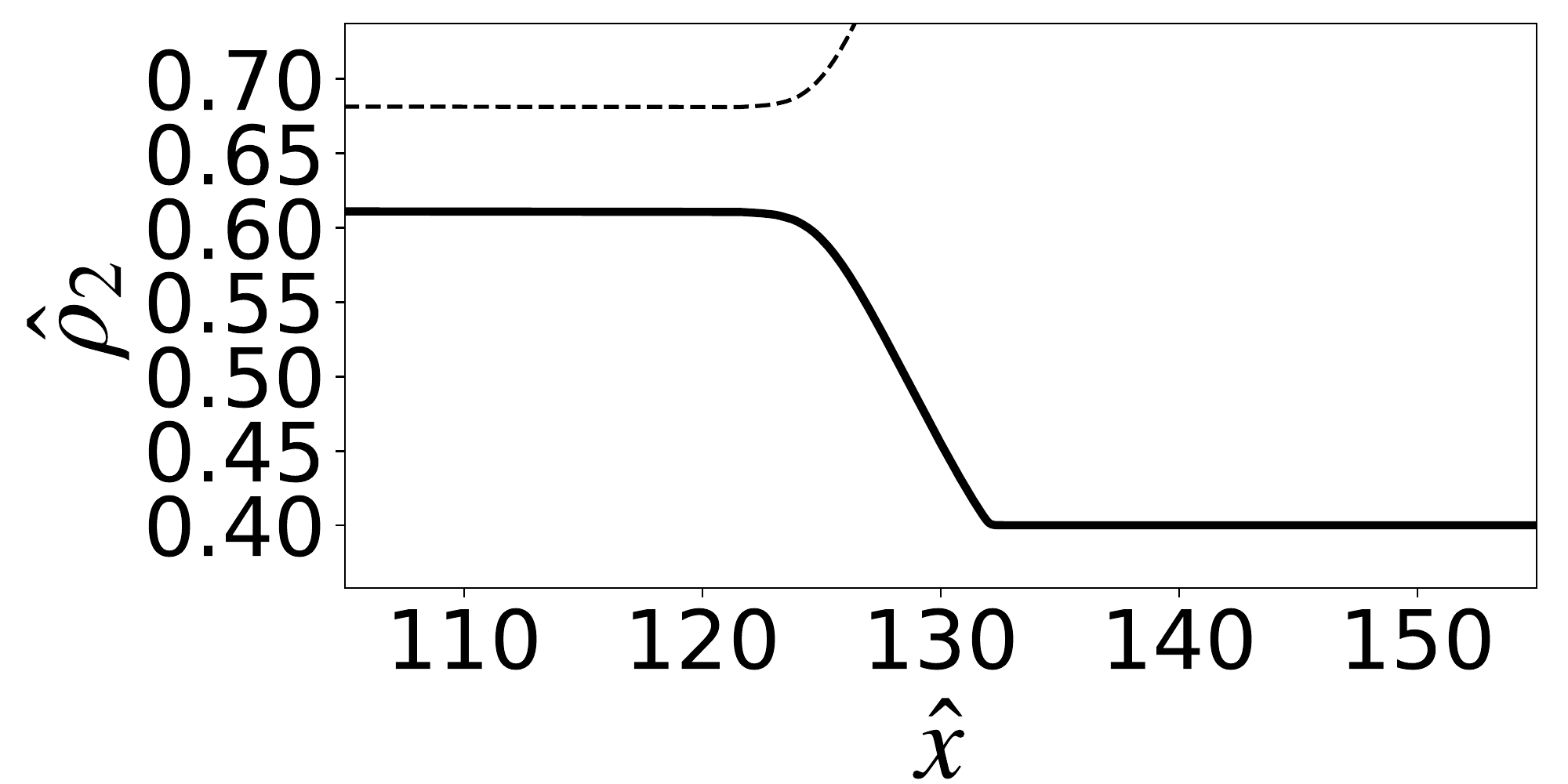}
		\includegraphics[width=0.45\linewidth]{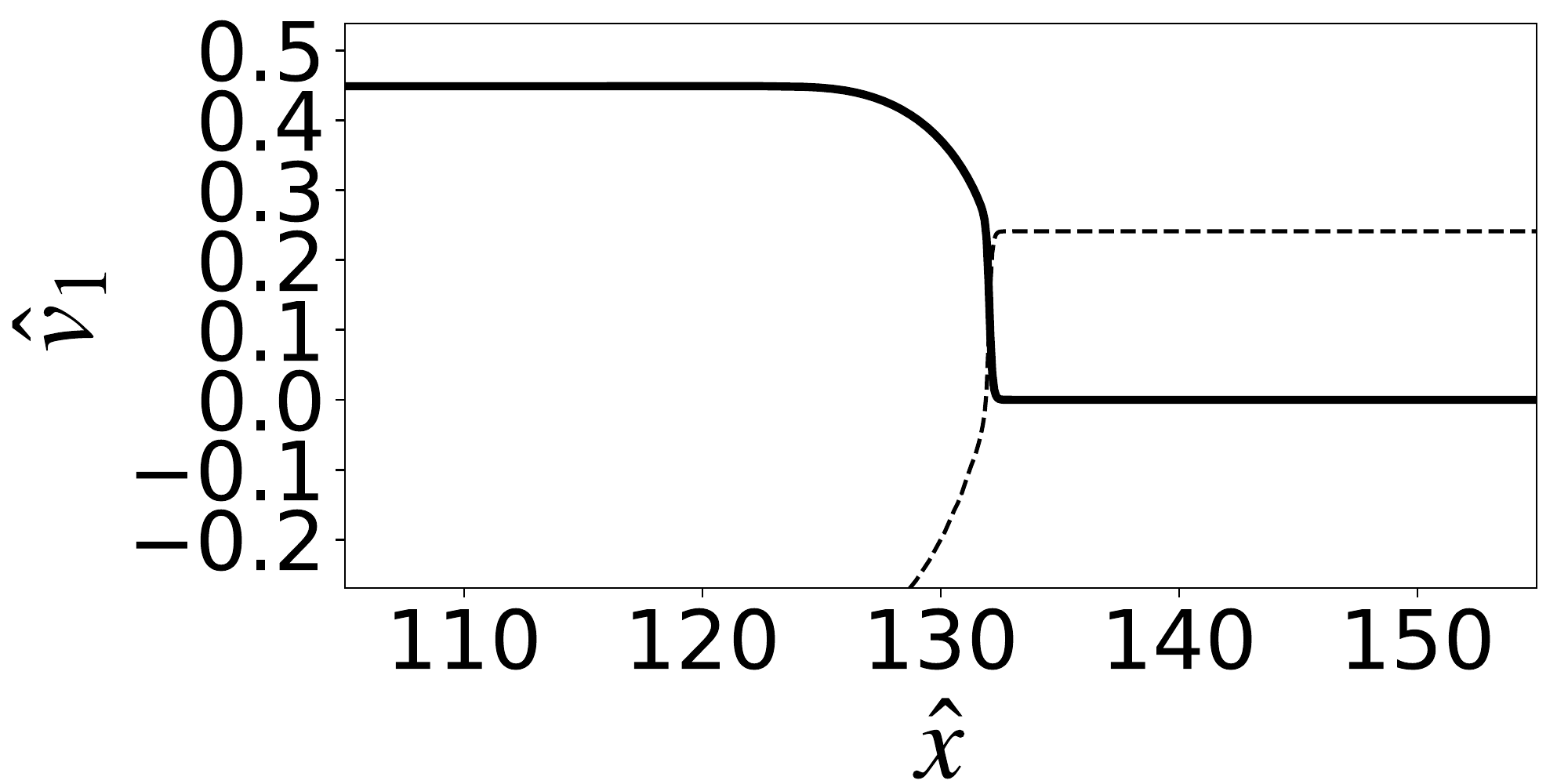} %
		\includegraphics[width=0.45\linewidth]{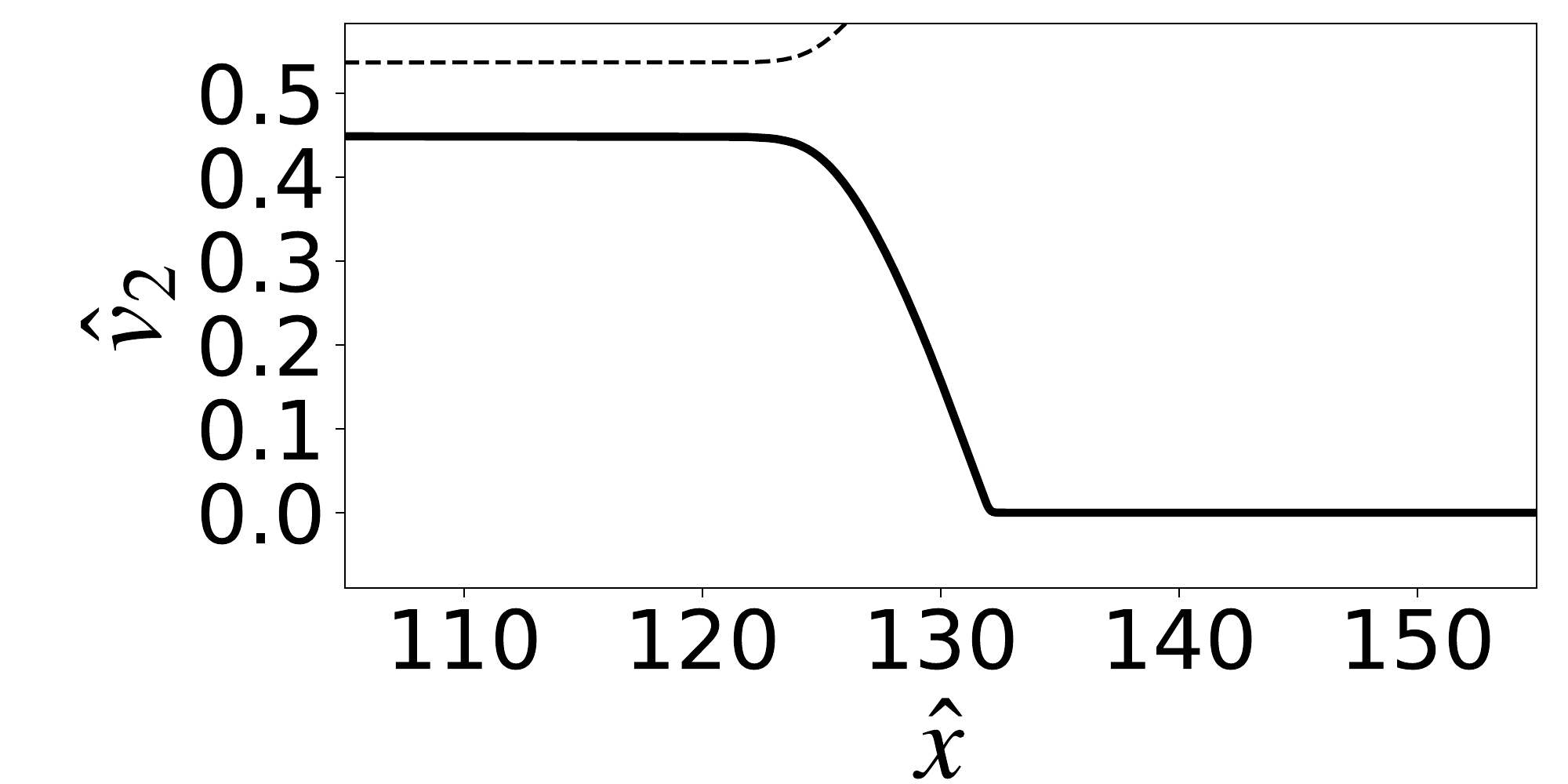}
		\includegraphics[width=0.45\linewidth]{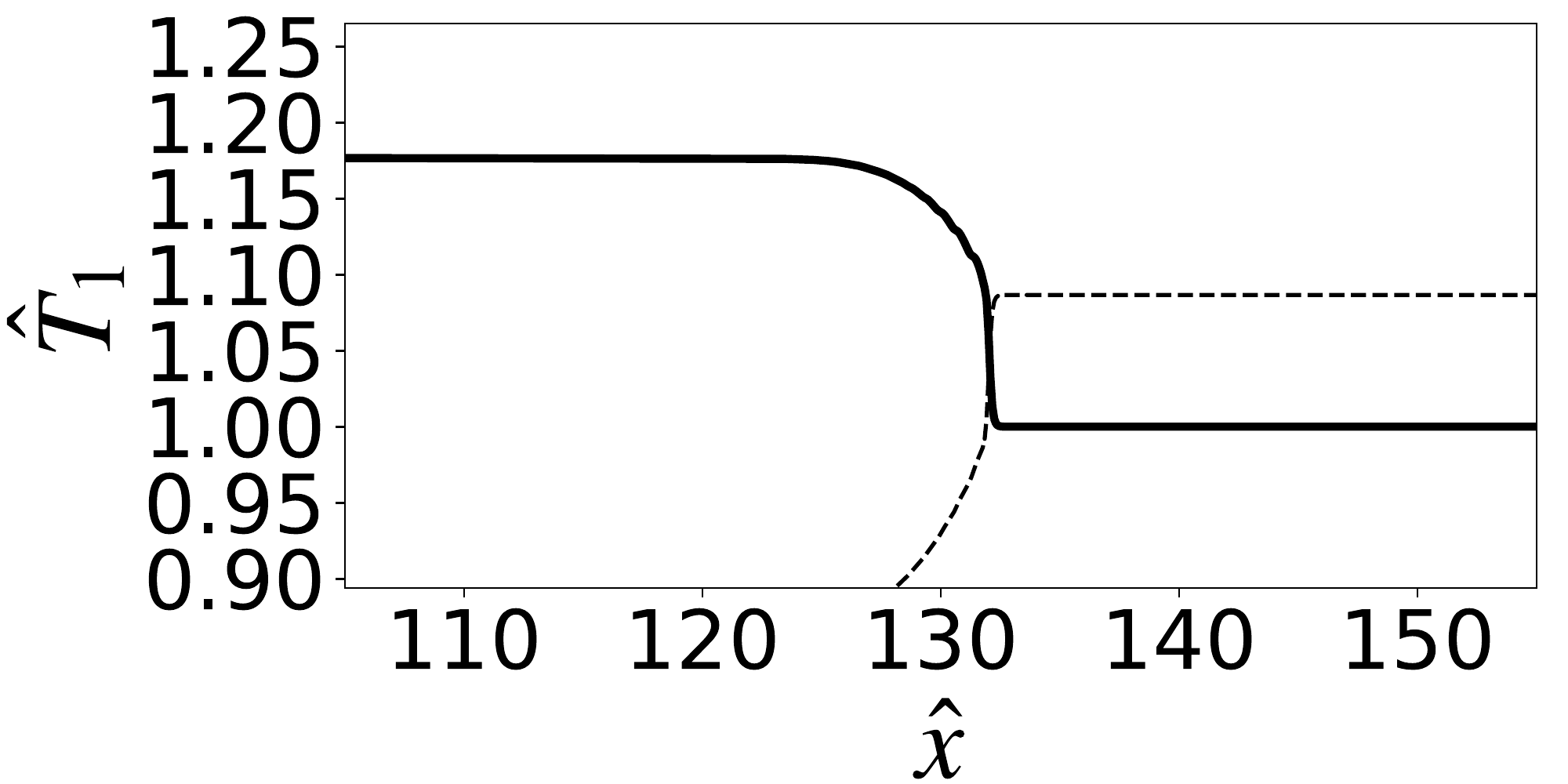} %
		\includegraphics[width=0.45\linewidth]{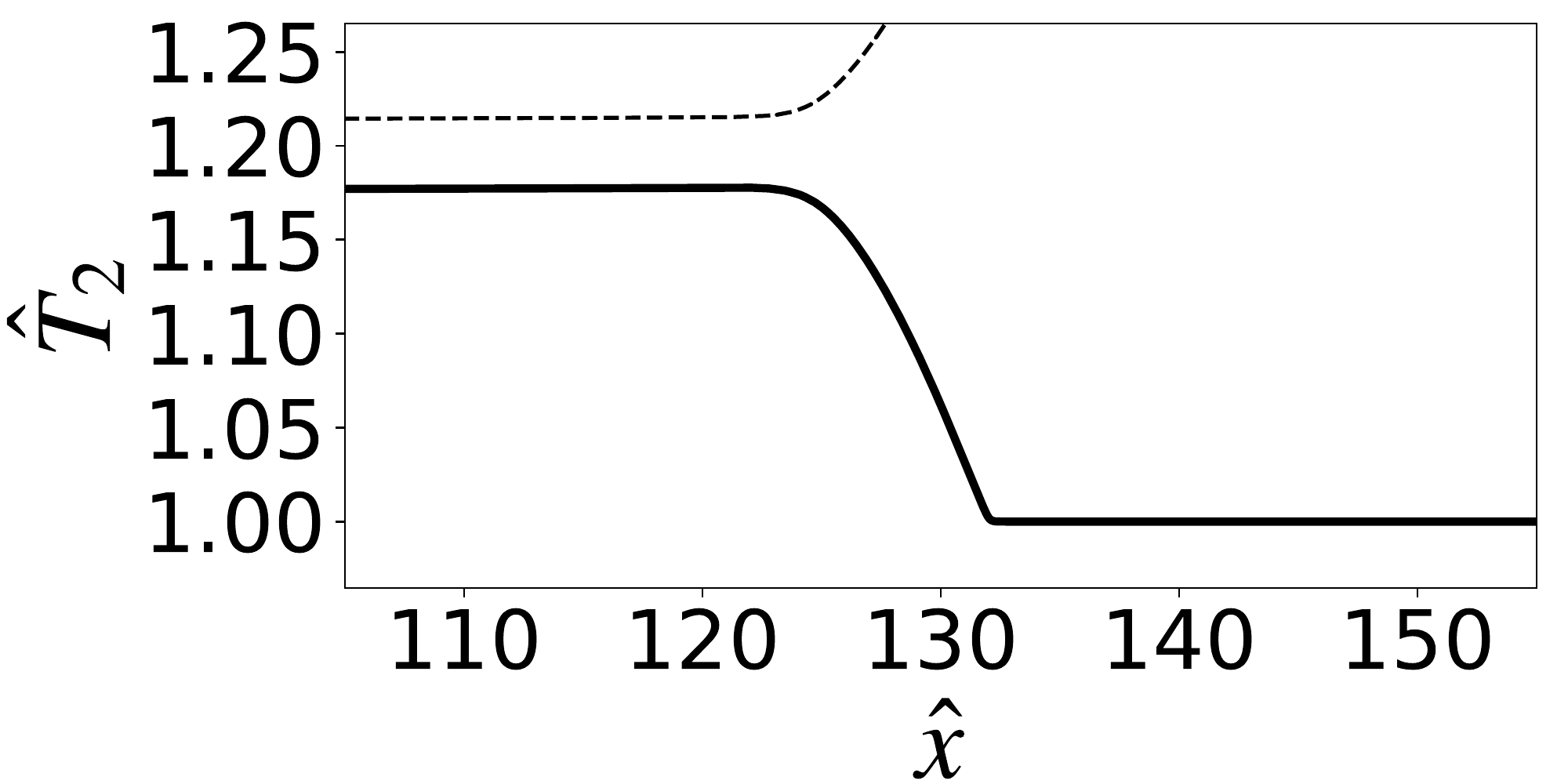}
	\end{center}
	\caption{Case A: Shock structure in a binary Eulerian mixture of polyatomic gases obtained at $\hat{t} = 100$. 
	The parameters are in Region III and correspond to the mark of No. 4 shown in Figure \ref{fig:subshockEuler_mu045}; 
	$\gamma_1 = 7/5$, $\gamma_2 = 9/7$, $\mu = 0.45$, $c_0 = 0.6$, and $M_0 = 1.3$. 
	The numerical conditions are $\hat{t} = 0.01$ and $\hat{x}=0.08$. }
	\label{fig:c06_M0-1_3}
\end{figure}

\begin{figure}
	\begin{center}
		\includegraphics[width=0.45\linewidth]{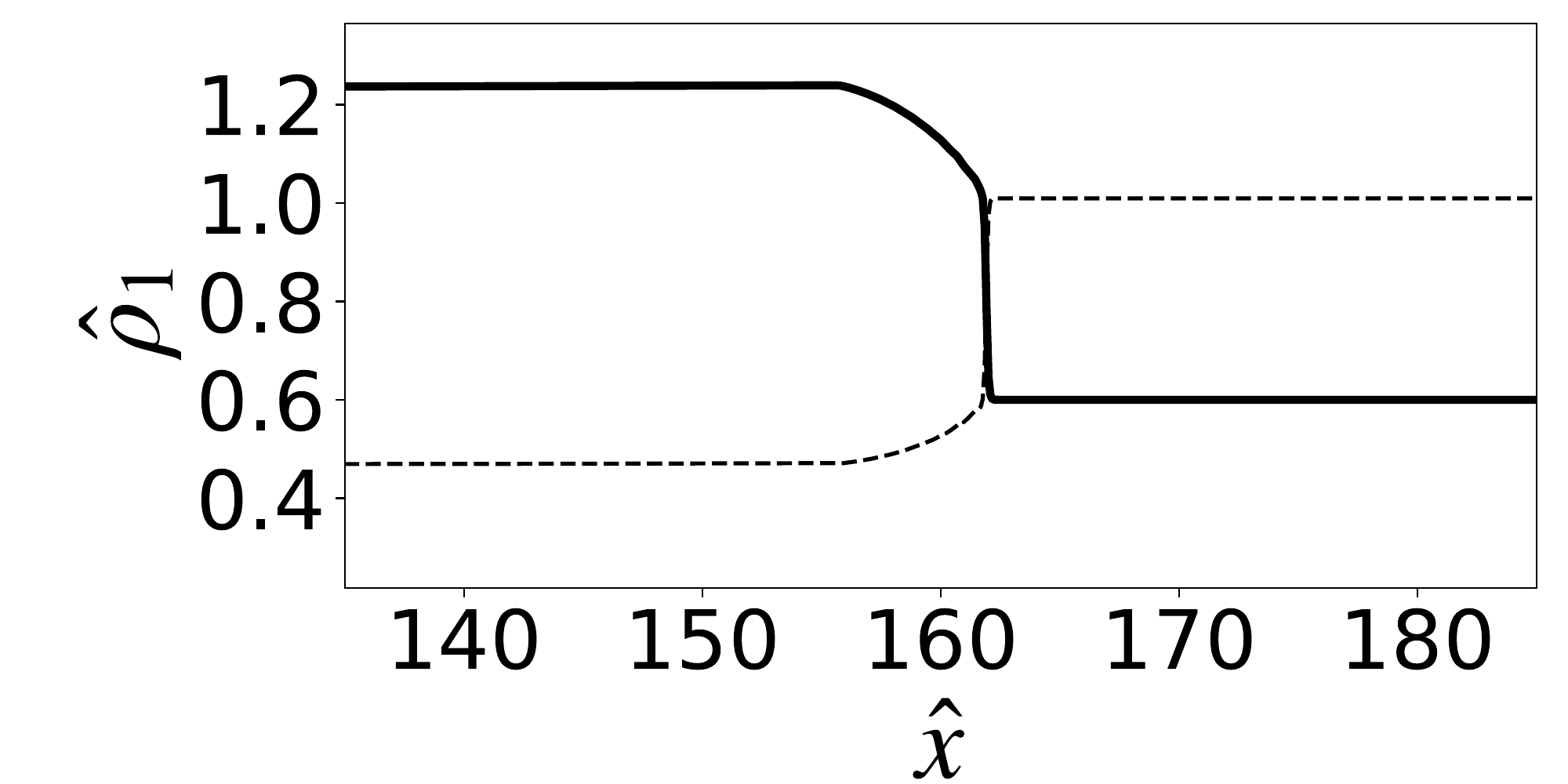} %
		\includegraphics[width=0.45\linewidth]{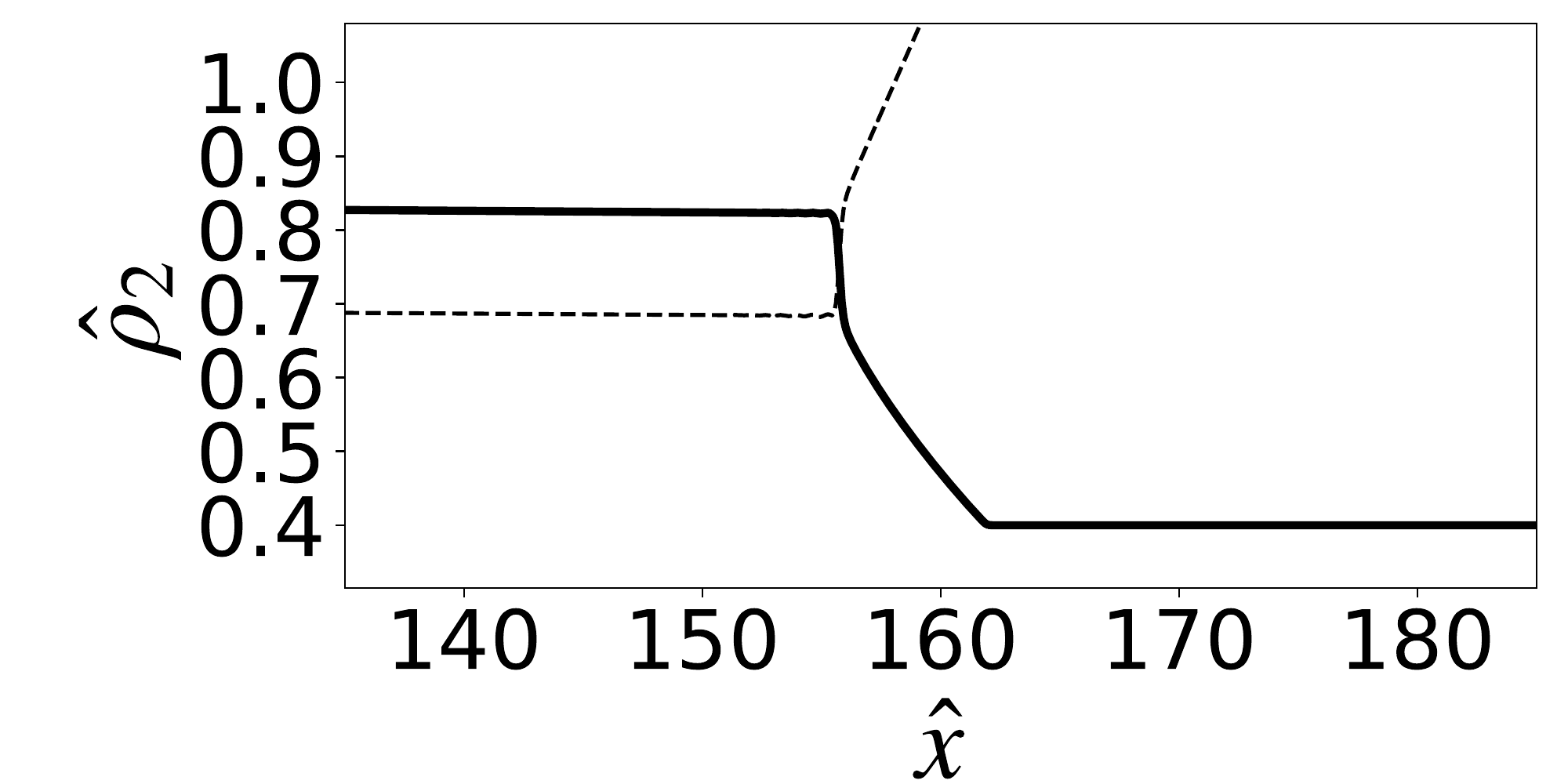}
		\includegraphics[width=0.45\linewidth]{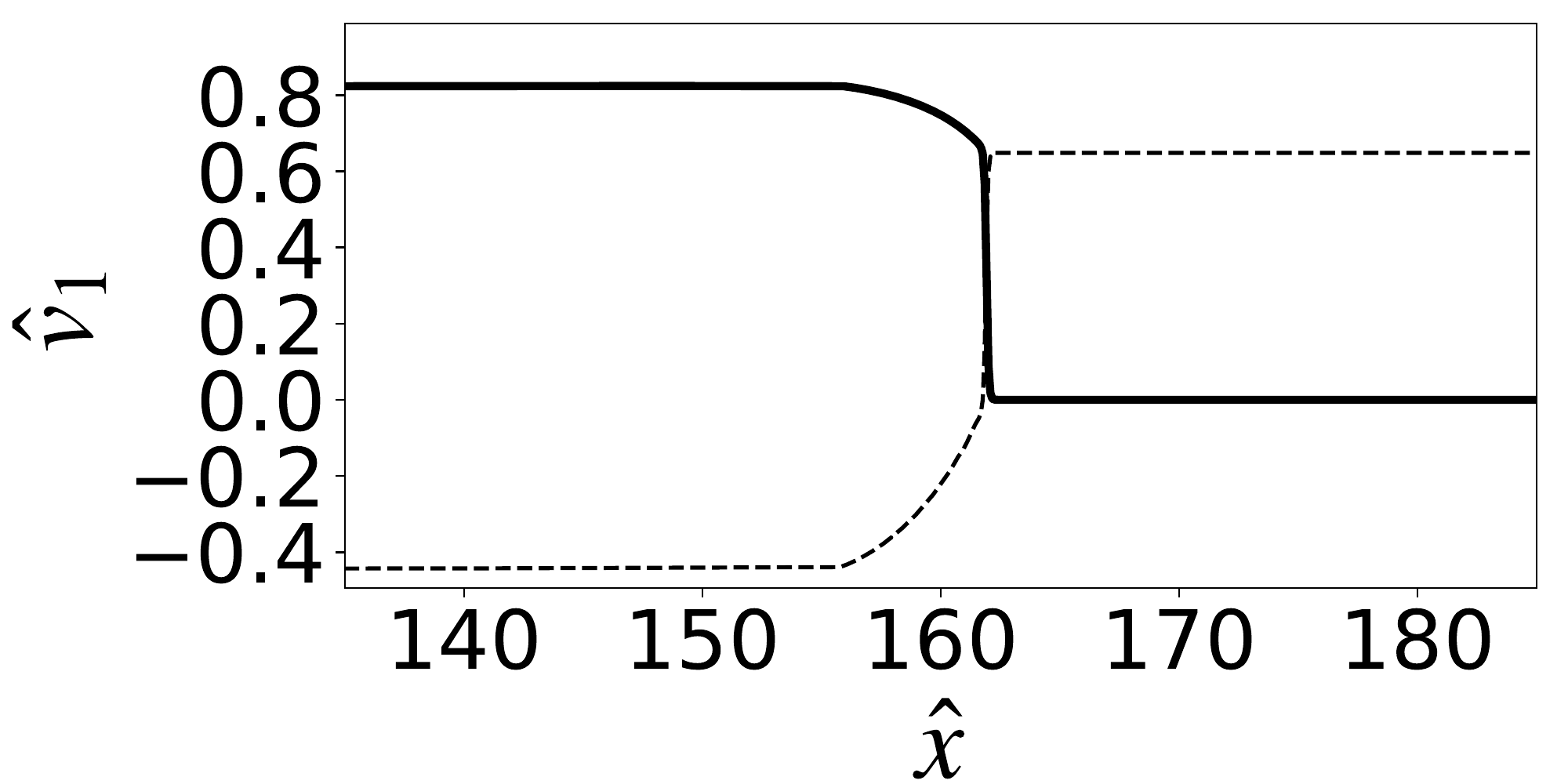} %
		\includegraphics[width=0.45\linewidth]{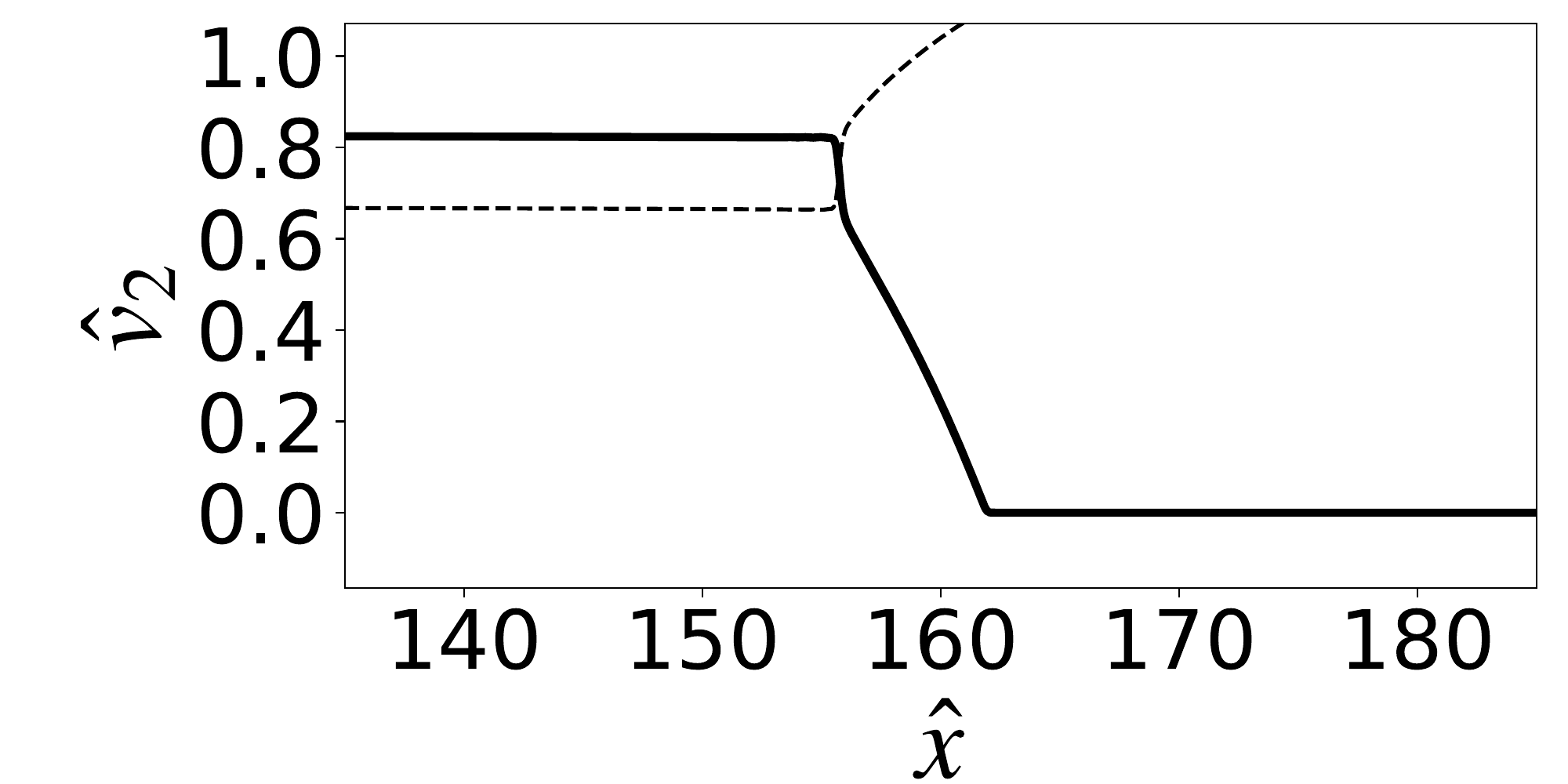}
		\includegraphics[width=0.45\linewidth]{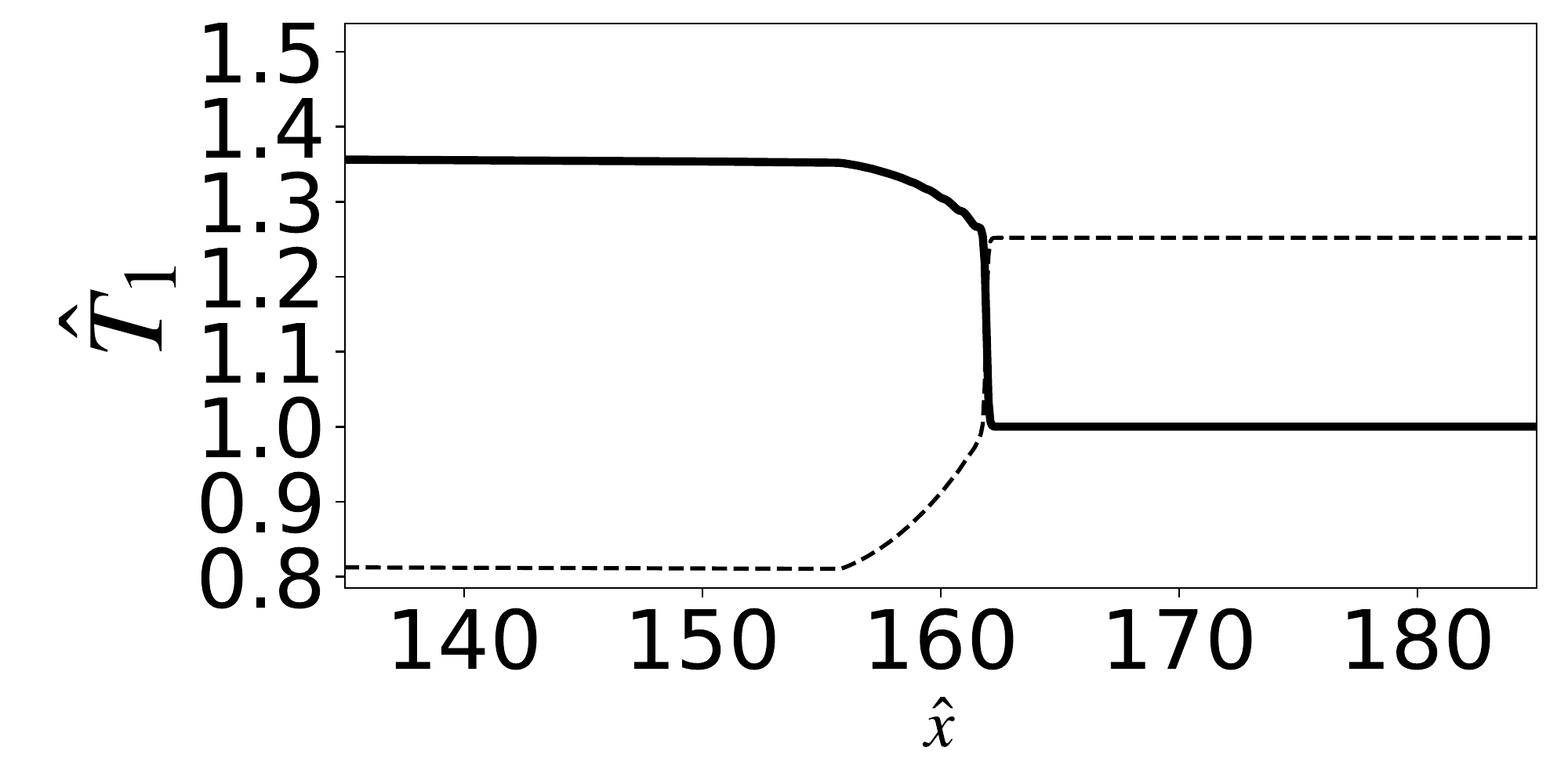} %
		\includegraphics[width=0.45\linewidth]{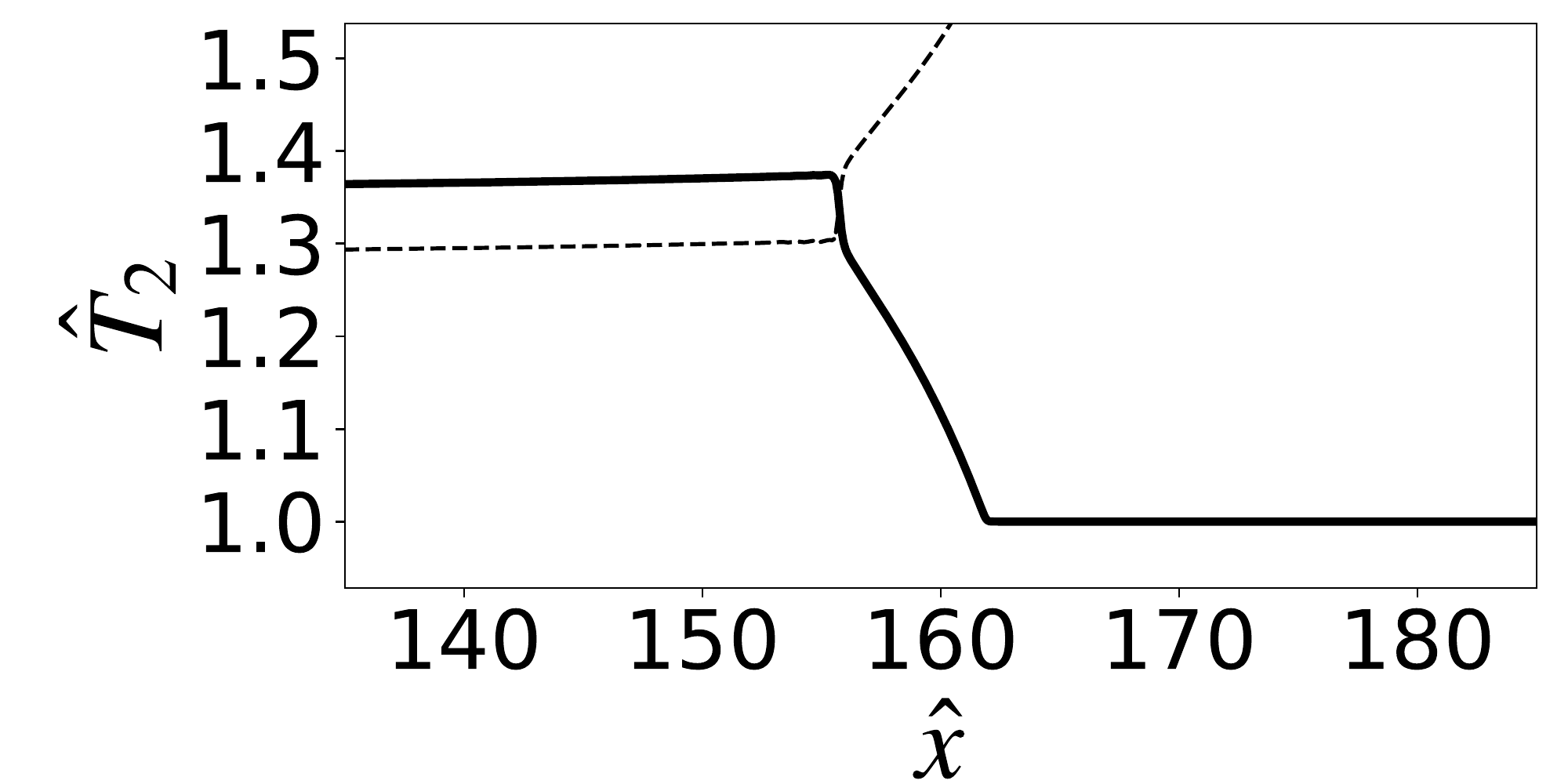}
	\end{center}
	\caption{Case A: Shock structure in a binary Eulerian mixture of polyatomic gases obtained at $\hat{t} = 100$.   
	The parameters are in Region IV and correspond to the mark of No. 5 shown in Figure \ref{fig:subshockEuler_mu045}; $\gamma_1 = 7/5$, $\gamma_2 = 9/7$, $\mu = 0.45$, $c_0 = 0.6$, and $M_0 = 1.6$. 
	The numerical conditions are $\Delta \hat{t} = 0.01$ and $\Delta \hat{x}=0.08$. }
	\label{fig:c06_M0-1_6}
\end{figure}

The shock structure for several Mach numbers with $c_0 = 0.6$ is shown in figures \ref{fig:c06_M0-1_3} and \ref{fig:c06_M0-1_6}.  
We understand that a sub-shock is formed for the constituent $1$ for $M_0 = 1.3$ in Region III from Figure \ref{fig:c06_M0-1_3}. 
From Figure \ref{fig:c06_M0-1_6} showing the case for $M_0 = 1.6$ in region IV, we see that multiple sub-shocks for both species $1$ and $2$ appear.  

To summarize the results obtained in the case of $\mu = 0.45$, $\gamma_1 = 7/5$, $\gamma_2 = 9/7$, from the viewpoint of the sub-shock formation, the predictions are similar to the ones observed in the context of the shock structure in a mixture of rarefied monatomic gases. 
Counter-example of the conjecture on the sub-shock formation can be constructed with the parameters in Region II, and multiple sub-shocks are also observed with the parameters in Region IV. 

\subsection{Shock structure in Case B$_1$}

\begin{figure}
	\begin{center}
		\includegraphics[width=0.49\linewidth]{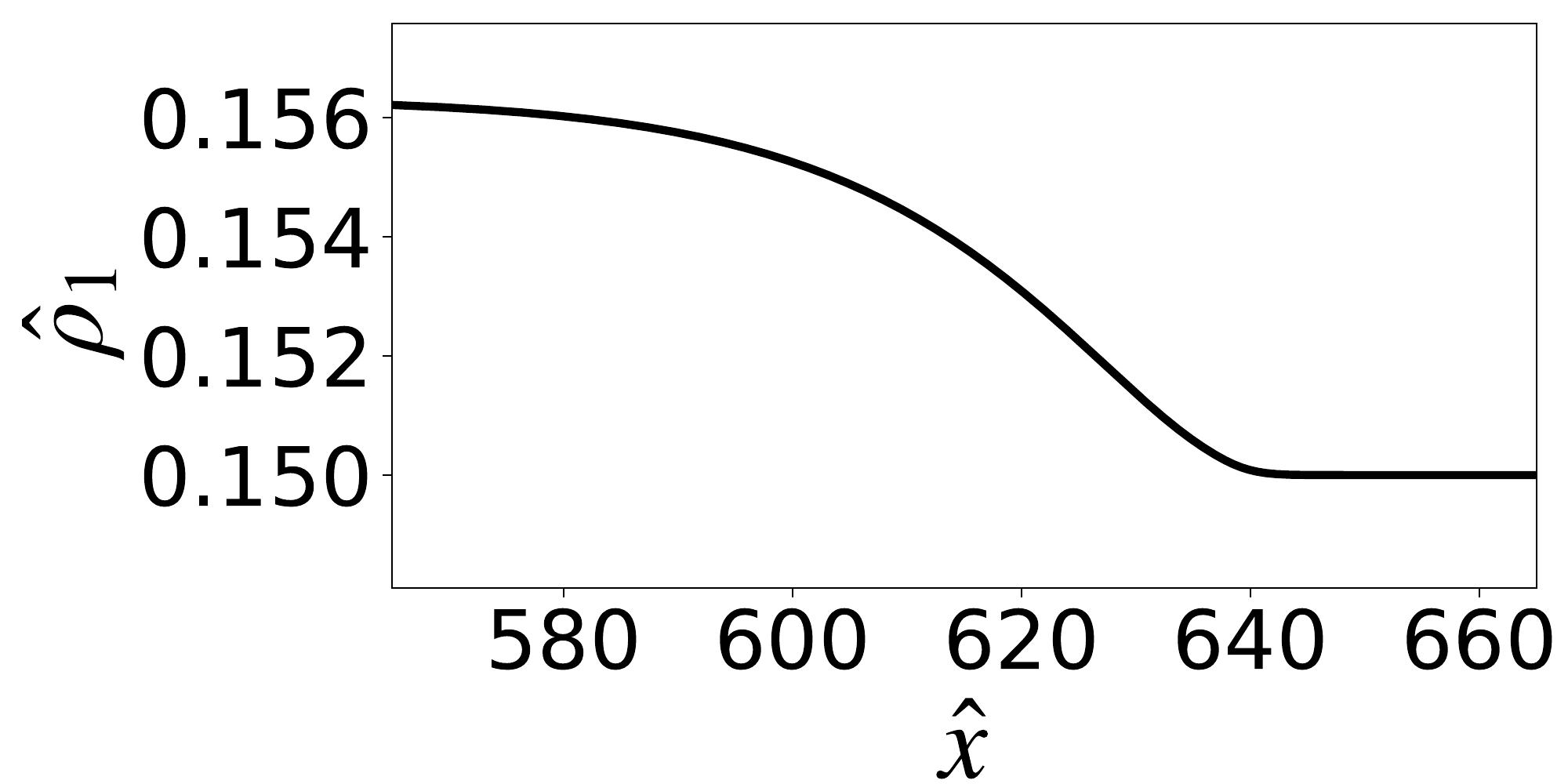} %
		\includegraphics[width=0.49\linewidth]{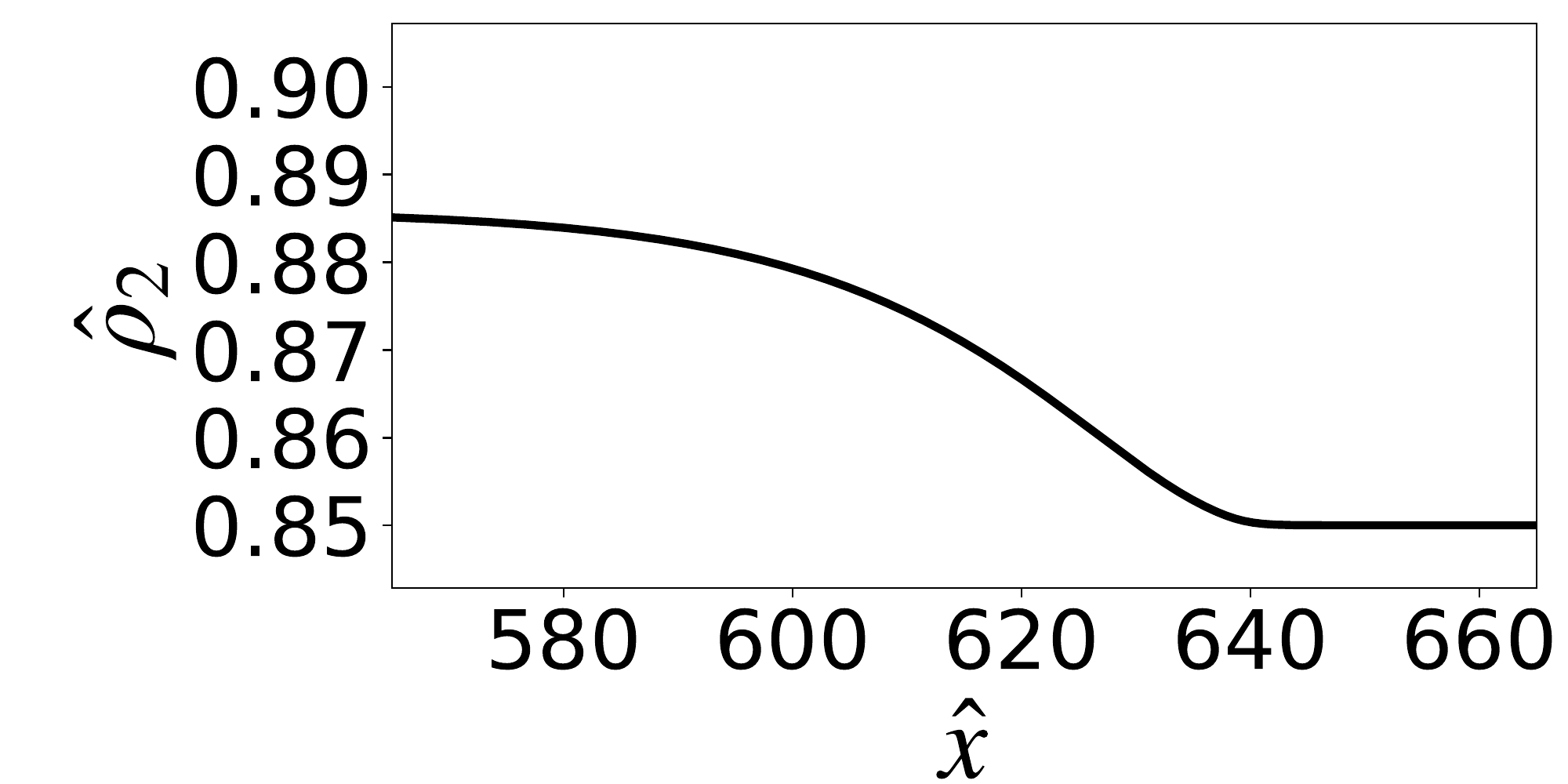}
		\includegraphics[width=0.49\linewidth]{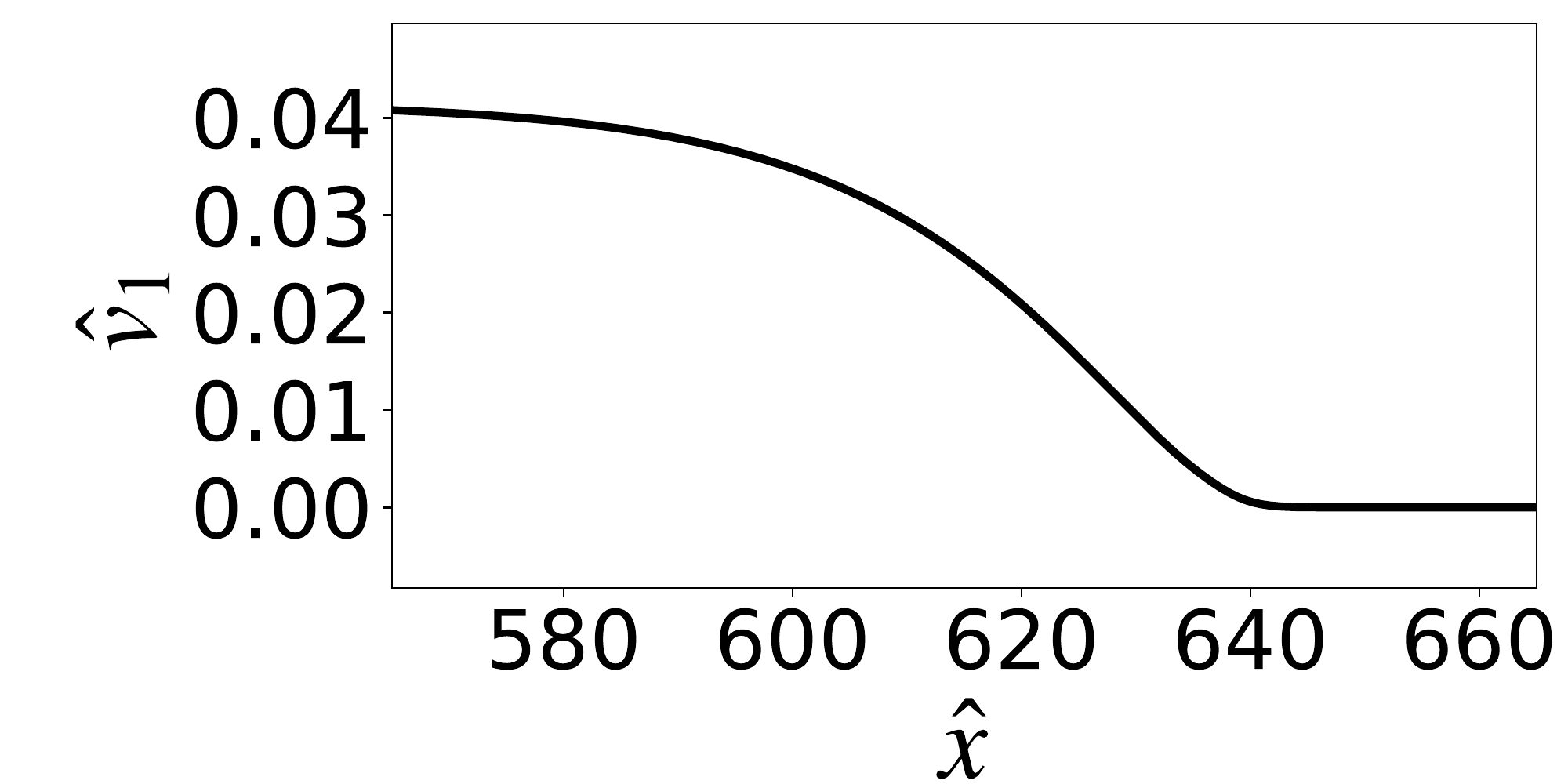} %
		\includegraphics[width=0.49\linewidth]{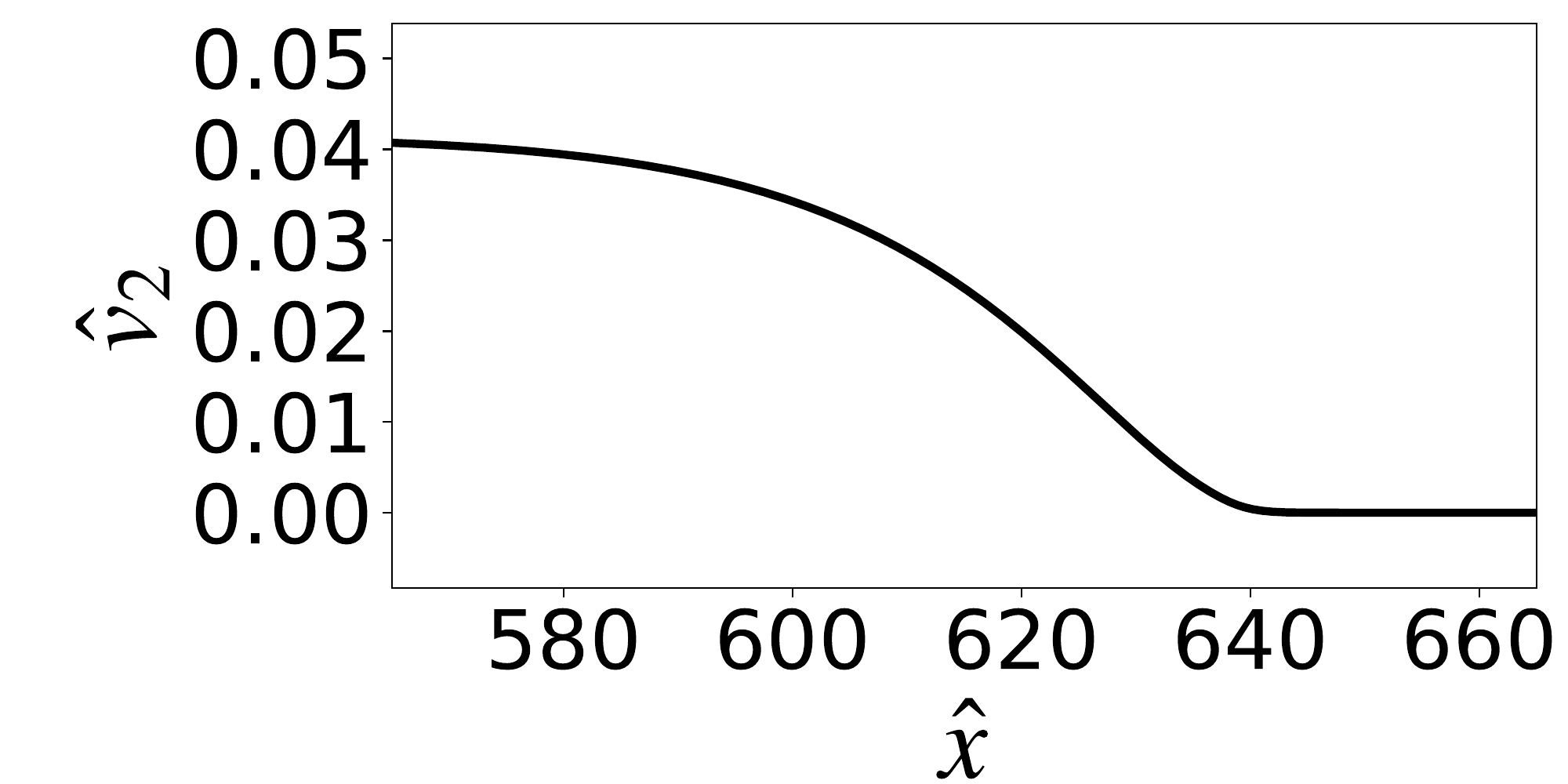}
		\includegraphics[width=0.49\linewidth]{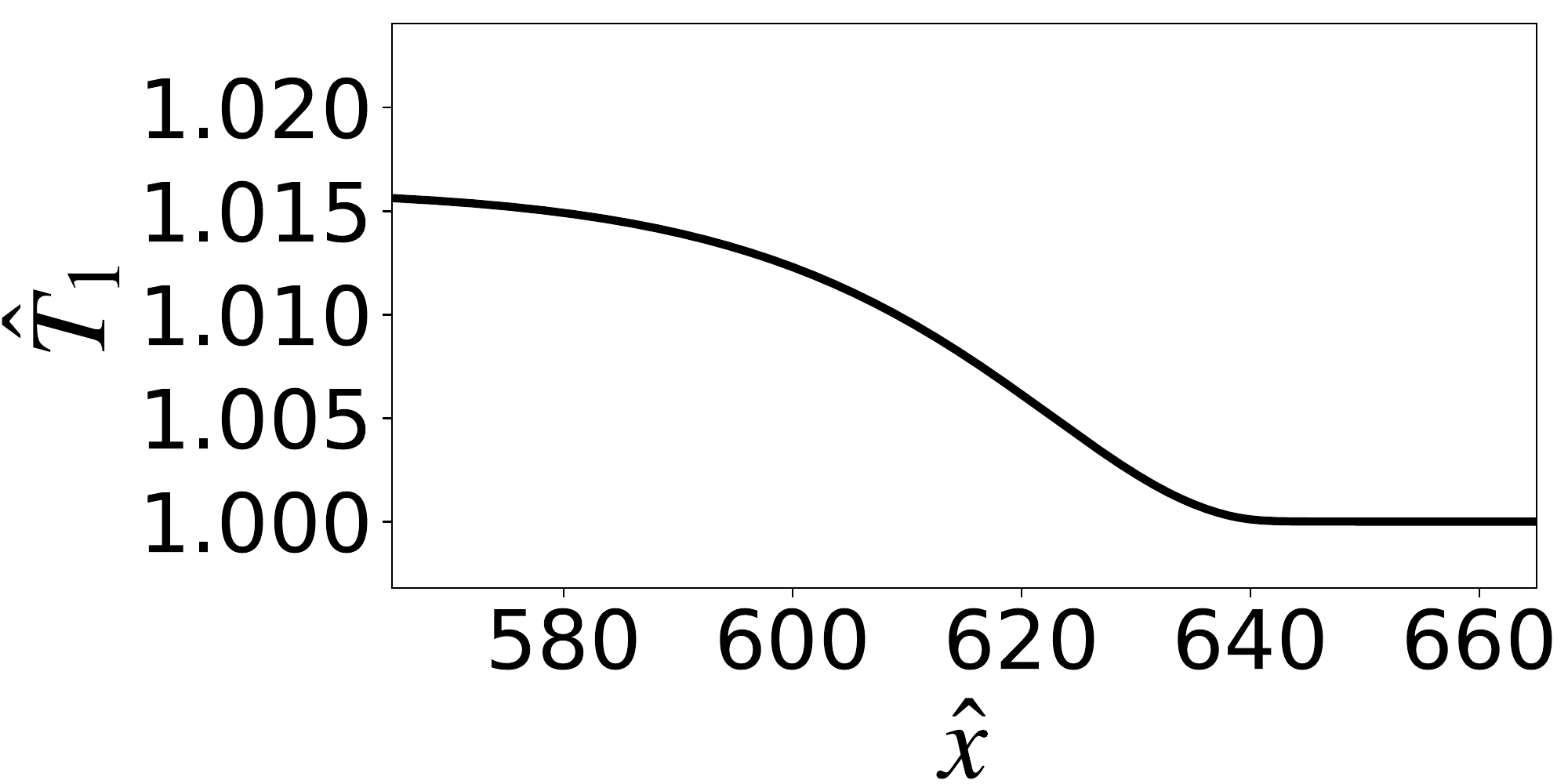} %
		\includegraphics[width=0.49\linewidth]{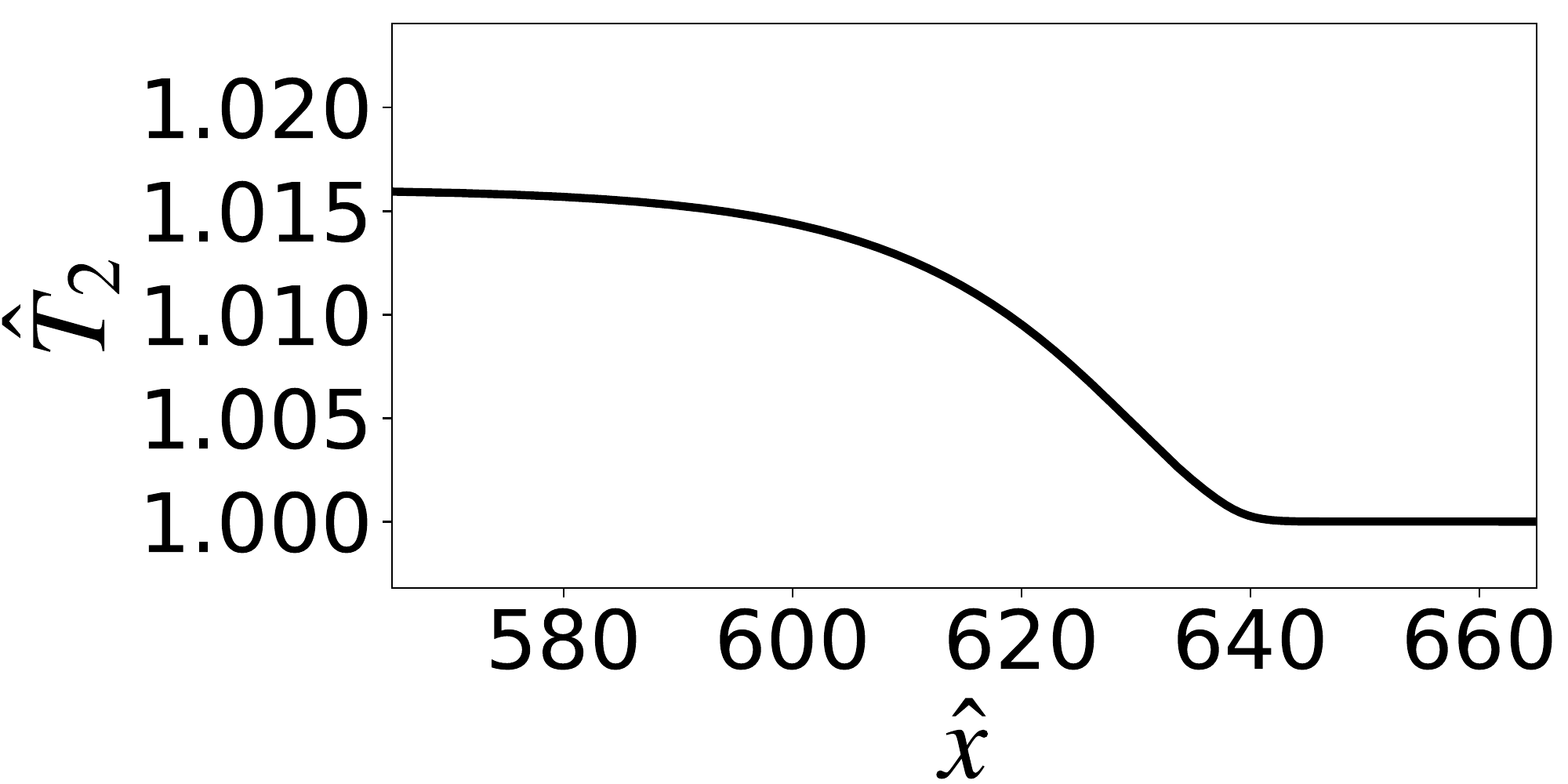}
	\end{center}
	\caption{Case B$_1$: Shock structure in a binary Eulerian mixture of polyatomic and monatomic gases obtained at $\hat{t} = 600$. 
	The parameters are in Region I and correspond to the mark of No. 1 shown in Figure \ref{fig:subshockEuler_mu055}; $\gamma_1 = 7/6$, $\gamma_2 = 5/3$, $\mu = 0.55$, $c_0 = 0.15$, and $M_0 = 1.025$
	The numerical conditions are $\Delta \hat{t} = 0.02$ and $\Delta \hat{x}=0.08$. }
	\label{fig:A_c015_M0-1_025}
\end{figure}

\begin{figure}
	\begin{center}
		\includegraphics[width=0.49\linewidth]{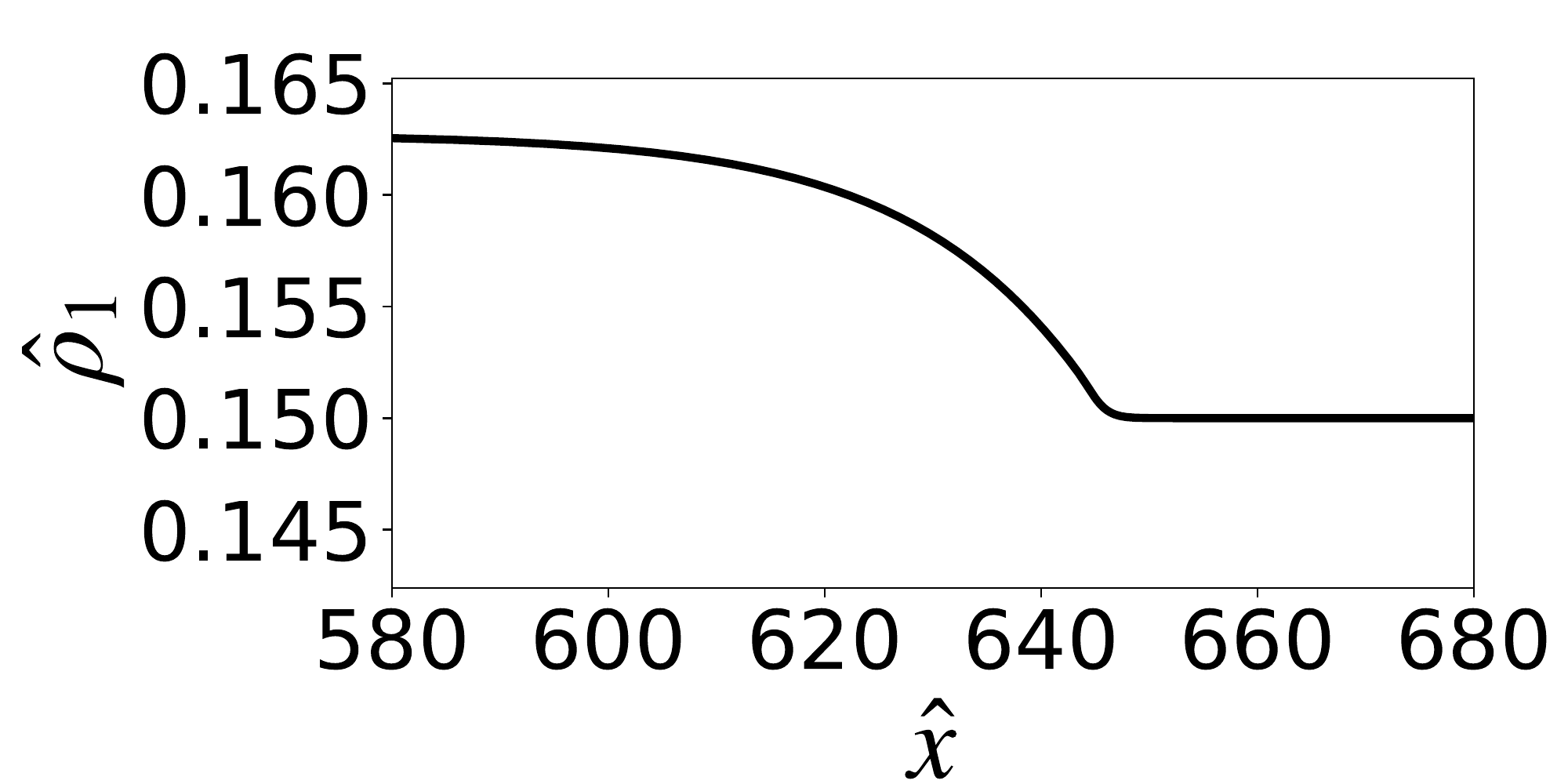} %
		\includegraphics[width=0.49\linewidth]{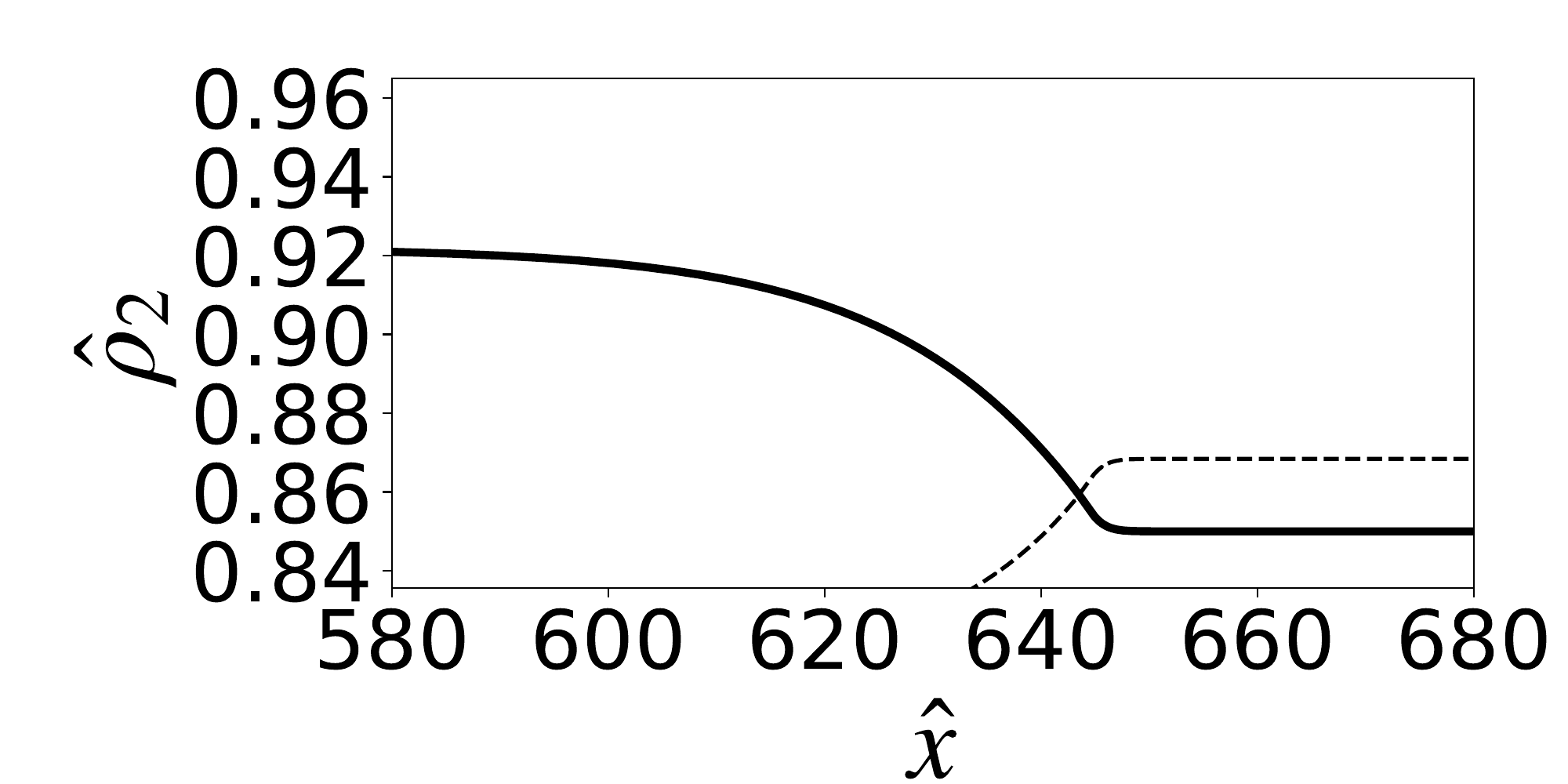}
		\includegraphics[width=0.49\linewidth]{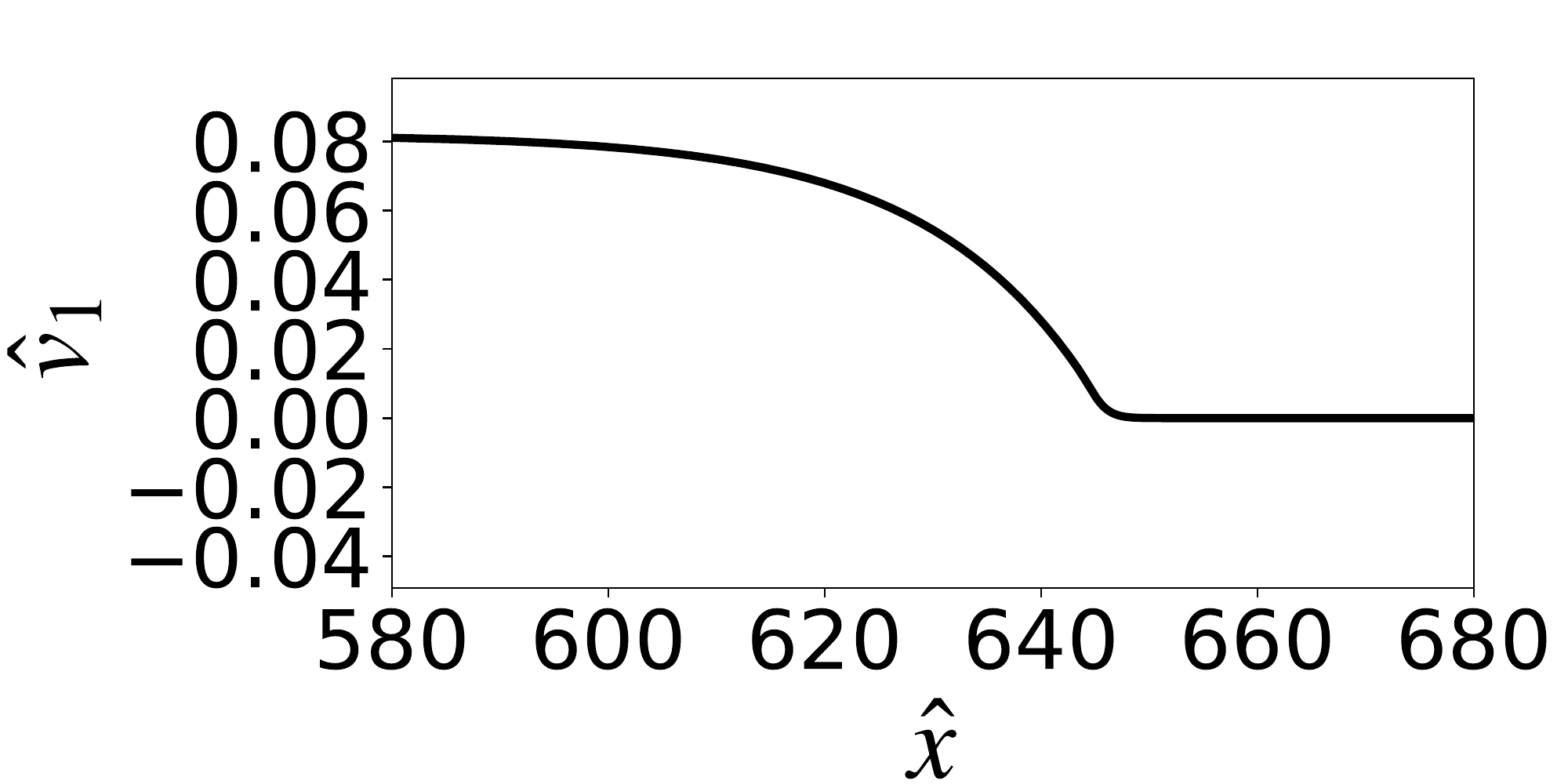} %
		\includegraphics[width=0.49\linewidth]{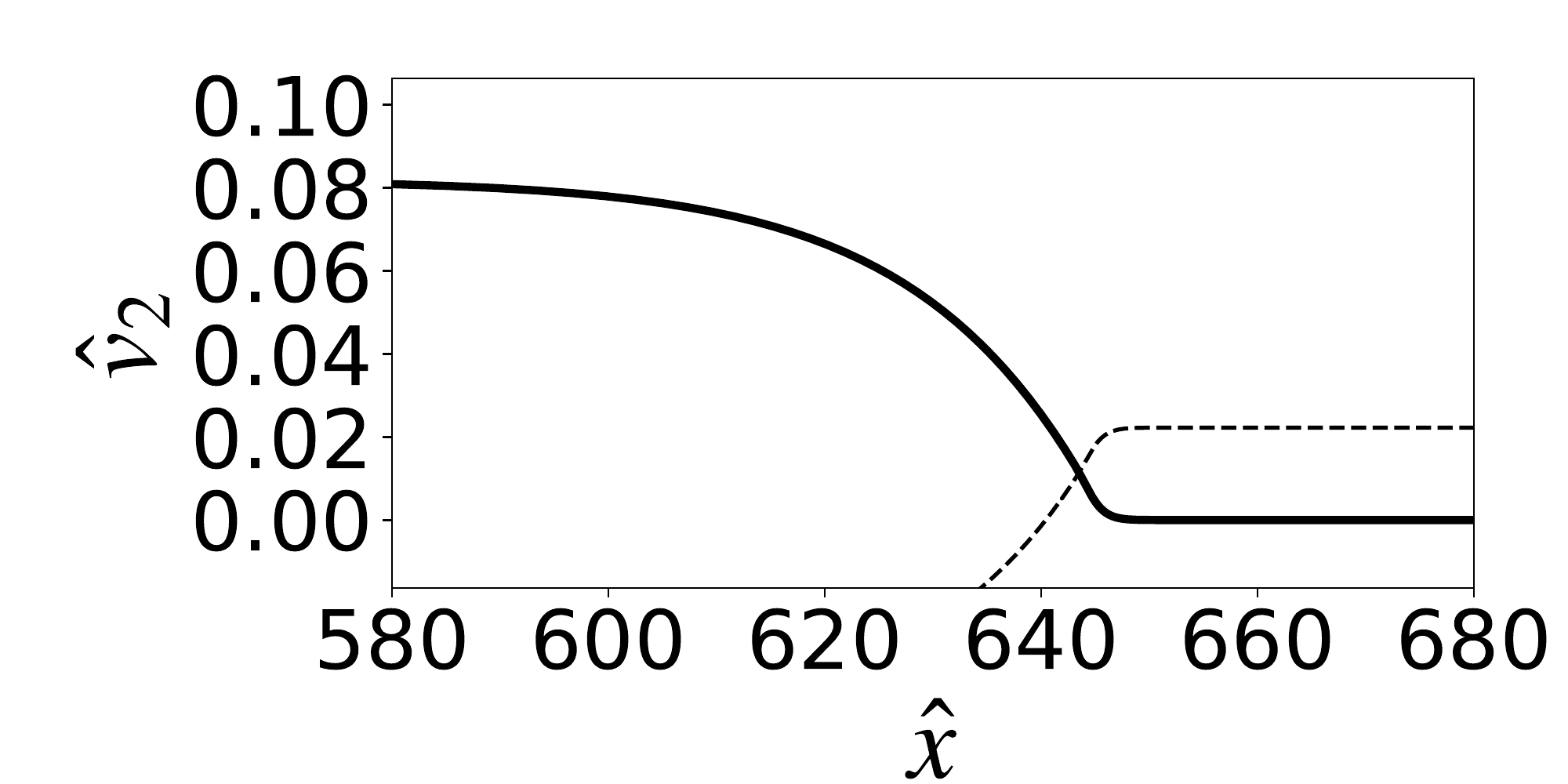}
		\includegraphics[width=0.49\linewidth]{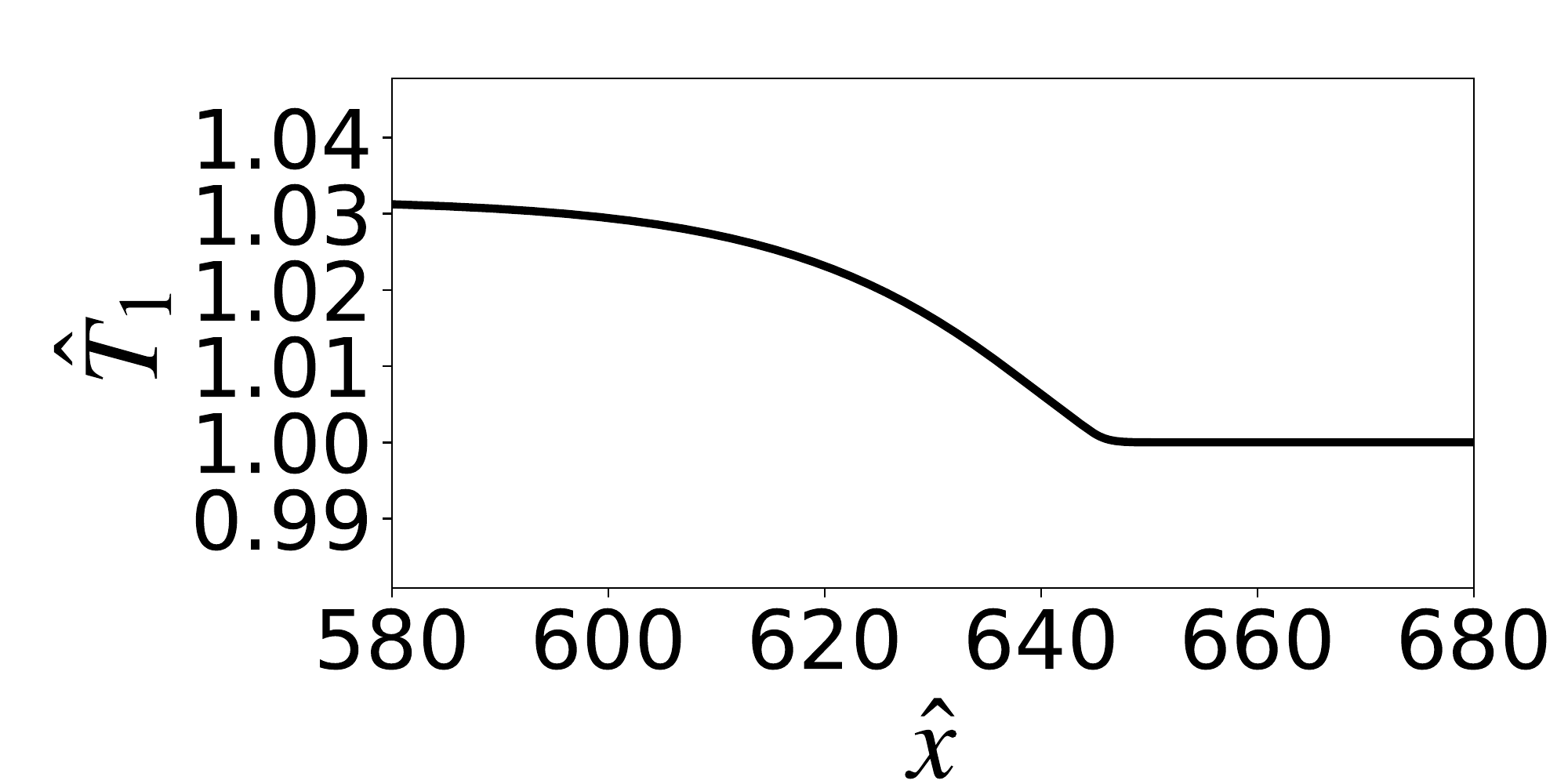} %
		\includegraphics[width=0.49\linewidth]{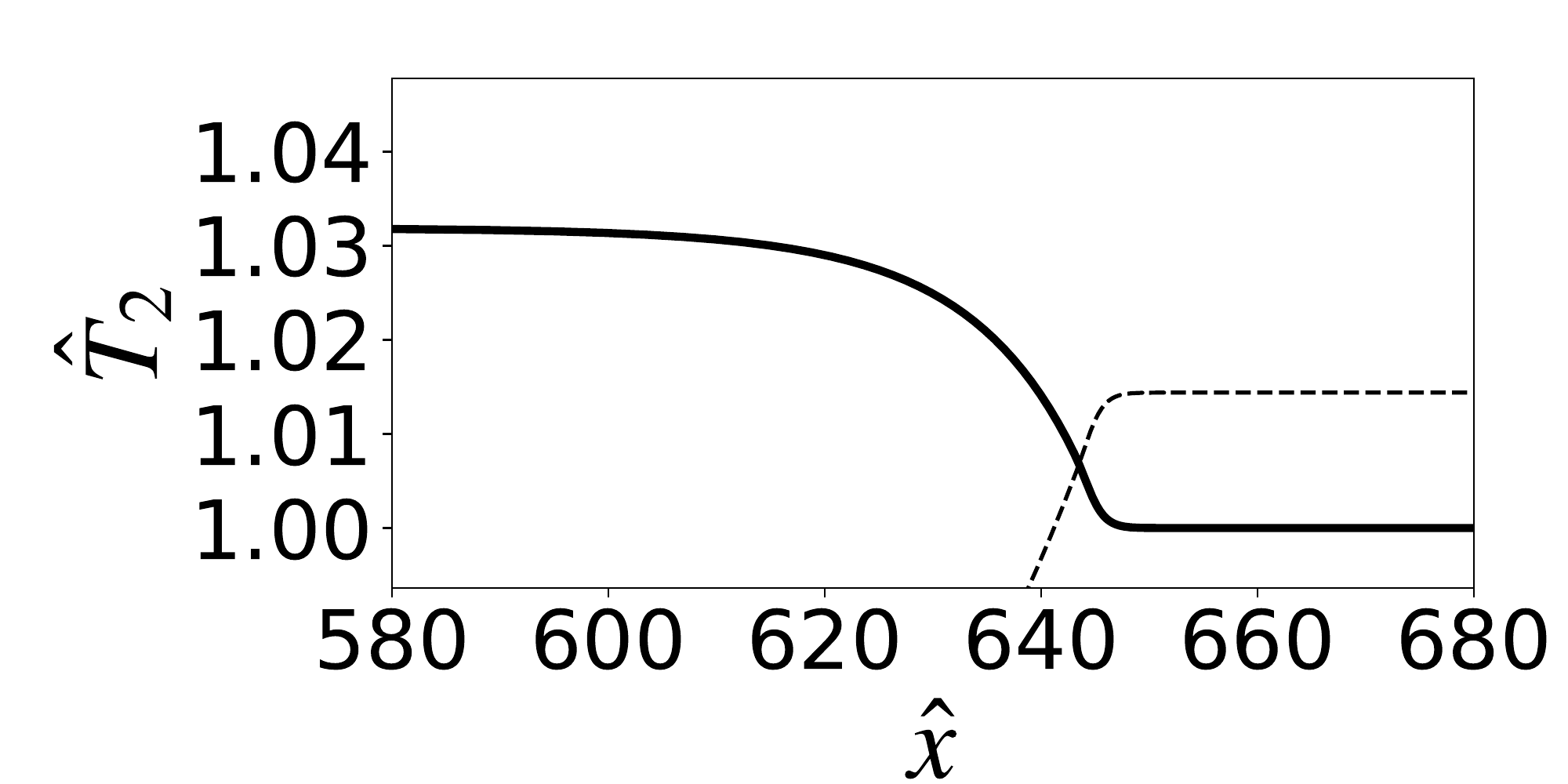}
	\end{center}
	\hspace{0.49\linewidth}
	\includegraphics[width=0.49\linewidth]{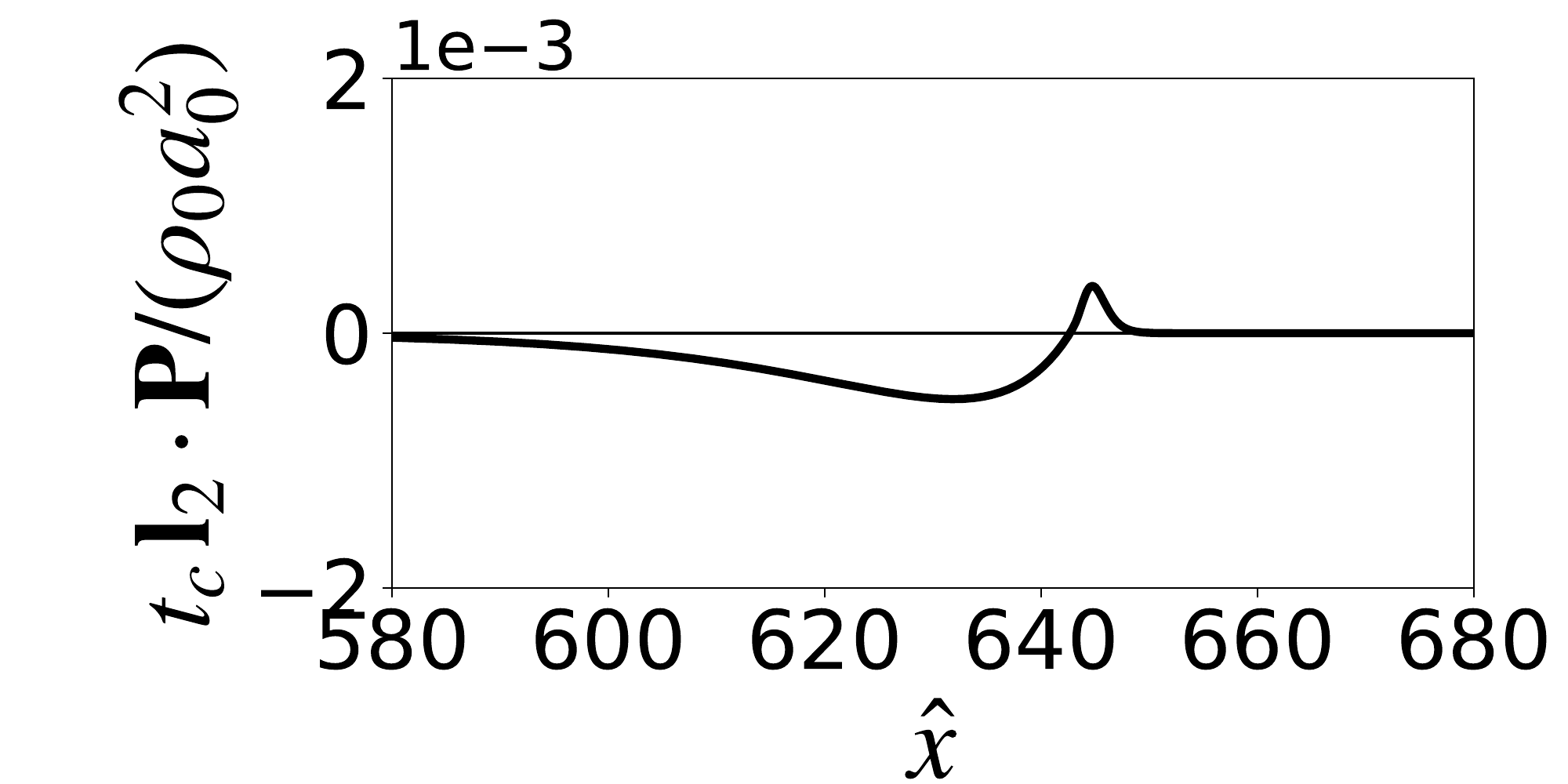}
	\caption{Case B$_1$: Shock structure in a binary Eulerian mixture of polyatomic and monatomic gases obtained at $\hat{t} = 600$. 
	The parameters are in Region II and correspond to the mark of No. 2 shown in Figure \ref{fig:subshockEuler_mu055}; $\gamma_1 = 7/6$, $\gamma_2 = 5/3$, $\mu = 0.55$, $c_0 = 0.15$, and $M_0 = 1.05$. 
	The numerical conditions are $\Delta \hat{t} = 0.02$ and $\Delta \hat{x}=0.08$. 
	The profiles of the potential sub-shock predicted by the RH conditions of the full system are also shown (dotted curves). }
	\label{fig:c015_M0-1_05}
\end{figure}

\begin{figure}
	\begin{center}
		\includegraphics[width=0.49\linewidth]{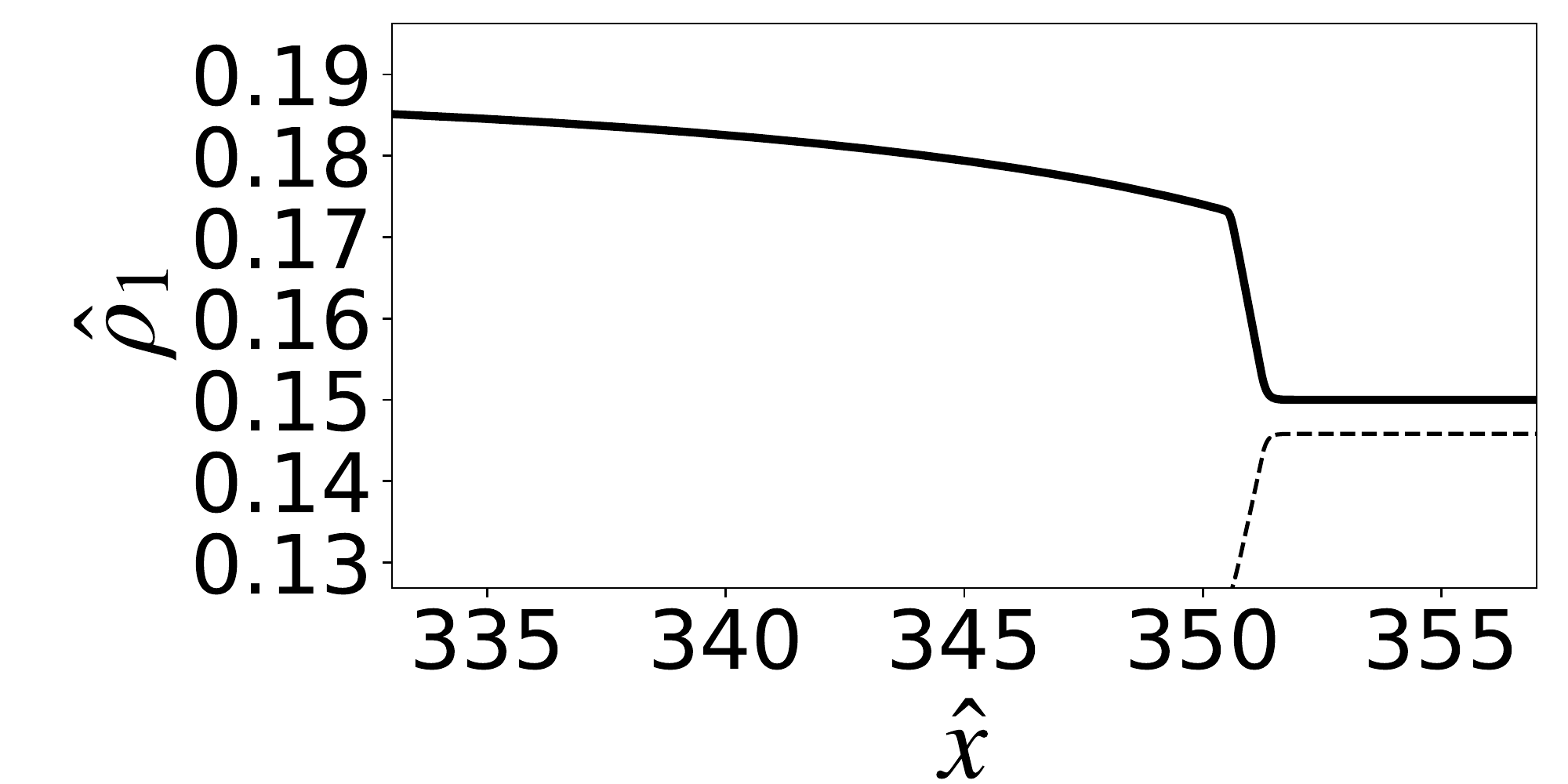} %
		\includegraphics[width=0.49\linewidth]{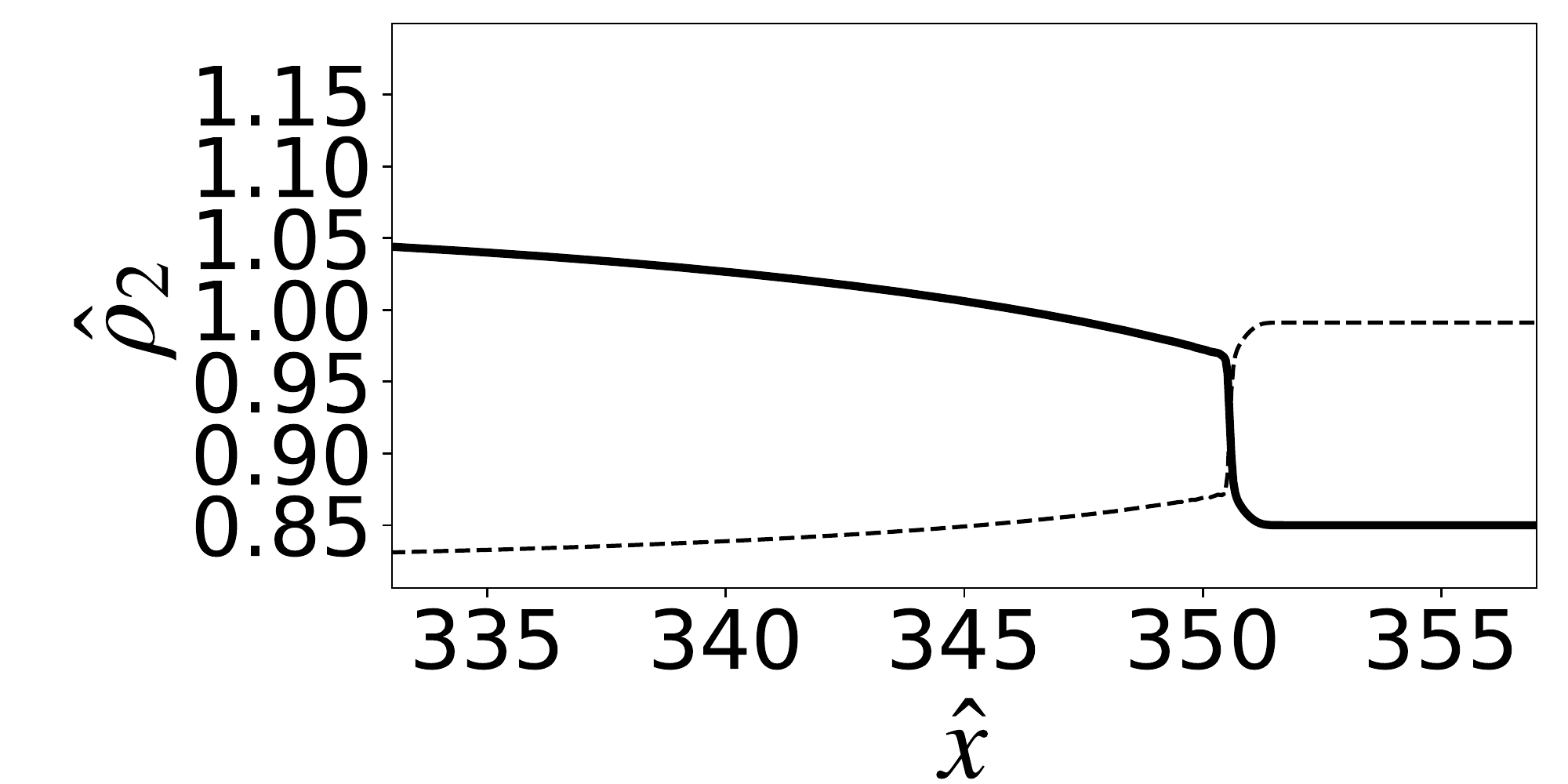}
		\includegraphics[width=0.49\linewidth]{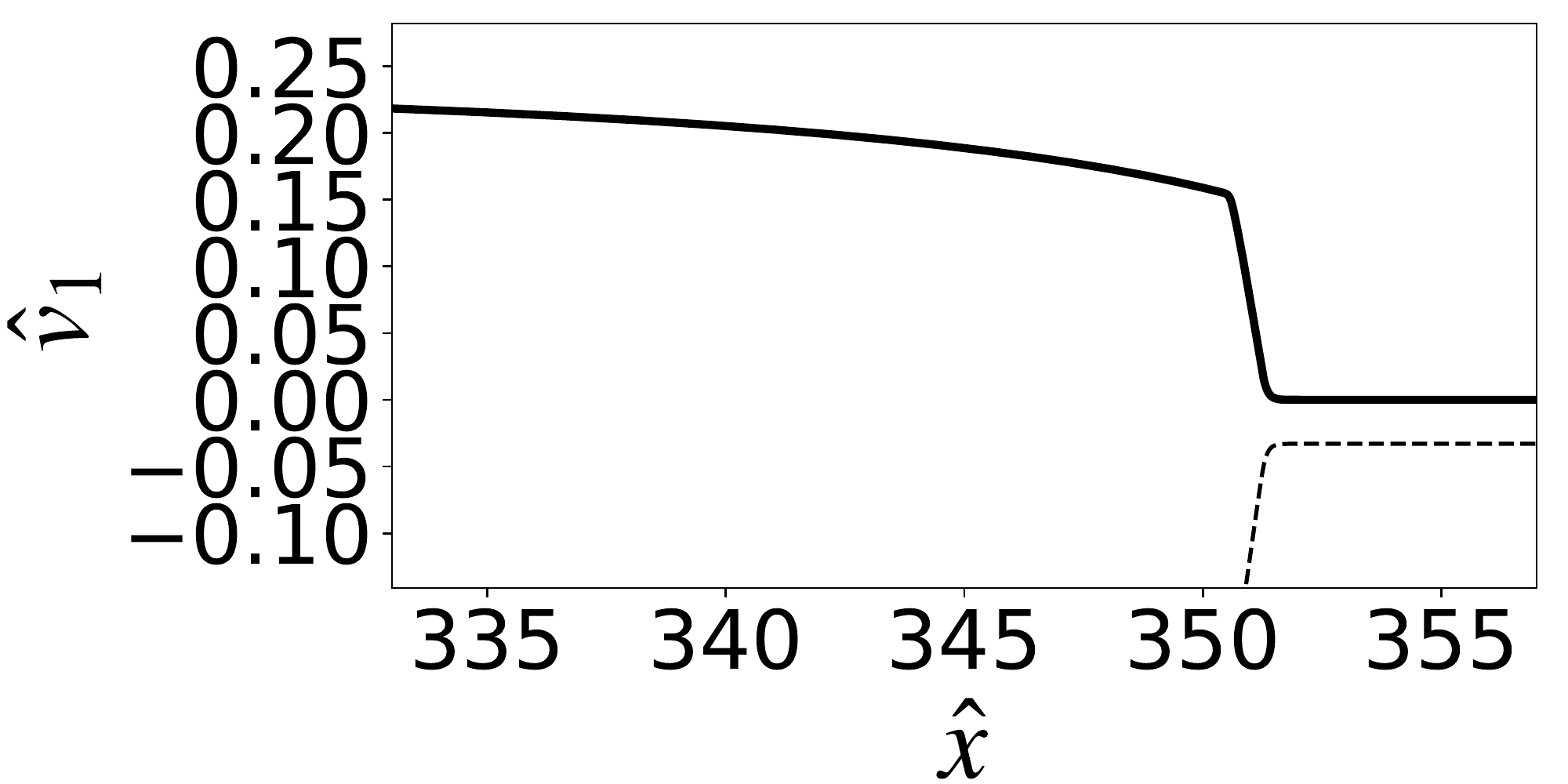} %
		\includegraphics[width=0.49\linewidth]{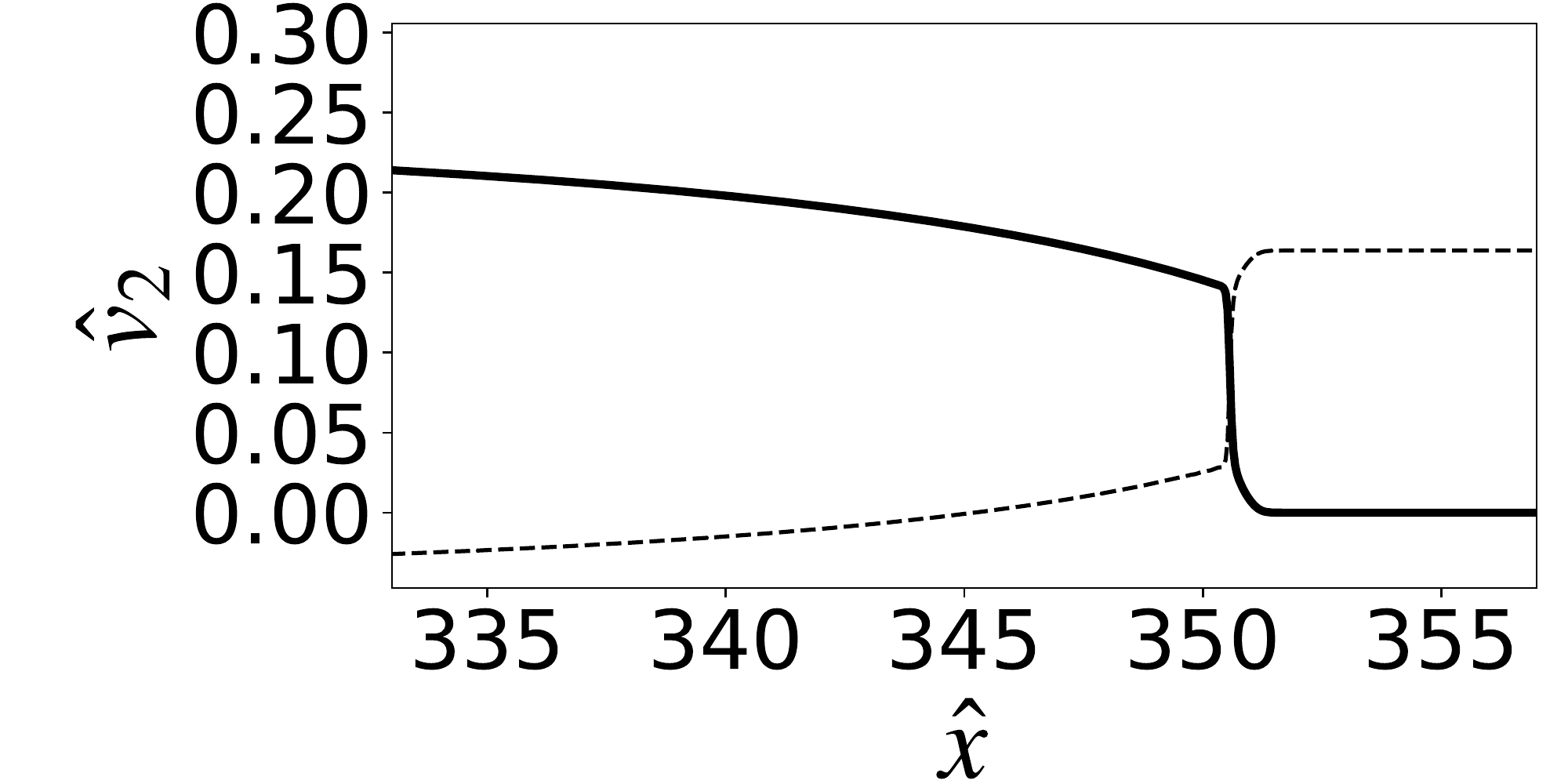}
		\includegraphics[width=0.49\linewidth]{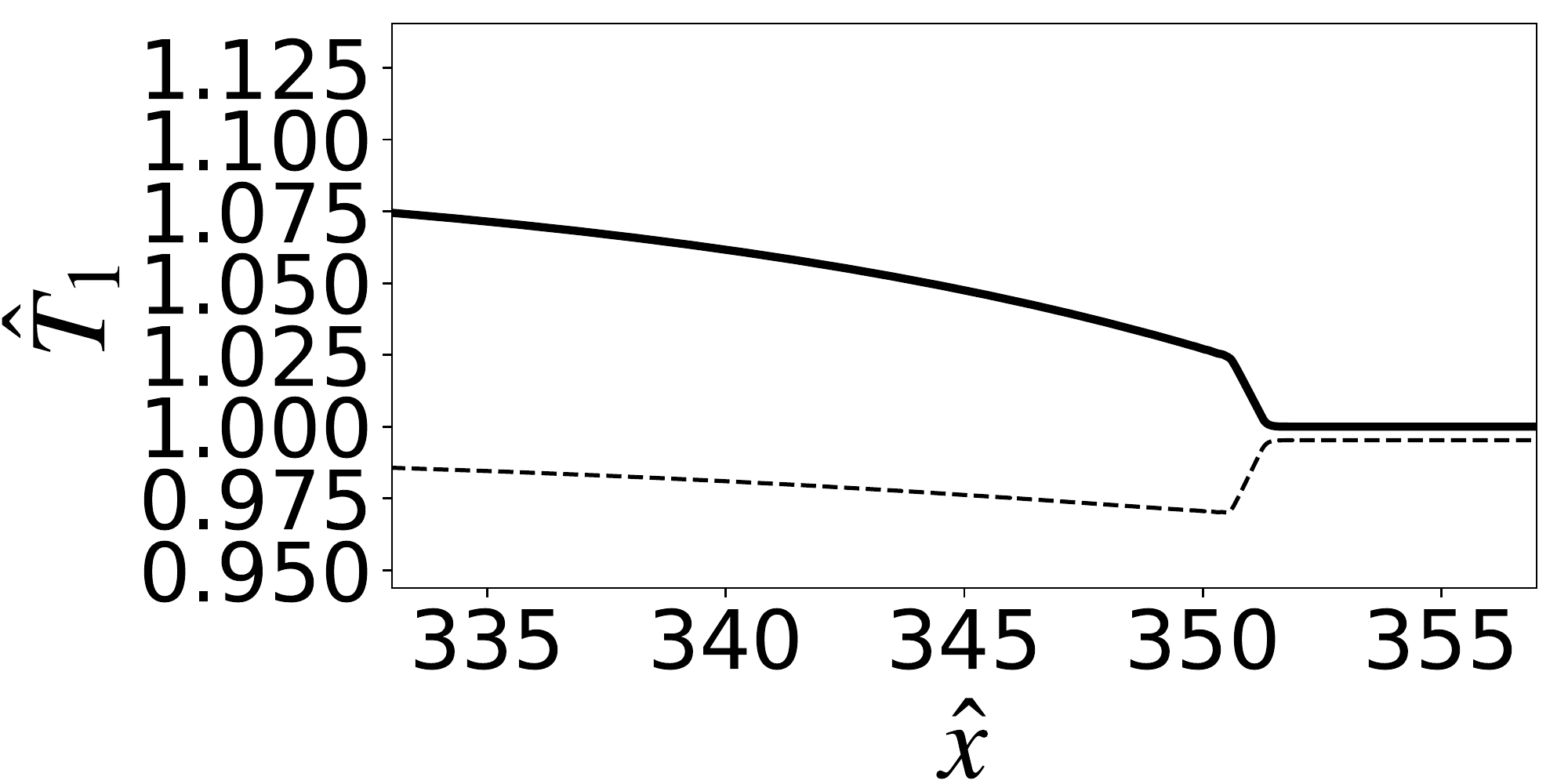} %
		\includegraphics[width=0.49\linewidth]{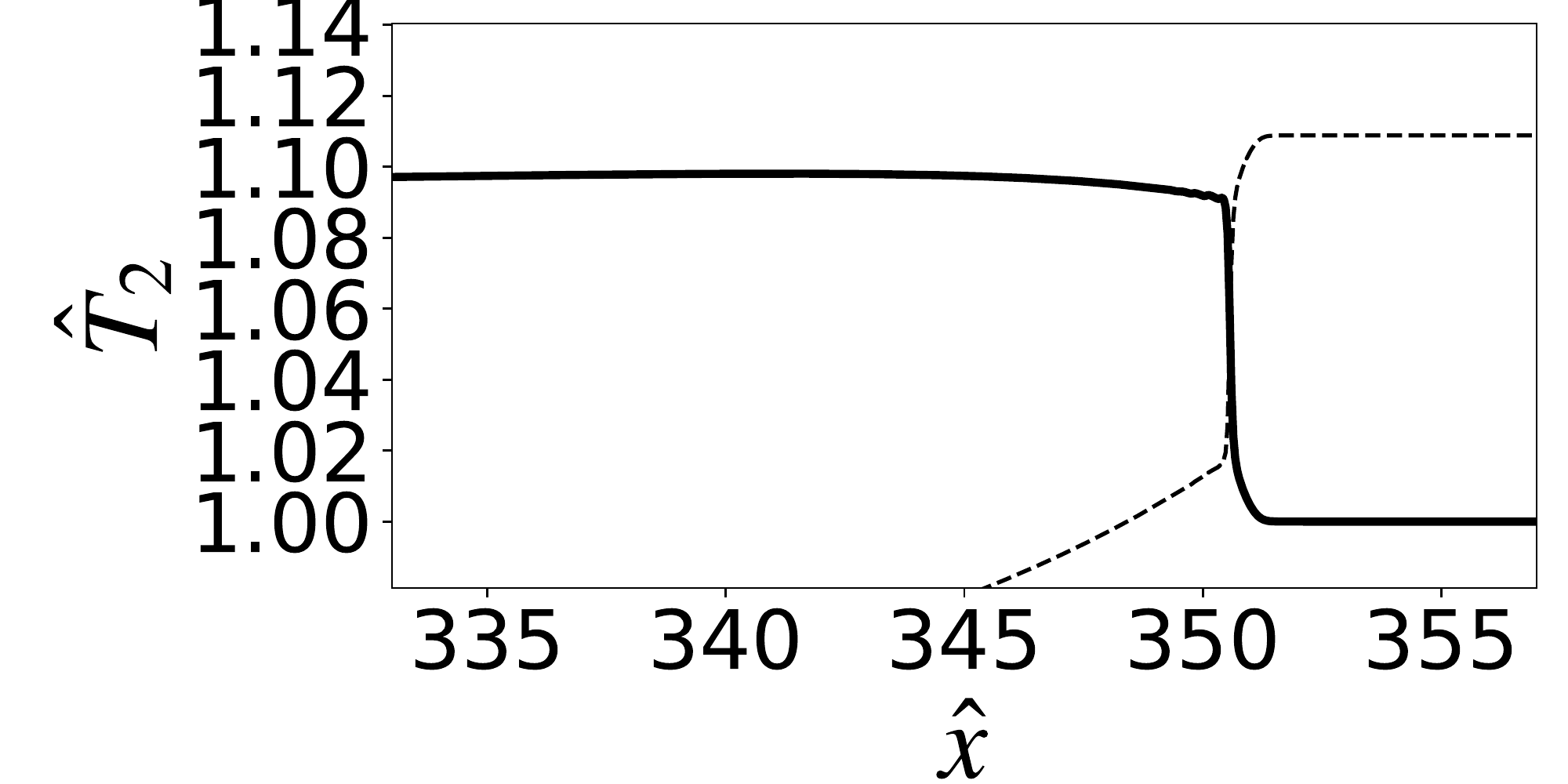}
	\end{center}
	\caption{Case B$_1$: Shock structure in a binary Eulerian mixture of polyatomic and monatomic gases obtained at $\hat{t} = 300$. 
	The parameters are in Region II and correspond to the mark of No. 3 shown in Figure \ref{fig:subshockEuler_mu055}; $\gamma_1 = 7/6$, $\gamma_2 = 5/3$, $\mu = 0.55$, $c_0 = 0.15$, and $M_0 = 1.15$. 
	The numerical conditions are $\Delta \hat{t} = 0.01$ and $\Delta \hat{x}=0.04$. }
	\label{fig:c015_M0-1_15}
\end{figure}

\begin{figure}
	\begin{center}
		\includegraphics[width=0.49\linewidth]{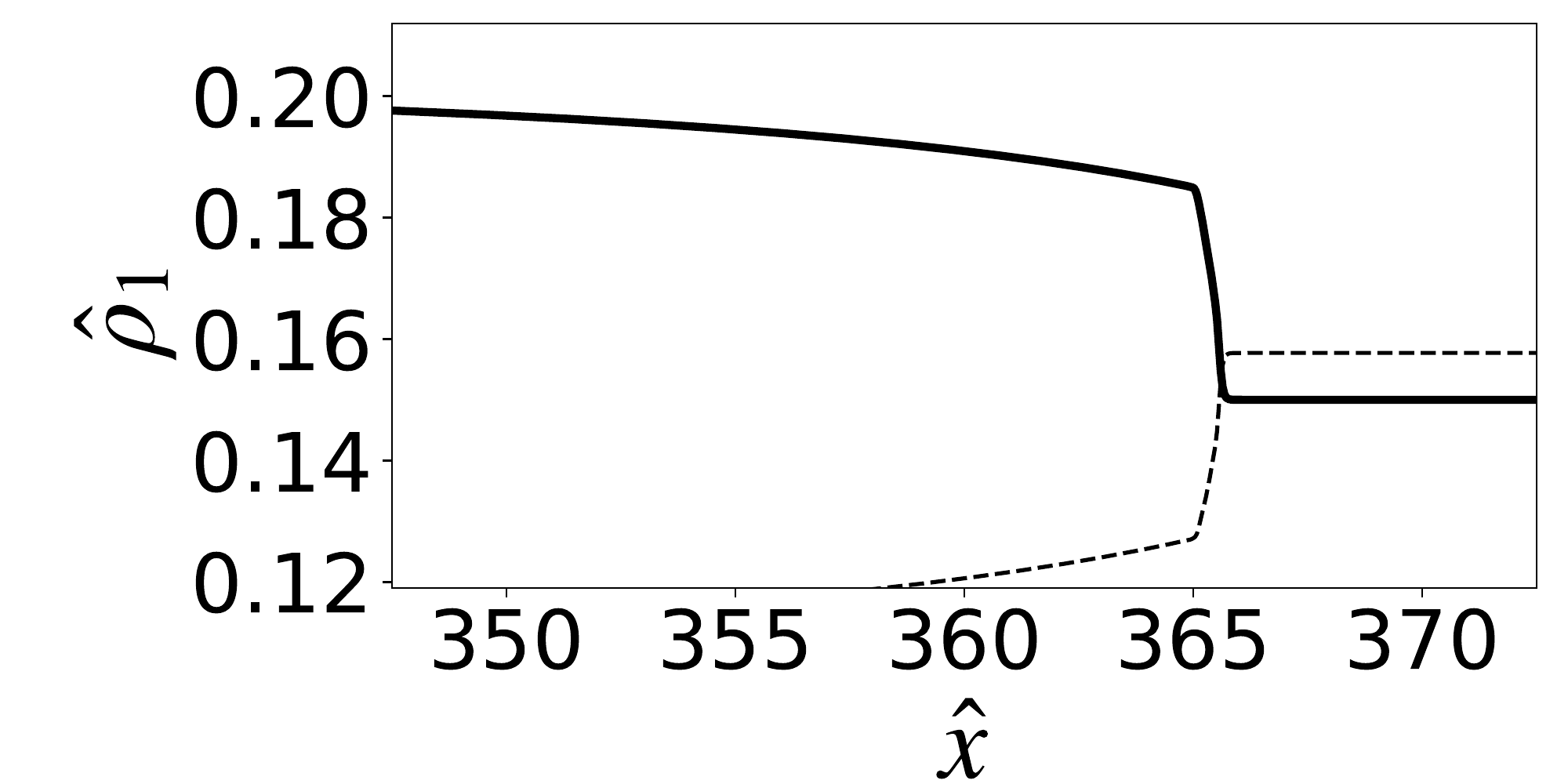} %
		\includegraphics[width=0.49\linewidth]{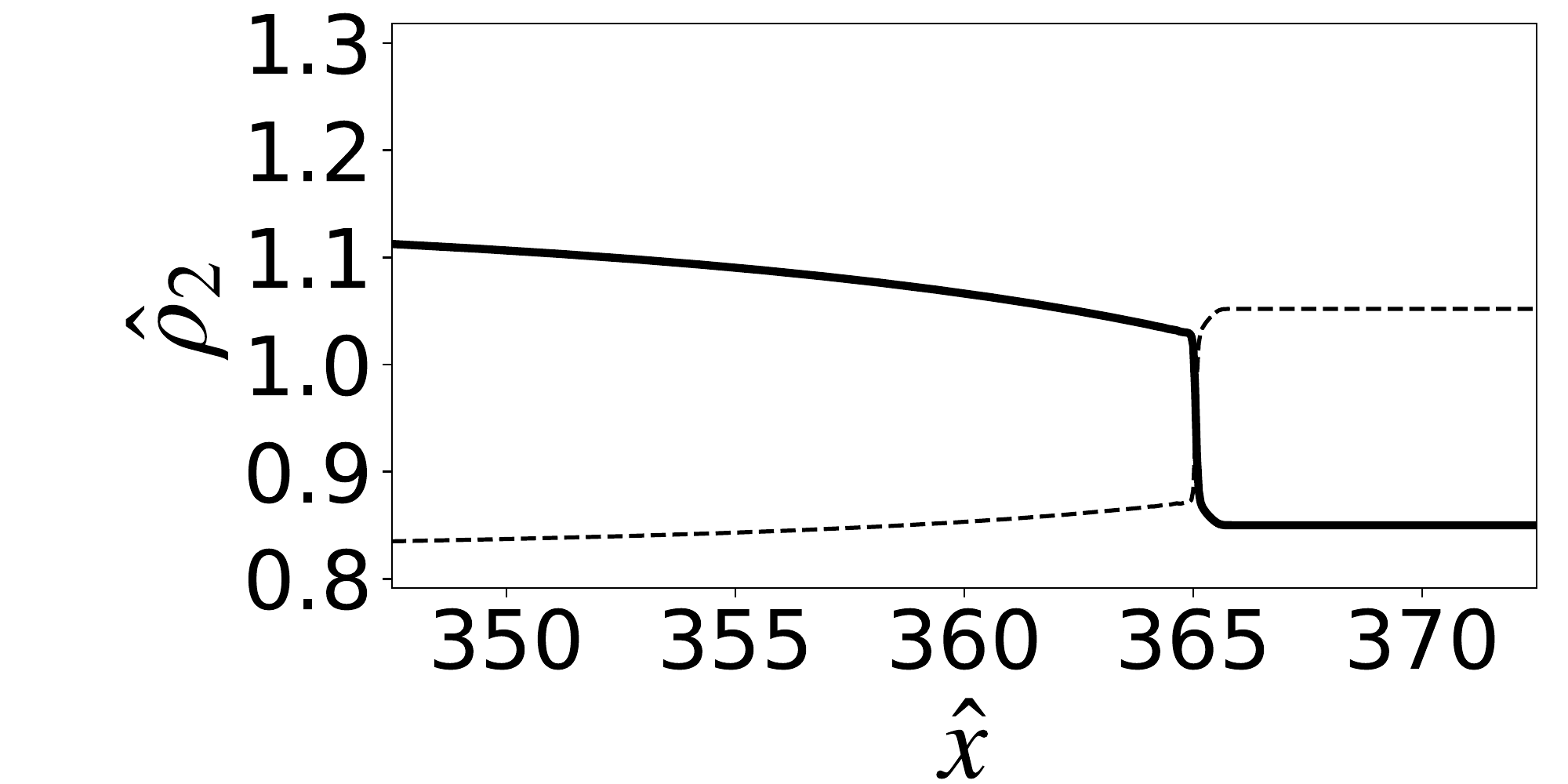}
		\includegraphics[width=0.49\linewidth]{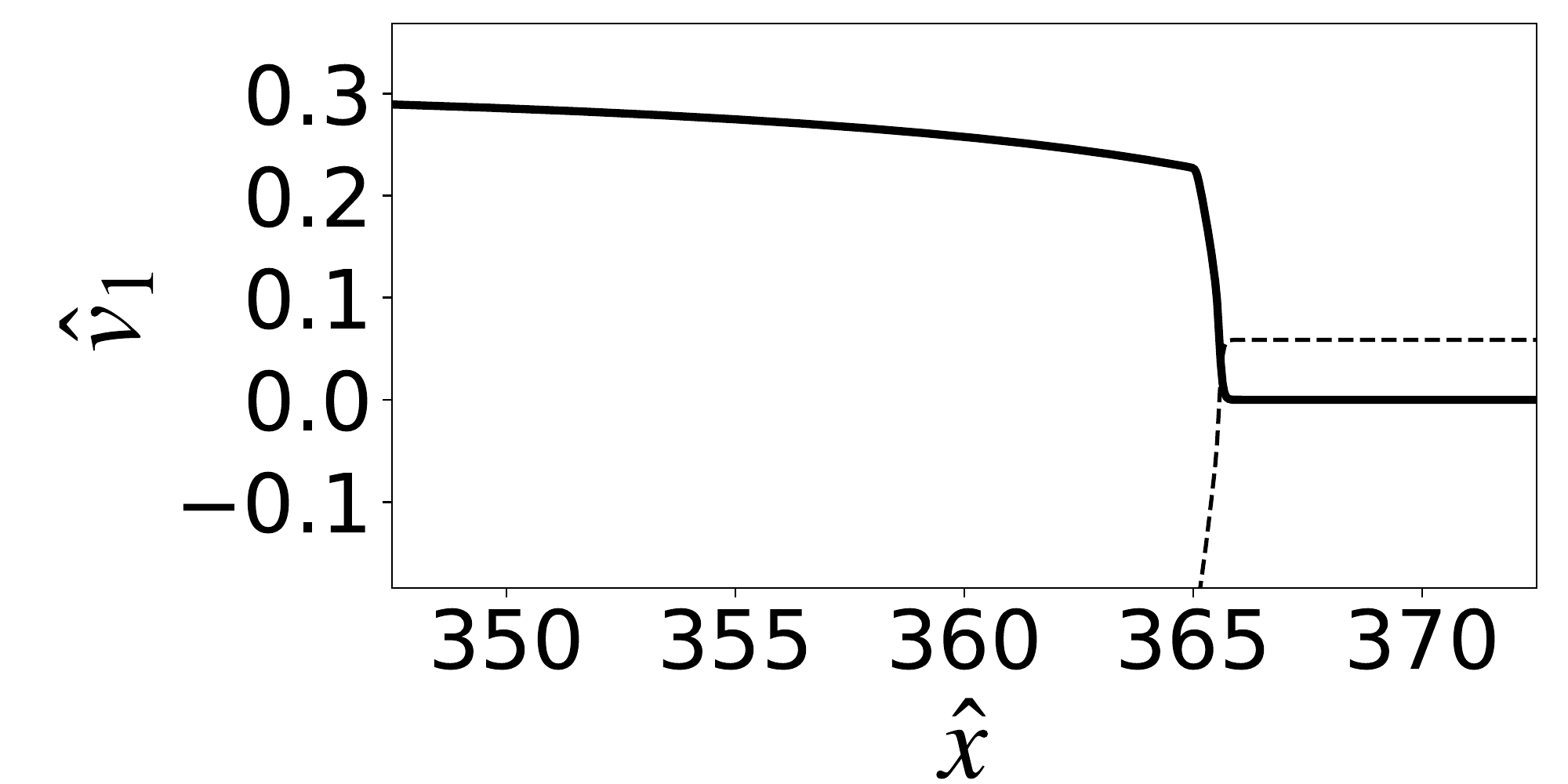} %
		\includegraphics[width=0.49\linewidth]{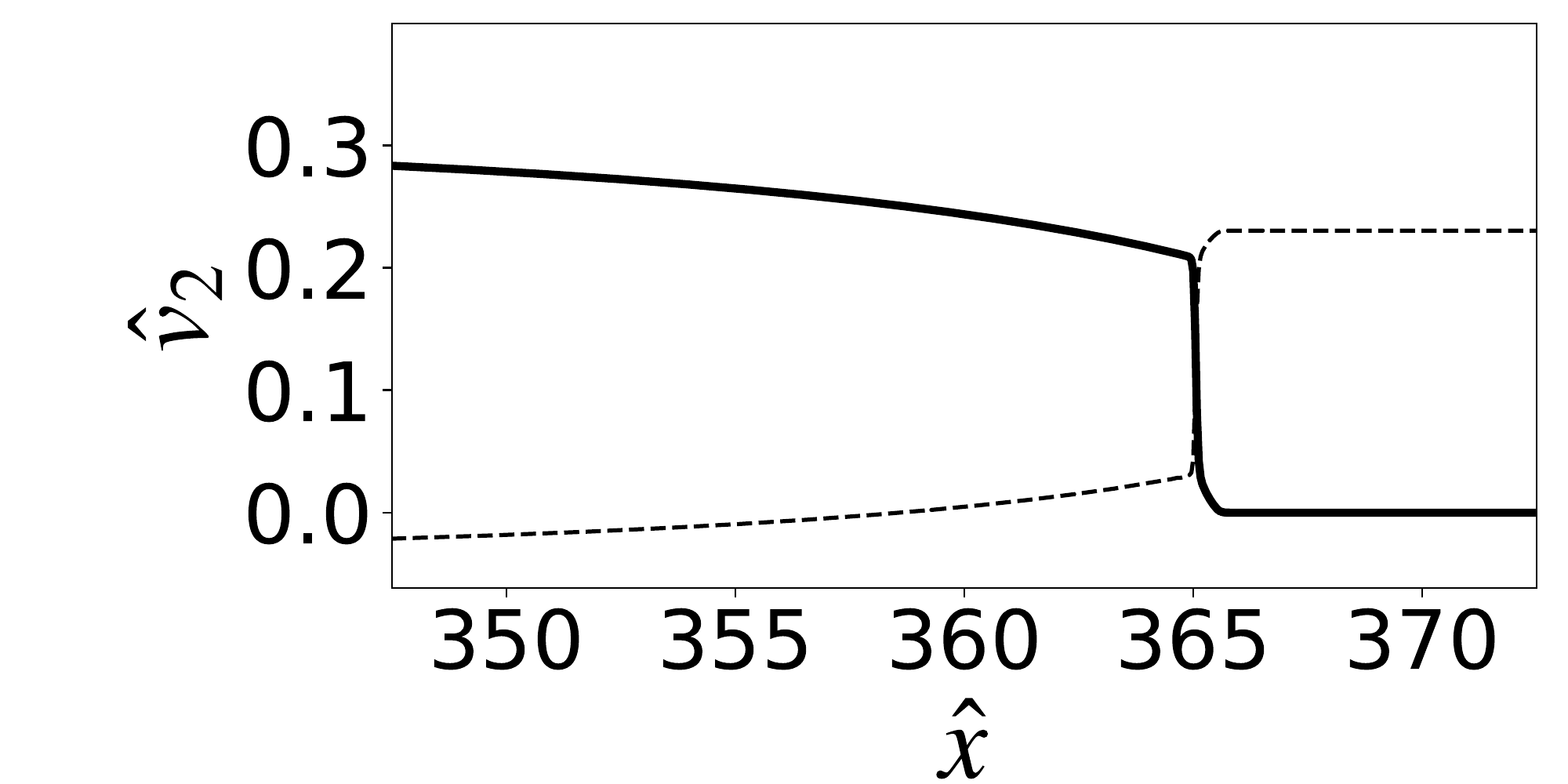}
		\includegraphics[width=0.49\linewidth]{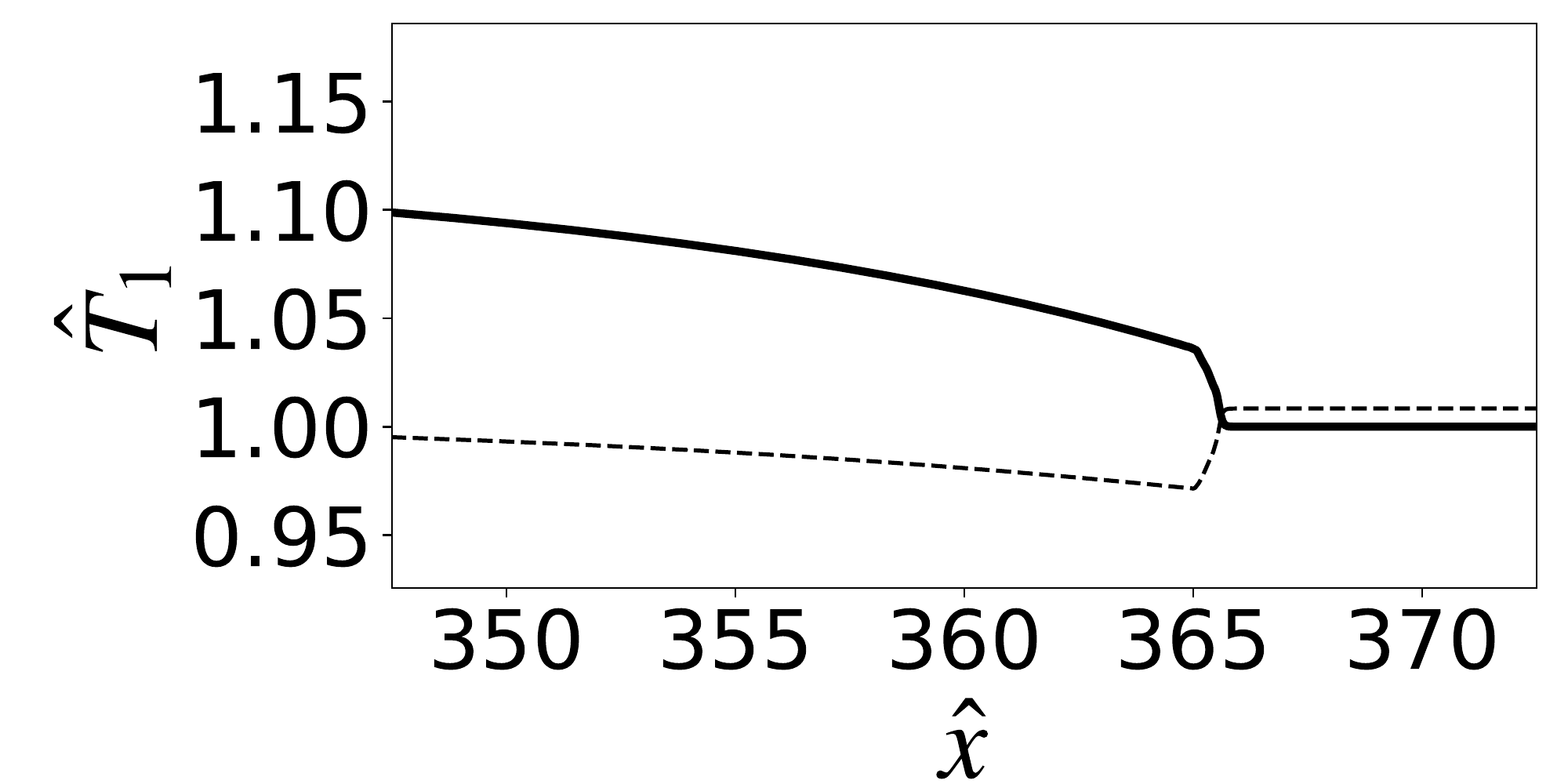} %
		\includegraphics[width=0.49\linewidth]{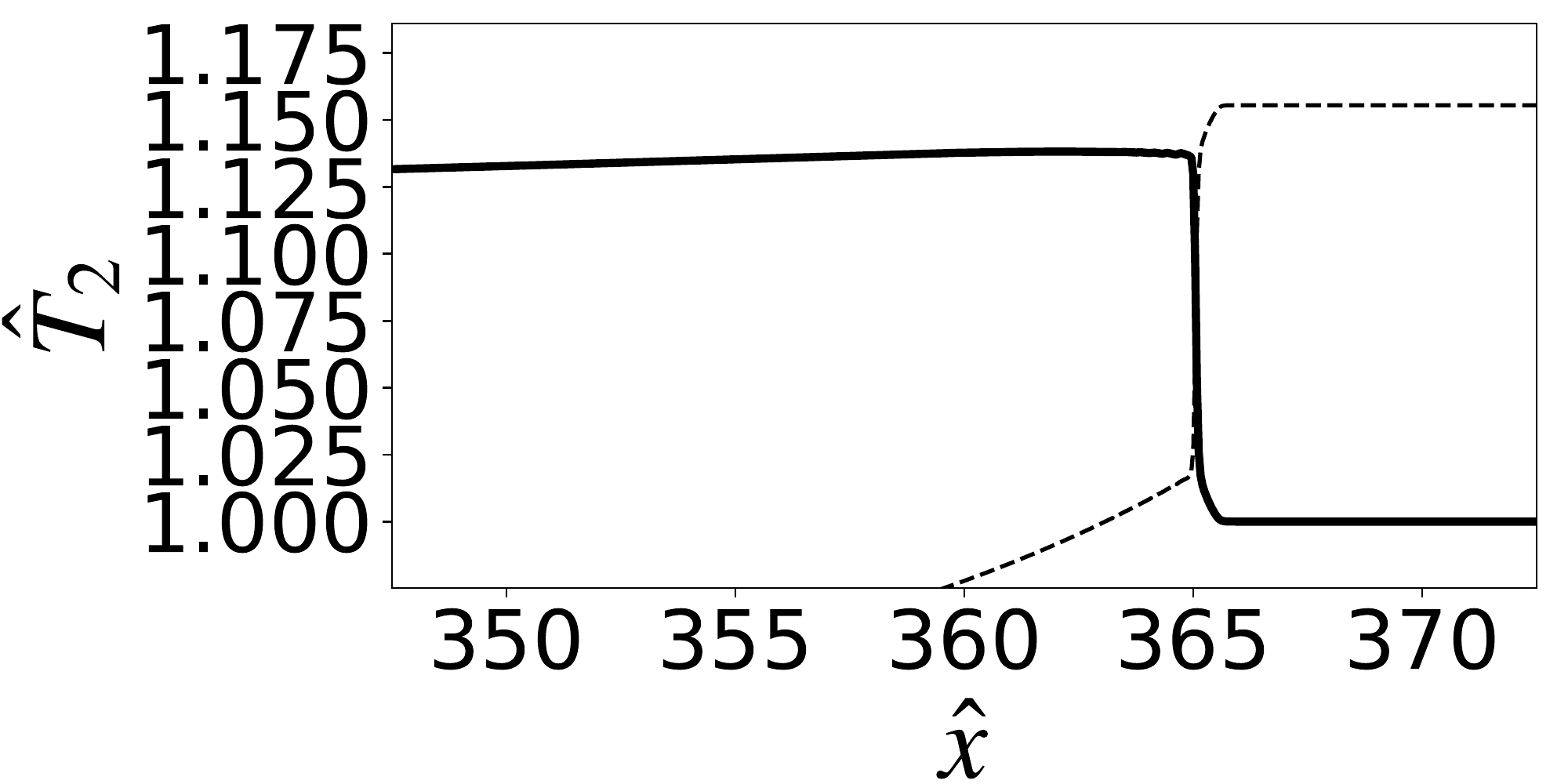}
	\end{center}
	\caption{Case B$_1$: Shock structure in a binary Eulerian mixture of polyatomic and monatomic gases obtained at $\hat{t} = 300$. 
	The parameters are in Region IV and correspond to the mark of No. 4 shown in Figure \ref{fig:subshockEuler_mu055}; $\gamma_1 = 7/6$, $\gamma_2 = 5/3$, $\mu = 0.55$, $c_0 = 0.15$, and $M_0 = 1.2$. 
	The numerical conditions are $\Delta \hat{t} = 0.01$ and $\Delta \hat{x}=0.04$. }
	\label{fig:A_c015_M0-1_2}
\end{figure}

Next let us analyze the shock structure with $\mu = 0.55$, $\gamma_1 = 7/6$, and $\gamma_2 = 5/3$ for different concentrations and the Mach numbers. 
In the case of $c_0 = 0.15$, we show the shock structure for $M_0 = 1.025, 1.05, 1.15$, and $1.2$ in Figures \ref{fig:A_c015_M0-1_025} -- \ref{fig:A_c015_M0-1_2}. 
Figure \ref{fig:c015_M0-1_05} for $M_0 = 1.05$ shows a surprising result that a sub-shock for constituent $1$ does not emerge even with $M_0 > M_{20}$. 
Although the curves of the potential sub-shock are intersected by the profile of the physical quantities, the shock structure is continuous and no sub-shock appears. 

Let us consider the behavior of $\mathbf{l}\cdot \mathbf{P}$ in the shock-structure solution for the constituent $2$, and we focus on the mode corresponding to $\lambda_2$. 
The left eigenvector $\mathbf{l}_2$ for the system \eqref{finale1} corresponding to $\lambda_2$ is 
\begin{equation*}
	\mathbf{l}_2 =
	\left[%
	\begin{array}{c}
		0 \\
		0 \\
		0 \\
		v_2 \left( v_2 - \sqrt{\gamma_2 \frac{k_{\mathrm{B}}}{m_{2}} T_2} \,\, \right) \\
		\sqrt{\gamma_2 \frac{k_{\mathrm{B}}}{m_{2}} T_2} - 2 v_2 \\
		1 \\
	\end{array}%
	\right]^T
\end{equation*}
and we depict its profile in the dimensionless form given by $t_c \, \mathbf{l}_2 \cdot \mathbf{P}/(\rho_0 a_0^2)$ in Figure \ref{fig:c015_M0-1_05}. 
From Figure \ref{fig:c015_M0-1_05}, it is confirmed that $\mathbf{l}_2 \cdot \mathbf{P}$ becomes zero at the intersection point between the profile and the curve of the potential sub-shock. 
Therefore we conclude that the shock structure for constituent $2$ is continuous without any sub-shock. 

The shock structure for $M_0 = 1.15$ is shown in Figure \ref{fig:c015_M0-1_15}, which lies in the same region in the case with $M_0 = 1.05$ in Figure \ref{fig:subshockEuler_mu055}. 
Again surprisingly, from Figure \ref{fig:c015_M0-1_15}, we see a sub-shock for constituent $2$ emerges for $M_0=1.15$. 
We conclude that the regular singular point corresponding to the second-fastest mode for constituent $2$ observed for $M_0 = 1.15$ is no more regular, and a sub-shock also for constituent $2$ emerges. 
This is also the first example that a singular point is a regular singular for the moderately large Mach number, and this regular singular point becomes singular for the larger Mach number. 

The fact that singular points in the shock-structure solution become regular is similar to the ones reported in the context of single fluids of rarefied monatomic and polyatomic gases~\cite{Weiss,IJNLM2017}. 
The singular point becoming regular is called a \emph{regular singular point}. 
Nevertheless, to the authors'  knowledge, the present result is the first example in which the singular point of the shock-structure solution becomes a regular singular point in the context of mixtures of rarefied gases. 

\begin{figure}
	\begin{center}
		\includegraphics[width=0.49\linewidth]{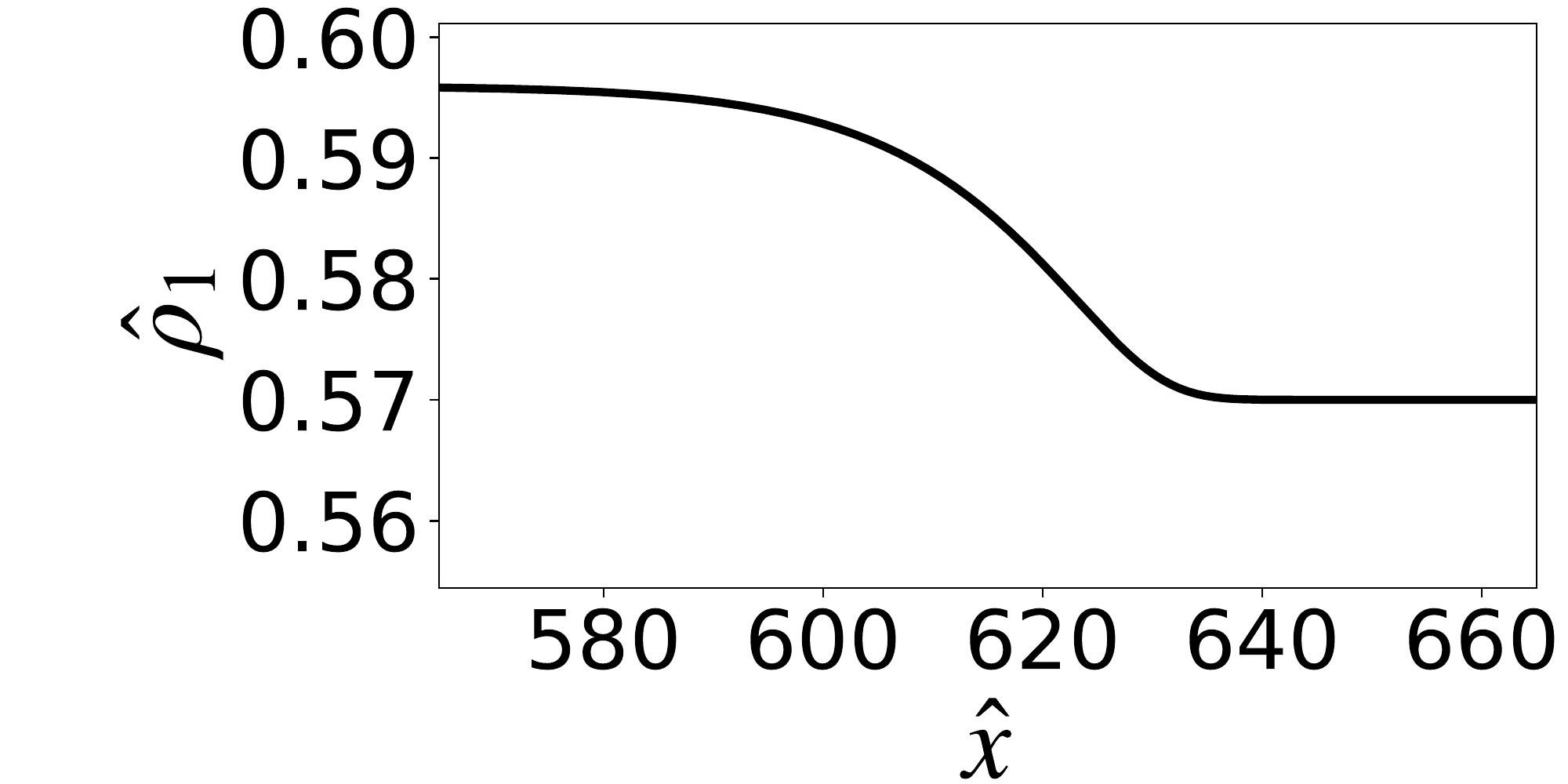} %
		\includegraphics[width=0.49\linewidth]{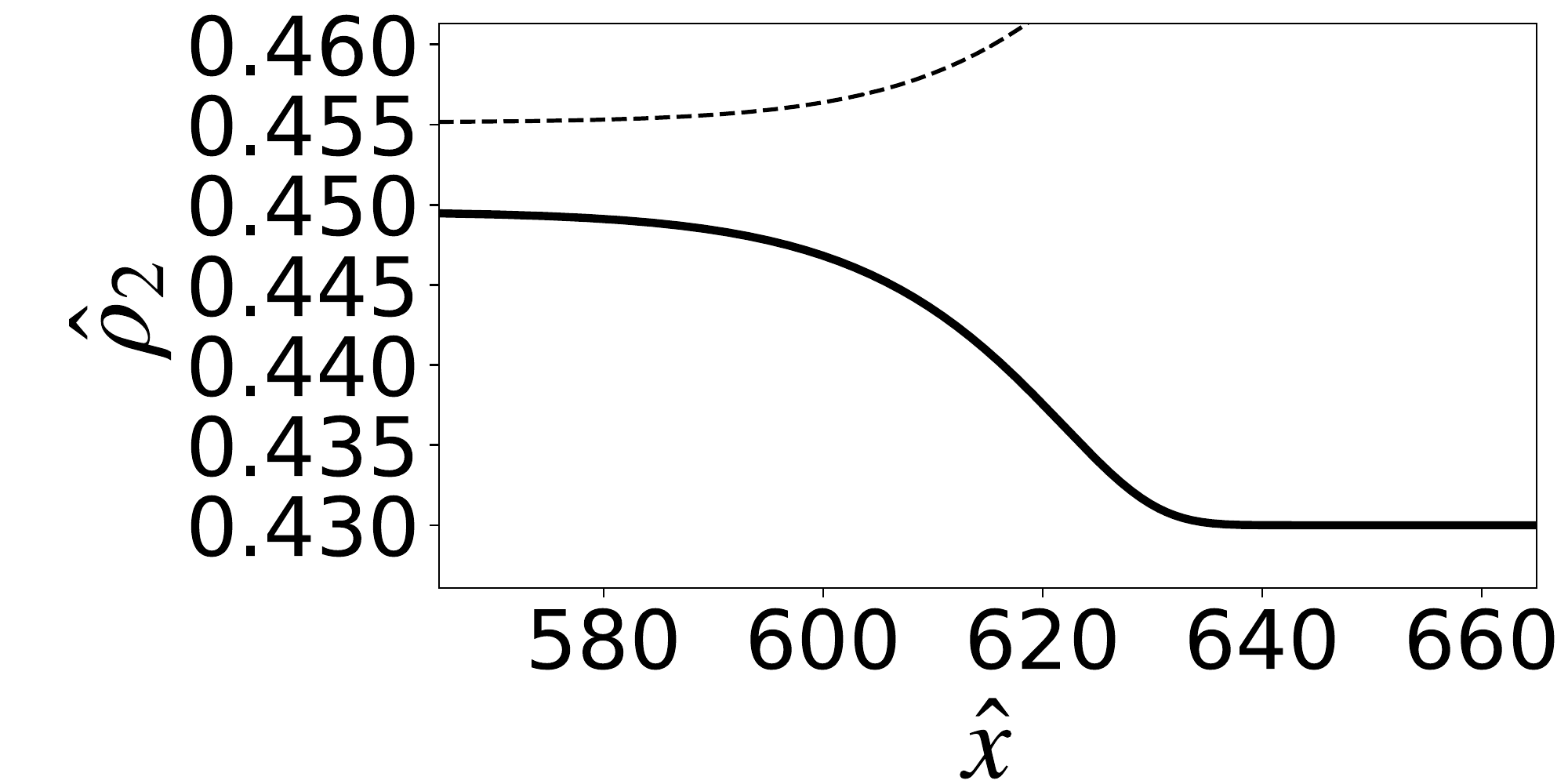}
		\includegraphics[width=0.49\linewidth]{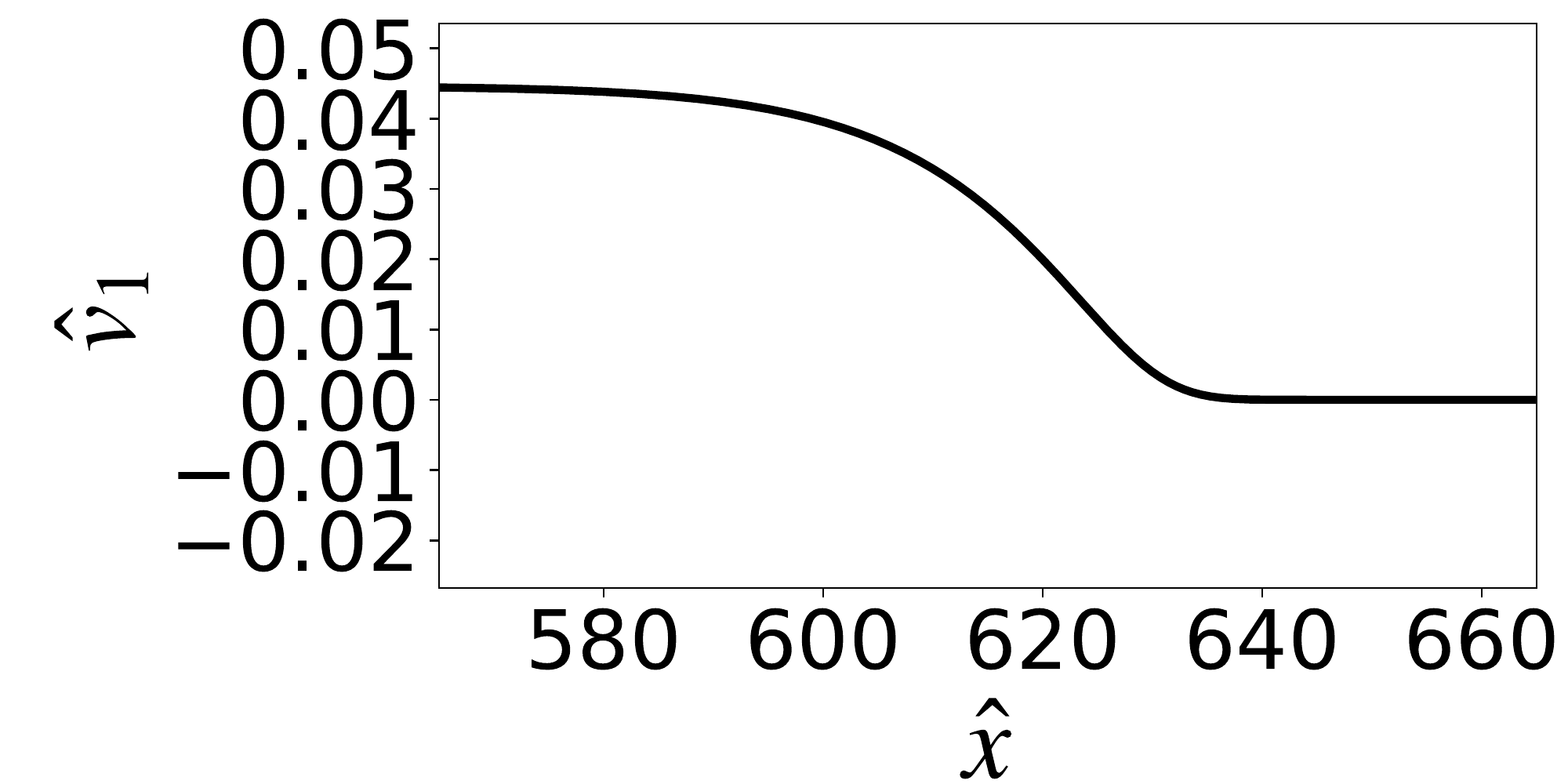} %
		\includegraphics[width=0.49\linewidth]{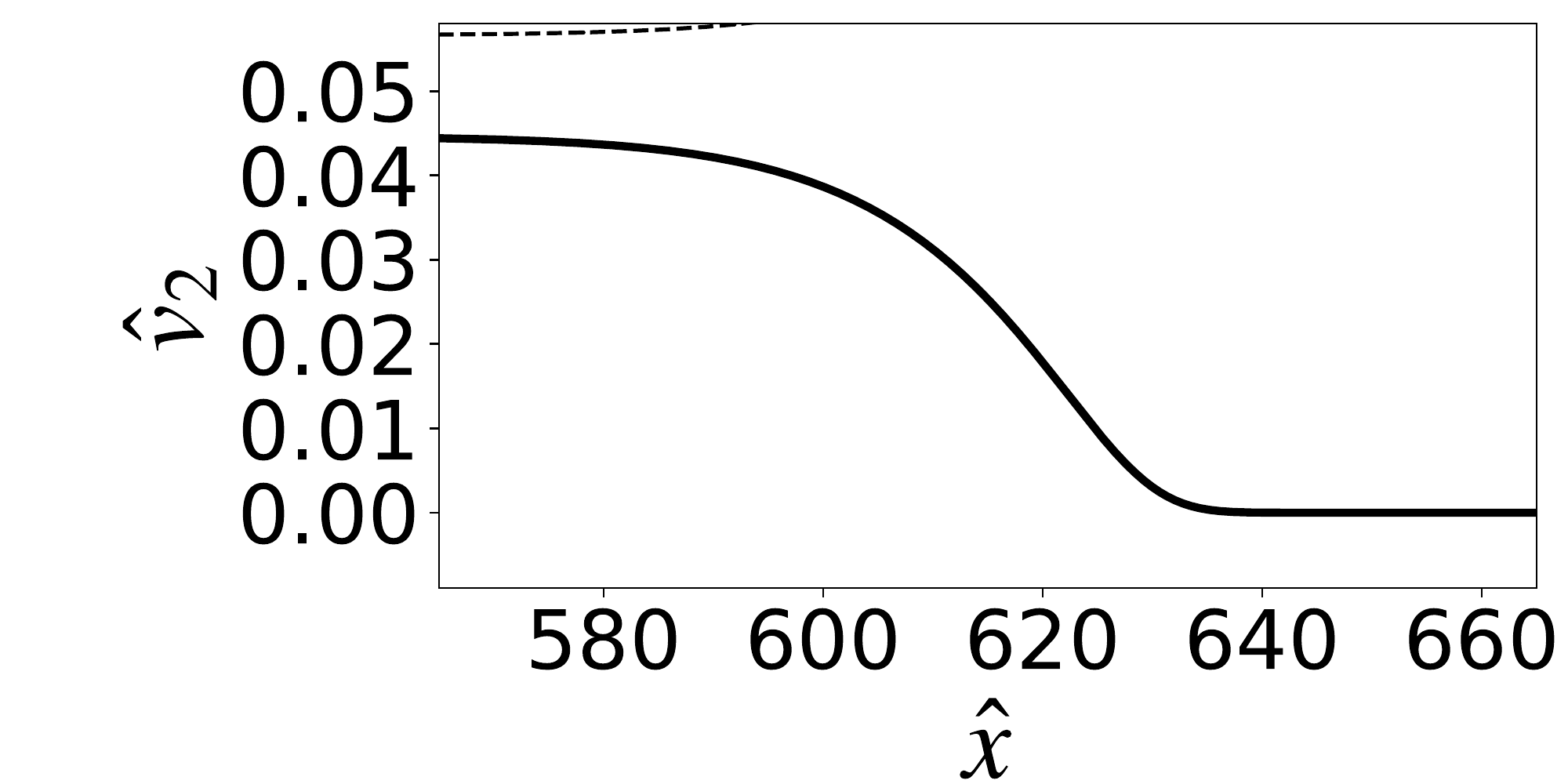}
		\includegraphics[width=0.49\linewidth]{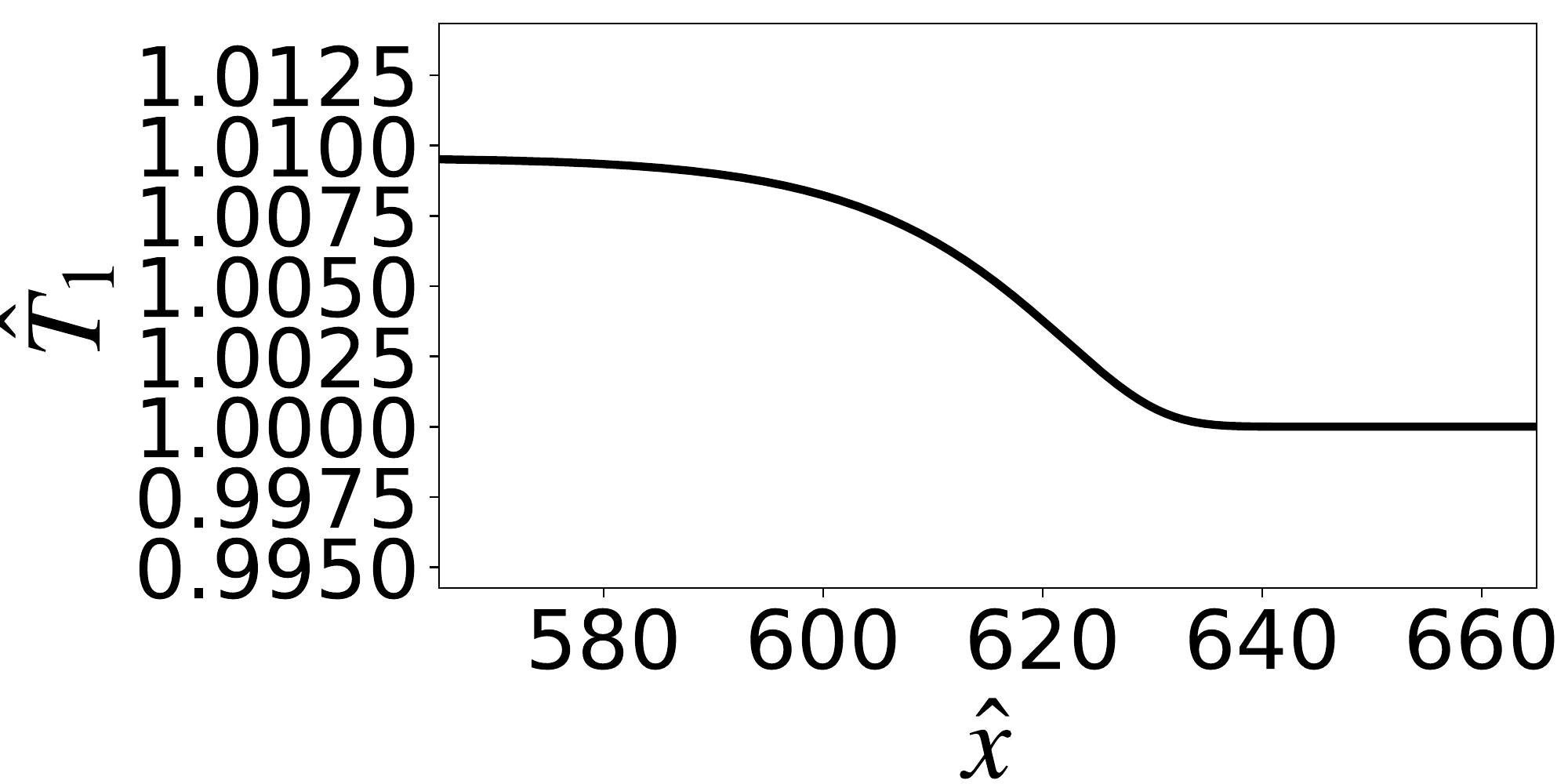} %
		\includegraphics[width=0.49\linewidth]{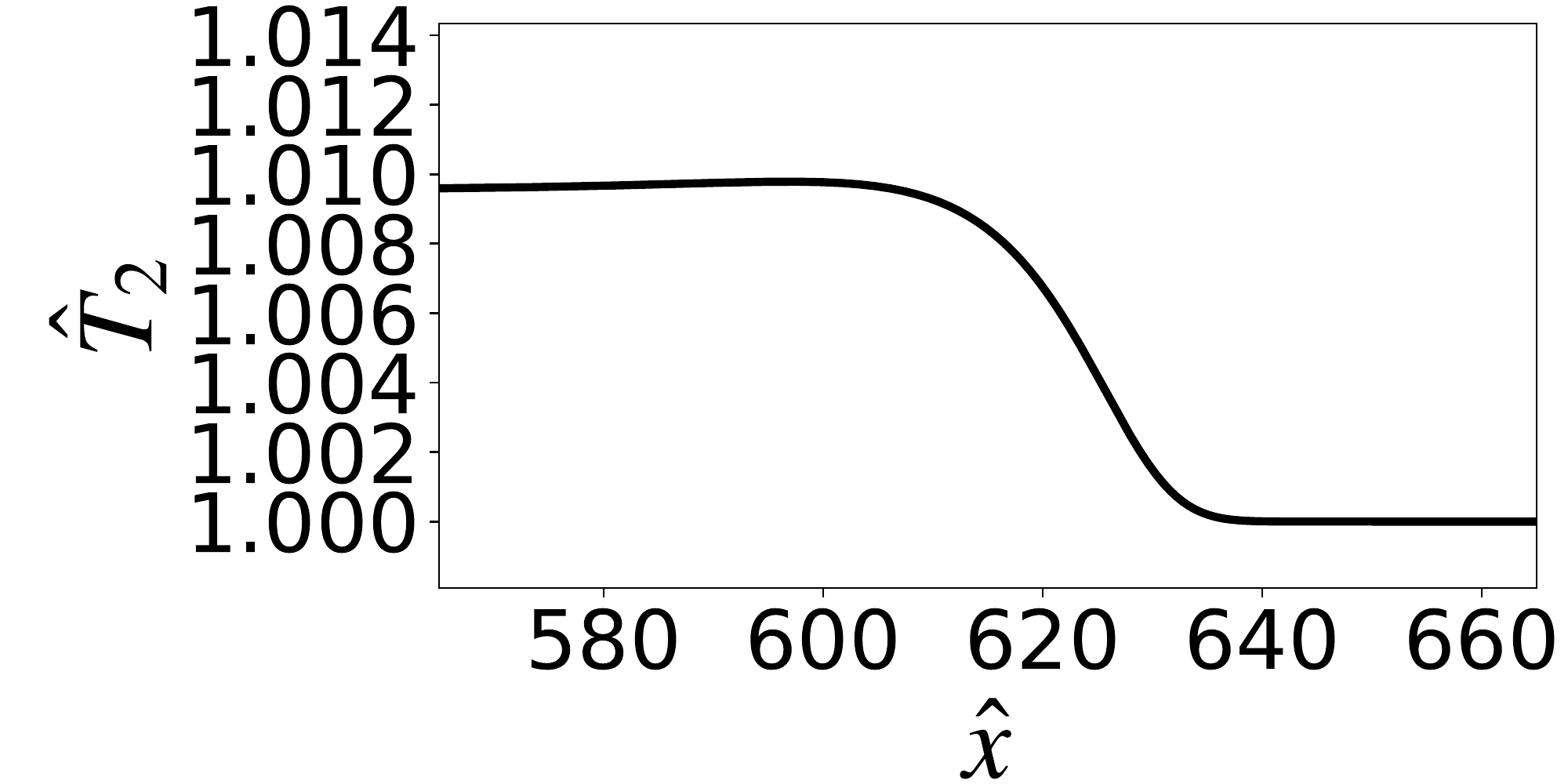}
	\end{center}
	\caption{Case B$_1$: Shock structure in a binary Eulerian mixture of polyatomic and monatomic gases obtained at $\hat{t} = 600$. 
	The parameters are in Region I and correspond to the mark of No. 5 shown in Figure \ref{fig:subshockEuler_mu055}; $\gamma_1 = 7/6$, $\gamma_2 = 5/3$, $\mu = 0.55$, $c_0 = 0.57$, and $M_0 = 1.025$. 
	The numerical conditions are $\Delta \hat{t} = 0.02$ and $\Delta \hat{x}=0.08$. }
	\label{fig:c057_M0-1_025}
\end{figure}

\begin{figure}
	\begin{center}
		\includegraphics[width=0.49\linewidth]{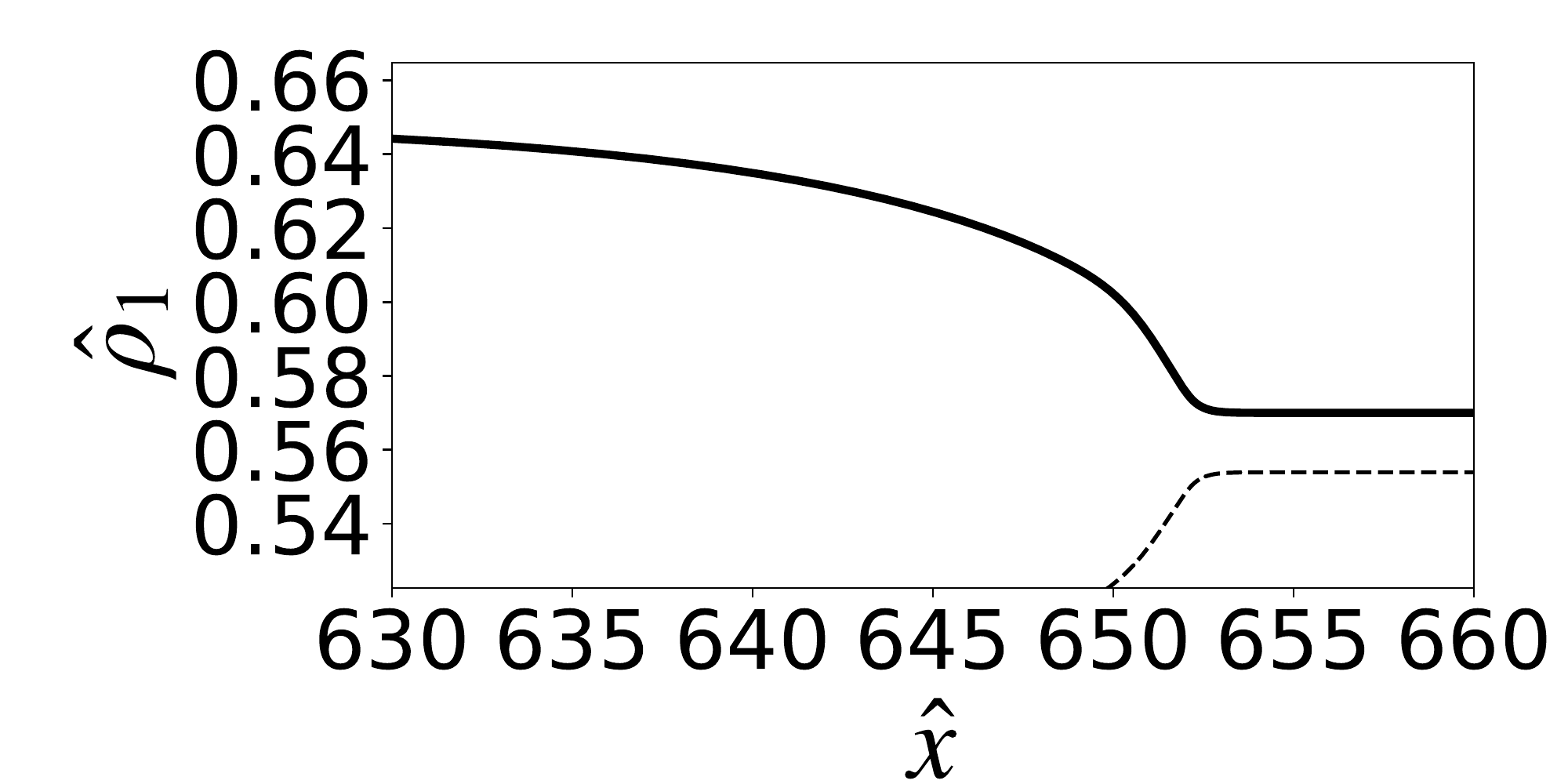} %
		\includegraphics[width=0.49\linewidth]{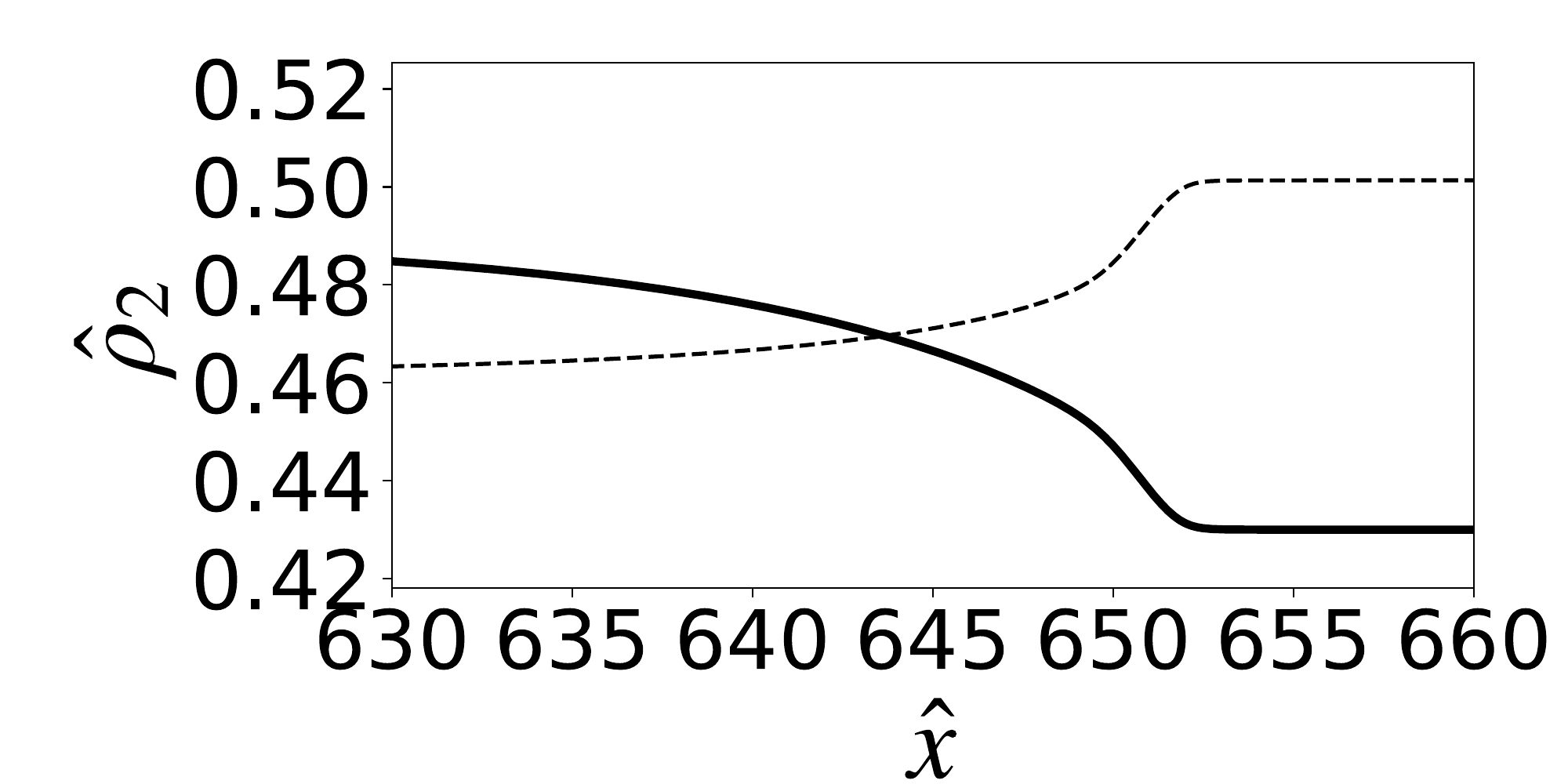}
		\includegraphics[width=0.49\linewidth]{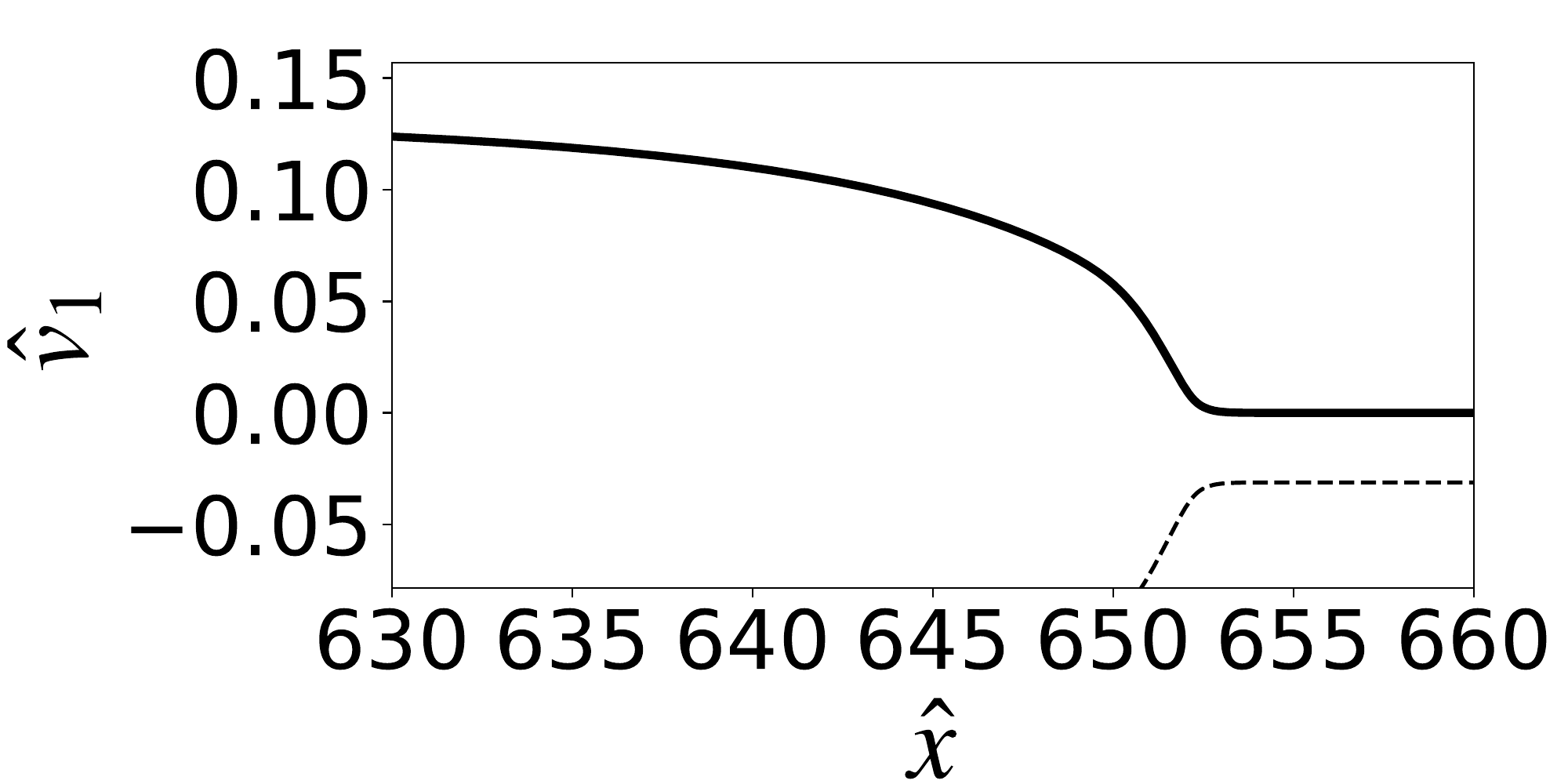} %
		\includegraphics[width=0.49\linewidth]{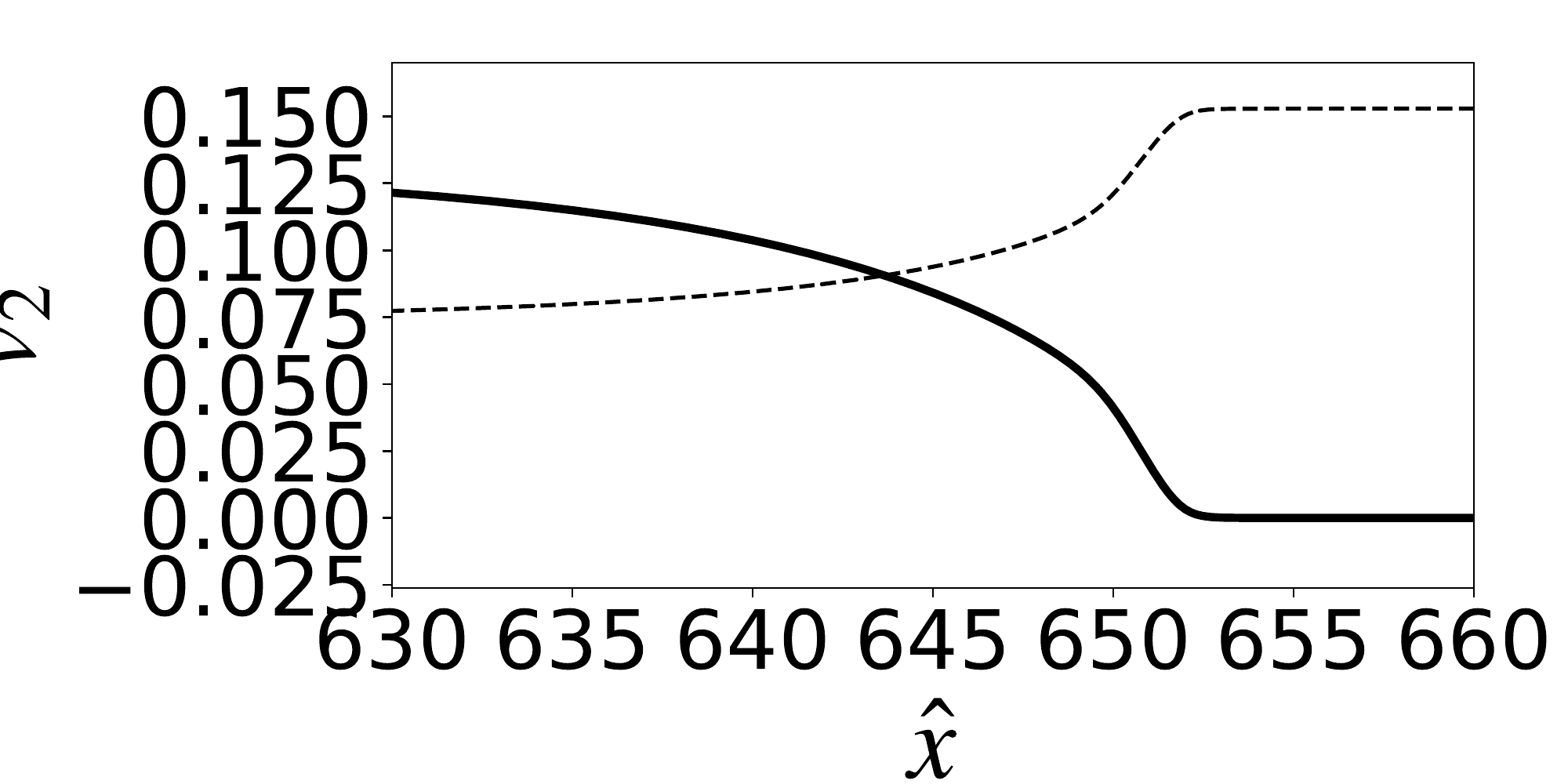}
		\includegraphics[width=0.49\linewidth]{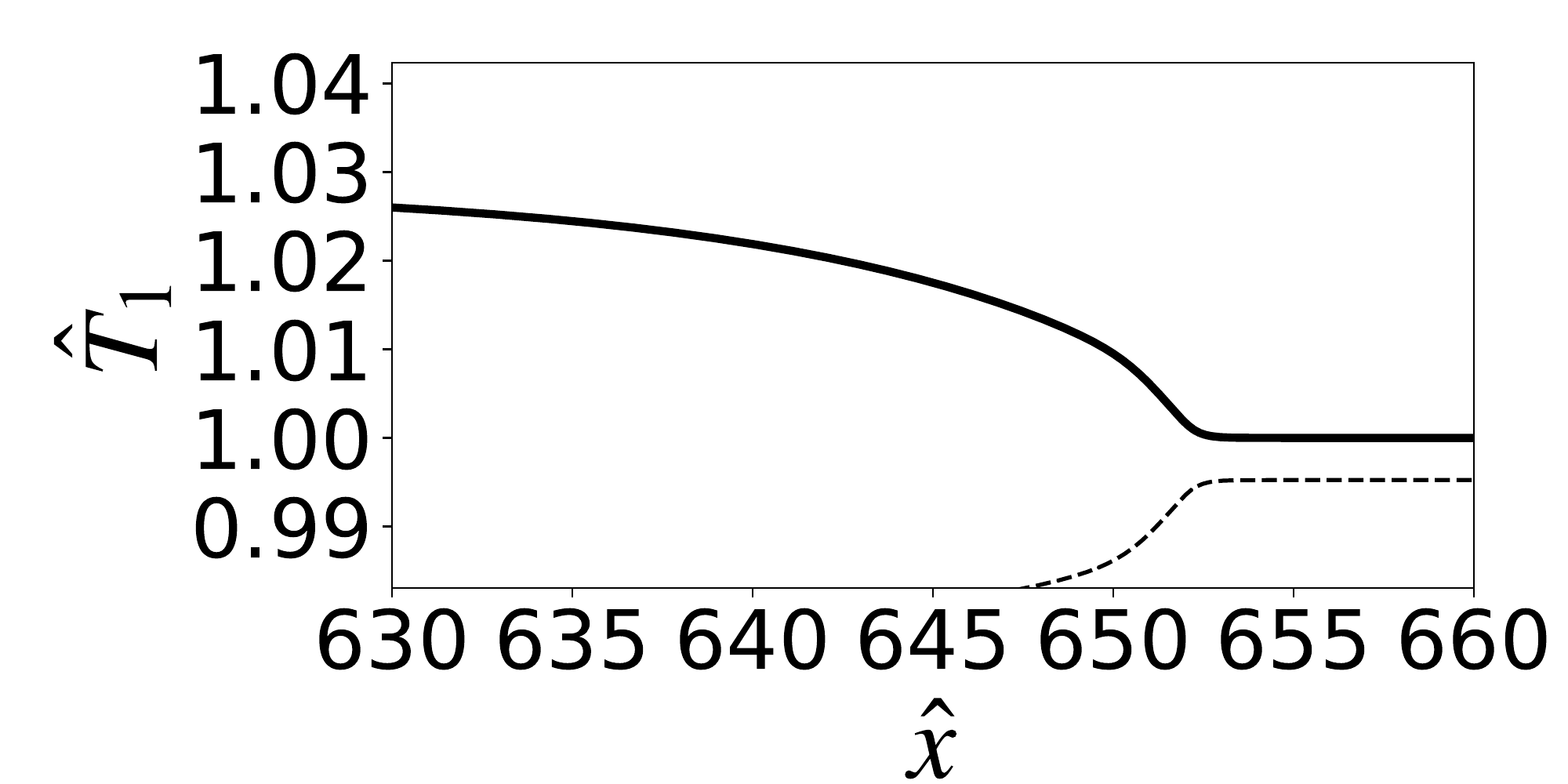} %
		\includegraphics[width=0.49\linewidth]{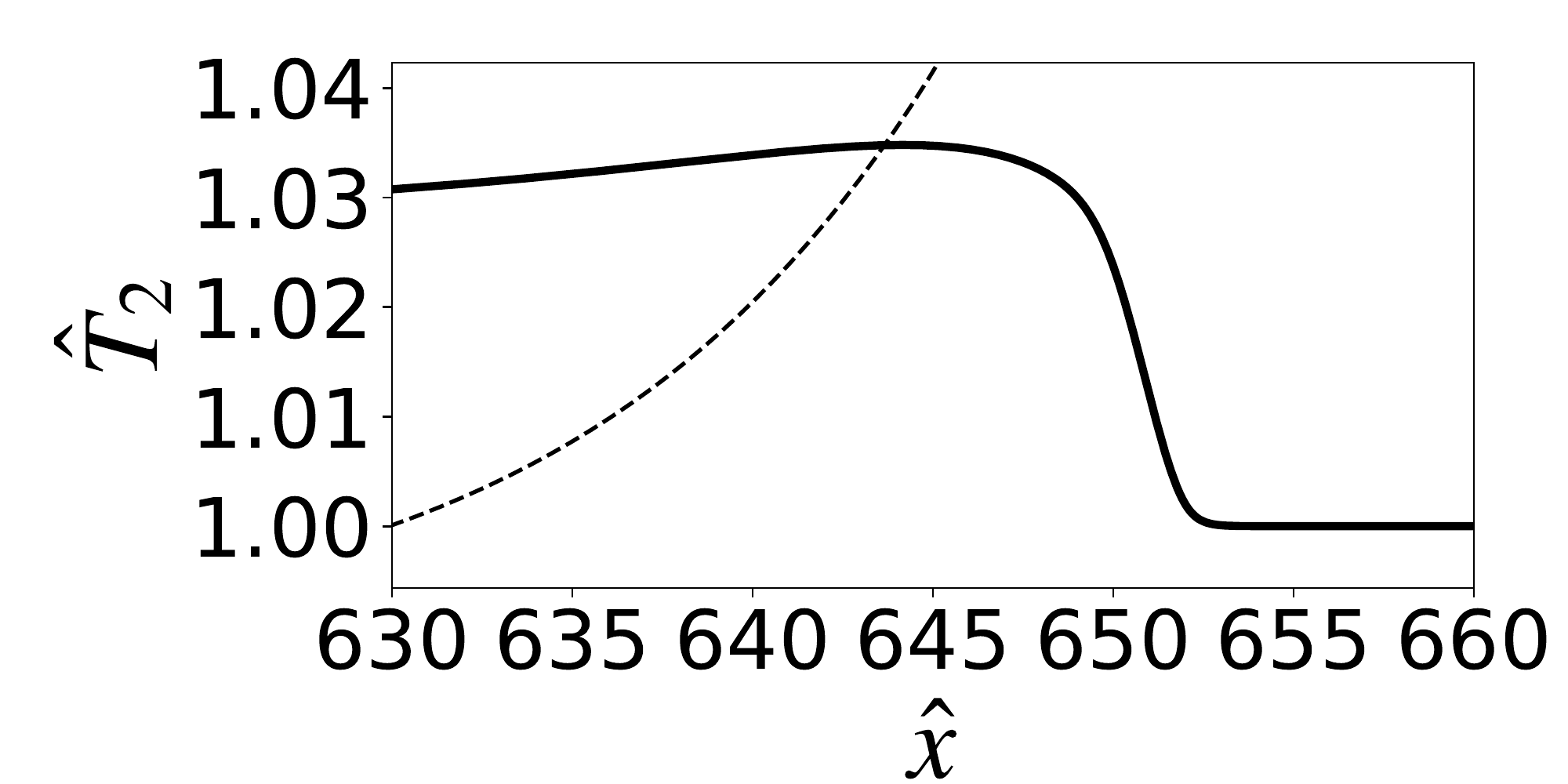}
	\end{center}
	\hspace{0.49\linewidth}
	\includegraphics[width=0.49\linewidth]{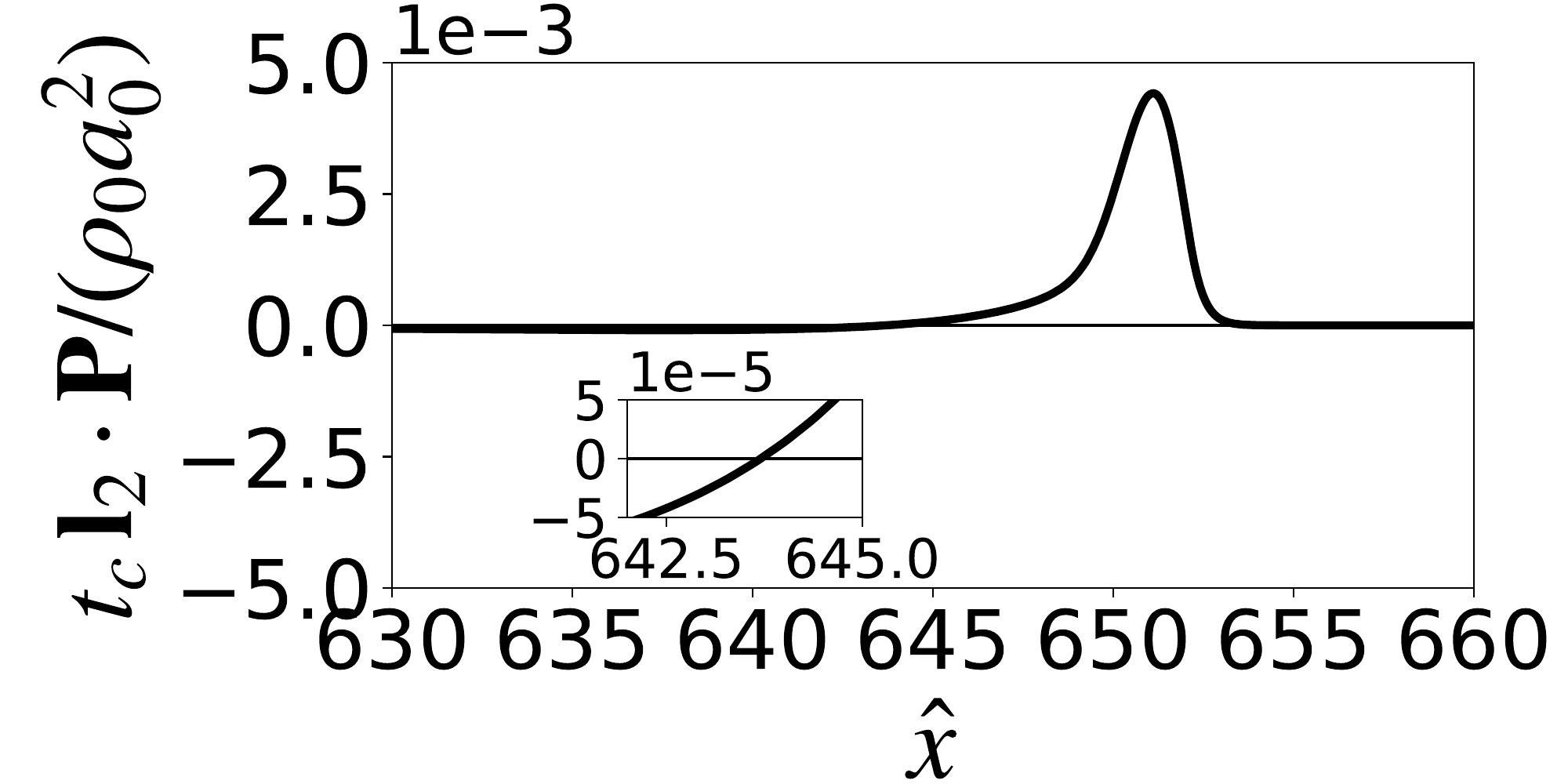}
	\caption{Case B$_1$: Shock structure in a binary Eulerian mixture of polyatomic and monatomic gases obtained at $\hat{t} = 600$. 
	The parameters are in Region II and correspond to the mark of No. 6 shown in Figure \ref{fig:subshockEuler_mu055}; $\gamma_1 = 7/6$, $\gamma_2 = 5/3$, $\mu = 0.55$, $c_0 = 0.57$, and $M_0 = 1.075$. 
	The numerical conditions are $\Delta \hat{t} = 0.02$ and $\Delta \hat{x}=0.08$. }
	\label{fig:c057_M0-1_075}
\end{figure}

\begin{figure}
	\begin{center}
		\includegraphics[width=0.49\linewidth]{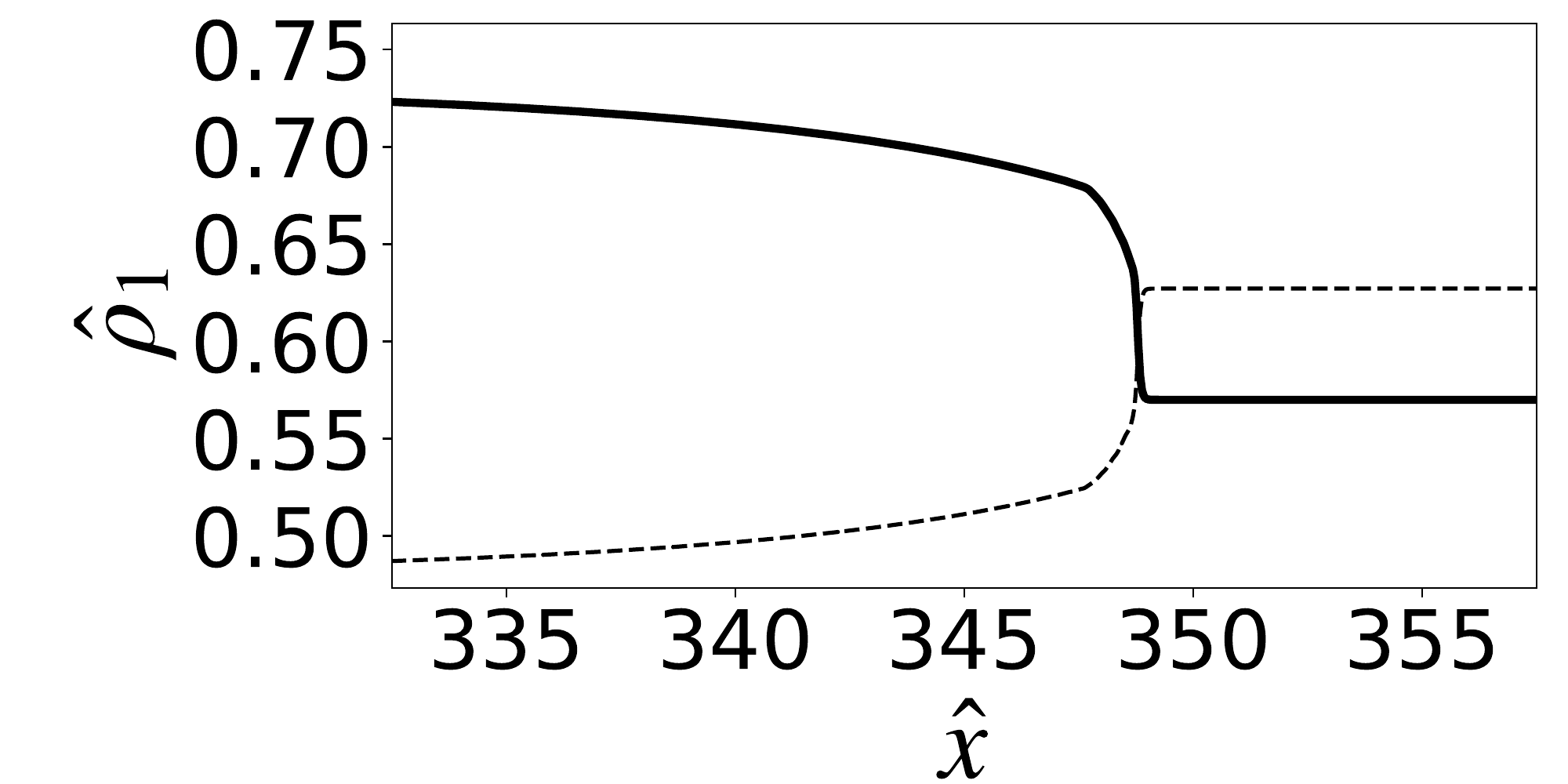} %
		\includegraphics[width=0.49\linewidth]{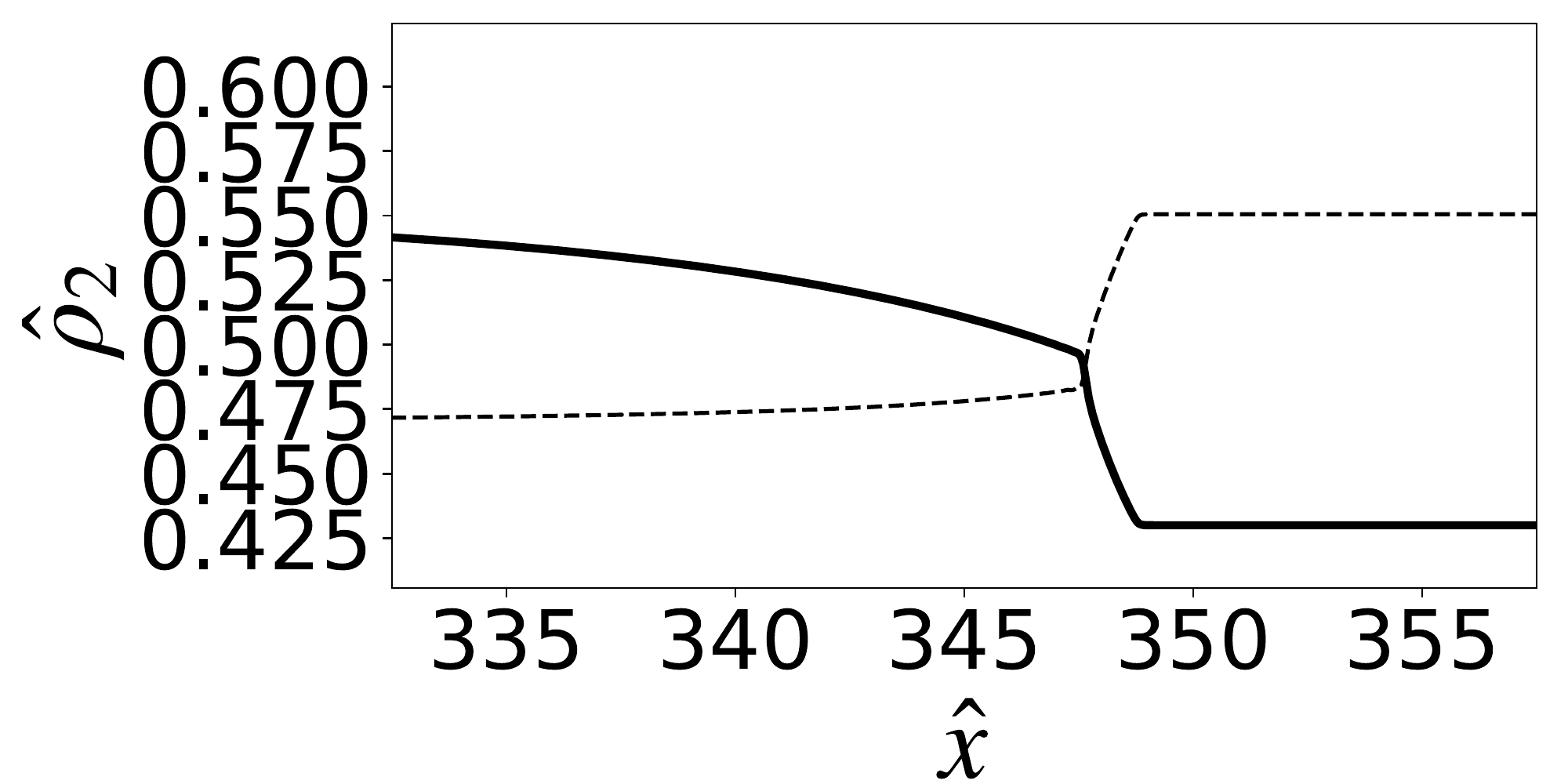}
		\includegraphics[width=0.49\linewidth]{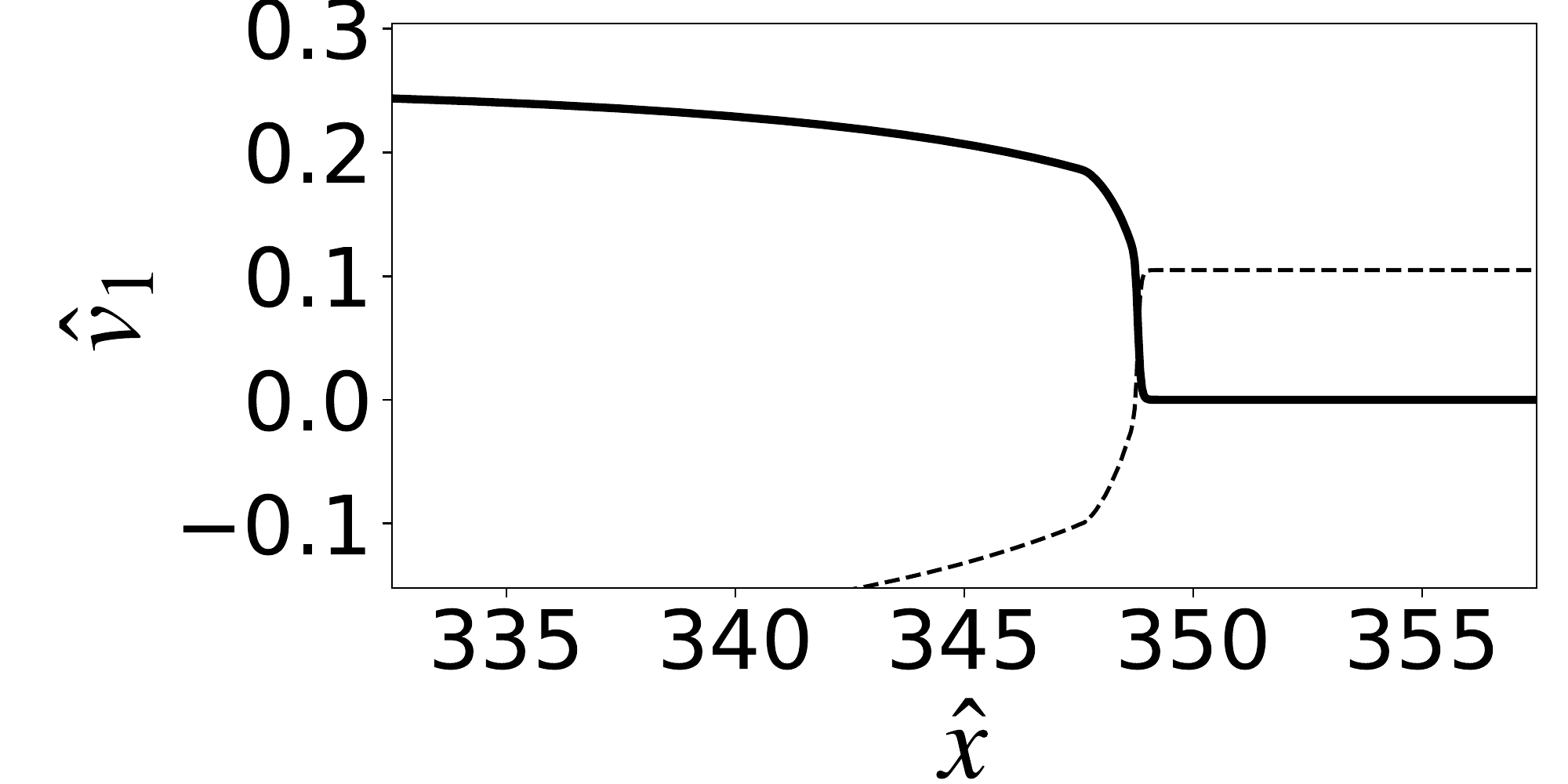} %
		\includegraphics[width=0.49\linewidth]{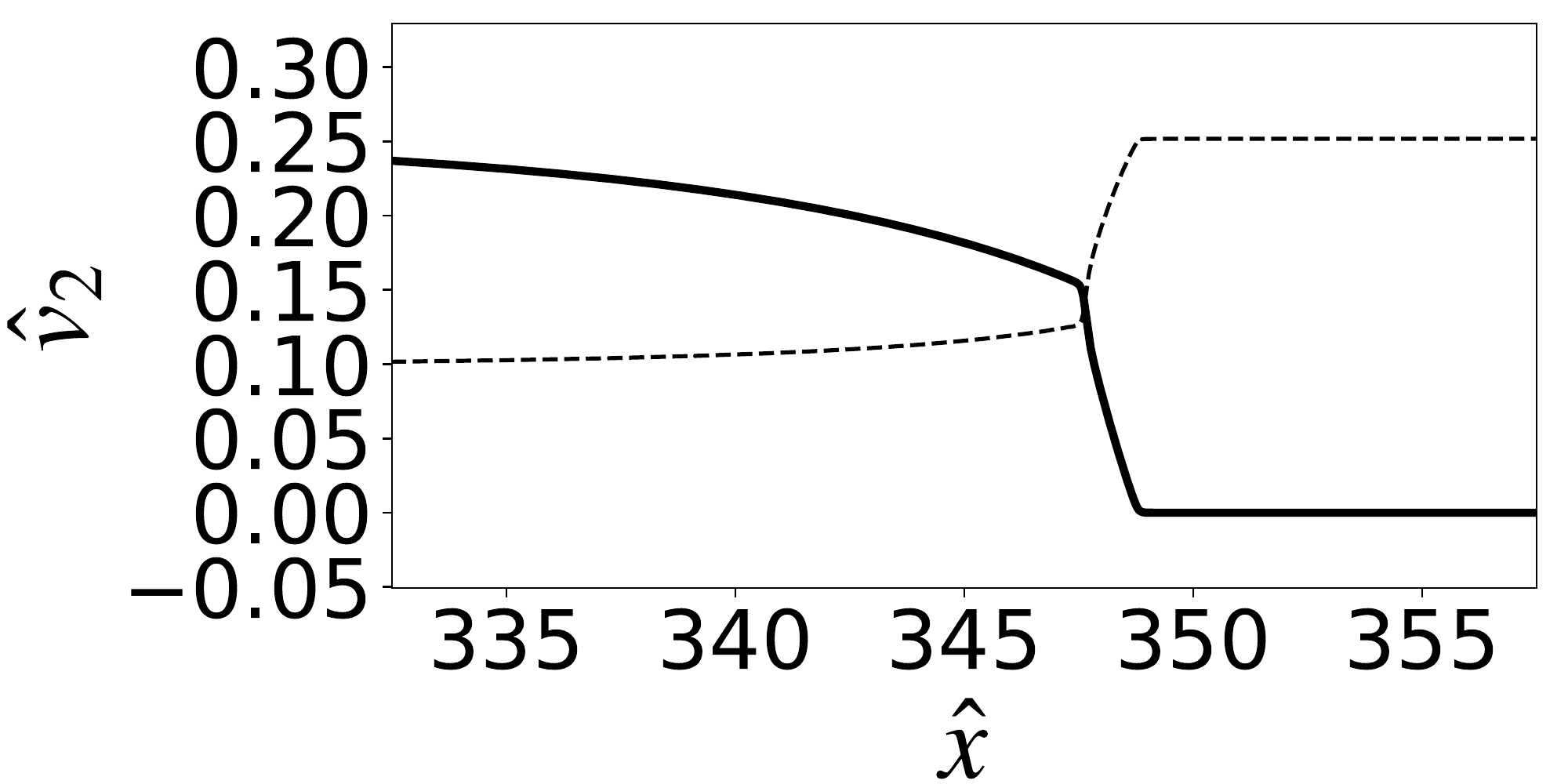}
		\includegraphics[width=0.49\linewidth]{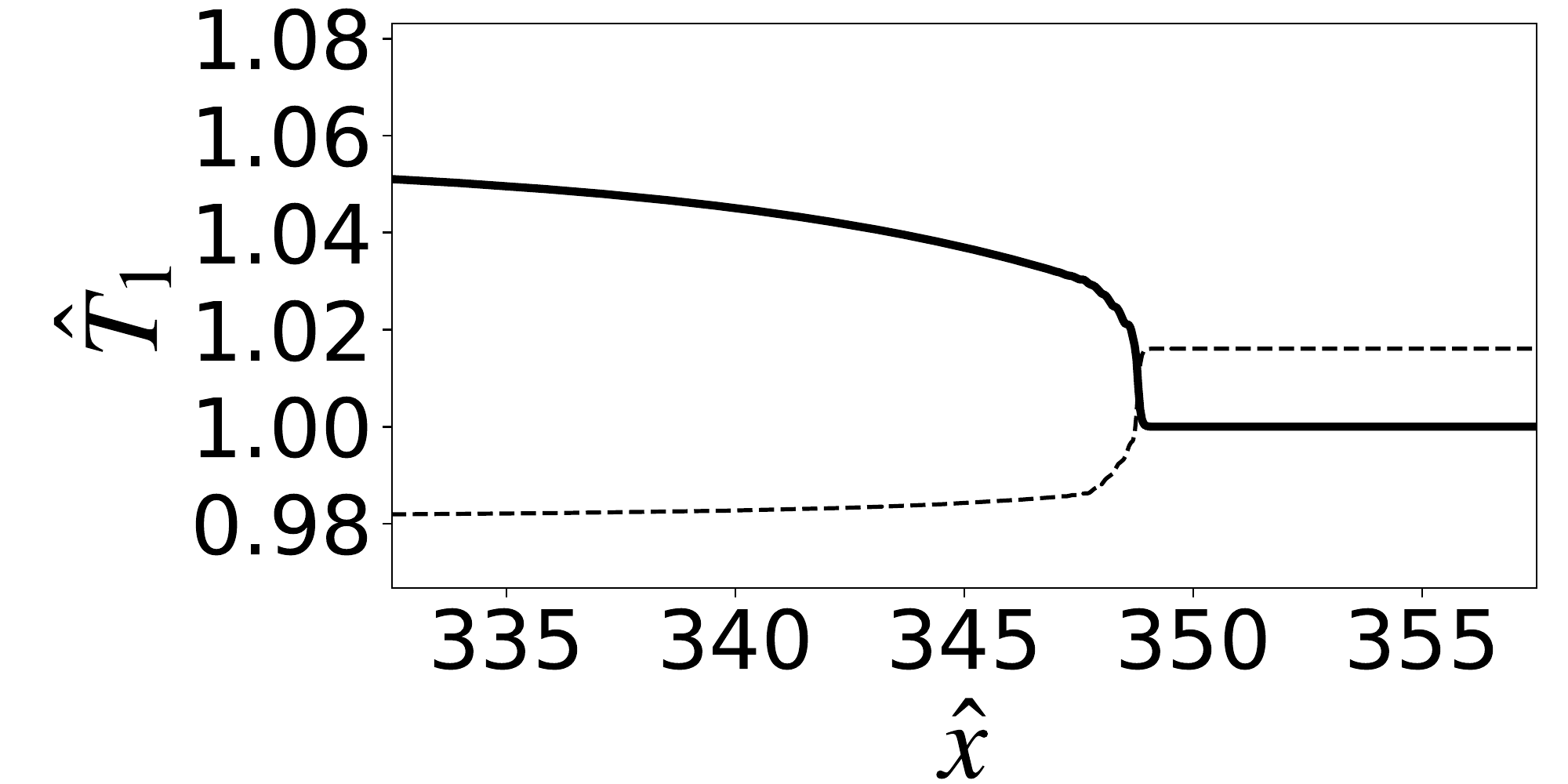} %
		\includegraphics[width=0.49\linewidth]{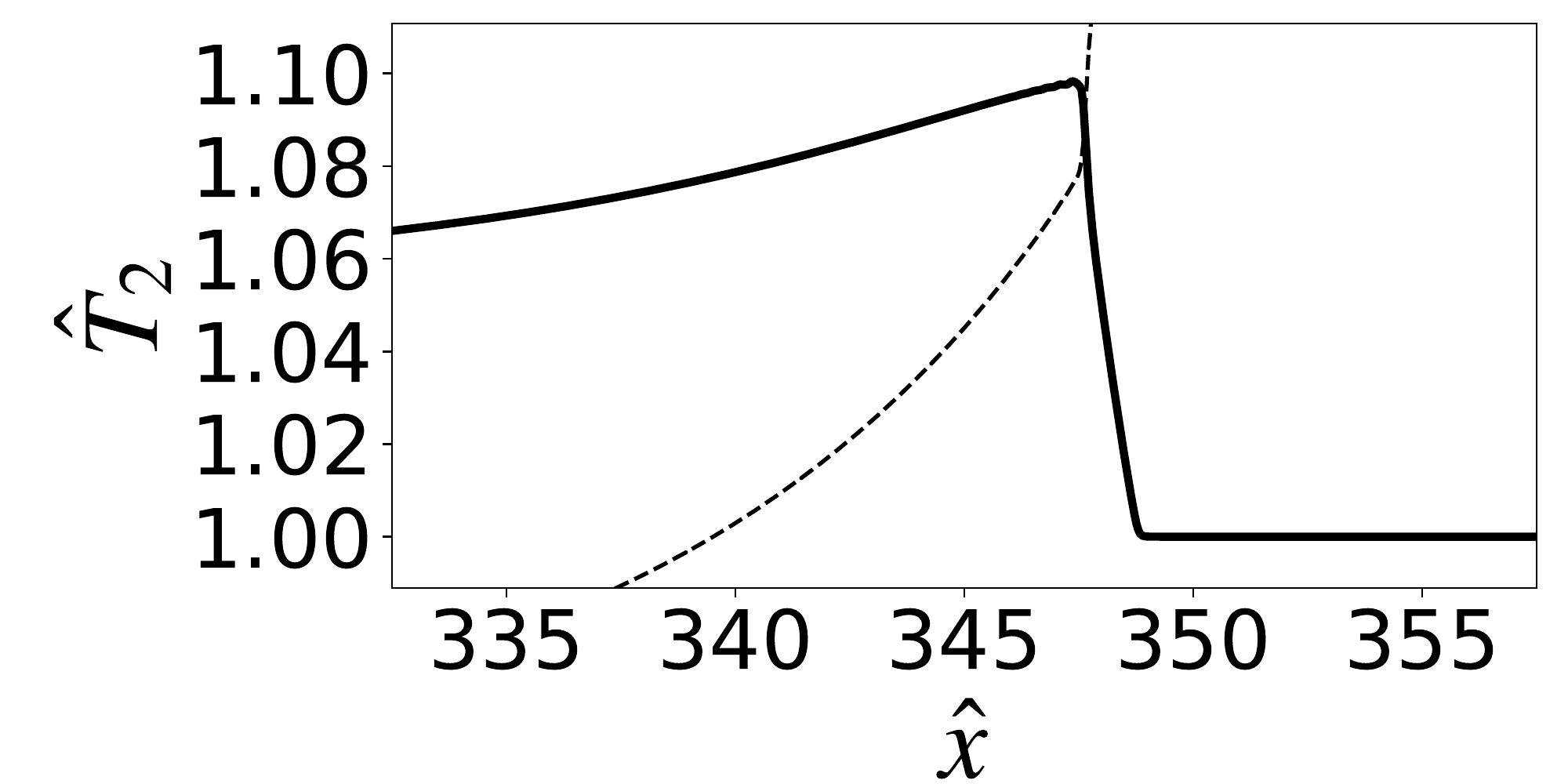}
	\end{center}
	\caption{Case B$_1$: Shock structure in a binary Eulerian mixture of polyatomic and monatomic gases obtained at $\hat{t} = 300$. 
	The parameters are in Region IV and correspond to the mark of No. 7 shown in Figure \ref{fig:subshockEuler_mu055}; $\gamma_1 = 7/6$, $\gamma_2 = 5/3$, $\mu = 0.55$, $c_0 = 0.57$, and $M_0 = 1.15$. 
	The numerical conditions are $\Delta \hat{t} = 0.01$ and $\Delta \hat{x}=0.04$. }
	\label{fig:c057_M0-1_15}
\end{figure}

In the case of $c_0 = 0.57$, we show the shock structure for $M_0 = 1.025, 1.075, 1.15$ in Figures \ref{fig:c057_M0-1_025} -- \ref{fig:c057_M0-1_15}. 
We confirm the similar behavior that continuous shock structure for $M_0 = 1.025$ becomes the shock structure with a regular singular point for $M_0 = 1.075$ and then the multiple sub-shocks appear for $M_0 = 1.15$. 

In summary, in the present case, we understand that the singular point may become regular for the moderately large Mach number. 
However, the corresponding singular point may become singular for the larger Mach number even with the same production terms. 
Therefore it is impossible to conclude the possibility of the sub-shock formation a priori even if we write down the production terms explicitly. 
The numerical analysis of the shock-structure solution is necessary to conclude whether the sub-shock forms or not if we adopt the parameters. 

By performing many numerical calculations of the shock structure, in principle, we obtain the boundary between the regular singular and singular points. 
For classification of the regions in more detail, the method of dynamical analysis of the ODE system is also useful. 
Please see Appendix \ref{sec:dynamical} for detail of the method.

\subsection{Shock structure in Case B$_2$}

\begin{figure}
	\begin{center}
		\includegraphics[width=0.49\linewidth]{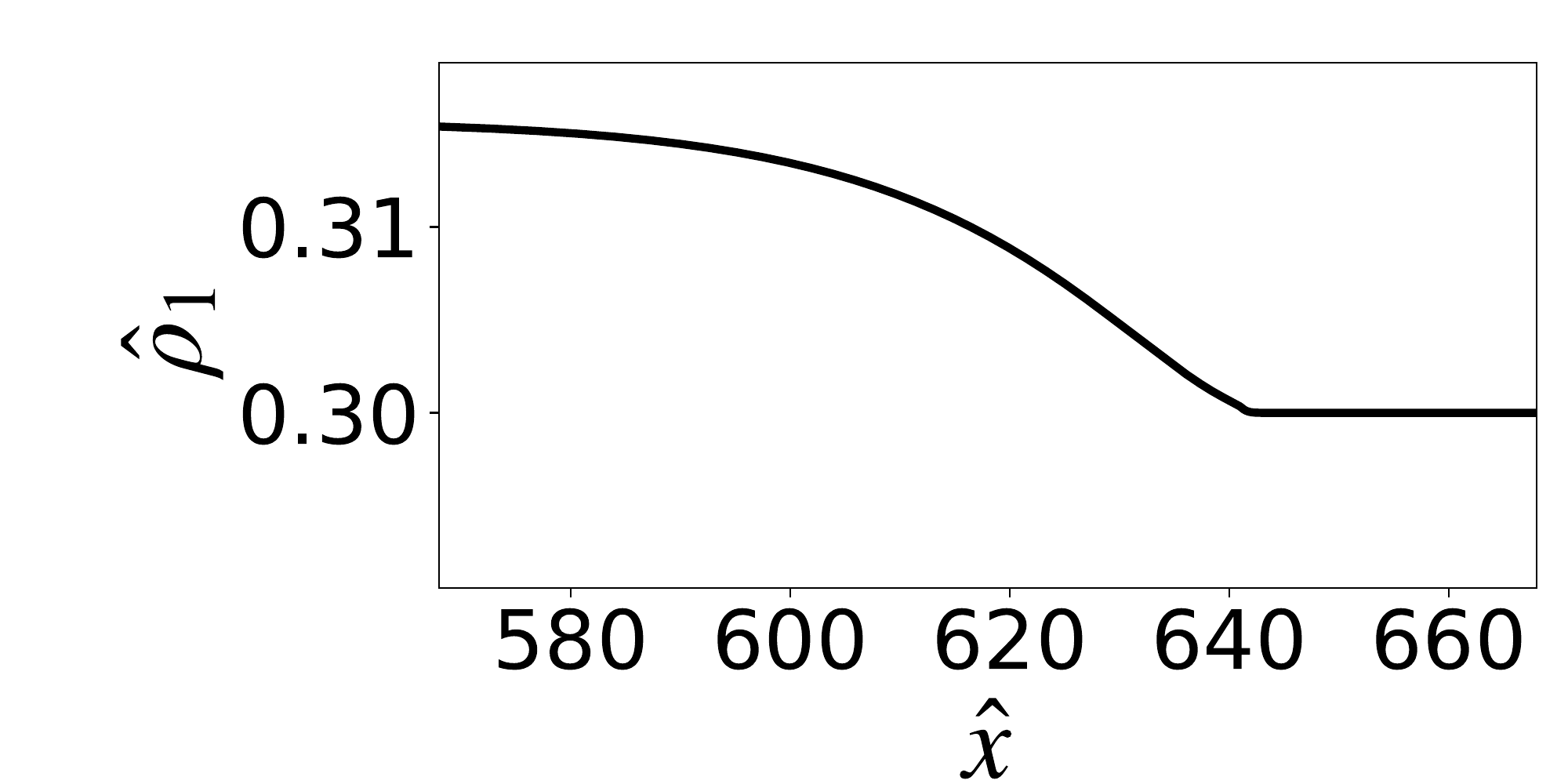} %
		\includegraphics[width=0.49\linewidth]{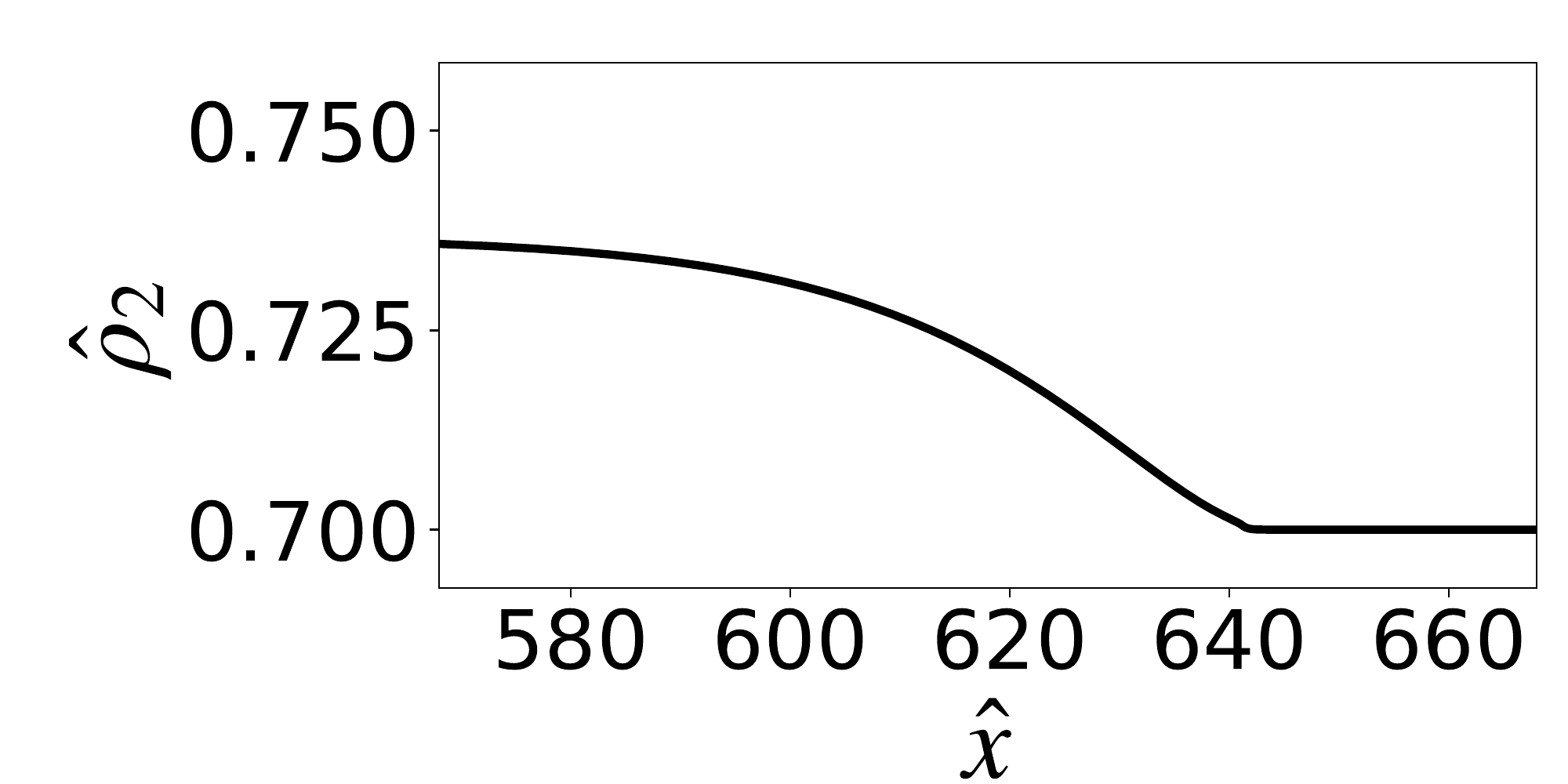}
		\includegraphics[width=0.49\linewidth]{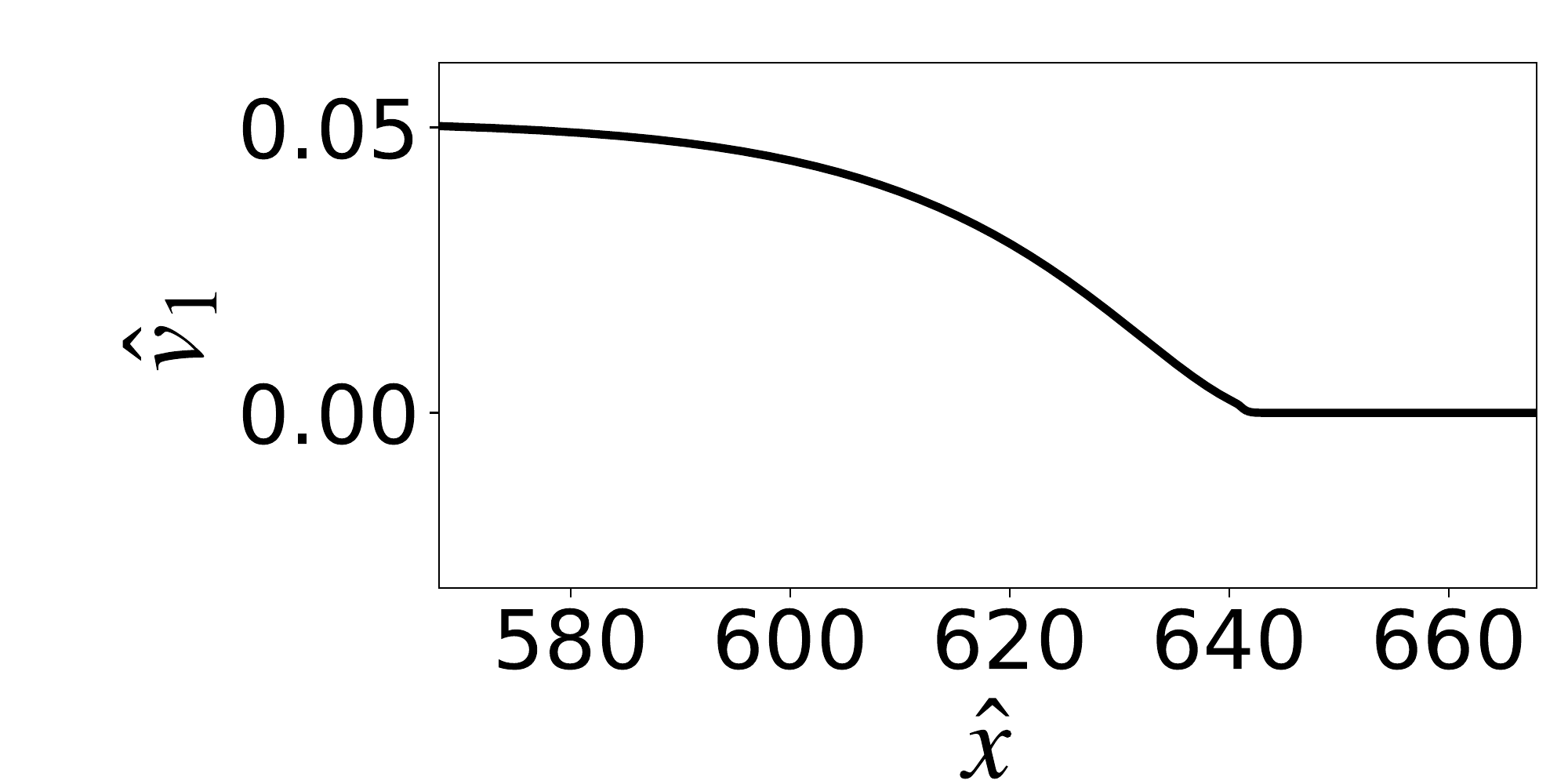} %
		\includegraphics[width=0.49\linewidth]{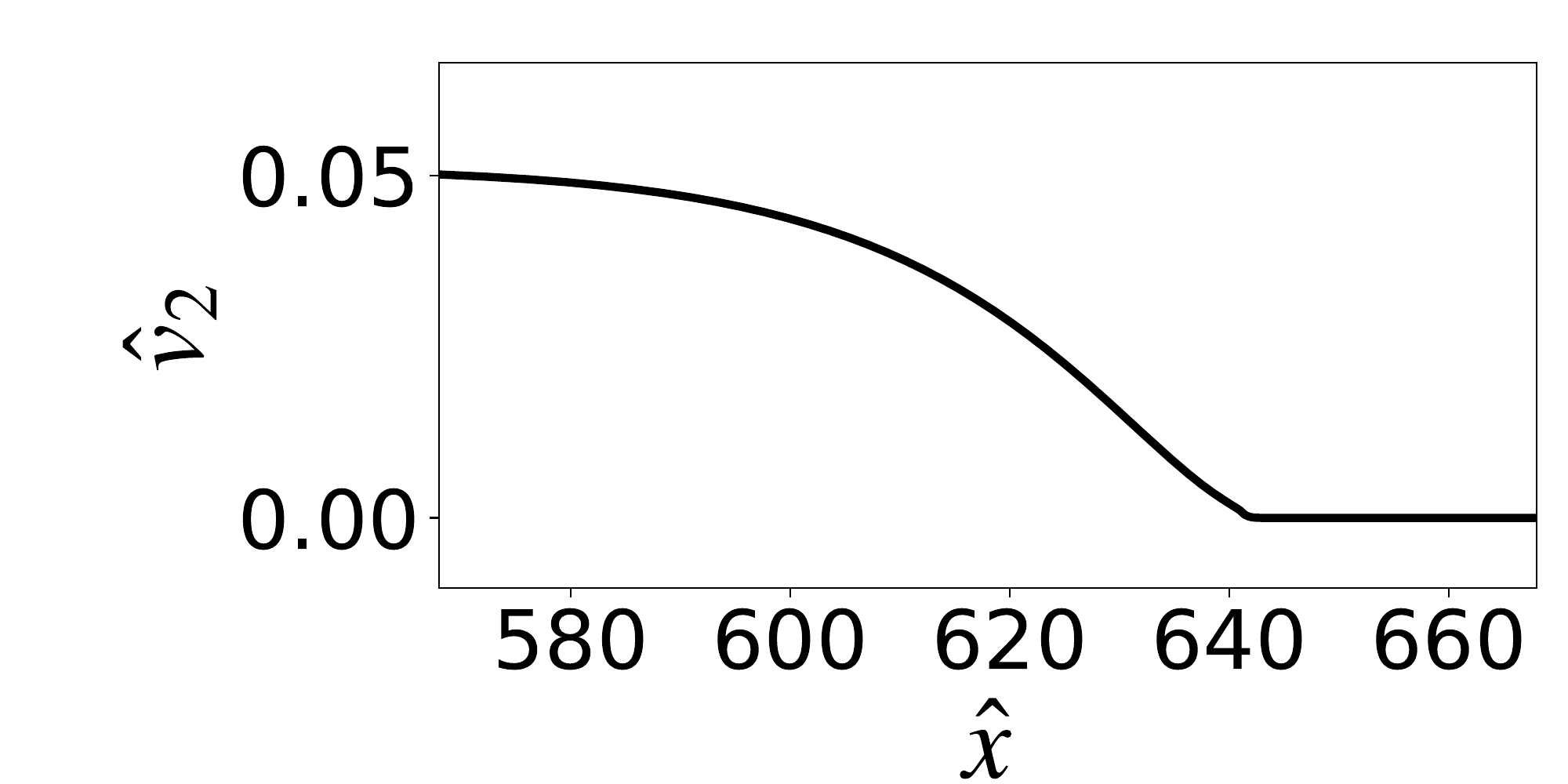}
		\includegraphics[width=0.49\linewidth]{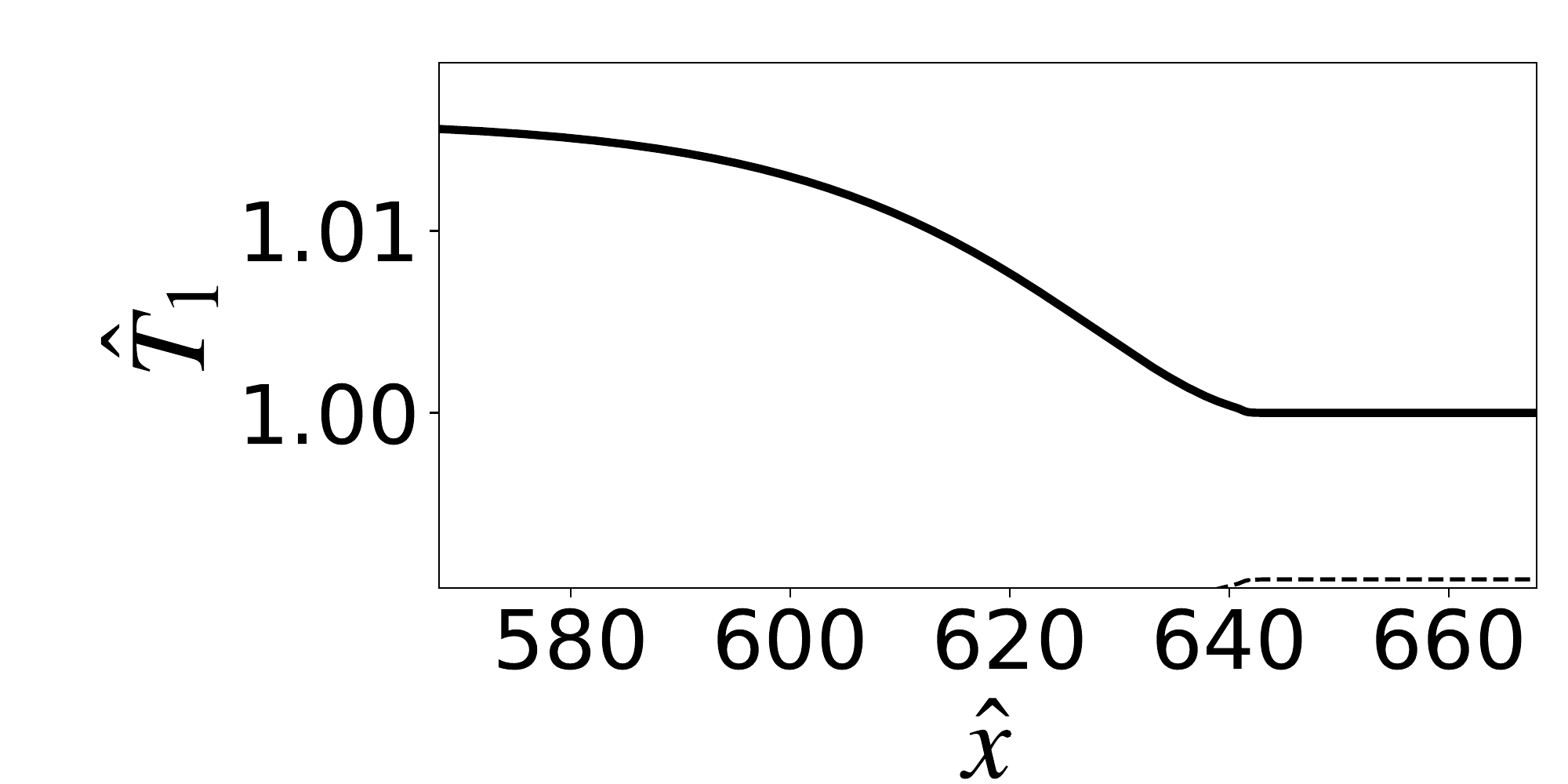} %
		\includegraphics[width=0.49\linewidth]{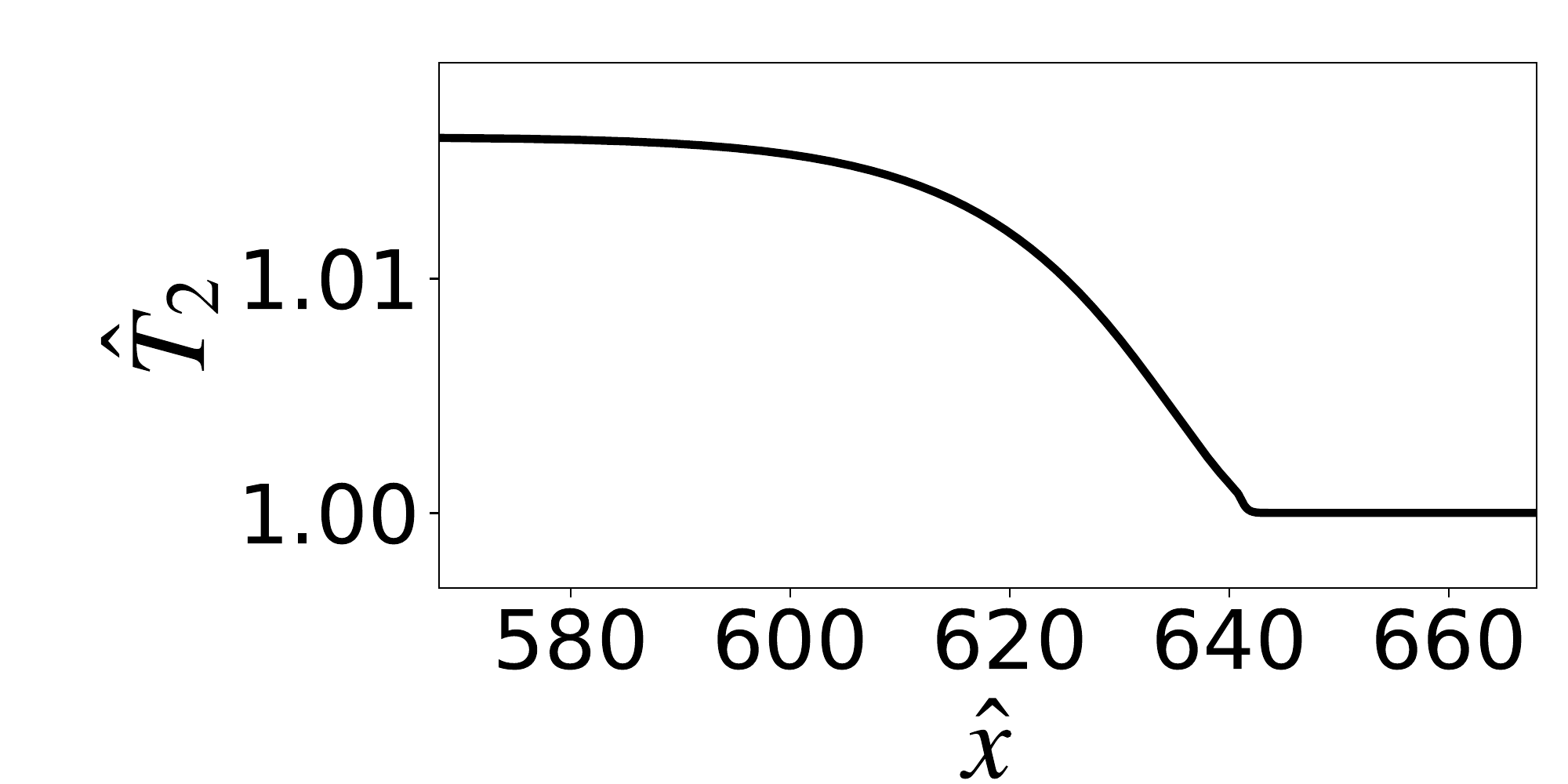}
	\end{center}
	\caption{Case B$_2$: Shock structure in a binary Eulerian mixture of polyatomic and monatomic gases obtained at $\hat{t} = 600$. 
	The parameters are in Region I and correspond to the mark of No. 1 shown in Figure \ref{fig:subshockEuler_mu07}; $\gamma_1 = 7/6$, $\gamma_2 = 5/3$, $\mu = 0.7$, $c_0 = 0.3$, and $M_0 = 1.03$. 
	The numerical conditions are $\Delta \hat{t} = 0.02$ and $\Delta \hat{x}=0.08$. }
	\label{fig:c03_M0-1_03}
\end{figure}

\begin{figure}
	\begin{center}
		\includegraphics[width=0.49\linewidth]{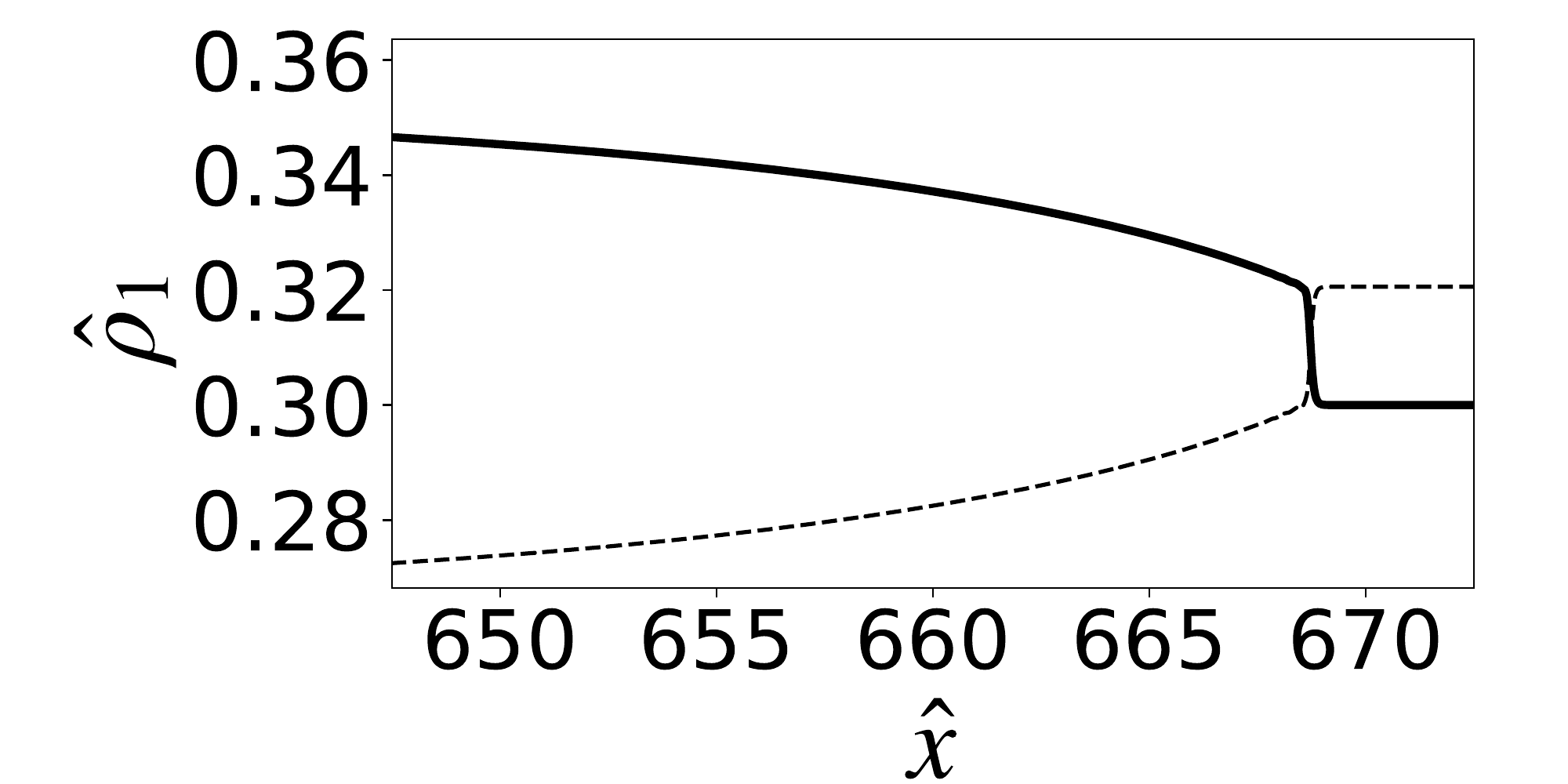} %
		\includegraphics[width=0.49\linewidth]{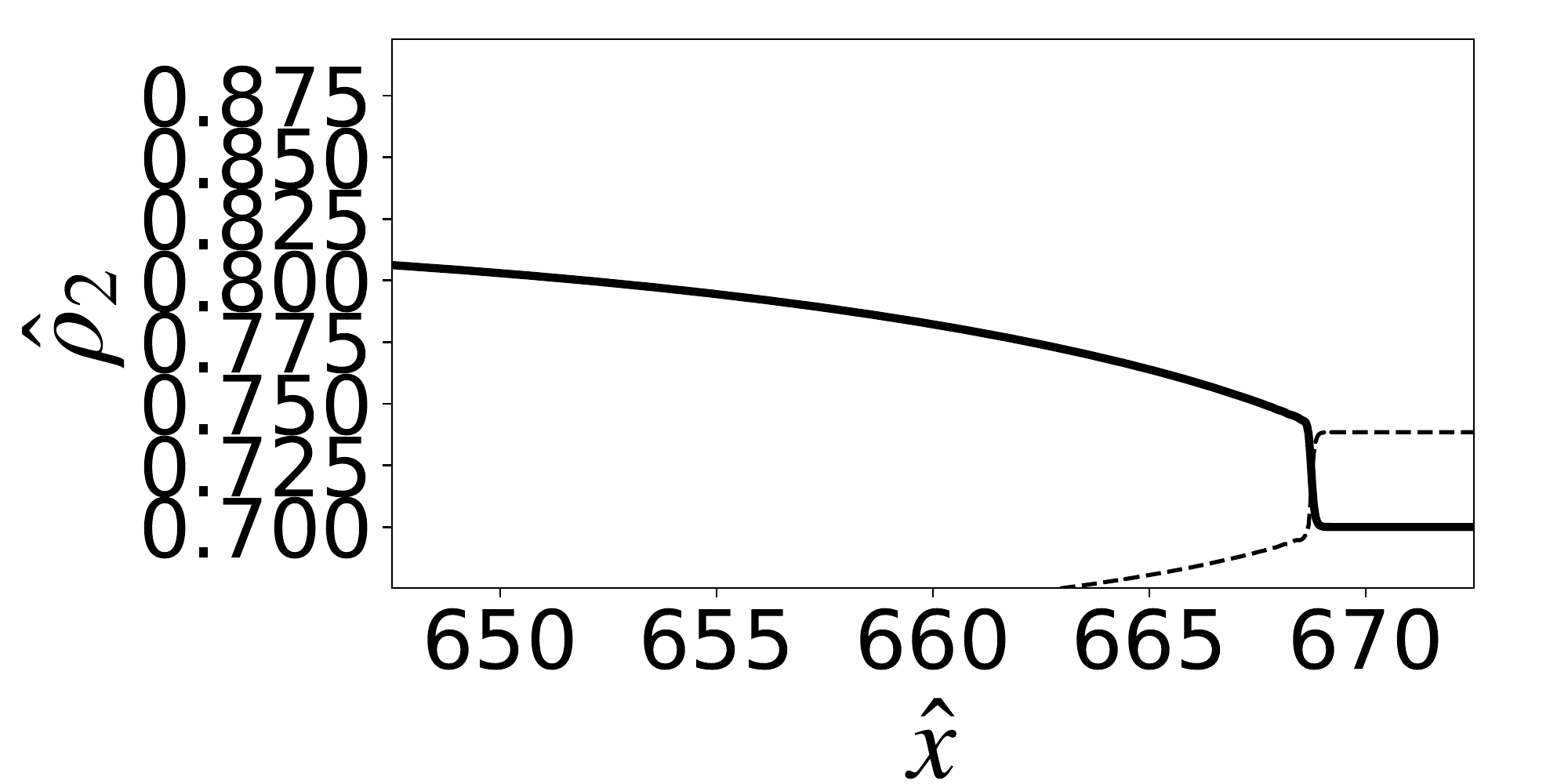}
		\includegraphics[width=0.49\linewidth]{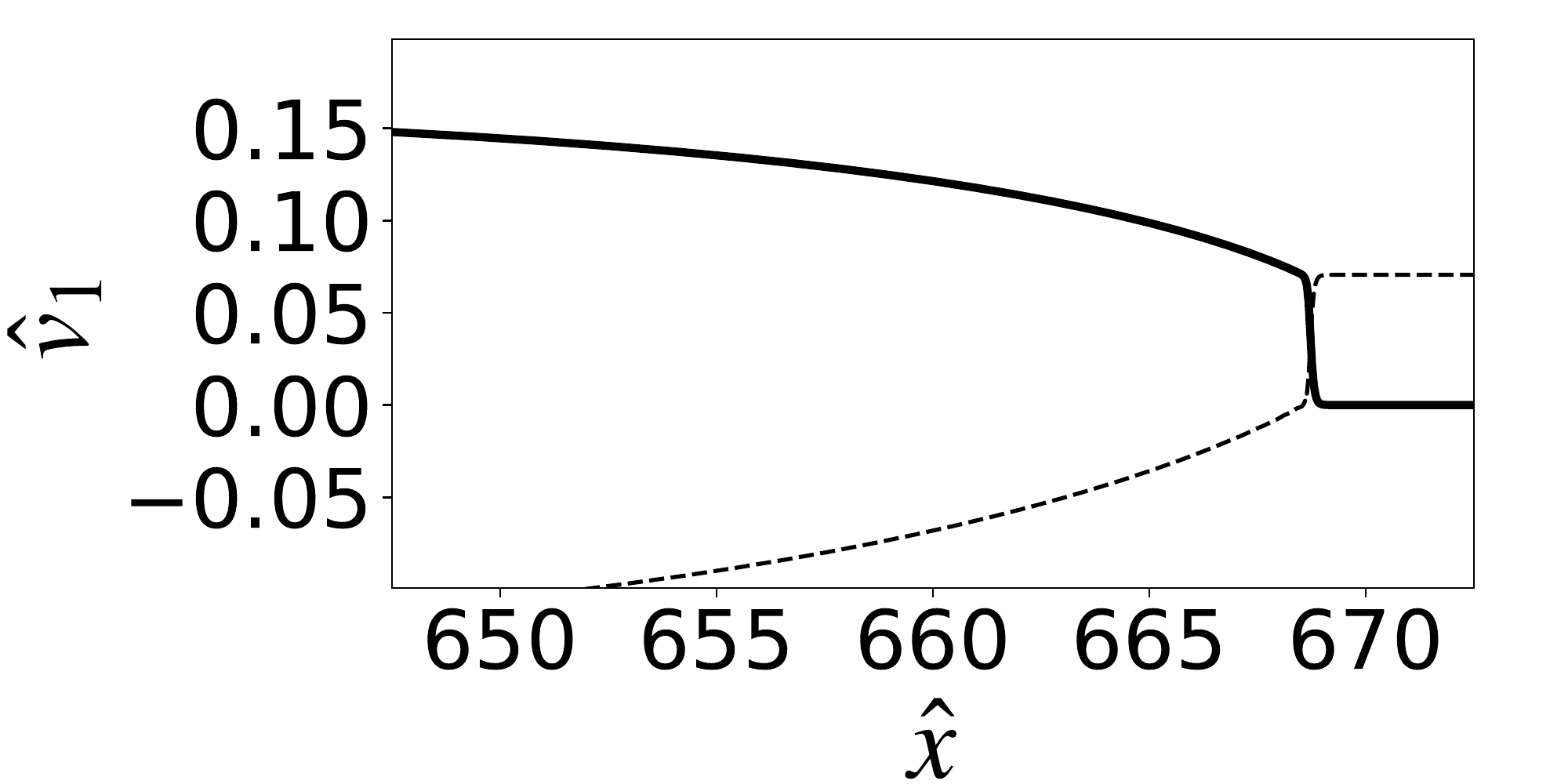} %
		\includegraphics[width=0.49\linewidth]{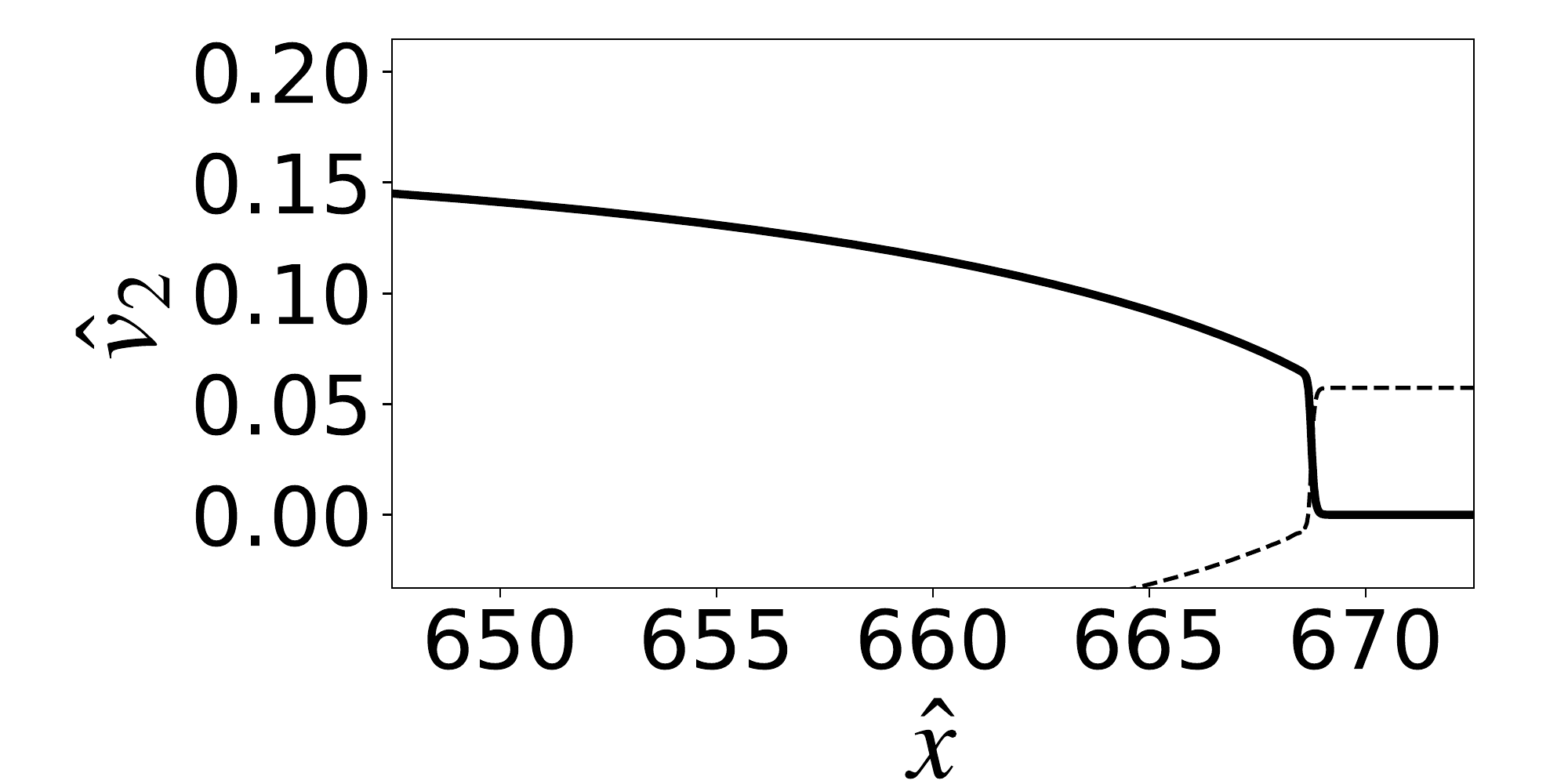}
		\includegraphics[width=0.49\linewidth]{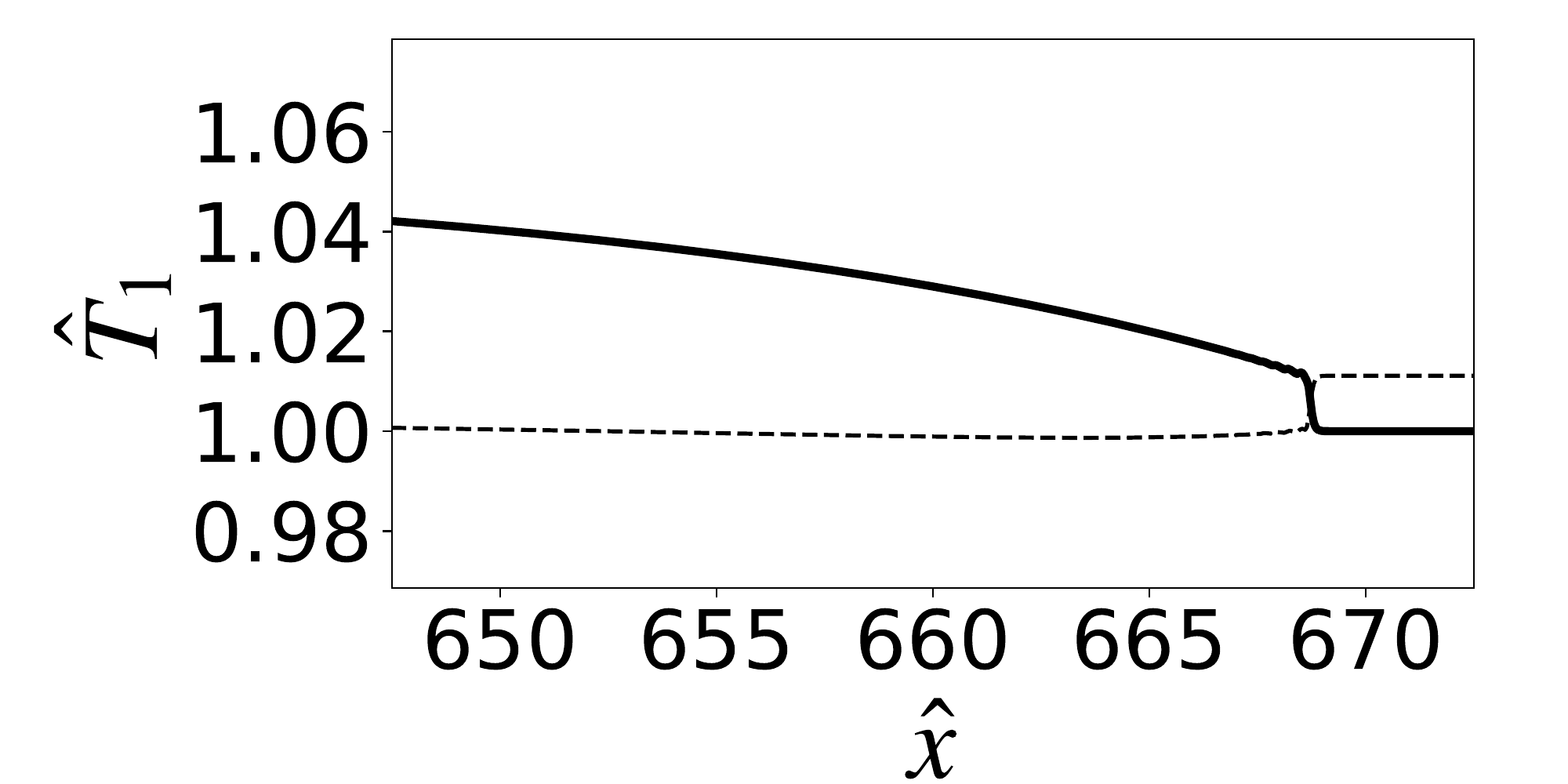} %
		\includegraphics[width=0.49\linewidth]{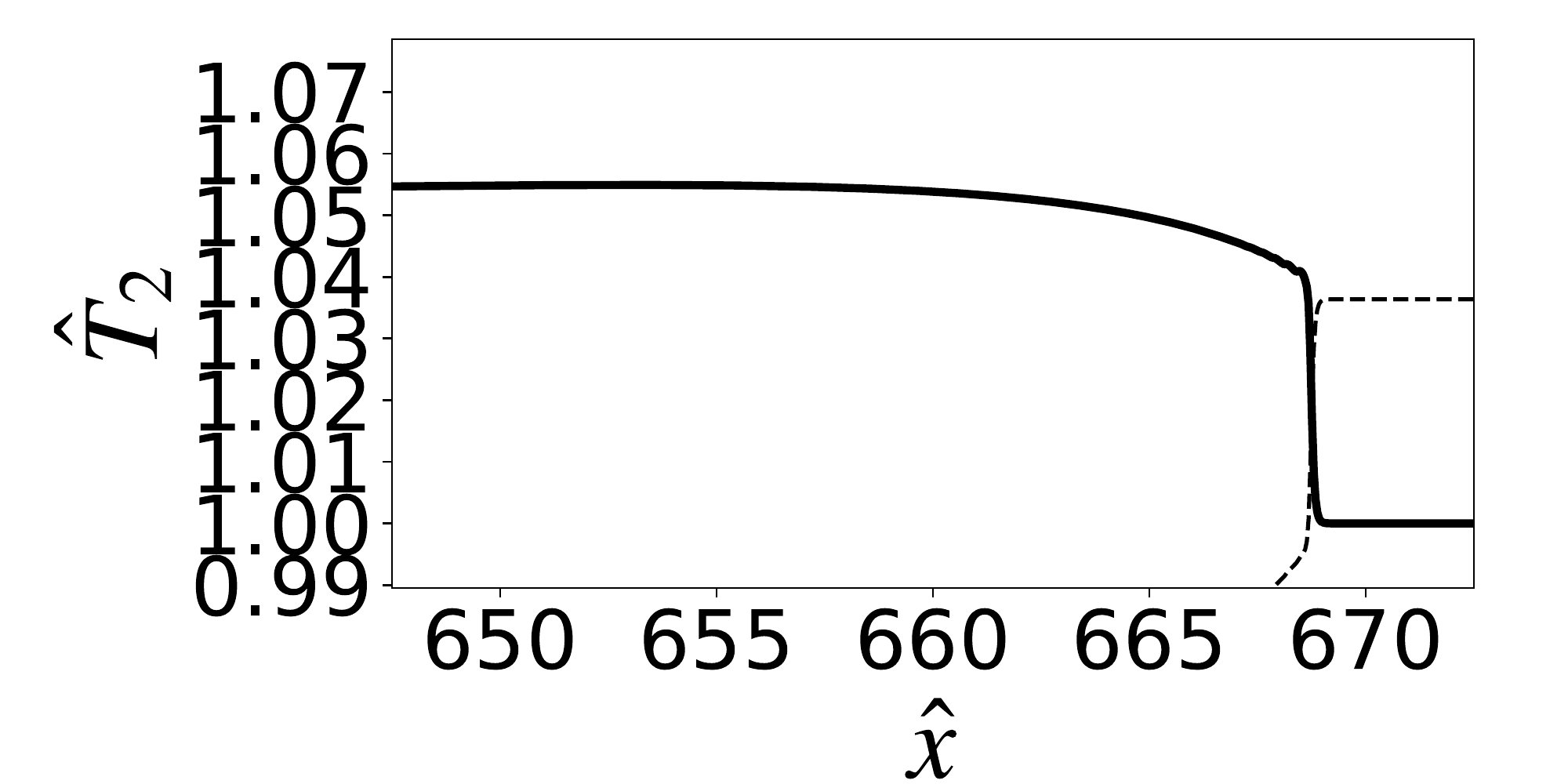}
	\end{center}
	\caption{Case B$_2$: Shock structure in a binary Eulerian mixture of polyatomic and monatomic gases obtained at $\hat{t} = 600$. 
	The parameters are in Region IV and correspond to the mark of No. 2 shown in Figure \ref{fig:subshockEuler_mu07}; $\gamma_1 = 7/6$, $\gamma_2 = 5/3$, $\mu = 0.7$, $c_0 = 0.3$, and $M_0 = 1.1$. 
	The numerical conditions are $\Delta \hat{t} = 0.01$ and $\Delta \hat{x}=0.04$. }
	\label{fig:c03_M0-1_1}
\end{figure}

We show the shock structure with $\gamma_1 = 7/6$, $\gamma_2 = 5/3$, $\mu = 0.7$, and $c_0 = 0.3$ for $M_0 = 1.03$ in Figure \ref{fig:c03_M0-1_03} and for $M_0 = 1.1$ in Figure \ref{fig:c03_M0-1_1}. 
As is expected, we observe a continuous shock structure for $M_0 = 1.03$ corresponding to the parameters in Region I and a shock structure with multiple sub-shocks for $M_0 = 1.1$ corresponding to the parameters in Region IV. 
It is noticeable that both sub-shocks for constituents $1$ and $2$ arise from the unperturbed state in contrast to other examples of the multiple sub-shock formation. 
Because both characteristic velocities in the unperturbed state have the common maximum value in this special case, both sub-shocks should be connected with the unperturbed state according to the theorem~\cite{Breakdown}. 

\subsection{Shock structure in Case B$_3$}

\begin{figure}
	\begin{center}
		\includegraphics[width=0.49\linewidth]{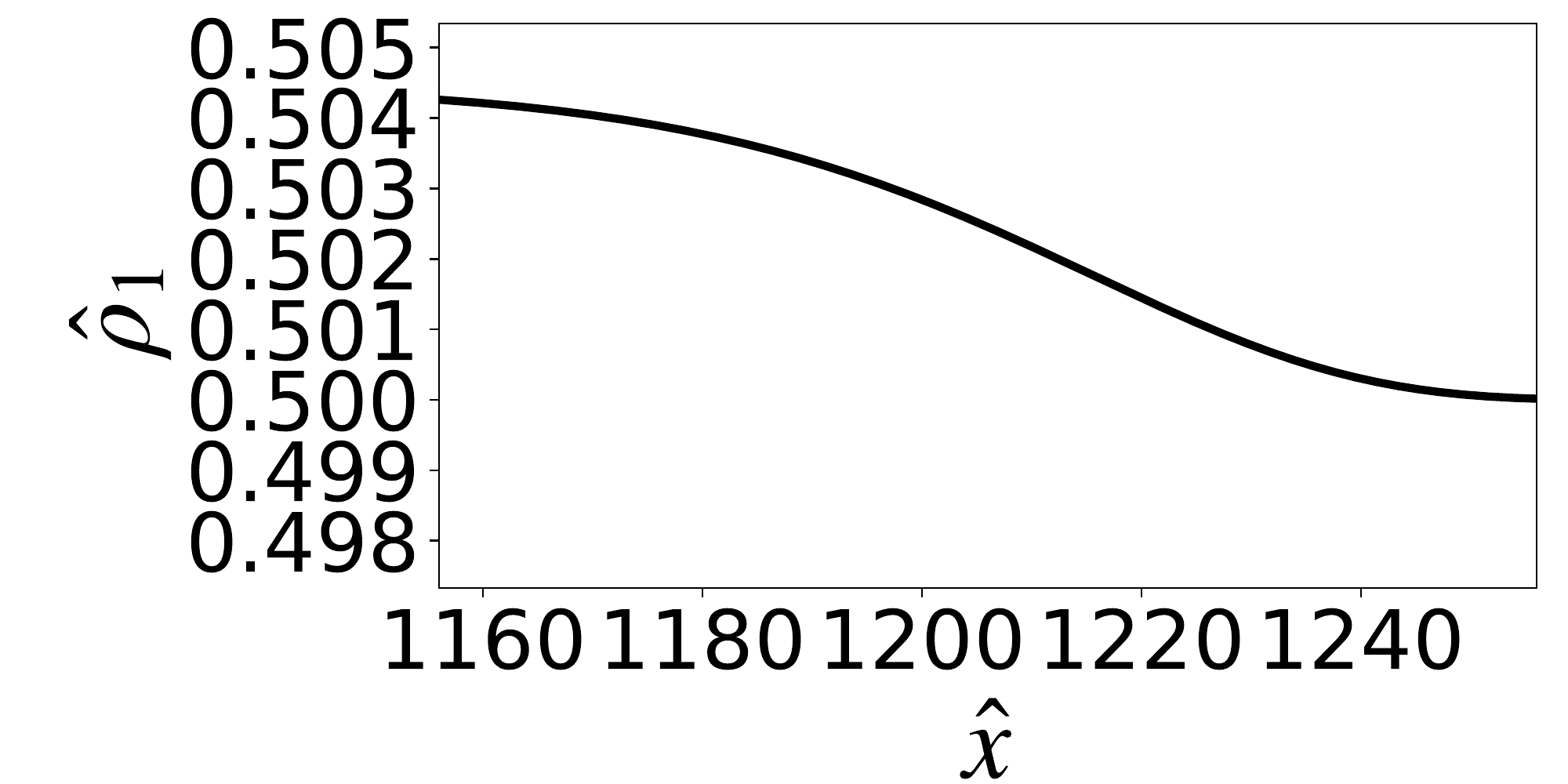} %
		\includegraphics[width=0.49\linewidth]{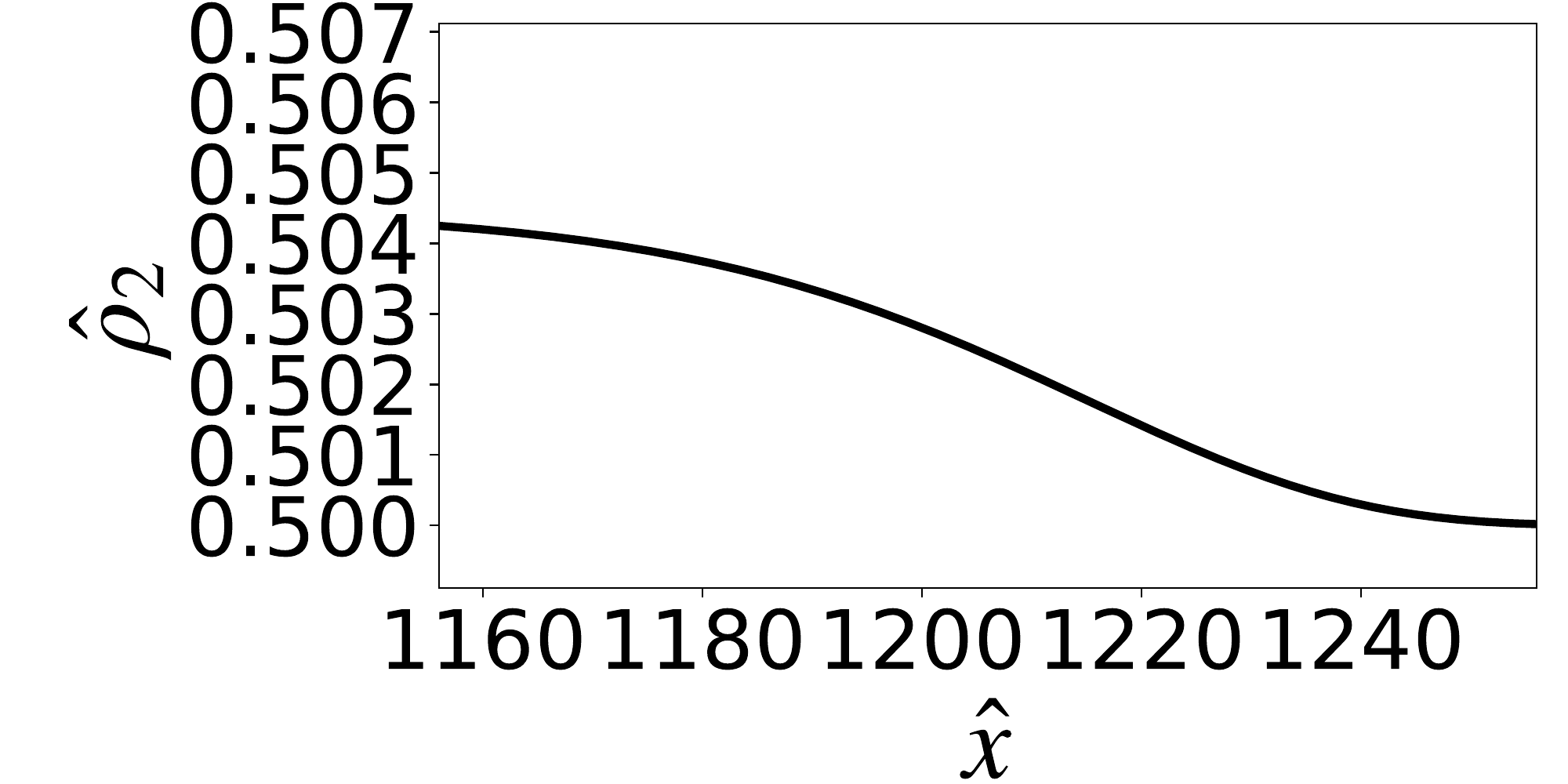}
		\includegraphics[width=0.49\linewidth]{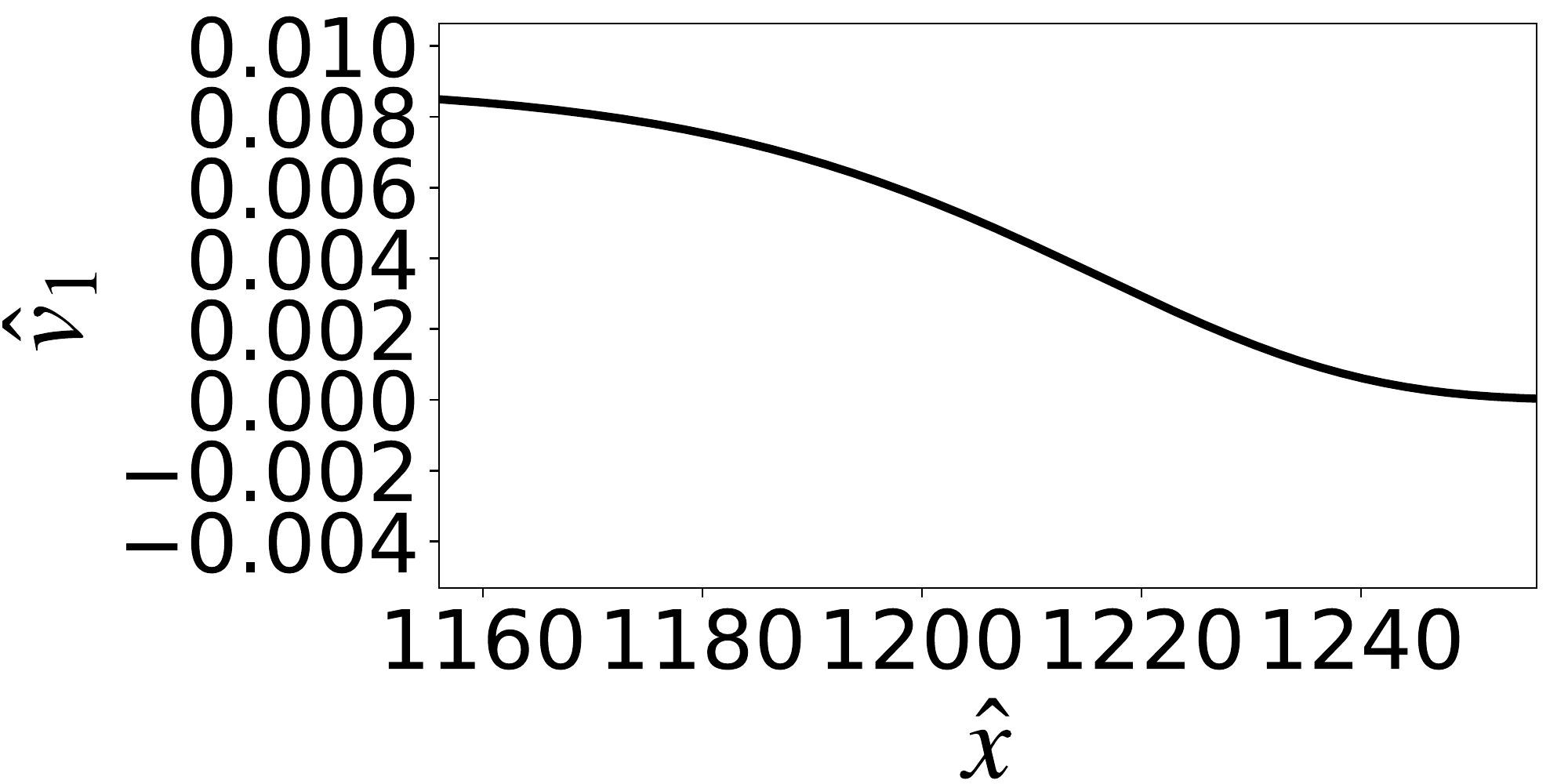} %
		\includegraphics[width=0.49\linewidth]{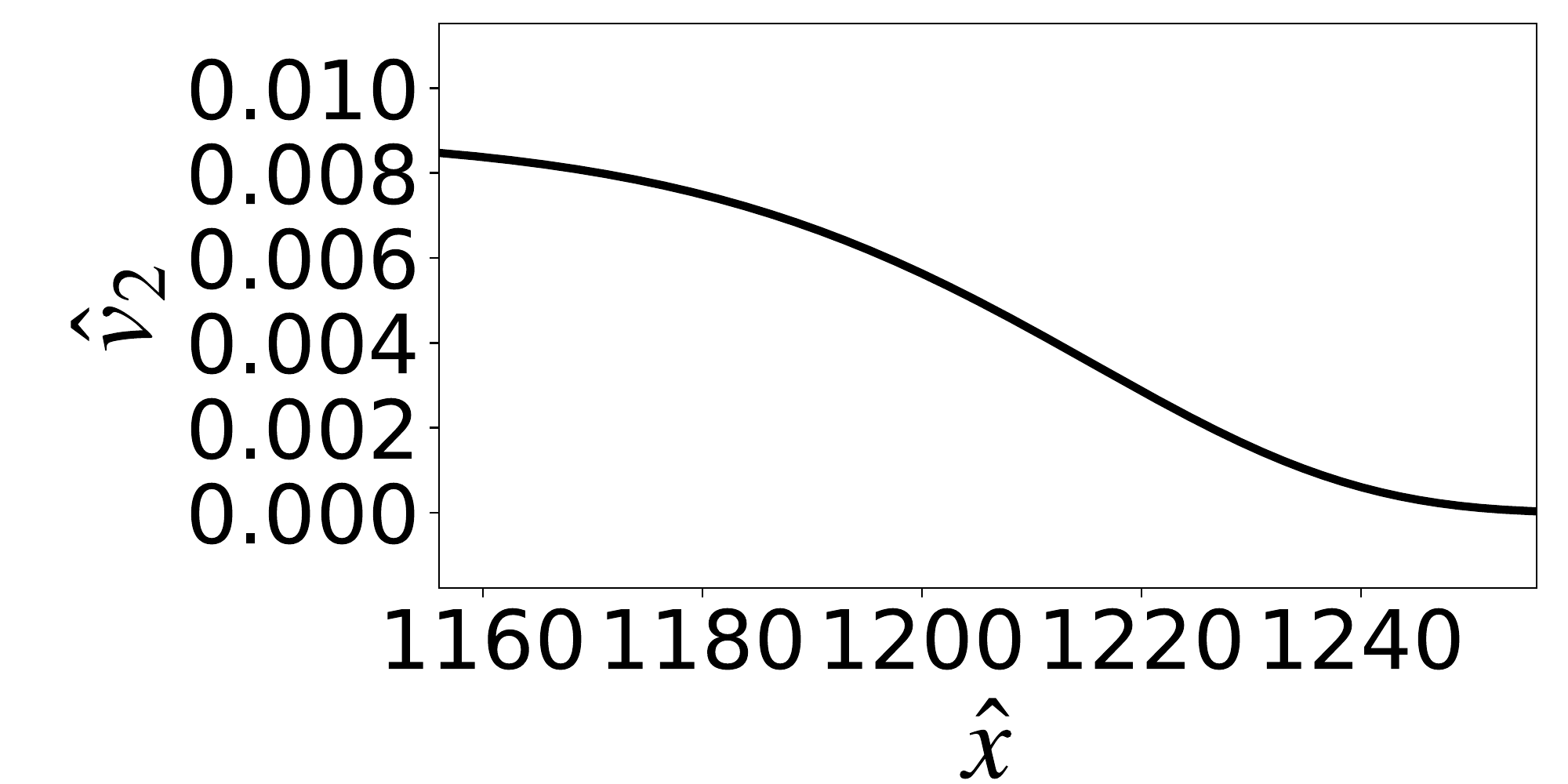}
		\includegraphics[width=0.49\linewidth]{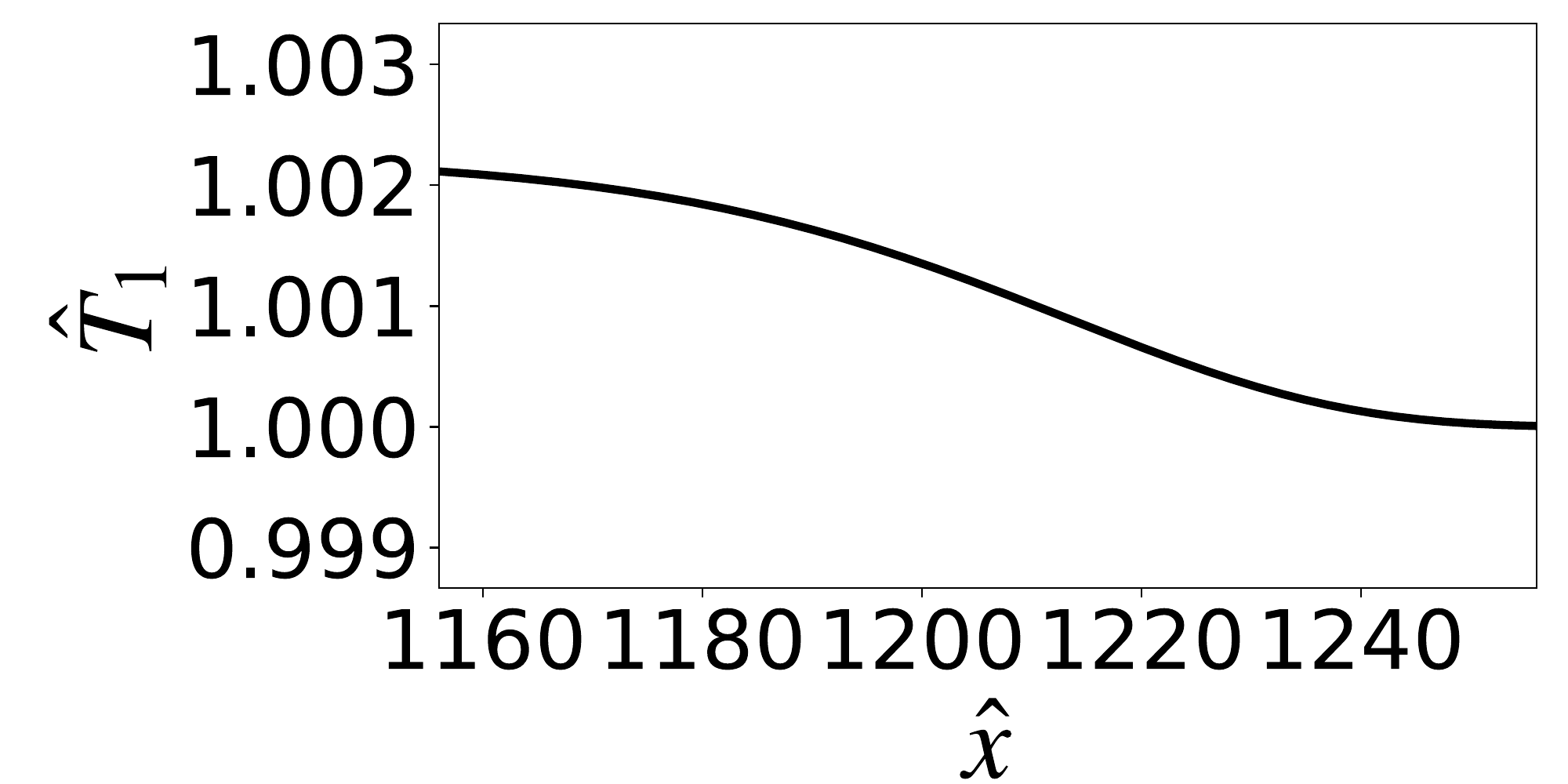} %
		\includegraphics[width=0.49\linewidth]{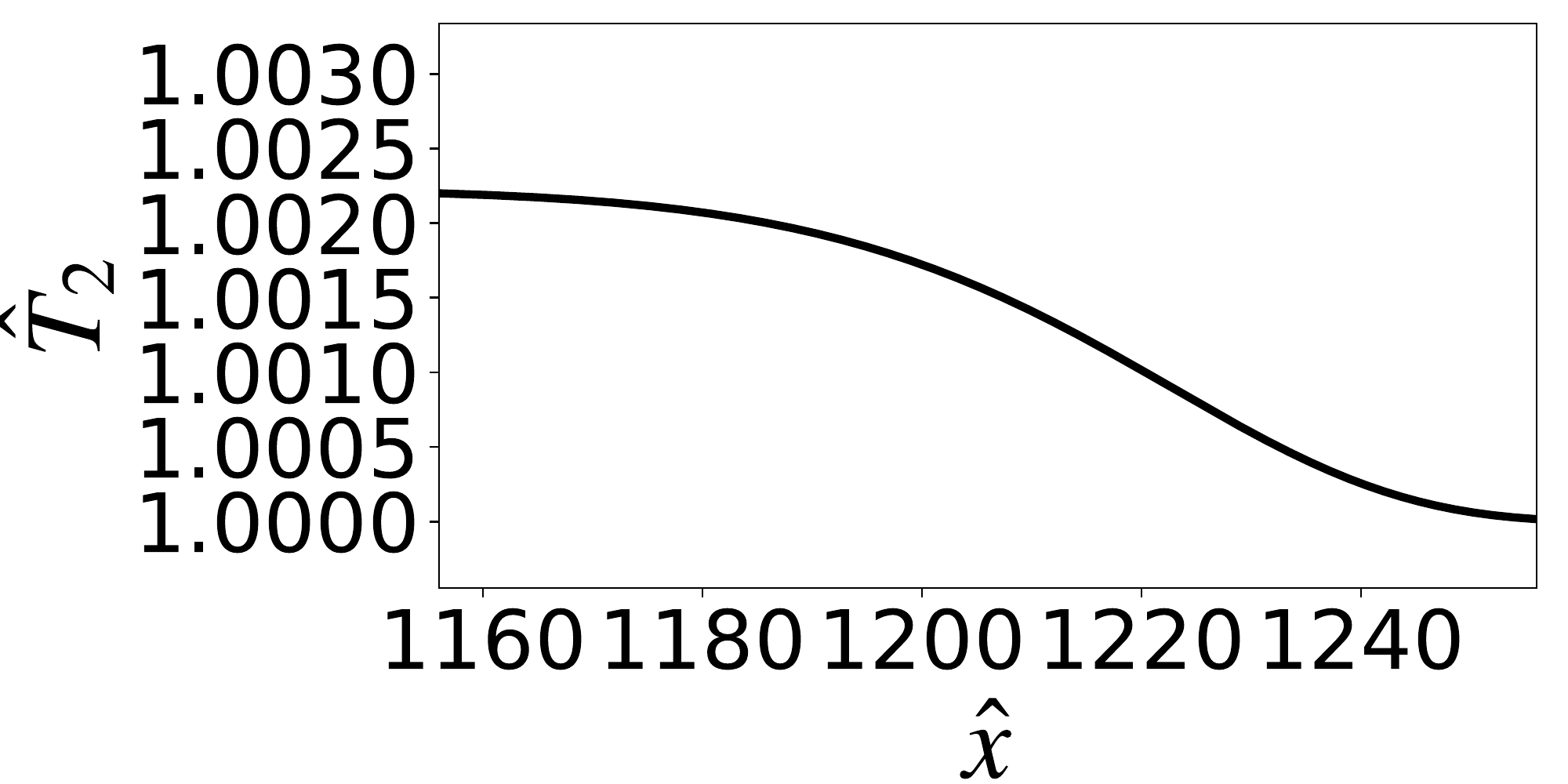}
	\end{center}
	\caption{Case B$_3$: Shock structure in a binary Eulerian mixture of polyatomic and monatomic gases obtained at $\hat{t} = 1200$. 
	The parameters are in Region I and correspond to the mark of No. 1 shown in Figure \ref{fig:subshockEuler_mu081}; $\gamma_1 = 7/6$, $\gamma_2 = 5/3$, $\mu = 0.81$, $c_0 = 0.5$, and $M_0 = 1.005$. 
	The numerical conditions are $\Delta \hat{t} = 0.04$ and $\Delta \hat{x}=0.16$. }
	\label{fig:c05_M0-1_005}
\end{figure}

\begin{figure}
	\begin{center}
		\includegraphics[width=0.49\linewidth]{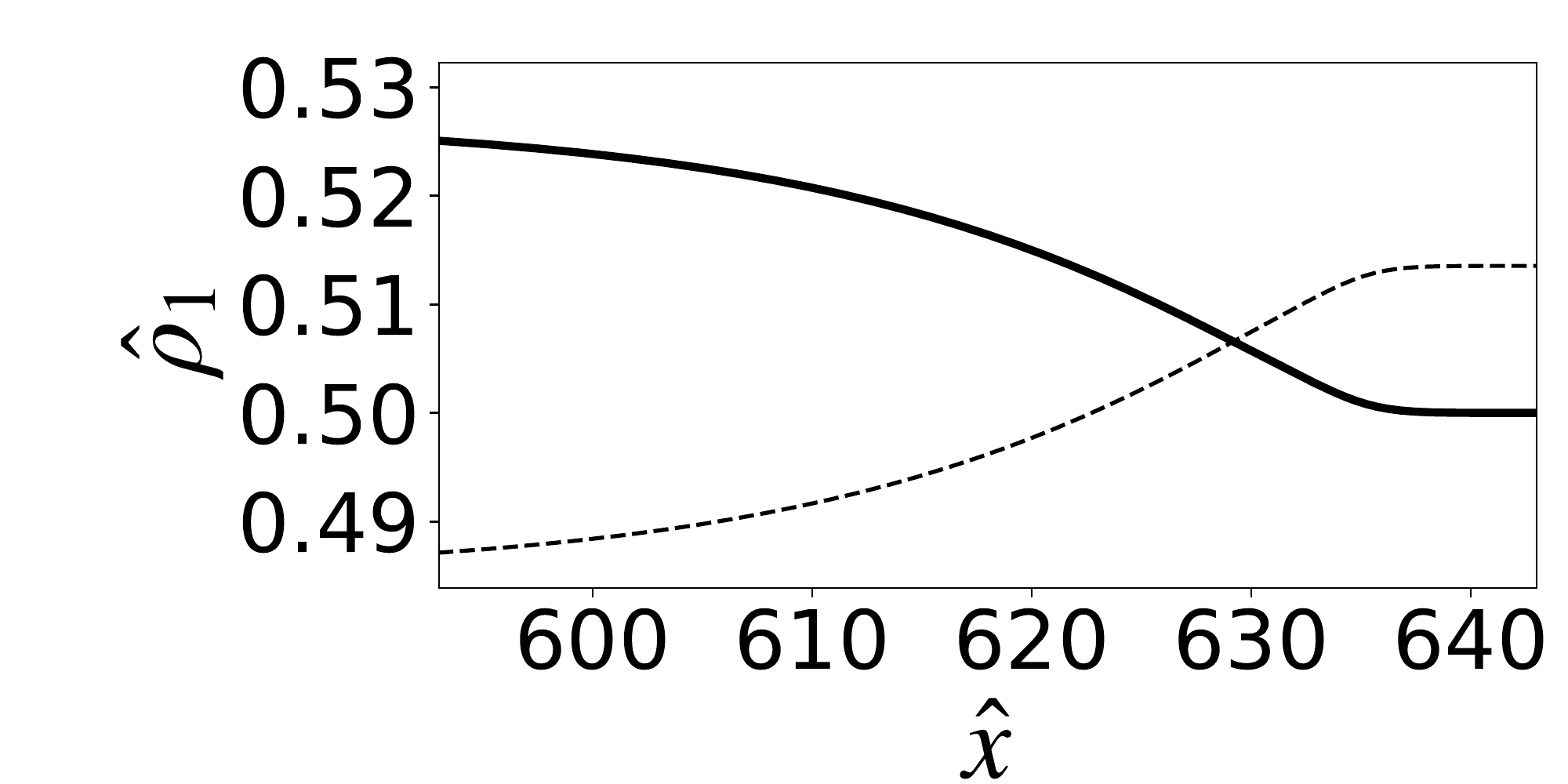} %
		\includegraphics[width=0.49\linewidth]{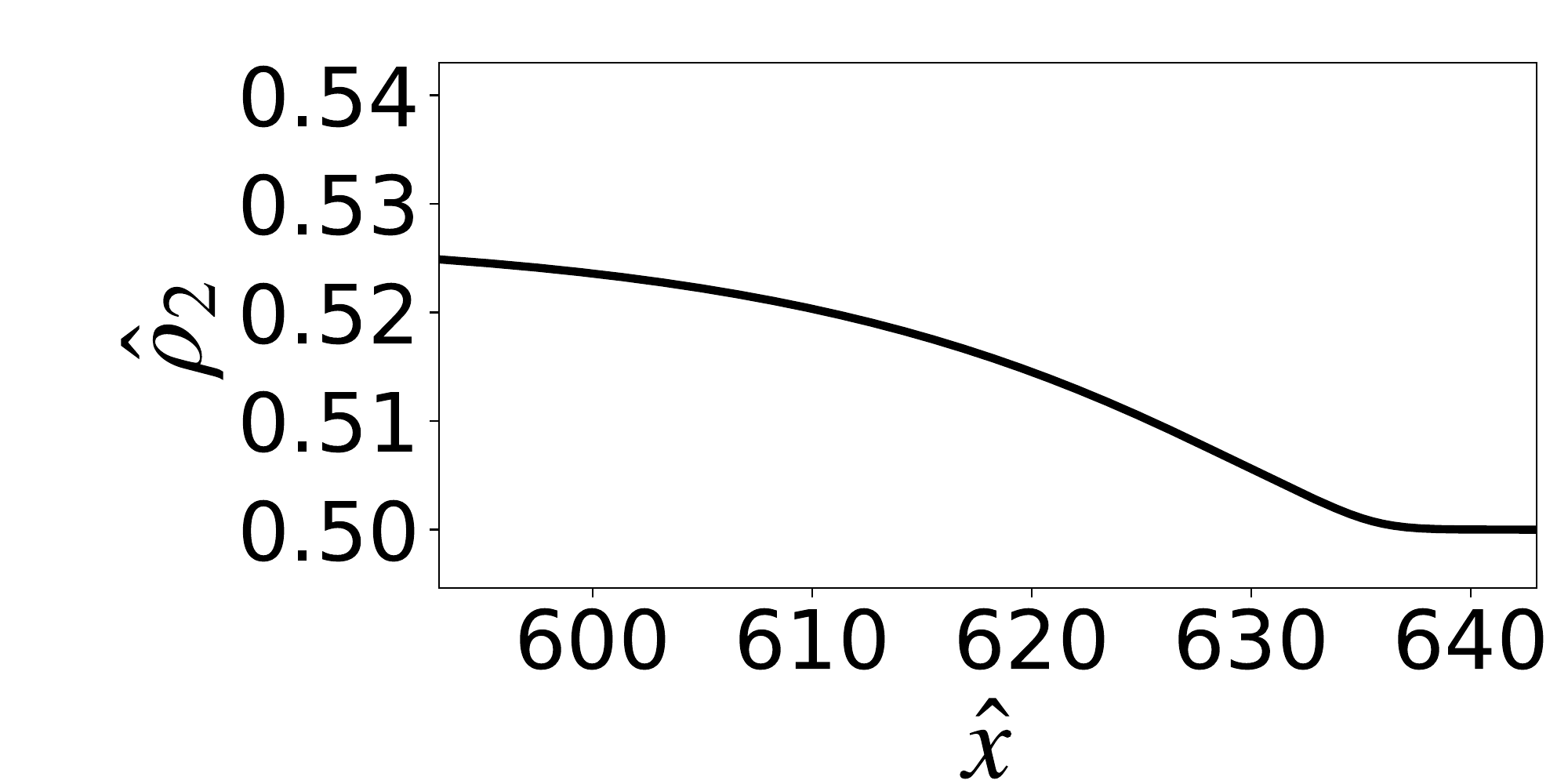}
		\includegraphics[width=0.49\linewidth]{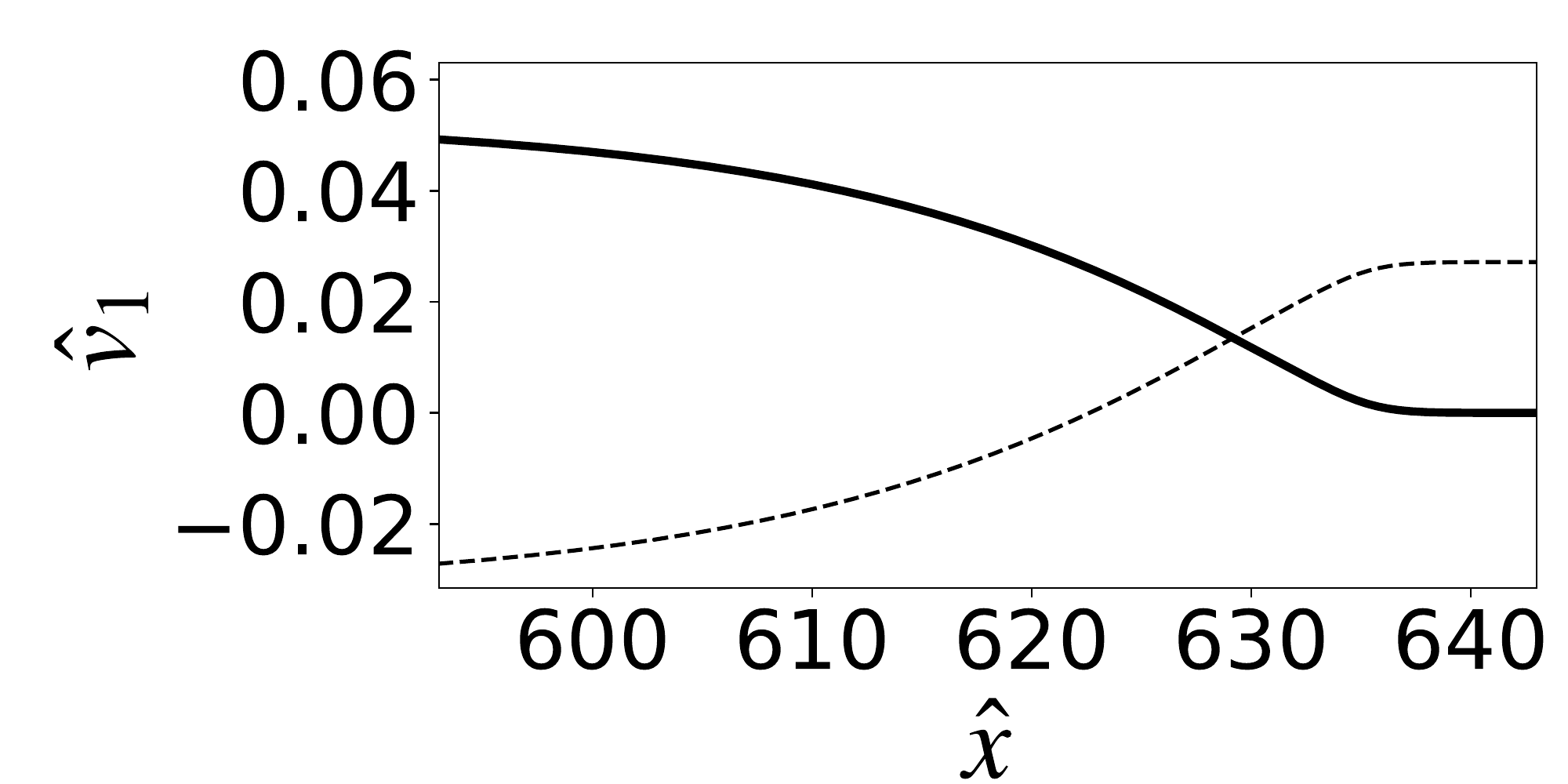} %
		\includegraphics[width=0.49\linewidth]{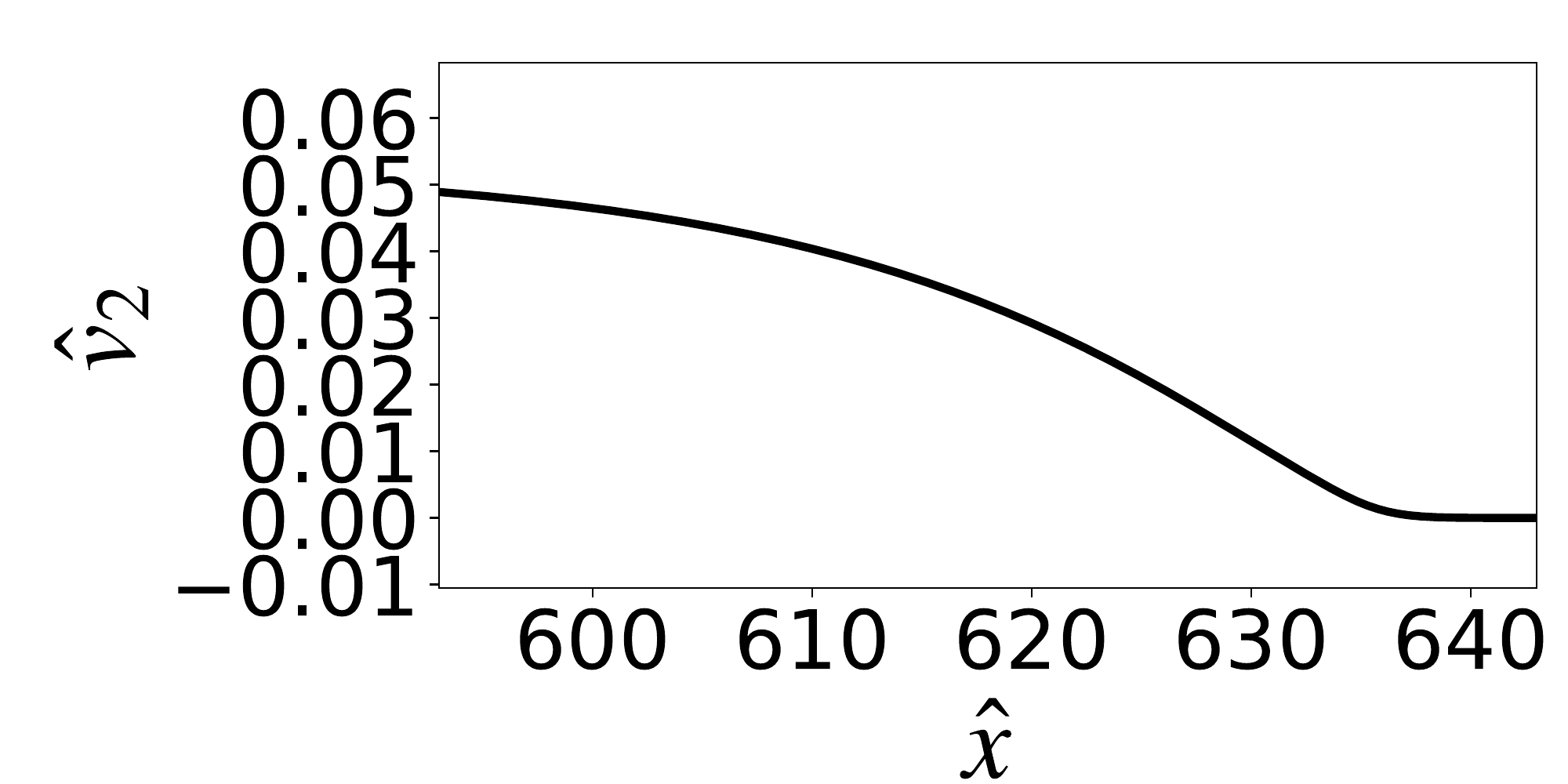}
		\includegraphics[width=0.49\linewidth]{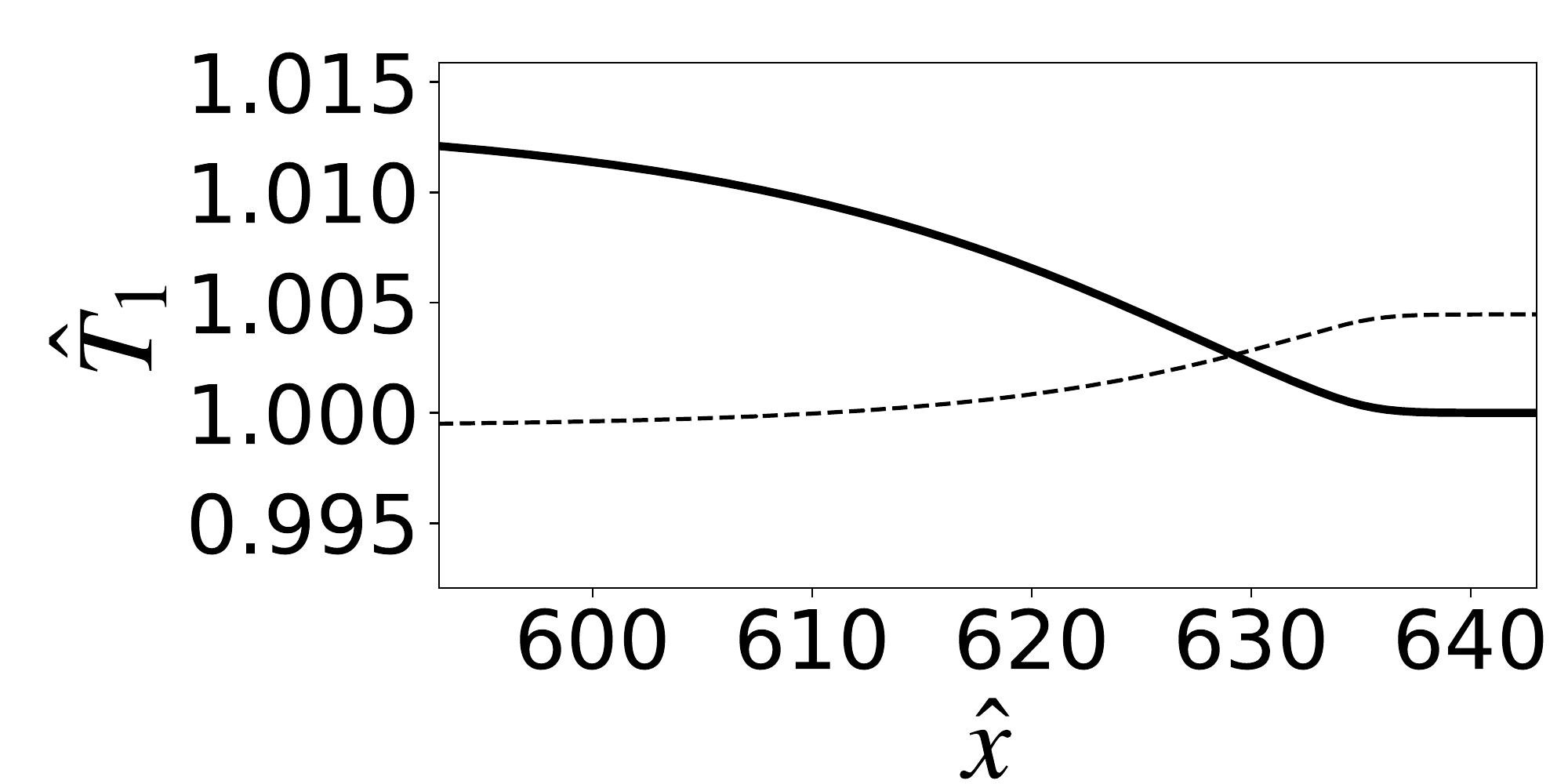} %
		\includegraphics[width=0.49\linewidth]{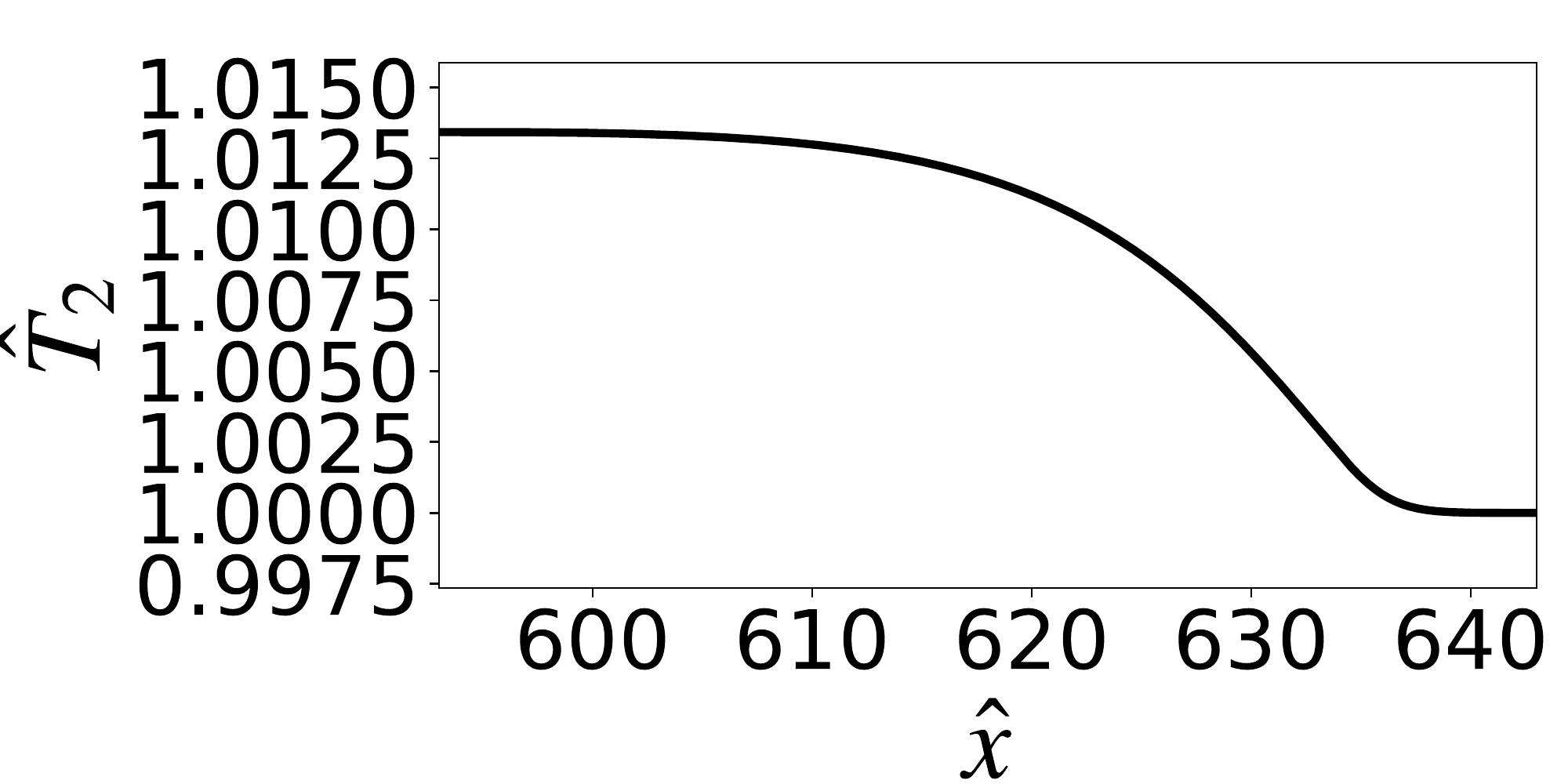}
	\end{center}
	\includegraphics[width=0.49\linewidth]{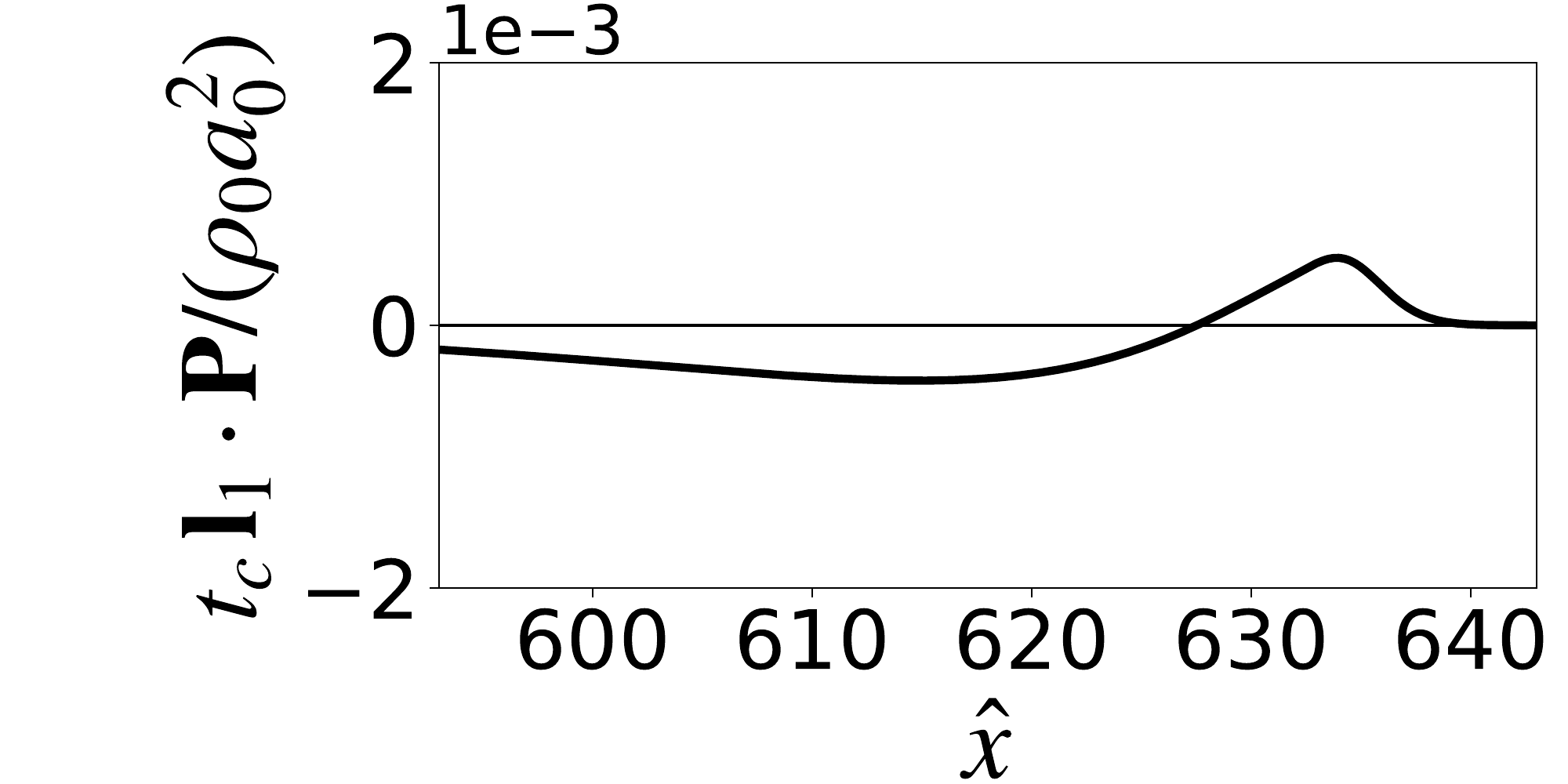}\hspace{0.49\linewidth}
	\caption{Case B$_3$: Shock structure in a binary Eulerian mixture of polyatomic and monatomic gases obtained at $\hat{t} = 600$. 
	The parameters are in Region IV and correspond to the mark of No. 2 shown in Figure \ref{fig:subshockEuler_mu081}; $\gamma_1 = 7/6$, $\gamma_2 = 5/3$, $\mu = 0.81$, $c_0 = 0.5$, and $M_0 = 1.03$. 
	The numerical conditions are $\Delta \hat{t} = 0.02$ and $\Delta \hat{x}=0.08$. }
	\label{fig:c05_M0-1_03}
\end{figure}

\begin{figure}
	\begin{center}
		\includegraphics[width=0.49\linewidth]{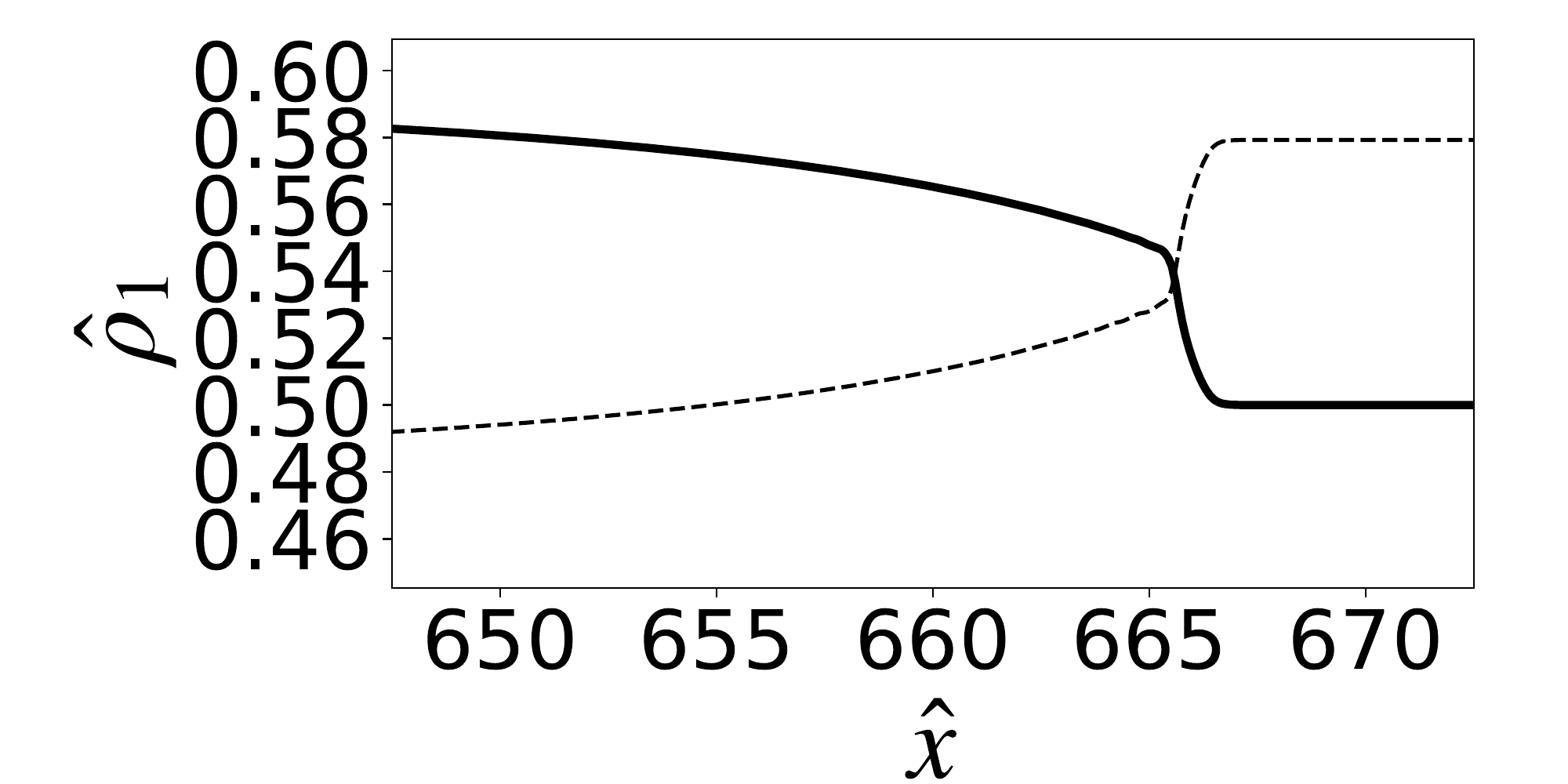} %
		\includegraphics[width=0.49\linewidth]{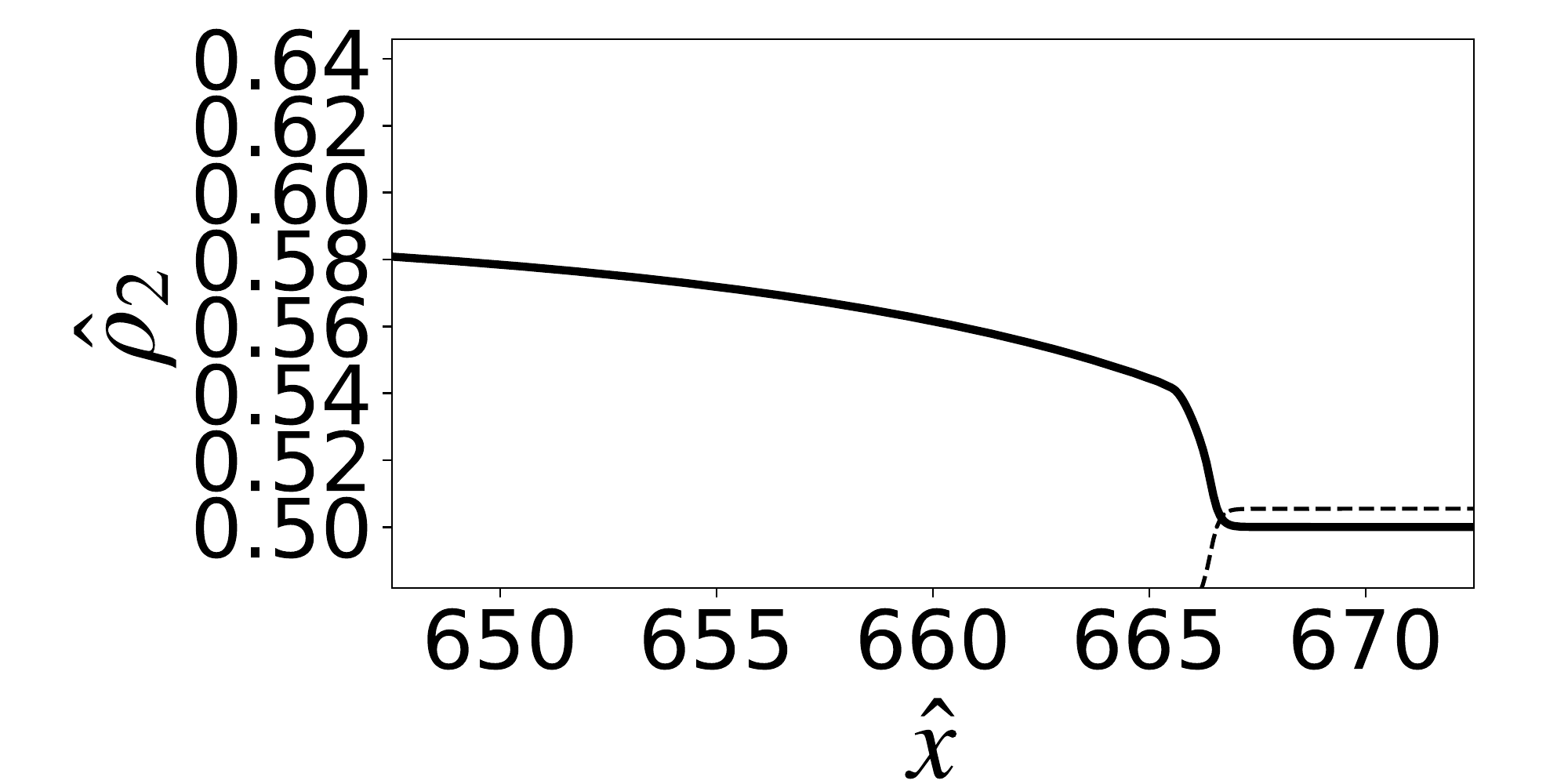}
		\includegraphics[width=0.49\linewidth]{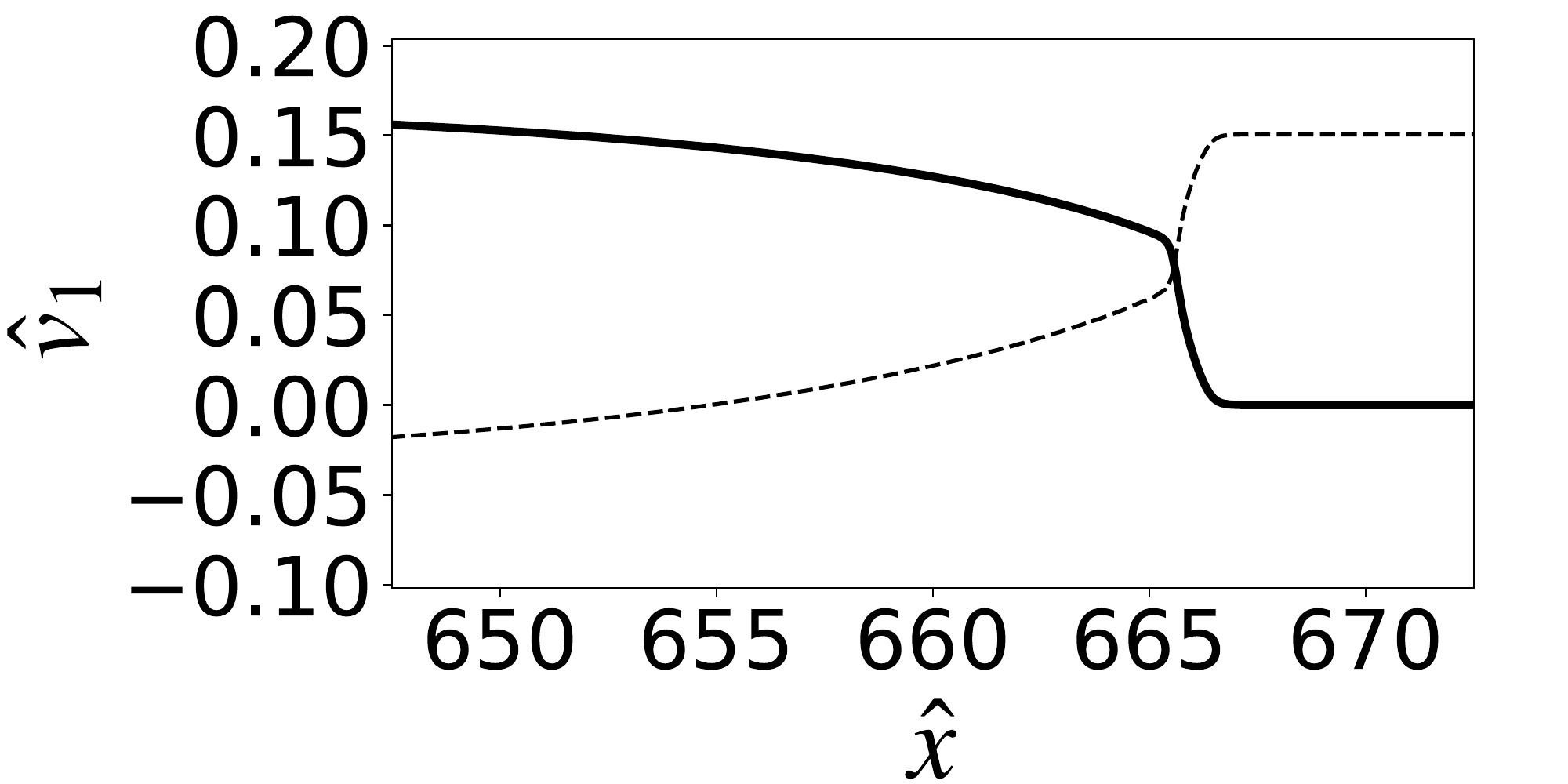} %
		\includegraphics[width=0.49\linewidth]{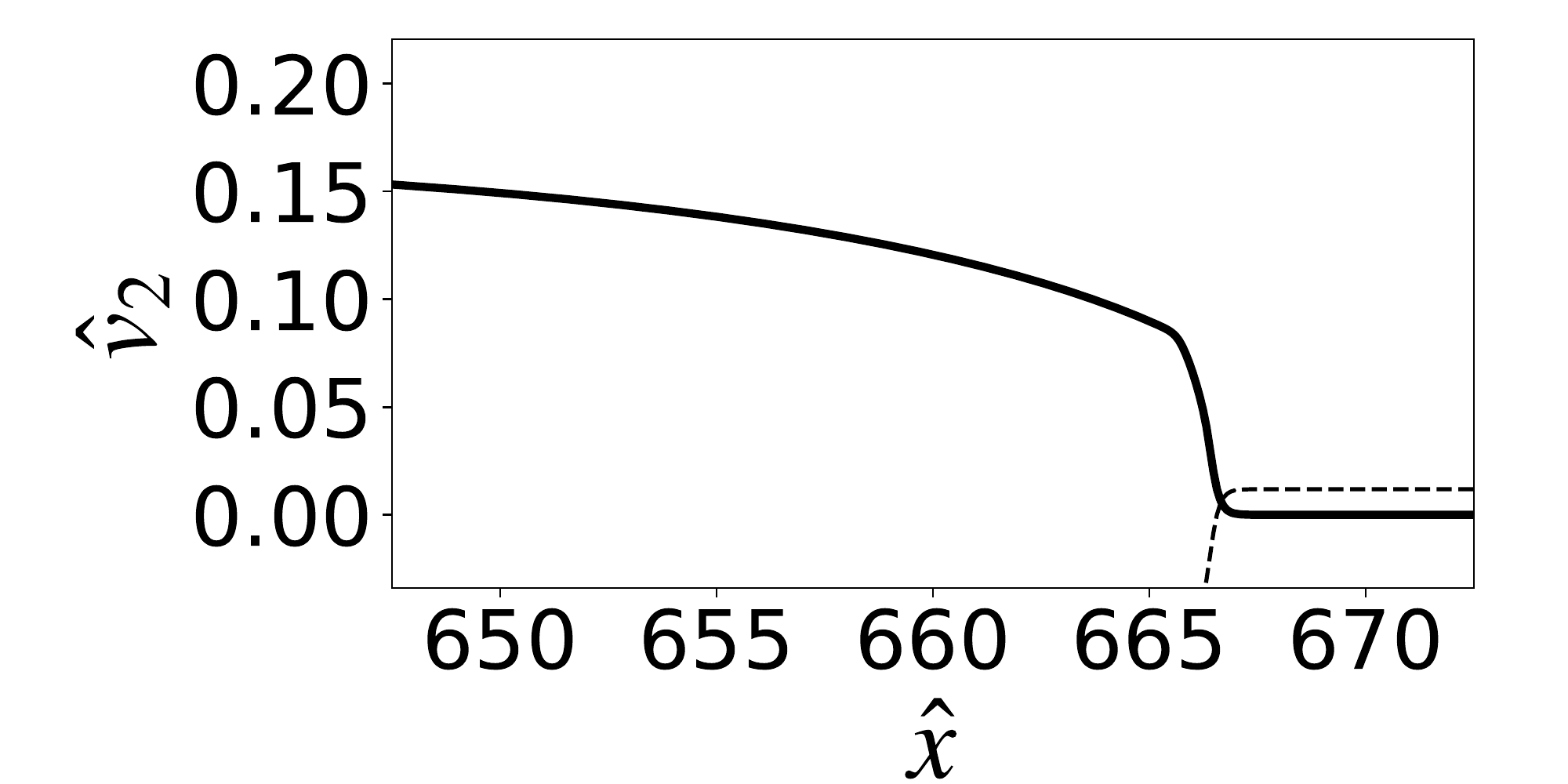}
		\includegraphics[width=0.49\linewidth]{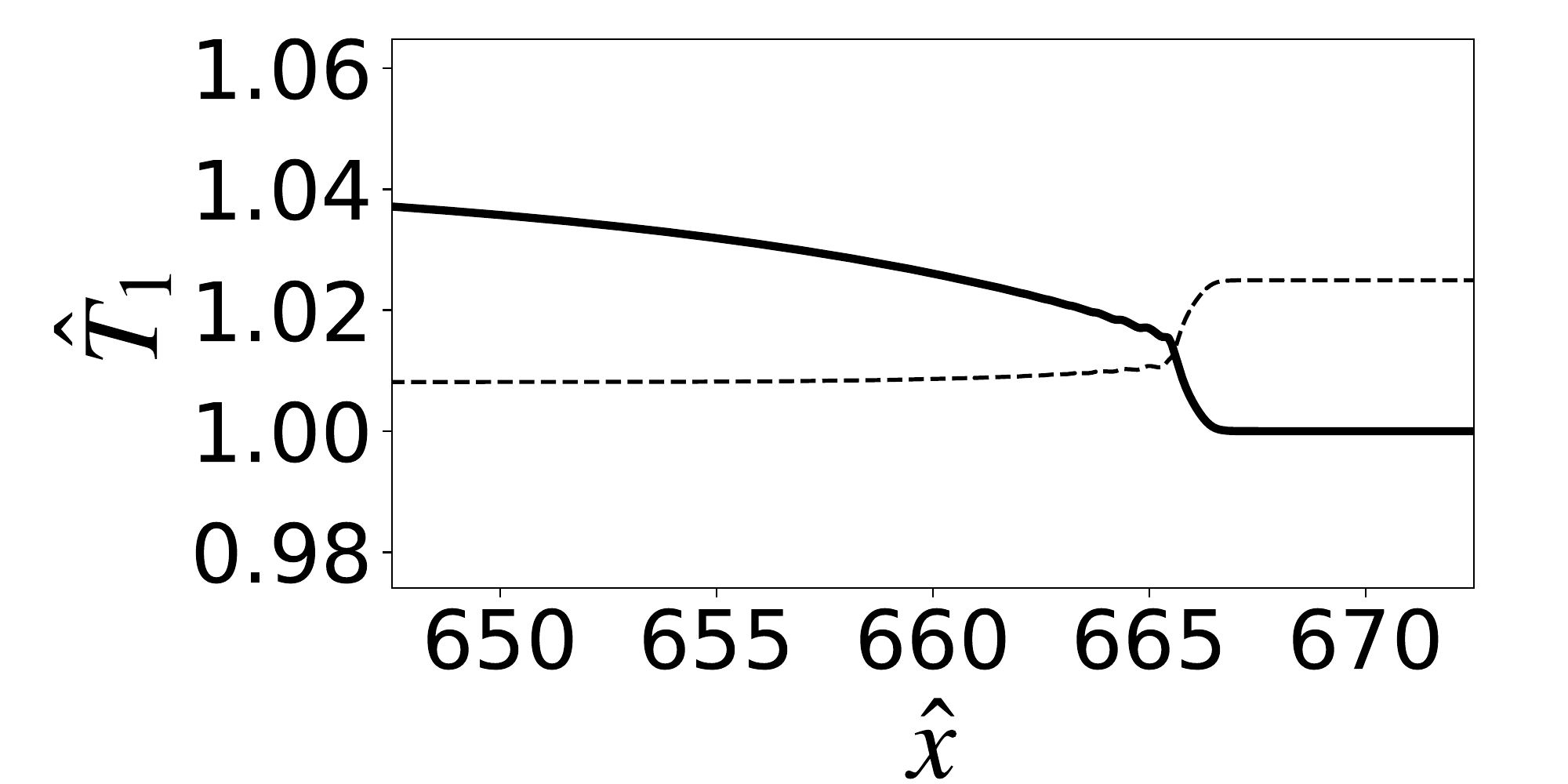} %
		\includegraphics[width=0.49\linewidth]{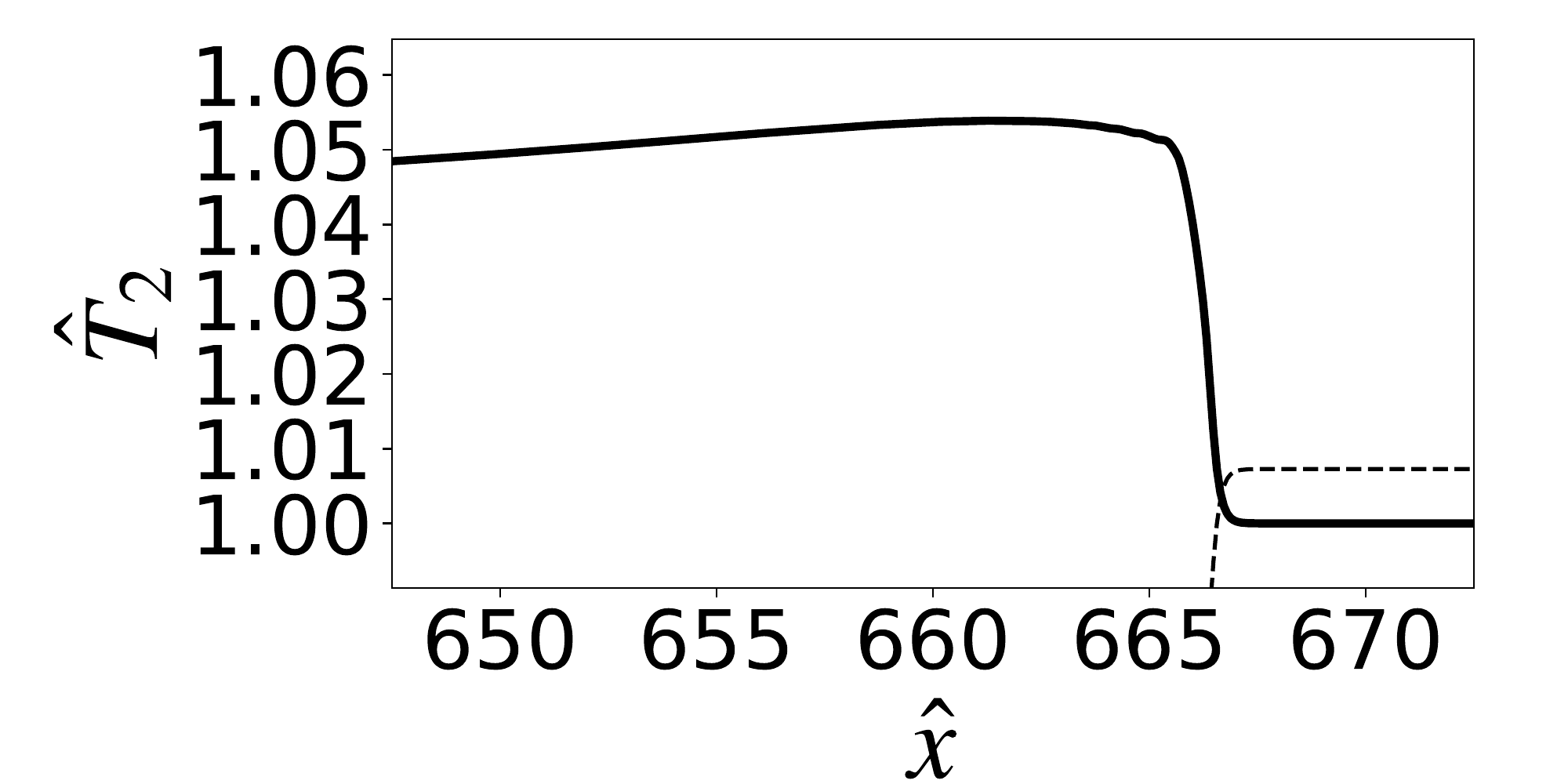}
	\end{center}
	\caption{Case B$_3$: Shock structure in a binary Eulerian mixture of polyatomic and monatomic gases obtained at $\hat{t} = 600$. 
	The parameters are in Region IV and correspond to the mark of No. 3 shown in Figure \ref{fig:subshockEuler_mu081}; $\gamma_1 = 7/6$, $\gamma_2 = 5/3$, $\mu = 0.81$, $c_0 = 0.5$, and $M_0 = 1.1$. 
	The numerical conditions are $\Delta \hat{t} = 0.02$ and $\Delta \hat{x}=0.08$. }
	\label{fig:c05_M0-1_1}
\end{figure}

In this subsection, in order to discuss the sub-shock formation in Case B$_3$, we adopt the following parameters; $\gamma_1 = 7/6$, $\gamma_2 = 5/3$ and $\mu = 0.81$. 
Figures \ref{fig:c05_M0-1_005} -- \ref{fig:c05_M0-1_1} show the shock structure for several Mach numbers in the case of $c_0 = 0.5$. 
We confirm from Figure \ref{fig:c05_M0-1_005} that the shock structure for $M_0 = 1.005$ is continuous and no sub-shock appears as expected. 

Again no sub-shock emerges in the shock structure for $M_0=1.03$ shown in Figure \ref{fig:c05_M0-1_03} although the shock velocity is greater than the characteristic velocity for the species $1$ evaluated in the unperturbed state. 
The left eigenvector $\mathbf{l}_1$ for the system \eqref{finale1} corresponding to $\lambda_1$ is 
\begin{equation*}
	\mathbf{l}_1 =
	\left[%
	\begin{array}{c}
		v_1 \left( v_1 - \sqrt{\gamma_1  \frac{k_{\mathrm{B}}}{m_{1}} T_1} \,\, \right) \\
		\sqrt{\gamma_1  \frac{k_{\mathrm{B}}}{m_{1}} T_1} - 2 v_1 \\
		1 \\
		0 \\
		0 \\
		0 \\
	\end{array}%
	\right]^T
\end{equation*}
From the profile of the $t_c \, \mathbf{l}_1 \cdot \mathbf{P}/(\rho_0 a_0^2)$ shown in Figure \ref{fig:c05_M0-1_03} we see that the singular point becomes regular singular and the shock-structure solution is continuous. 
Furthermore, this regular singular point becomes singular and a sub-shock for constituent $1$ appears if we increase the Mach number more and we observe multiple sub-shocks for $M_0=1.1$ shown in Figure \ref{fig:c05_M0-1_1}.

\subsection{Discontinuity wave induced by a sub-shock}
From the profiles of the shock structure with a sub-shock, we notice that the derivative of the physical quantities for a constituent jumps when a sub-shock appears in the other constituent. 
In other words, a sub-shock for a constituent induces a discontinuity wave for the other constituent. 
In this section, we derive the relationship between these waves explicitly.  

For analyzing the discontinuity wave, it is convenient to rewrite the system of field equations \eqref{finale1} in the following form: 
\begin{align}\label{eq:material_time_derivative}
	\begin{split}
		& \dot{\rho}_1 + \rho_1 \frac{\partial v_1}{\partial x} = 0,  \\ 
		& \rho_1 \dot{v}_1 + \frac{\partial p_1}{\partial x} = \hat{m}_1, \\
		& \rho_1 \dot{\varepsilon}_1 + p_1 \frac{\partial v_1}{\partial x} 
		= \hat{e}_1 - \hat{m}_1 u_1, \\
		& \dot{\rho}_2 + \rho_2 \frac{\partial v_2}{\partial x} = 0,  \\ 
		& \rho_2 \dot{v}_2 + \frac{\partial p_2}{\partial x} = - \hat{m}_1, \\
		& \rho_2 \dot{\varepsilon}_2 + p_2 \frac{\partial v_2}{\partial x} 
		= - \hat{e}_1 + \hat{m}_1 u_2,  
	\end{split}
\end{align}
where a dot on physical quantities represents the material time derivative and in the present case $\dot{X}_\alpha = (\partial X_\alpha  / \partial t) + v_\alpha (\partial X_\alpha  / \partial x)$ holds with $X_\alpha $ being a generic quantity for the constituent $\alpha=1,2$. 
We introduce $[\![ X_\alpha  ]\!]$ to represent the jump of a generic variable $X_\alpha $ across a sub-shock. 
Let us consider the case that a sub-shock arise in the constituent $2$ ($[\![ X_2 ]\!] \neq 0$) and calculate the jump of the derivative of the physical quantities of the constituent $1$ ($[\![ (\partial X_1)/(\partial \varphi) ]\!] \neq 0$). 
We introduce $\delta = [\![\partial/\partial \varphi]\!]$ and by taking the fact that the acceleration wave propagates with the velocity of the shock $s$ into account, the formal substitutions are given by
\begin{equation}\label{eq:substitution}
	\begin{split}
	&[\![ X_1 ]\!] = 0, \qquad
	\left[\!\!\left[ \frac{\partial X_1}{\partial t} \right]\!\!\right] \rightarrow -s \delta X_1, \\
	&\left[\!\!\left[ \frac{\partial X_1}{\partial x} \right]\!\!\right] \rightarrow \delta X_1, \qquad 
	\left[\!\!\left[ \dot{X_1} \right]\!\!\right] \rightarrow - (s - v_1)\delta X_1. 
	\end{split}
\end{equation}
Subtracting both sides of \eqref{eq:material_time_derivative}$_{1-3}$ and applying the formal substitutions \eqref{eq:substitution}, we have
\begin{equation}\label{eq:discontinuity1}
	\begin{split}
		&\delta \rho_1 = - \frac{m_1 \left[  (\gamma_1 - 1) [\![\hat{e}_1]\!] - (\gamma_1 - 1) [\![\hat{m}_1 u_1] \!] +  (s - v_1) [\![\hat{m}_1] \!] \right]}
		{m_1 (s - v_1)^3 - \gamma_1 k_B T_1 (s - v_1)}, \\
		&\delta v_1 = - \frac{m_1 \left[  (\gamma_1 - 1) [\![\hat{e}_1]\!] - (\gamma_1 - 1) [\![\hat{m}_1 u_1] \!] +  (s - v_1) [\![\hat{m}_1] \!] \right]}
		{\rho_1 \left[ m_1 (s - v_1)^2 -  k_B \gamma_1 T_1 \right]}, \\
		&\delta T_1 = - \frac{m_1 (\gamma_1 - 1)}{k_B \rho_1 (s - v_1) \left[ m_1 (s - v_1)^2 -  k_B \gamma_1 T_1 \right]}\\
		& \times \left\{ \left[ m_1 (s - v_1)^2 - k_B T_1 \right] \left( [\![\hat{e}_1]\!] - [\![\hat{m}_1 u_1]\!] \right) 
		+ k_B T_1 (s - v_1) [\![\hat{m}_1] \!] \right\}. 
	\end{split}
\end{equation}
We confirm the validity of the expressions \eqref{eq:discontinuity1} in the following way. 
First, we estimate the strength of the sub-shock appearing in constituent $2$ from the shock structure obtained numerically with the use of \eqref{eq:subshockRH}. 
Then, by inserting the estimated values of the jump of the physical quantities in the constituent $2$ into \eqref{eq:discontinuity1}, we obtain the derivative for the constituent $1$ and plot the theoretical line in addition to the shock structure. 
In this step, we need to pay attention that, according to the conjecture on the large-time behavior of the Riemann problem, the derivative with respect to $x$ obtained by the numerical calculation after a long time is the same with the derivative with respect to $\varphi$. 

\begin{figure}
	\begin{center}
		\includegraphics[width=0.49\linewidth]{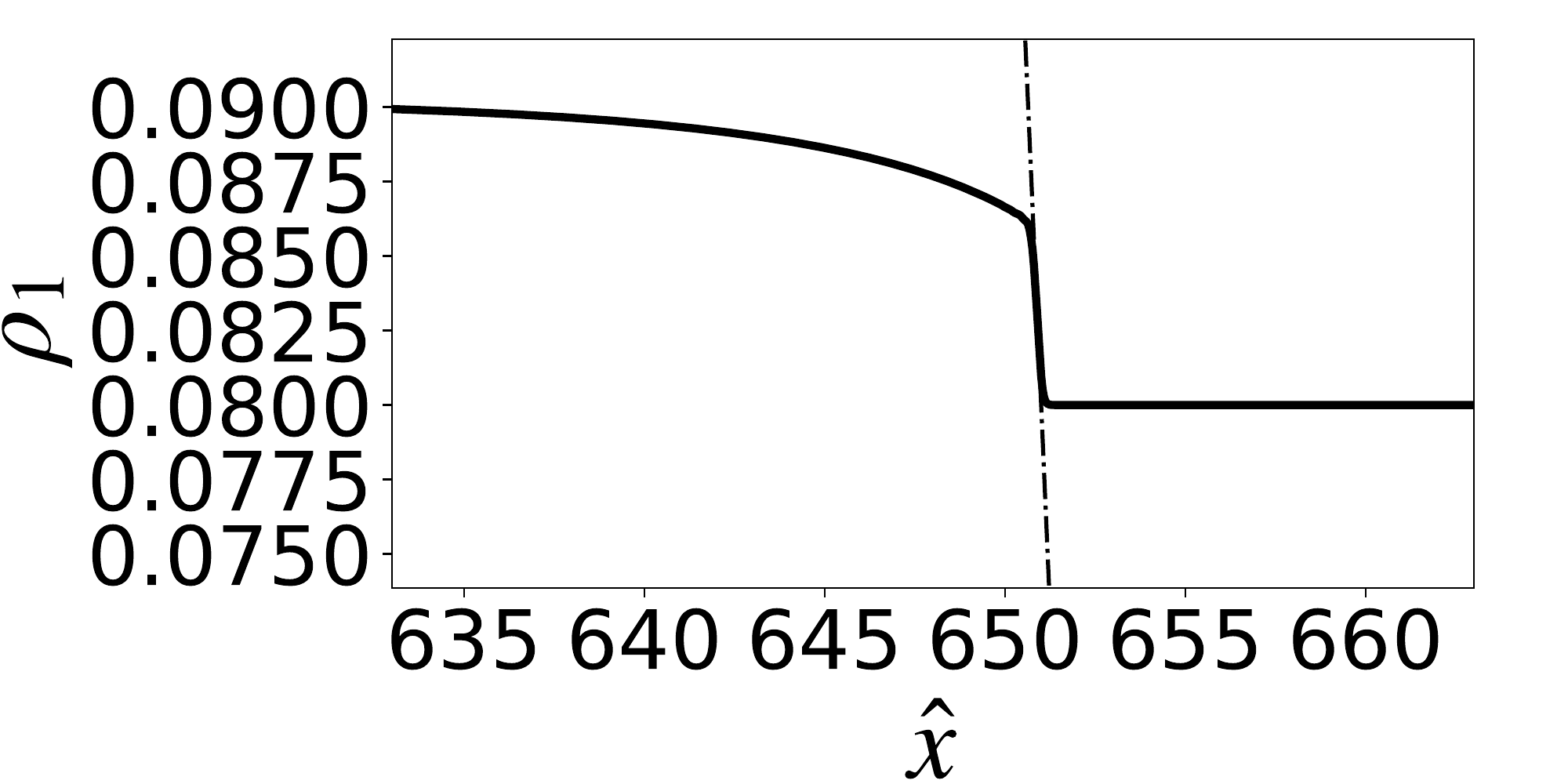} %
		\includegraphics[width=0.49\linewidth]{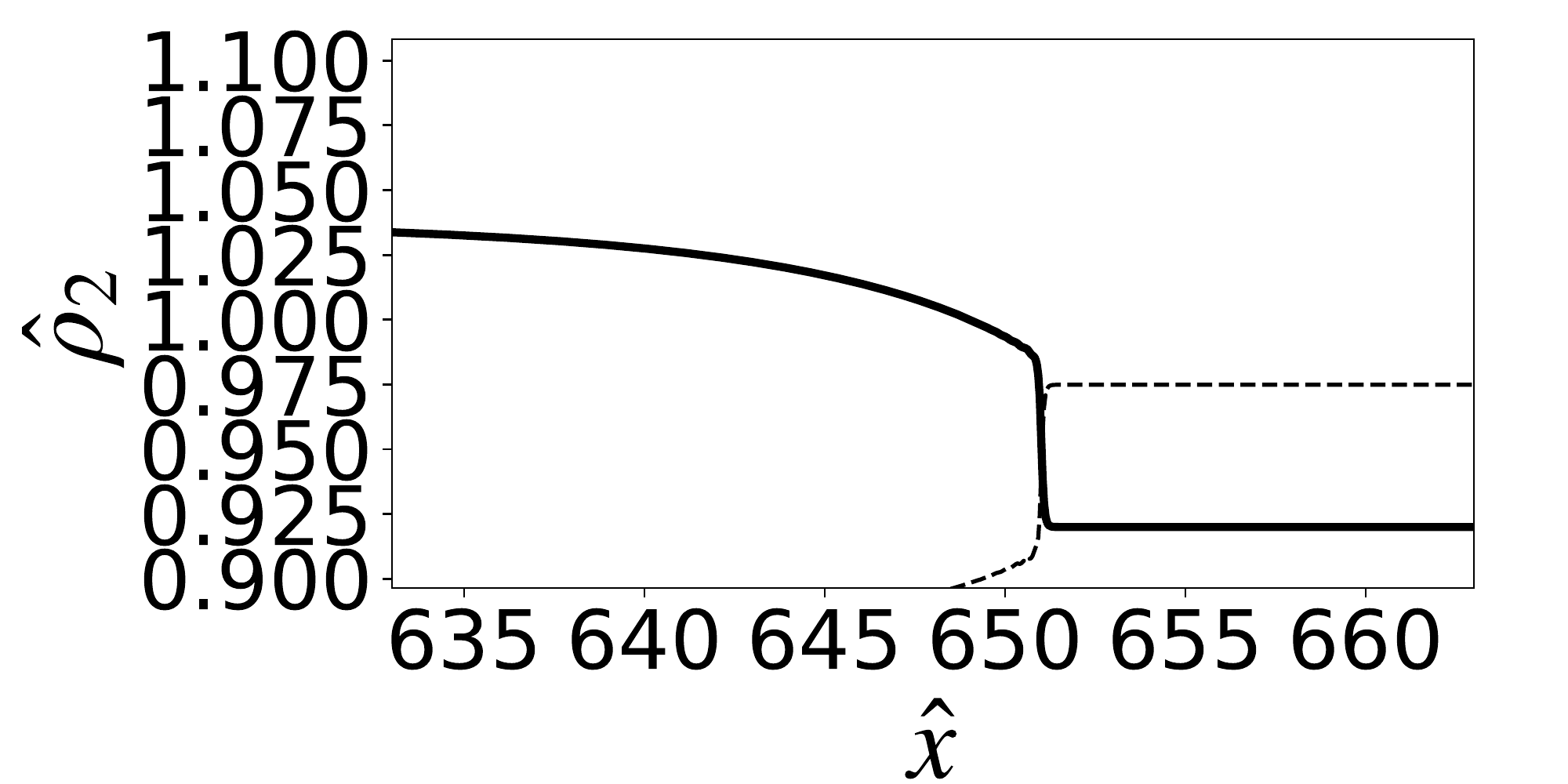} %
		\includegraphics[width=0.49\linewidth]{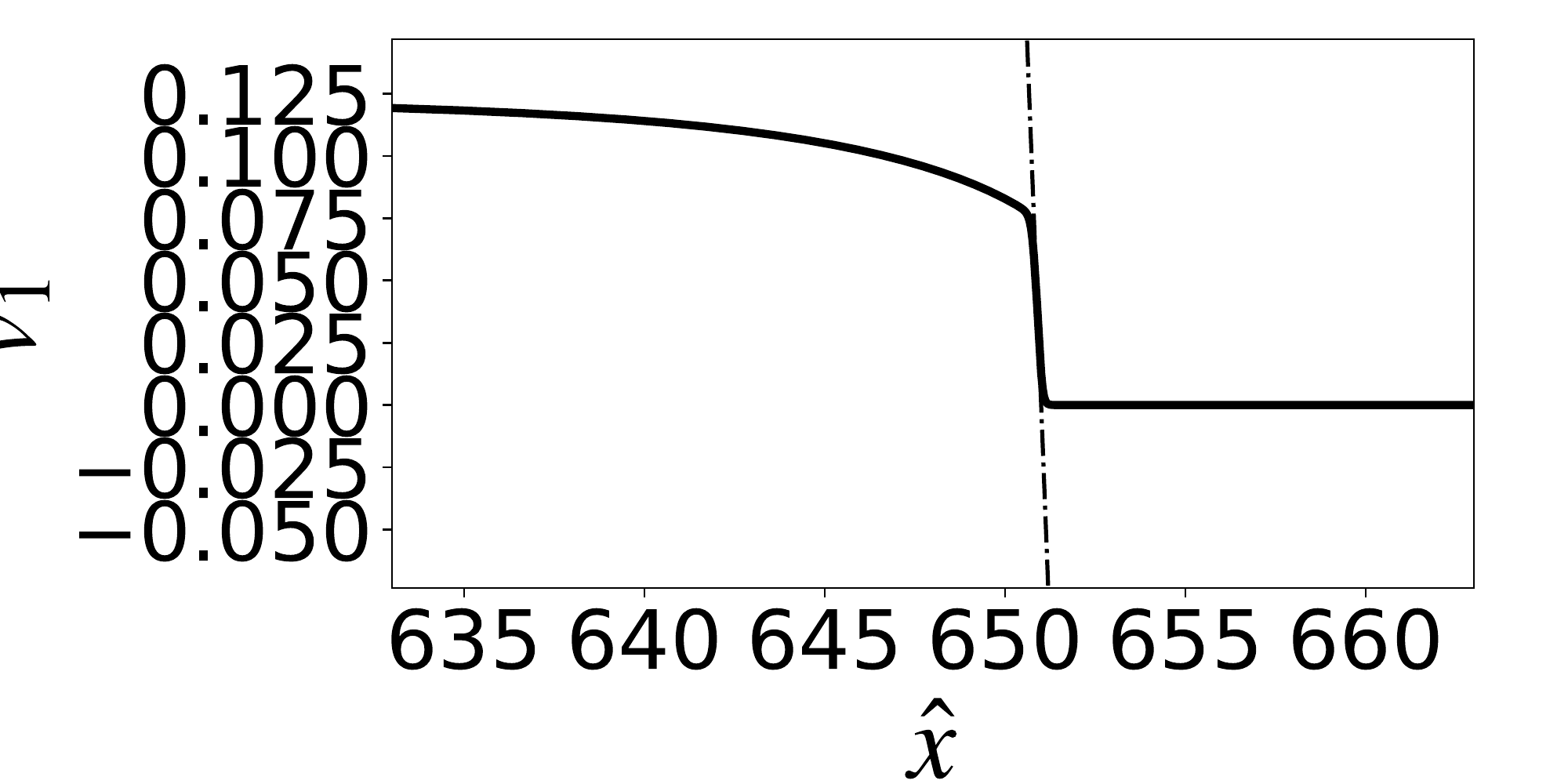} %
		\includegraphics[width=0.49\linewidth]{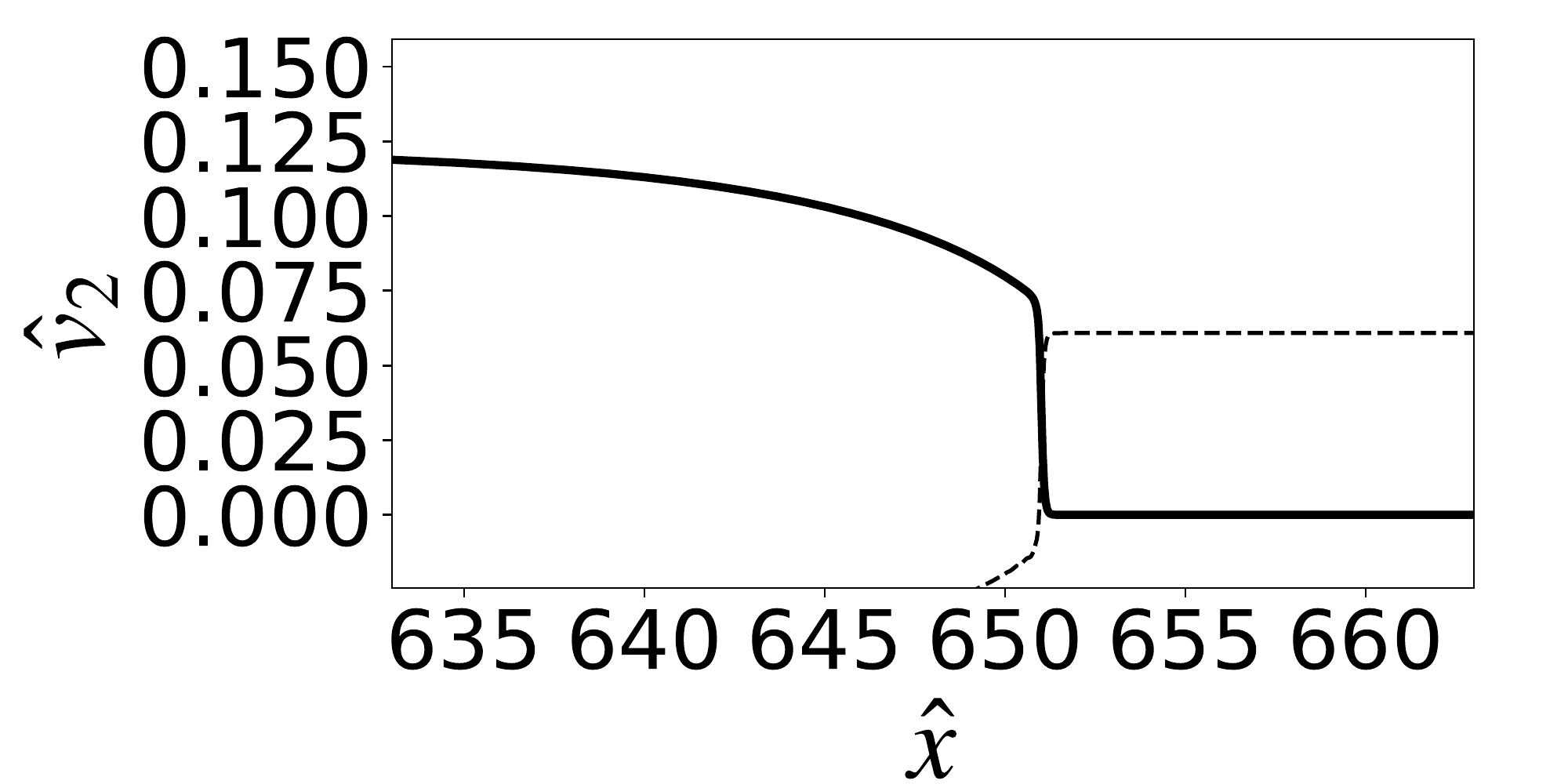} %
		\includegraphics[width=0.49\linewidth]{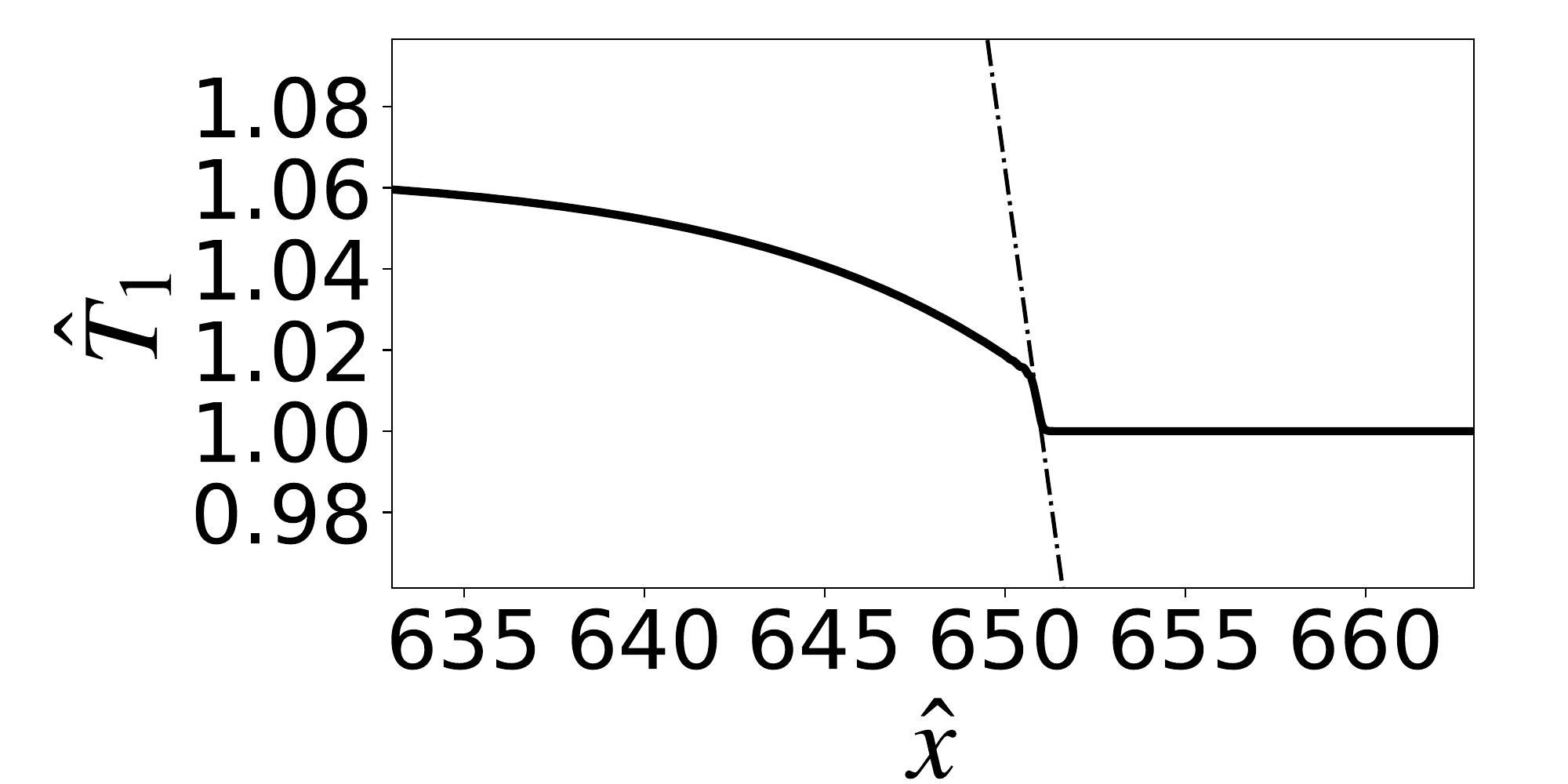} %
		\includegraphics[width=0.49\linewidth]{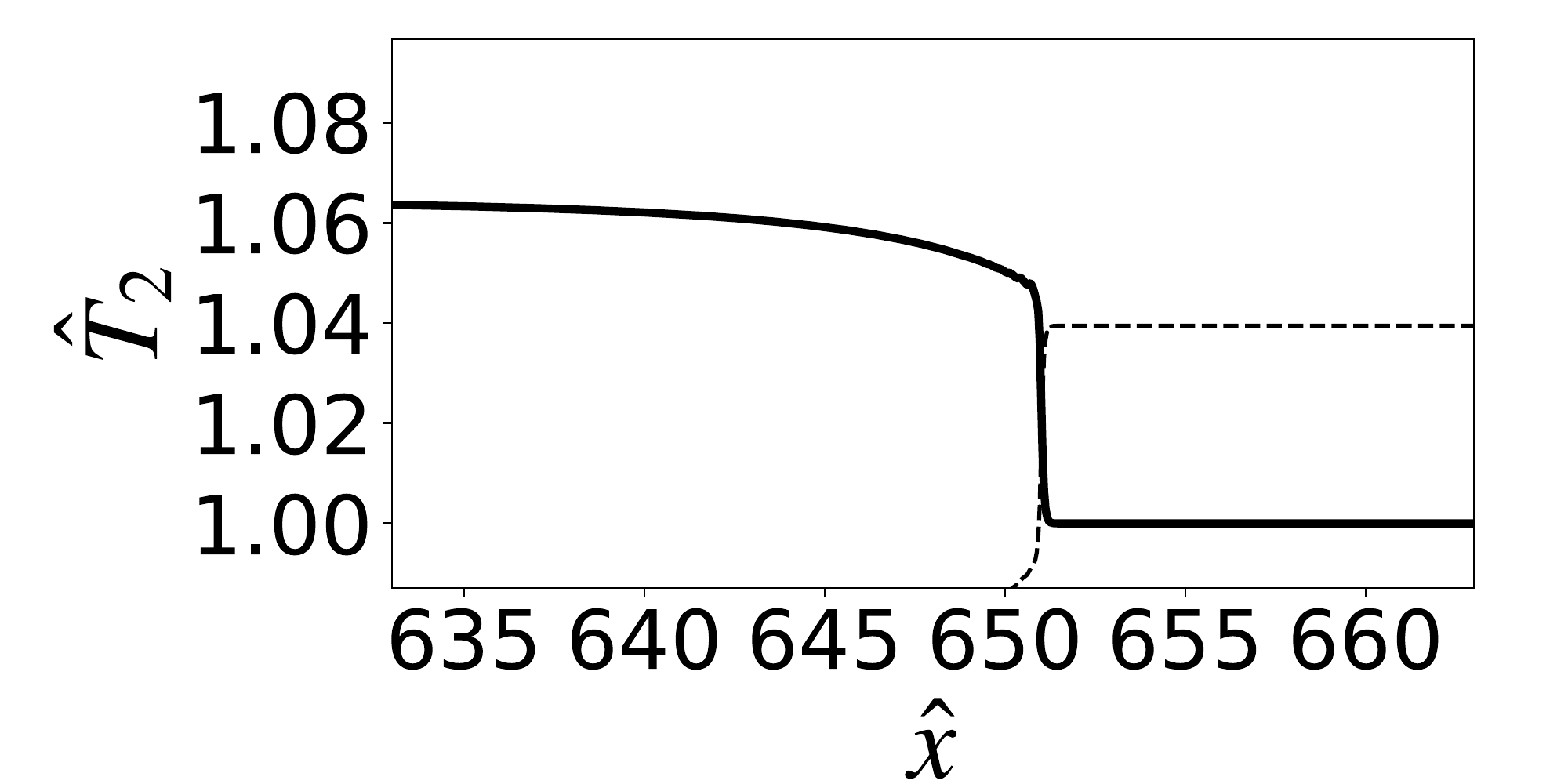} %
	\end{center}
	\caption{Shock structure obtained at $\hat{t} = 600$. 
	$\gamma_1 = 7/6$, $\gamma_2 = 5/3$, $\mu = 0.81$, $c_0 = 0.08$, $M_0 = 1.08$, $\Delta \hat{t} = 0.02$, and $\Delta \hat{x}=0.08$. 
	The dashed line shown in the profile for the constituent $1$ represent the theoretical predictions of the derivative of the discontinuity wave. 
	The dotted curves for the constituent $2$ are the potential sub-shock predicted any point of the shock structure.}
	\label{fig:c008_M0-1_08_discontinuity}
\end{figure}

Figure \ref{fig:c008_M0-1_08_discontinuity} shows a typical example of the discontinuity wave for the constituent $1$ induced by a sub-shock emerging in the constituent $2$. 
We see that the jump of the derivative of the constituent $1$ can be explained by the relationship \eqref{eq:discontinuity1}. 
To the author's knowledge, although the discontinuity waves can be observed numerically in other systems reported in previous papers on the shock structure~\cite{FMR,Bisi1,Bisi2,IJNLM2017,subshock2}, this is the first case to explain it quantitatively. 
For completeness, we summarize the expression of the jump of the derivative in the discontinuous wave for constituent $2$ in terms of the jump of the physical quantities due to the sub-shock for constituent $1$ obtained in a similar way: 
\begin{equation*}
	\begin{split}
		&\delta \rho_2 = \frac{m_2 \left[  (\gamma_2 - 1) [\![\hat{e}_1]\!] - (\gamma_2 - 1) [\![\hat{m}_1 u_2] \!] +  (s - v_2) [\![\hat{m}_1] \!] \right]}
		{m_2 (s - v_2)^3 - \gamma_2 k_B T_2 (s - v_1)}, \\
		&\delta v_2 =  \frac{m_2 \left[  (\gamma_2 - 1) [\![\hat{e}_1]\!] - (\gamma_2 - 1) [\![\hat{m}_1 u_2] \!] +  (s - v_2) [\![\hat{m}_1] \!] \right]}
		{\rho_2 \left[ m_2 (s - v_2)^2 -  k_B \gamma_2 T_2 \right]}, \\
		&\delta T_2 = \frac{m_2 (\gamma_2 - 1)}{k_B \rho_2 (s - v_2) \left[ m_2 (s - v_2)^2 -  k_B \gamma_2 T_2 \right]}\\
		& \times \left\{ \left[ m_2 (s - v_2)^2 - k_B T_2 \right] \left( [\![\hat{e}_1]\!] - [\![\hat{m}_1 u_2]\!] \right) 
		+ k_B T_2 (s - v_2) [\![\hat{m}_1] \!] \right\}. \\
	\end{split}
\end{equation*}

\subsection{Profiles of the global quantities and average temperature}
\begin{figure}
	\begin{center}
		\includegraphics[width=0.49\linewidth]{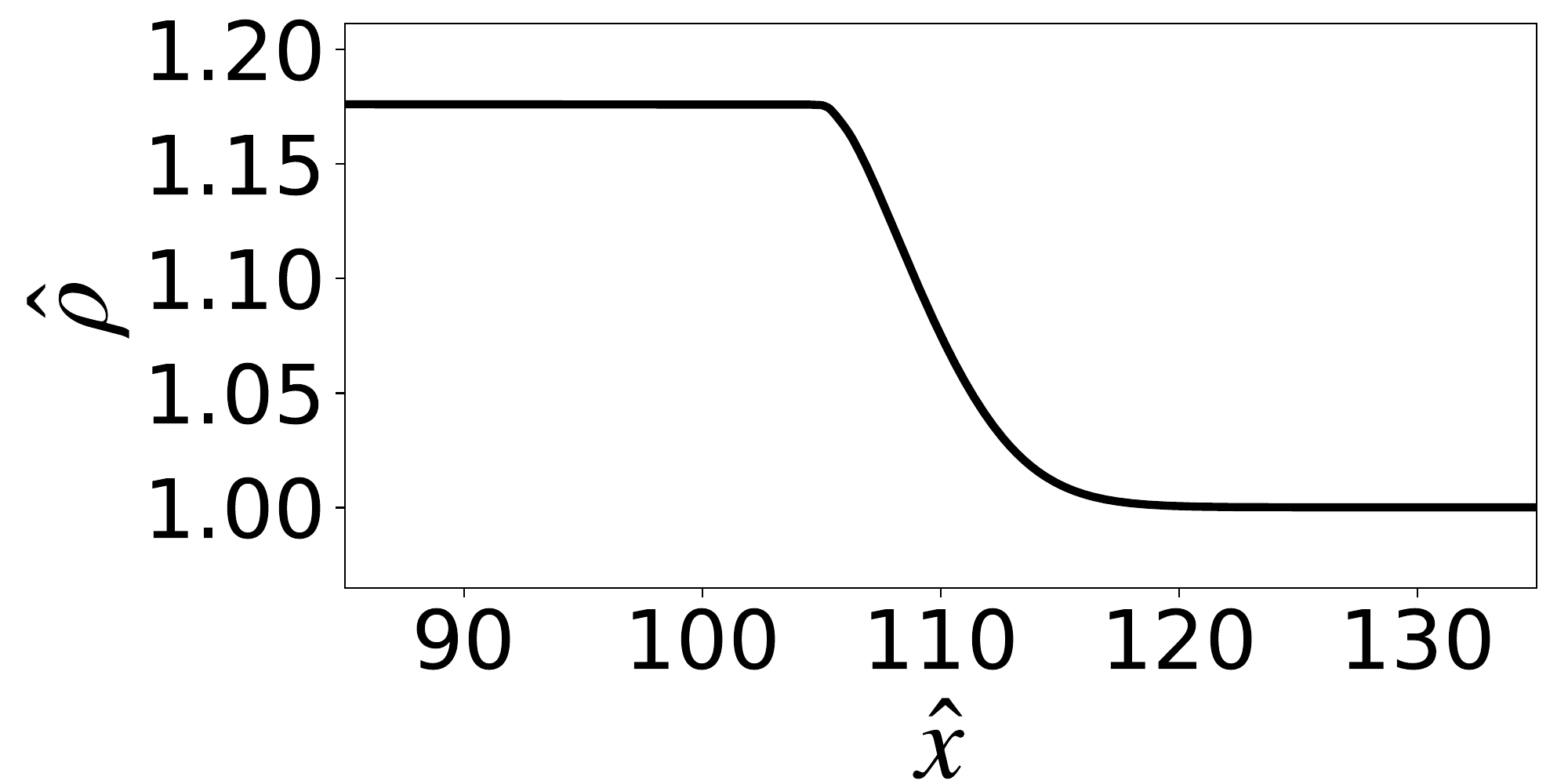} %
		\includegraphics[width=0.49\linewidth]{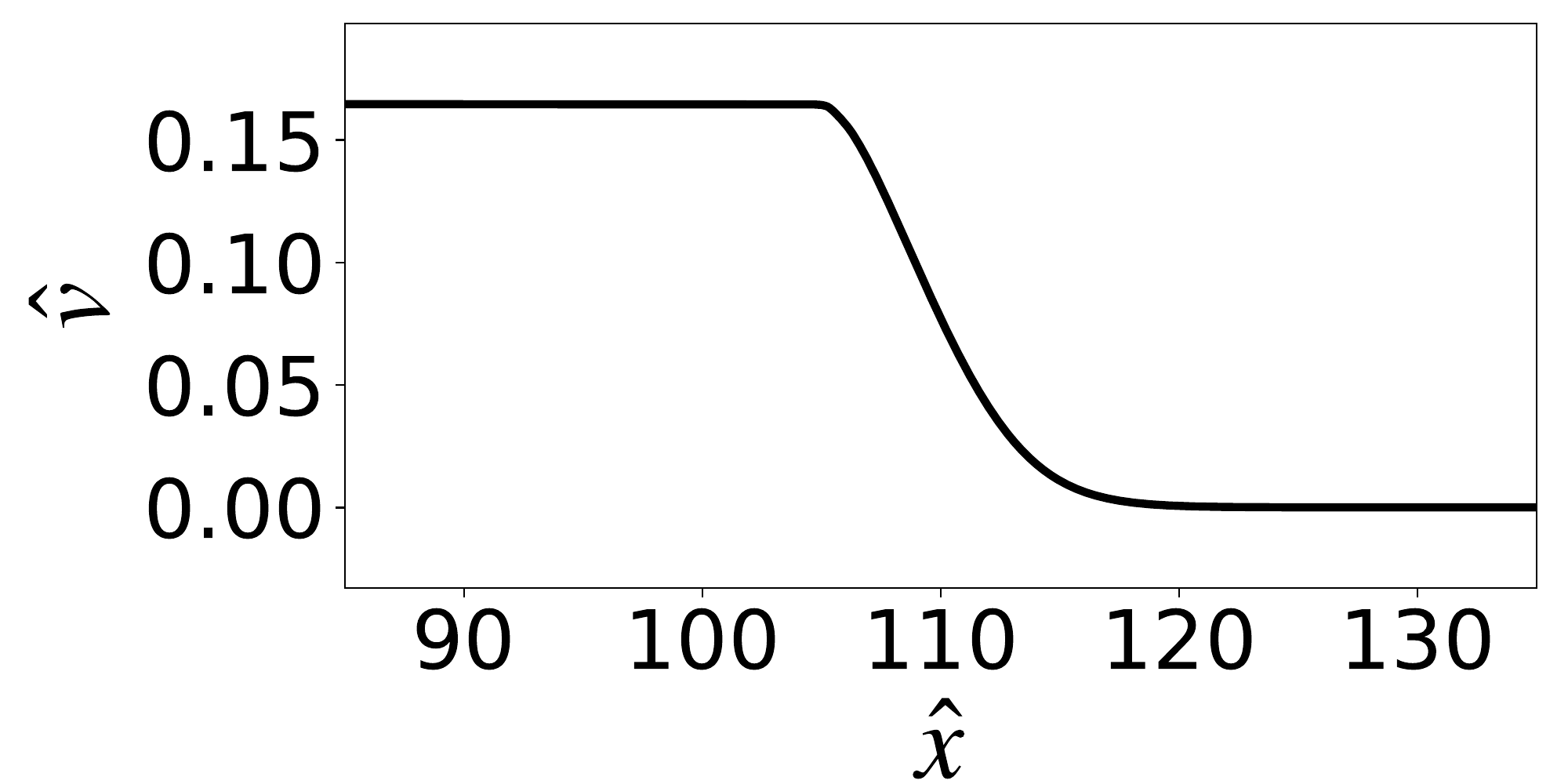} %
		\includegraphics[width=0.49\linewidth]{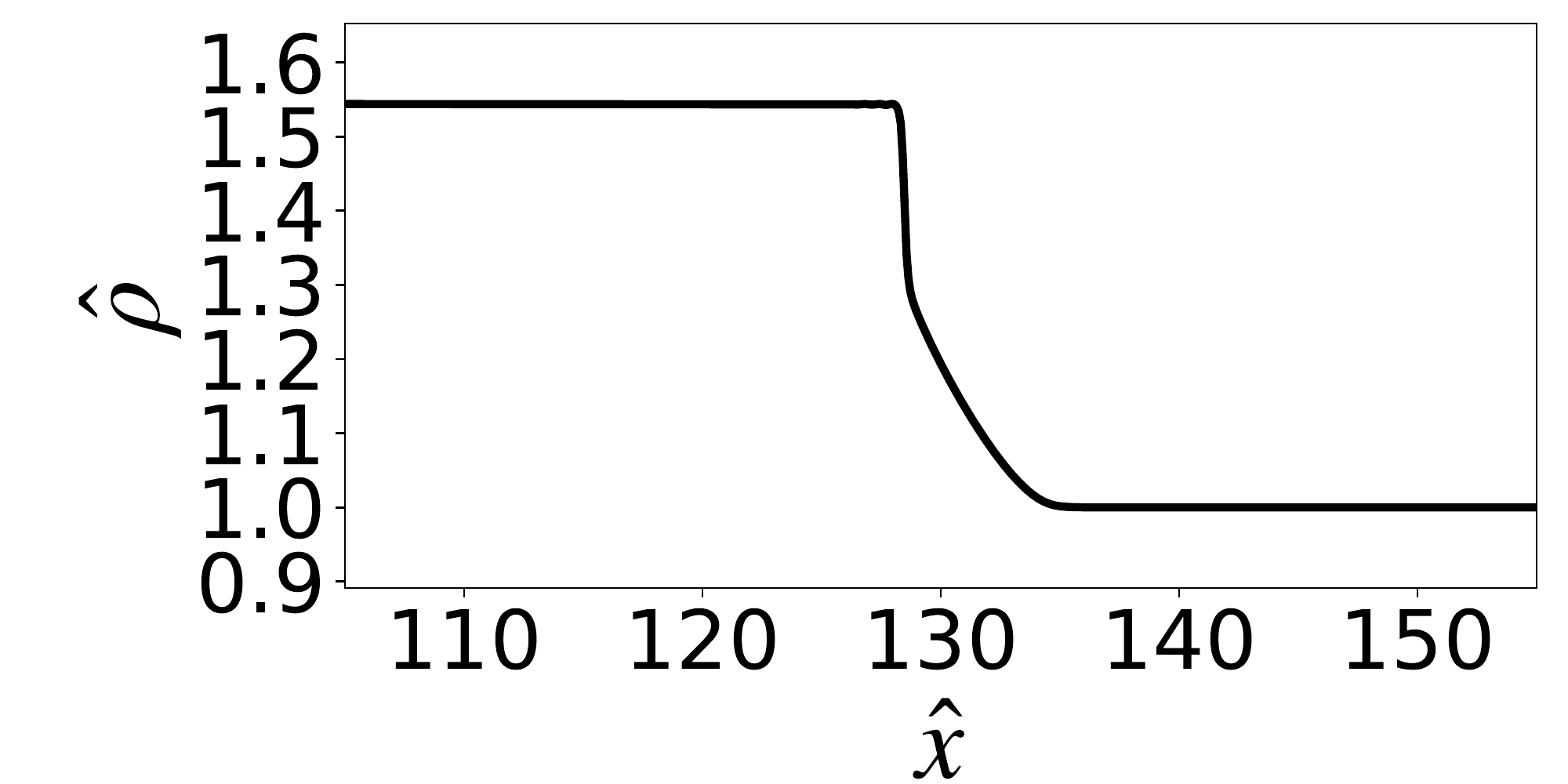} %
		\includegraphics[width=0.49\linewidth]{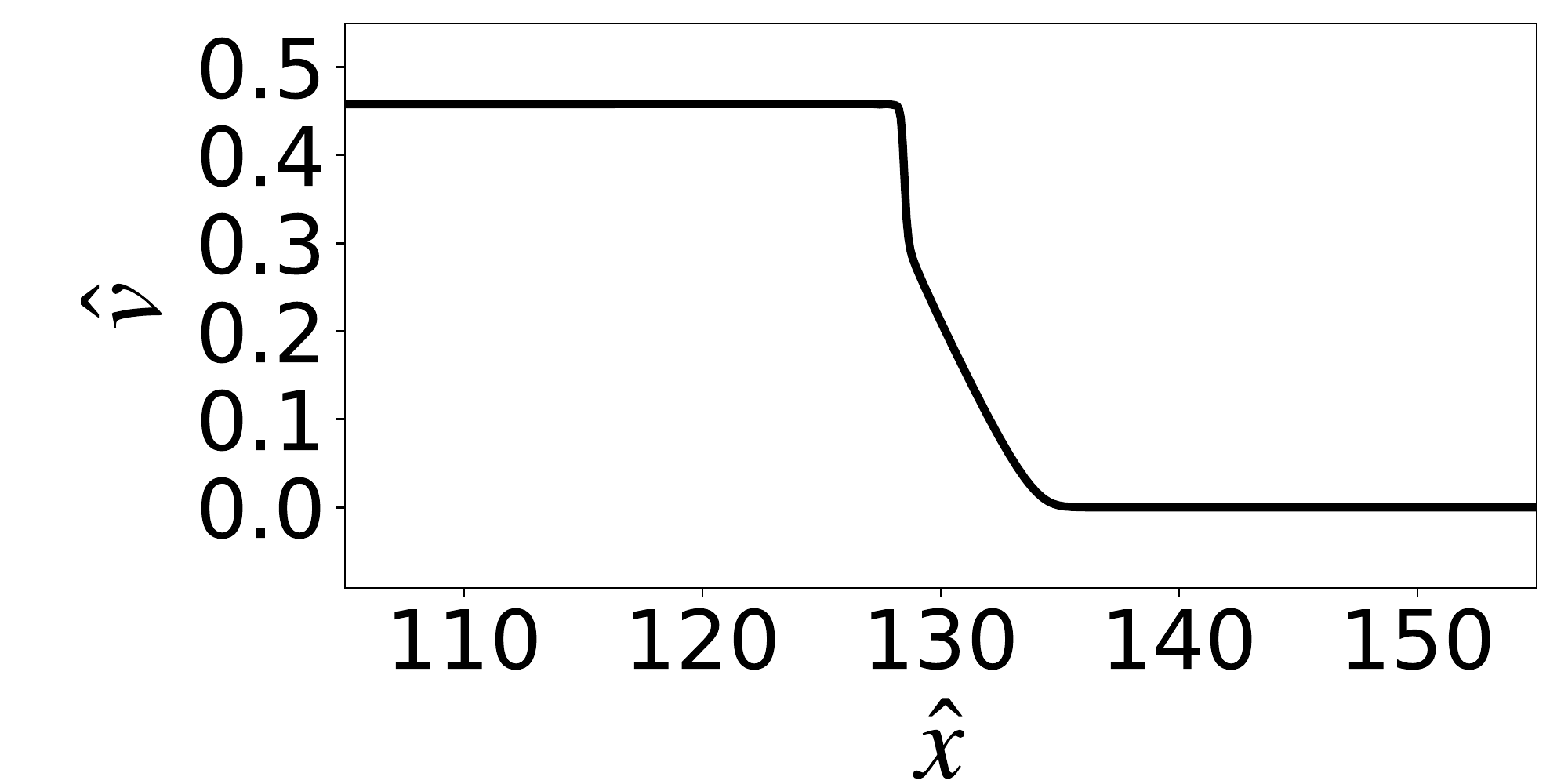} %
		\includegraphics[width=0.49\linewidth]{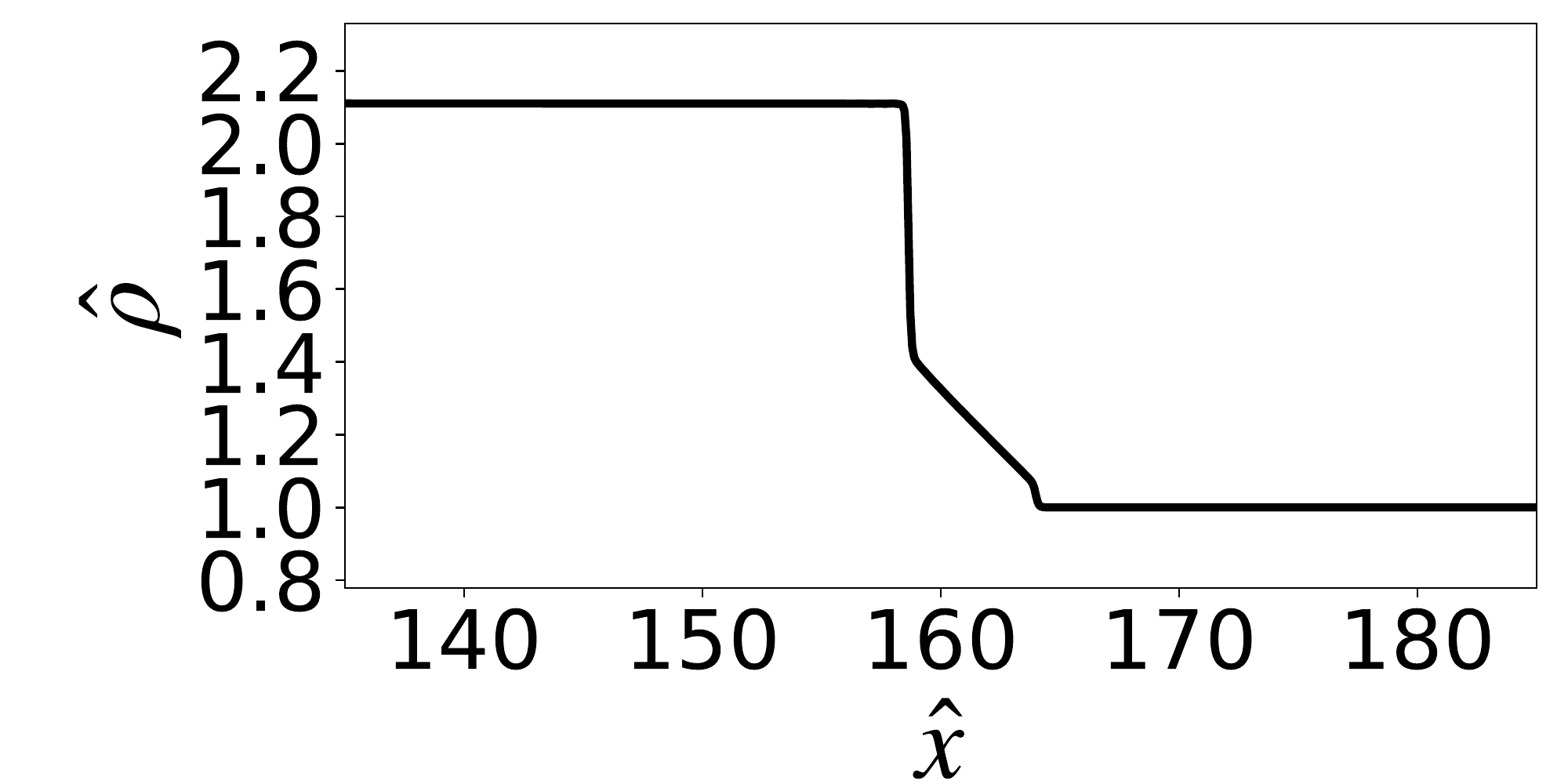} %
		\includegraphics[width=0.49\linewidth]{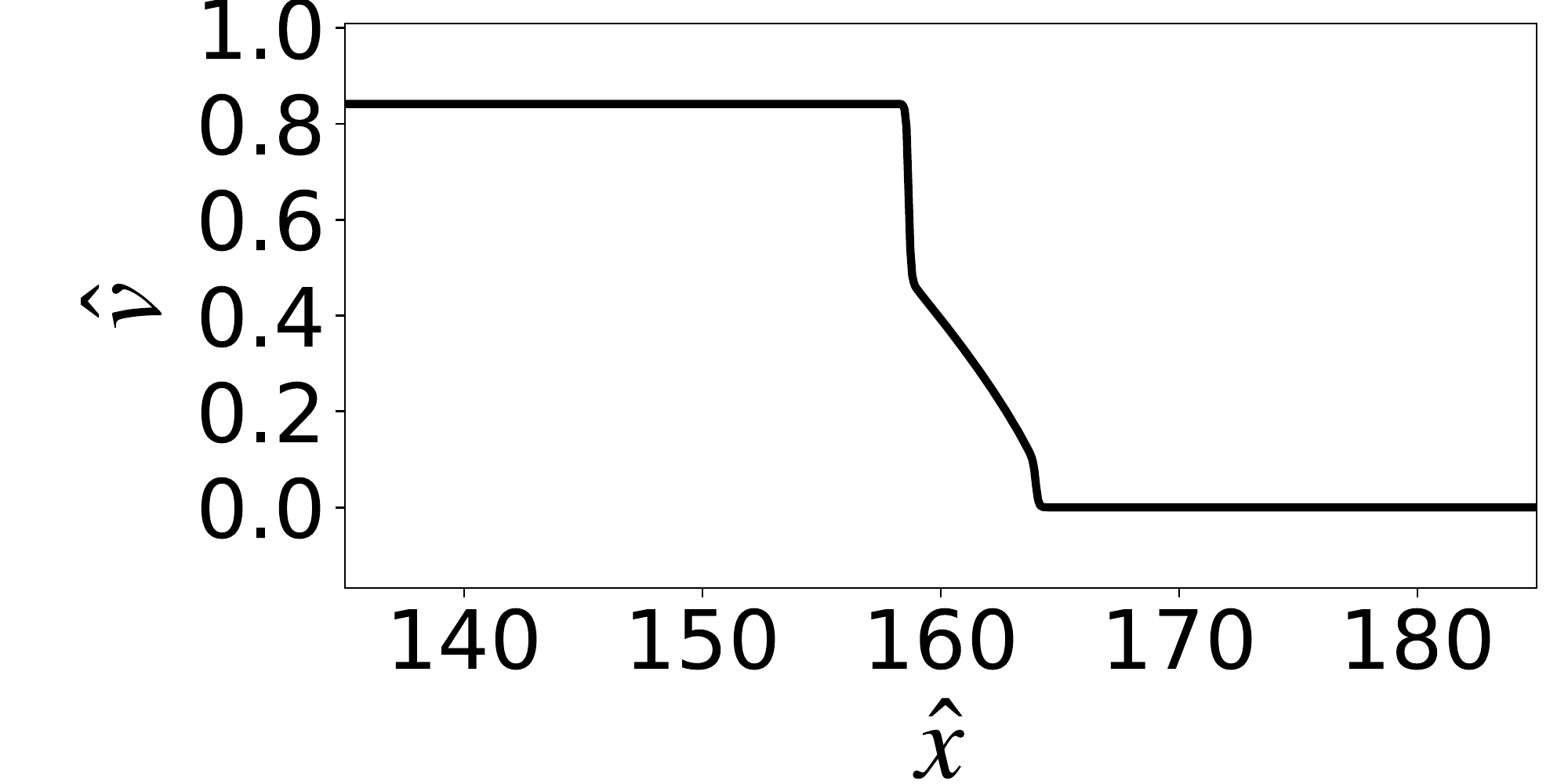} %
	\end{center}
	\caption{Case A: Profiles of the dimensionless total mass density and the mixture velocity.  
	The parameters are $\gamma_1 = 7/5$, $\gamma_2 = 9/7$, $\mu = 0.45$, $c_0 = 0.2$. 
	The Mach numbers are $M_0 = 1.1$ (top), $M_0 = 1.3$ (middle), and $M_0 = 1.6$ (bottom). }
	\label{fig:totalrhov}
\end{figure}

Because it is quite difficult to measure the profile for each constituent independently, the profiles of averaged or total quantities are often observed experimentally. 
To understand the shock structure for averaged or total quantities, we need to analyze how the sub-shocks appearing for each constituent affect the profile of these quantities. 
By using the definitions \eqref{eq:totalrho} and \eqref{def:v}, we can easily obtain the profiles of the total mass density and the mixture velocity from the ones of the mass densities and the velocity for both constituents. 
Figure \ref{fig:totalrhov} shows the total mass density and the mixture velocity profile corresponding to the parameters adopted in Figures \ref{fig:c02_M0-1_1}--\ref{fig:c02_M0-1_6}. 
Now the meaning of the multiple sub-shocks becomes clear because the two jumps appear in a profile of the total mass density or the mixture velocity for $M_0 = 1.6$ corresponding to Region IV. 

 Ruggeri and Simi\'c in the paper ~\cite{RuggeriSimic2008proceedings} proposed two possible definition of average temperature. The first is based on the  idea  that the global equation of the energy \eqref{finale}$_3$ as in single fluids govern the evolution of the average temperature and  for homogeneous solutions (independent of $x$), the internal energy $\varepsilon$ together with  $\Theta$ become constant (see ref. \cite{RuggeriSimic2008proceedings,Ha_Ruggeri}). Therefore taking into account the expression of the global internal energy \eqref{campiG}$_3$ we have  
   the average temperature $\Theta$  implicitly defined by:
 \begin{equation}\label{eq:def_AveT}
 	\begin{split}
 		&\rho_1 \varepsilon_1 (\rho_1, \Theta) + \rho_2 \varepsilon_2 (\rho_2, \Theta)\\
 		&= \rho_1 \varepsilon_1 (\rho_1 , T_1) + \rho_2 \varepsilon_2 (\rho_2, T_2) + \frac{1}{2} \rho_1 u_1^2 + \frac{1}{2} \rho_2 u_2^2. 
 	\end{split}
 \end{equation}
 
 Another possible definition is to consider the identification of the average temperature $\mathcal{T}$ such that the intrinsic internal energy $\varepsilon_I$ of the multi-temperature mixture resembles the structure of intrinsic internal energy of a single-temperature one. Therefore, the following implicit definition of an average temperature is adopted:
  \begin{equation}\label{eq:def_AveTL}
 	\begin{split}
 		\rho \varepsilon_I 
		 &= \rho_1 \varepsilon_1 (\rho_1, \mathcal{T}) + \rho_2 \varepsilon_2 (\rho_2, \mathcal{T})\\
 		&= \rho_1 \varepsilon_1 (\rho_1 , T_1) + \rho_2 \varepsilon_2 (\rho_2, T_2) . 
 	\end{split}
 \end{equation}
If the diffusion is not big so that the process is not far from equilibrium, we may neglect the non-linear terms in \eqref{eq:def_AveT} and have $\Theta \simeq \mathcal{T}$.
The definition of $\mathcal{T}$ was reconsidered in successive papers \cite{GouinRuggeri,RuggeriSimic2009,Ruggeri_Lou} and, in particular, in \cite{RuggeriSimic2009}, it is proved that $\mathcal{T}$ is the natural definition such that the entropy density is maximum in an equilibrium state.
 
By inserting the caloric equation of state \eqref{coneq}$_2$ into \eqref{eq:def_AveT}, we obtain the explicit expression of $\Theta$: 
\begin{equation}
	\begin{split}
		\Theta = \frac{\frac{k_B }{m_1} \frac{\rho_1 T_1}{\qty(\gamma_1 - 1)} + \frac{k_B }{m_2} \frac{\rho_2 T_2}{\qty(\gamma_2 - 1)} + \frac{1}{2} \rho_1 u_1^2 + \frac{1}{2} \rho_2 u_2^2}
		{k_B \qty{ \frac{\rho_1}{m_1 \qty(\gamma_1 - 1)} + \frac{\rho_2}{m_2 \qty(\gamma_2 - 1)} }}. 
	\end{split}
\end{equation}
or
\begin{equation}
	\Theta = \frac{\frac{k_B }{m_1}\frac{\rho_1 T_1}{\qty(\gamma_1 - 1)} + \frac{k_B }{m_2}\frac{\rho_2 T_2}{\qty(\gamma_2 - 1)} + \frac{1}{2}\qty(\Pi - \sigma)}
	{k_B \qty{ \frac{\rho_1}{m_1 \qty(\gamma_1 - 1)} + \frac{\rho_2}{m_2 \qty(\gamma_2 - 1)} }}. 
\end{equation}
While $\mathcal{T}$ becomes: 
\begin{equation}
	\mathcal{T} = \frac{\frac{1}{m_1} \frac{\rho_1 T_1}{\qty(\gamma_1 - 1)} + \frac{1}{m_2} \frac{\rho_2 T_2}{\qty(\gamma_2 - 1)}}
		{\frac{\rho_1}{m_1 \qty(\gamma_1 - 1)} + \frac{\rho_2}{m_2 \qty(\gamma_2 - 1)}}.
\end{equation}

For convenience, we introduce the dimensionless average temperature $\hat{\Theta} \equiv \Theta/T_0$: 
\begin{equation}
	\begin{split}
		&\hat{\Theta} = \frac{\frac{\hat{\rho}_1 \hat{T}_1}{\qty(\gamma_1 - 1)} + \mu\frac{\hat{\rho}_2 \hat{T}_2}{\qty(\gamma_2 - 1)} 
		+ \frac{1}{2}\gamma_0\frac{m_1}{m_0}\qty{\hat{\rho}_1 (\hat{v}_1 - \hat{v})^2 + \hat{\rho}_2 (\hat{v}_2 - \hat{v})^2}}
		{\frac{\hat{\rho}_1}{\qty(\gamma_1 - 1)} + \mu\frac{\hat{\rho}_2}{\qty(\gamma_2 - 1)} }, 
	\end{split}
\end{equation}
where taking into account \eqref{equ:avgTemp} $m_1/m_0$ can be written as
\begin{equation}
	\begin{split}
		\frac{m_1}{m_0} = c_0 + \mu (1 - c_0). 
	\end{split}
\end{equation}
The dimensionless form of $\hat{\mathcal{T}} \equiv \mathcal{T}/T_0$ is given by
\begin{equation}
	\hat{\mathcal{T}}
	= \frac{\frac{\hat{\rho}_1 \hat{T}_1}{\qty(\gamma_1 - 1)} + \mu\frac{\hat{\rho}_2 \hat{T}_2}{\qty(\gamma_2 - 1)}}
	{\qty{ \frac{\hat{\rho}_1}{\qty(\gamma_1 - 1)} + \mu\frac{\hat{\rho}_2}{\qty(\gamma_2 - 1)}}}. 
\end{equation}

\begin{figure}[h]
	\begin{center}
		\includegraphics[width=0.49\linewidth]{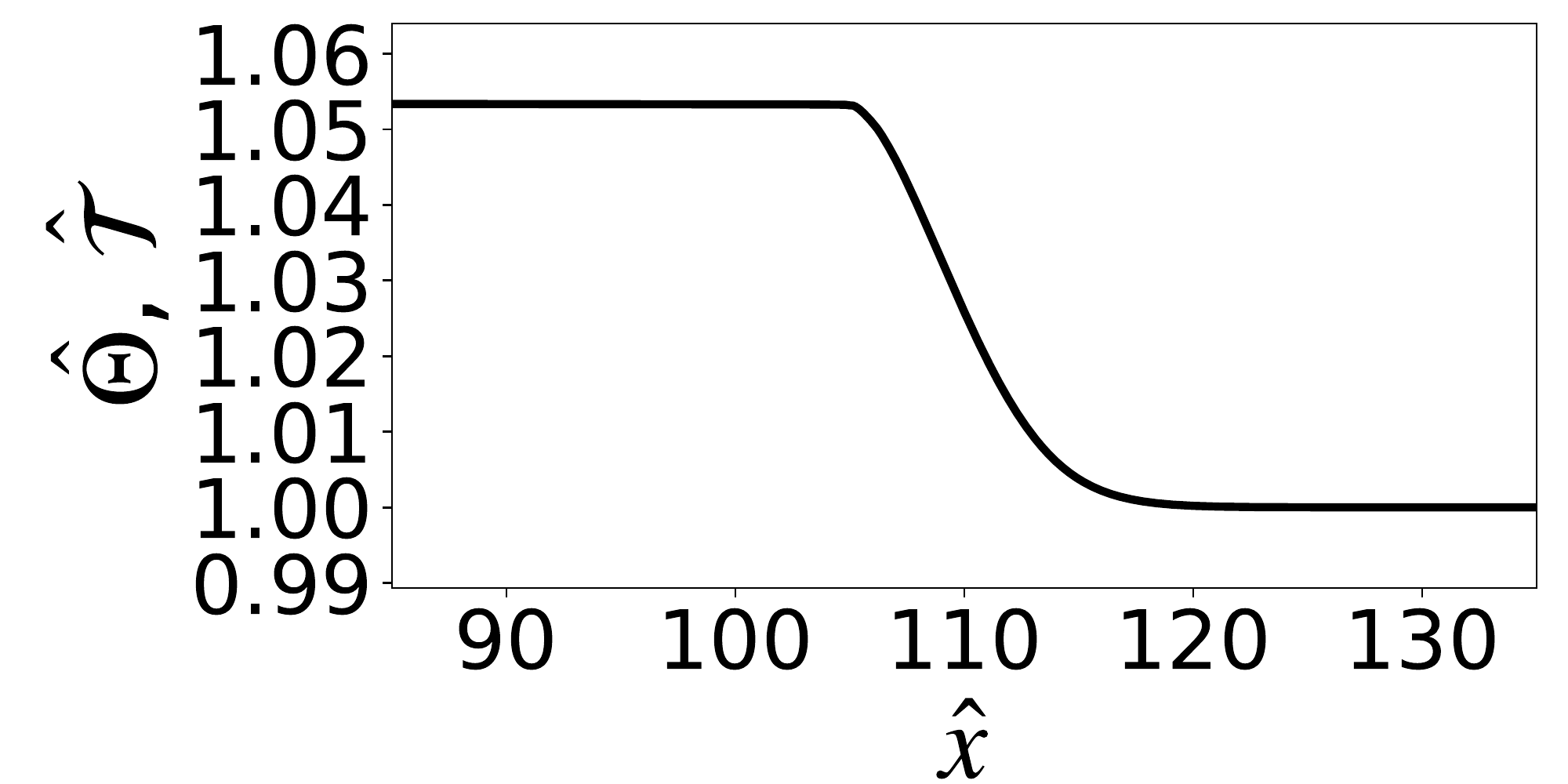} 	
		\includegraphics[width=0.49\linewidth]{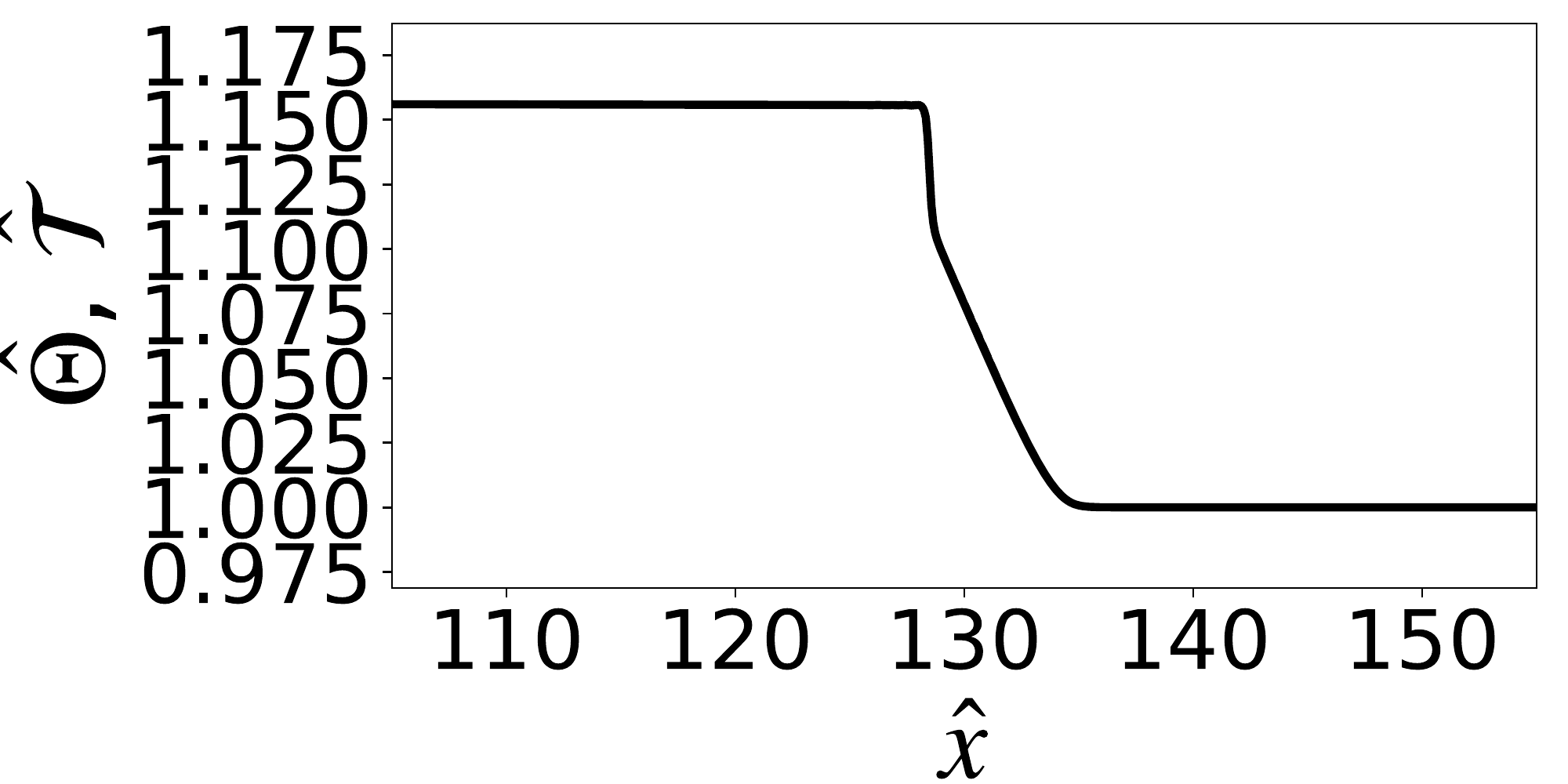} %
		\includegraphics[width=0.49\linewidth]{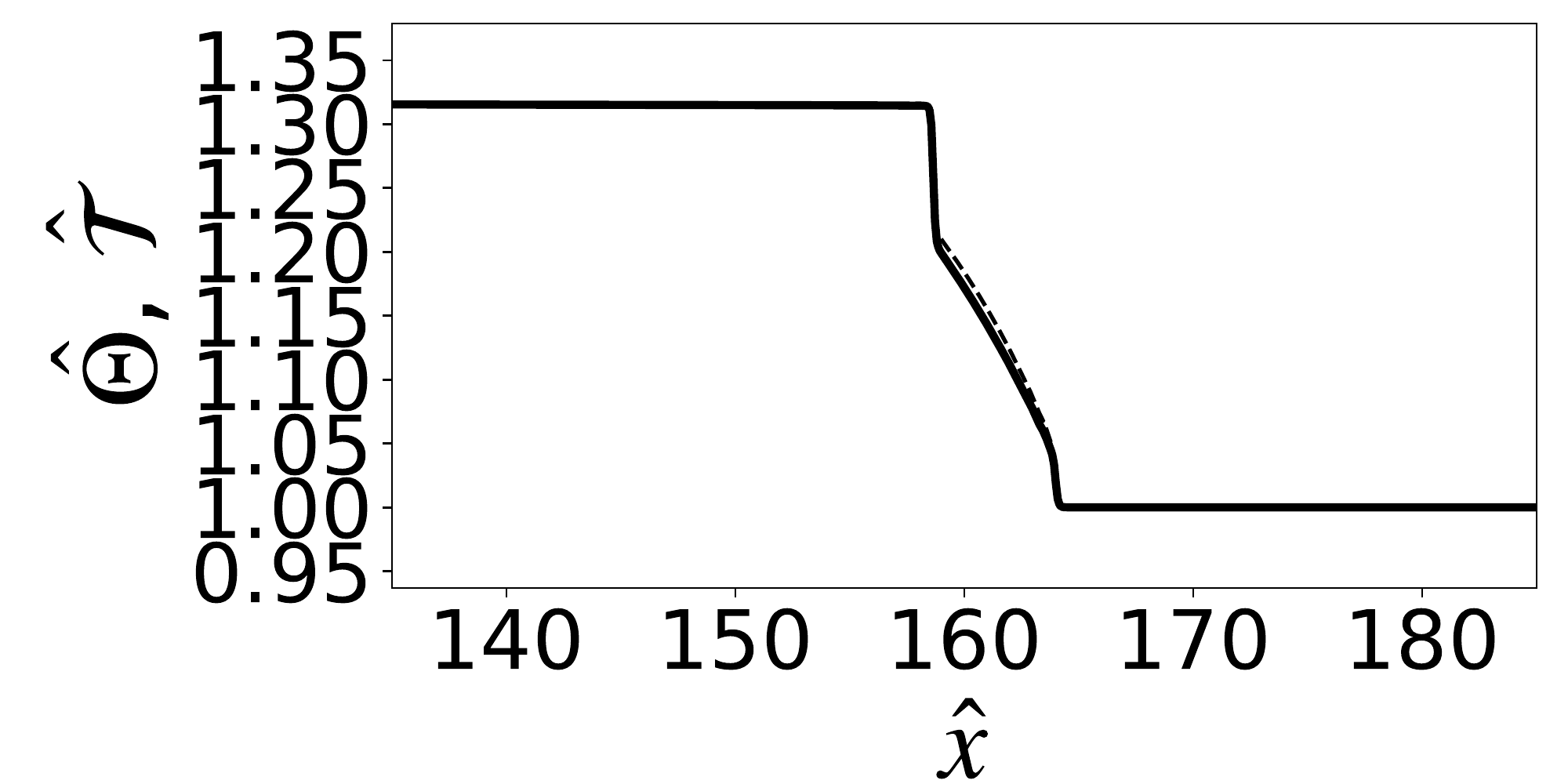} %
	\end{center}
	\caption{Profiles of the dimensionless average temperature $\hat{\Theta}$ (solid curve) and $\hat{\mathcal{T}}$ (dotted curve). 
	$\gamma_1 = 7/5$, $\gamma_2 = 9/7$, $\mu = 0.45$, and $c_0 = 0.2$. 
	The Mach numbers are $M_0 = 1.1$ (above left), $M_0 = 1.3$ (above right), and $M_0 = 1.6$ (below). }
	\label{fig:aveT_Euler_mu045_c02}
\end{figure}

\begin{figure}[h]
	\begin{center}
		\includegraphics[width=0.49\linewidth]{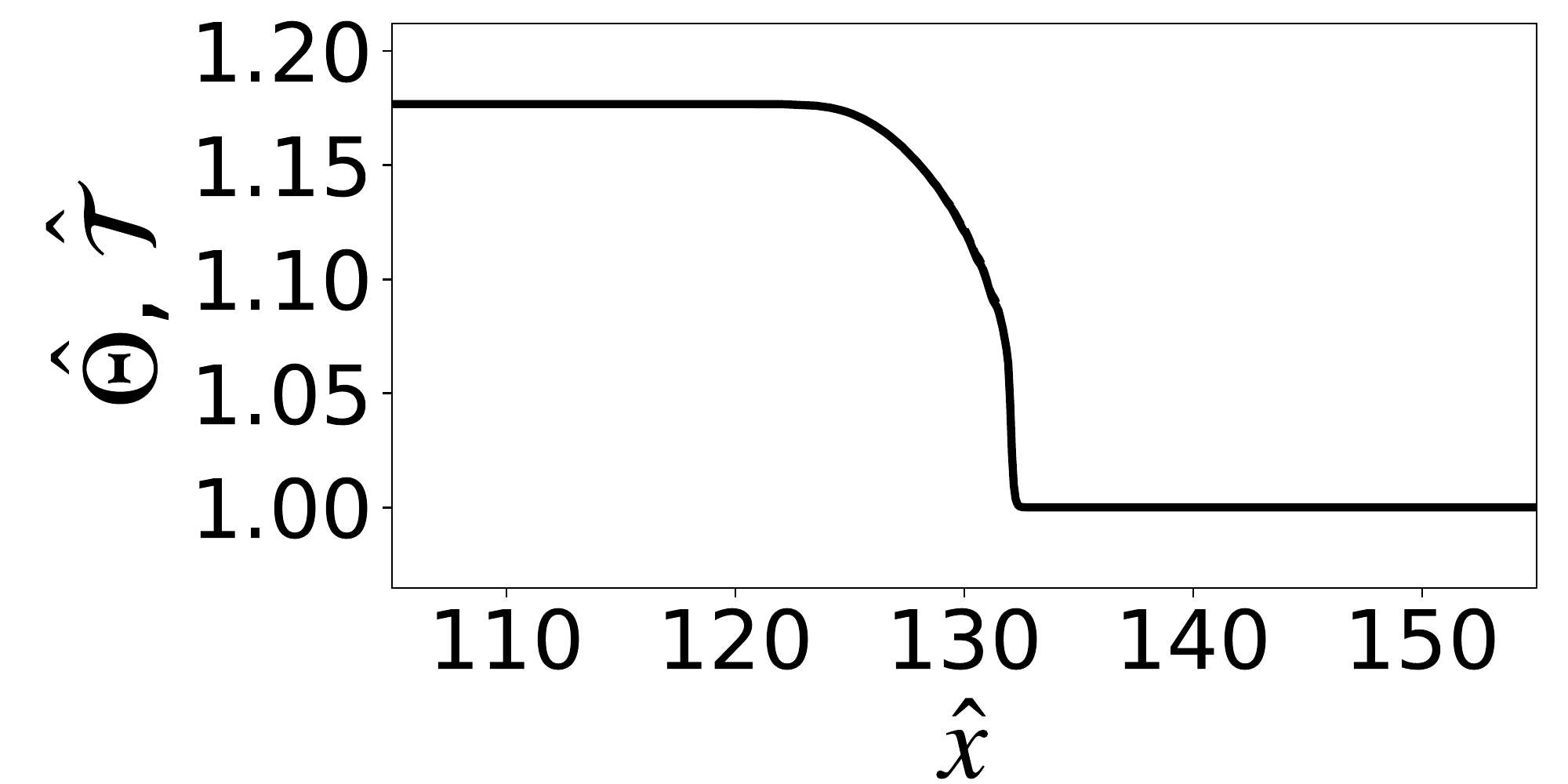} %
		\includegraphics[width=0.49\linewidth]{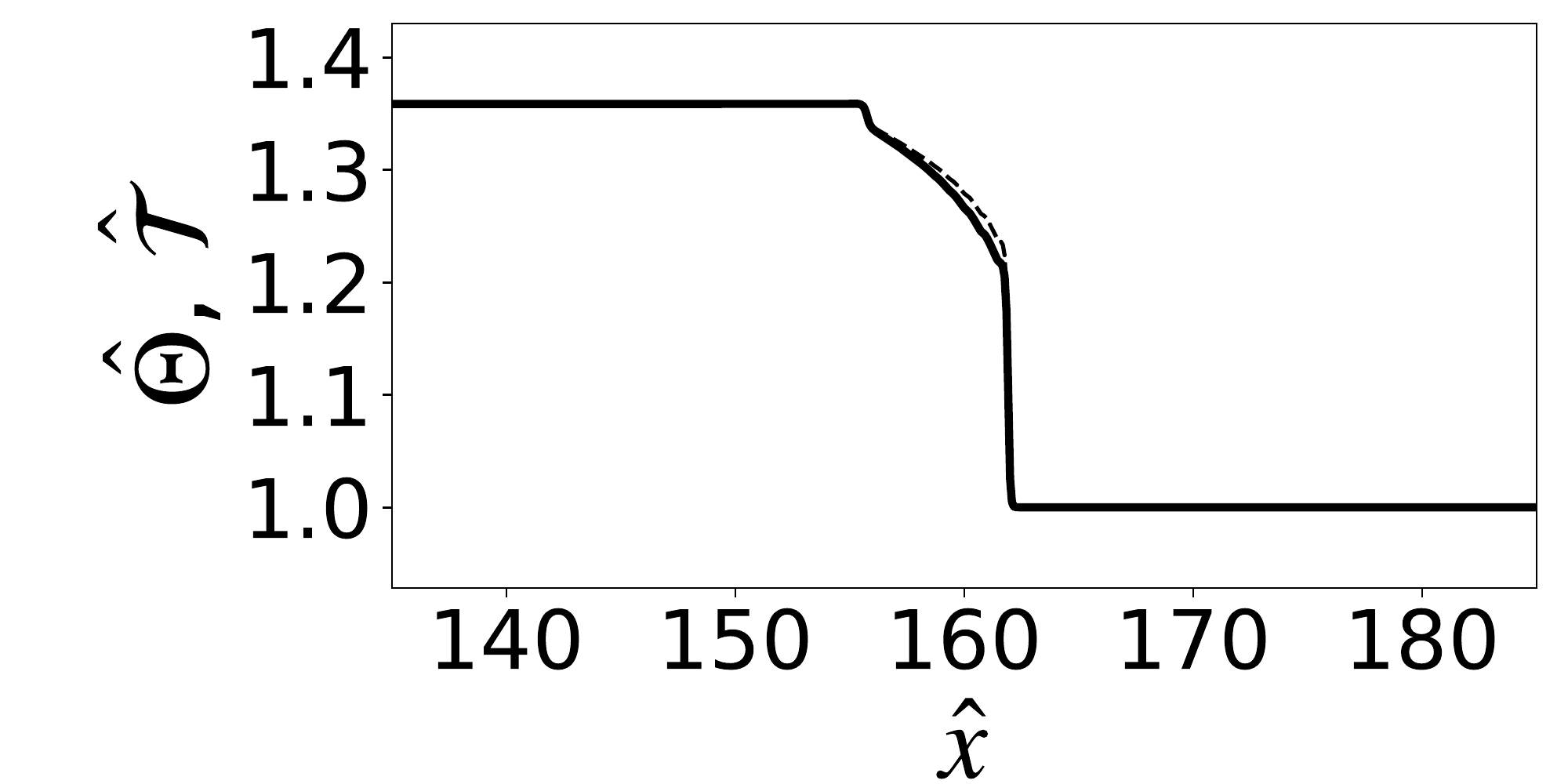} %
	\end{center}
	\caption{Profiles of the dimensionless average temperature $\hat{\Theta}$ (solid curve) and $\hat{\mathcal{T}}$ (dotted curve). 
	$\gamma_1 = 7/5$, $\gamma_2 = 9/7$, $\mu = 0.45$, and $c_0 = 0.6$. 
	The Mach numbers are 
	$M_0 = 1.3$ (left) and $M_0 = 1.6$ (right). }
	\label{fig:aveT_Euler_mu045_c06}
\end{figure}

We show the profile of the average temperature $\Theta$ and $\mathcal{T}$ in Figure \ref{fig:aveT_Euler_mu045_c02} with the parameters corresponding to the ones in Figures \ref{fig:c02_M0-1_1}-\ref{fig:c02_M0-1_6} and also in Figure \ref{fig:aveT_Euler_mu045_c06} with the parameters corresponding to the ones in Figures \ref{fig:c06_M0-1_3} and \ref{fig:c06_M0-1_6}. 
It is confirmed that the sub-shock formation can also be detected through the measurement of the average temperature. 
While the difference between $\Theta$ and $\mathcal{T}$ is negligible for small Mach numbers, the nonlinear terms may play a role in the profile of the average temperature for large Mach numbers. 

\bigskip

\section{Summary and concluding remarks}
\label{sec:summary}

In the present paper, we have analyzed the shock structure in a binary Eulerian mixture of gases with different degrees of freedom by using the framework of the multi-temperature model. 
In the first part, by considering the necessary conditions of the sub-shock formation, we have classified the possible regions for the sub-shock formation into four parts in terms of the concentration and the Mach number. 
It is found that the topology of the regions can be quite different from the one obtained in a binary mixture of monatomic gases if the degrees of freedom of a lighter molecule is larger than the one of a heavier molecule. 
In the second part, we have numerically solved the system of field equations and have made the possibility of the sub-shock formation clear for all cases of the parameters in the classified regions. 
Moreover, we have found that a regular singular point for moderate Mach numbers may become singular for larger Mach numbers. 
The relationship between the sub-shock for a constituent and the acceleration wave for the other constituent is explicitly derived. 
From the viewpoint of possible experimental observation, we have also shown the shock structure for the global or average quantities, in particular, the one for the average temperature. 

The remarks of the present study are summarized as follows: 

(I) We need to pay attention to the fact that the present analysis is based on the Eulerian mixture of polyatomic gases in which the viscosity and heat conductivity of each constituent are neglected. 
In reality, the sub-shock (discontinuous surface) does not exist, and it is replaced by a steep but continuous change in the physical quantity. Nevertheless, the thickness of the shock is for many gases, particularly for monatomic gas of the order of mean free path and therefore negligible at the macroscopic scale. 
The previous observation agrees that the shock wave theory based on the system of Euler equations in a single fluid has vast literature. 
We could not find experimental data for the binary mixture to compare our results. 
For possible future experimental verification of the corresponding phenomena found in the present paper, it is essential to choose a pair of gases with appropriate degrees of freedom and the ratio of the masses. 
We summarize the values of $\gamma_1$, $\gamma_2$, and $\mu$ for possible pairs of gases classified into cases A and B in Tables \ref{tbl:caseA} and \ref{tbl:caseB}. 
\begin{table}[htbp]
	\begin{center}
		\caption{Values of the ratio of the specific heats at 298.15 [K] and the ratio of the masses~\cite{CRChandbook} for typical binary mixtures of gases in Case A. 
		}
		\begin{tabular}{c | c c c | c}\label{tbl:caseA}
			& $\gamma_1$ & $\gamma_2$ & $\mu$ & $\mu^{*}$  \\ \hline
			\mbox{He / CO$_2$} & 5/3 & 1.29 & 0.091 & 0.259 \\ 
			\mbox{H$_2$ / CO$_2$} & 1.41 & 1.29 & 0.046 & 0.157 \\ 
			\mbox{N$_2$ / CO$_2$} & 1.40 & 1.29 & 0.637 & 0.155 \\ 
		\end{tabular}
	\end{center}
\end{table}
\begin{table}[htbp]
	\begin{center}
		\caption{Values of the ratio of the specific heats at 298.15 [K] and the ratio of the masses~\cite{CRChandbook} for typical binary mixtures of gases in Case B. 
		}
		\begin{tabular}{c | c c c | c c}\label{tbl:caseB}
			& $\gamma_1$ & $\gamma_2$ & $\mu$ & $ g $ & $\mu^{**}$  \\ \hline
			\mbox{N$_2$ / Ar} & 1.40 & 5/3 & 0.701 & 0.840 & 0.897\\ 
			\mbox{O$_2$ / Ar} & 1.39 & 5/3 & 0.801 & 0.837 & 0.897\\ 
			\mbox{HCl / Ar} & 1.40 & 5/3 & 0.913 & 0.836 & 0.897\\ 
		\end{tabular}
	\end{center}
\end{table}

(II) The sub-shock formation plays an essential role in the context of the shock structure in a single fluid of some rarefied polyatomic gases that have a large bulk viscosity~\cite{ET6shock,NLET6shock,ET6shockFukuoka}. 
The theory of mixtures of polyatomic gases with large bulk viscosity based on the RET theory has been proposed~\cite{GMixture} and  
we will consider in a future paper the shock structure in a binary mixture incorporating the effect of the dynamic pressure.  

\begin{acknowledgments}
	This work is partially supported by GNFM- INdAM (TR) and by JSPS KAKENHI Grant Number JP19K04204 (ST).
\end{acknowledgments}	

\section*{Author Declarations}

\section*{Conflict of Interest}

The authors have no conflicts to disclose. 

\section*{Data Availability}

Data sharing is not applicable to this article as no new data were created or analyzed in this study.

\bigskip

\appendix

\section{Dynamical analysis of the ODE system}
\label{sec:dynamical}
\begin{figure}
	\begin{center}
		\includegraphics[width=0.45\linewidth]{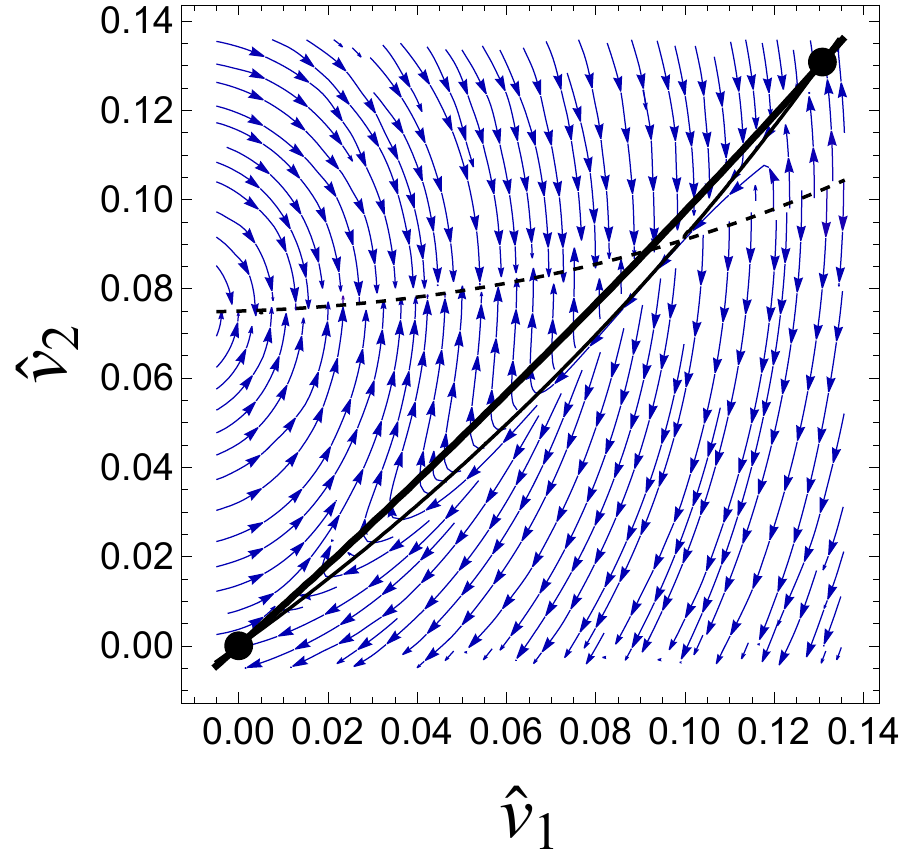} %
		\includegraphics[width=0.45\linewidth]{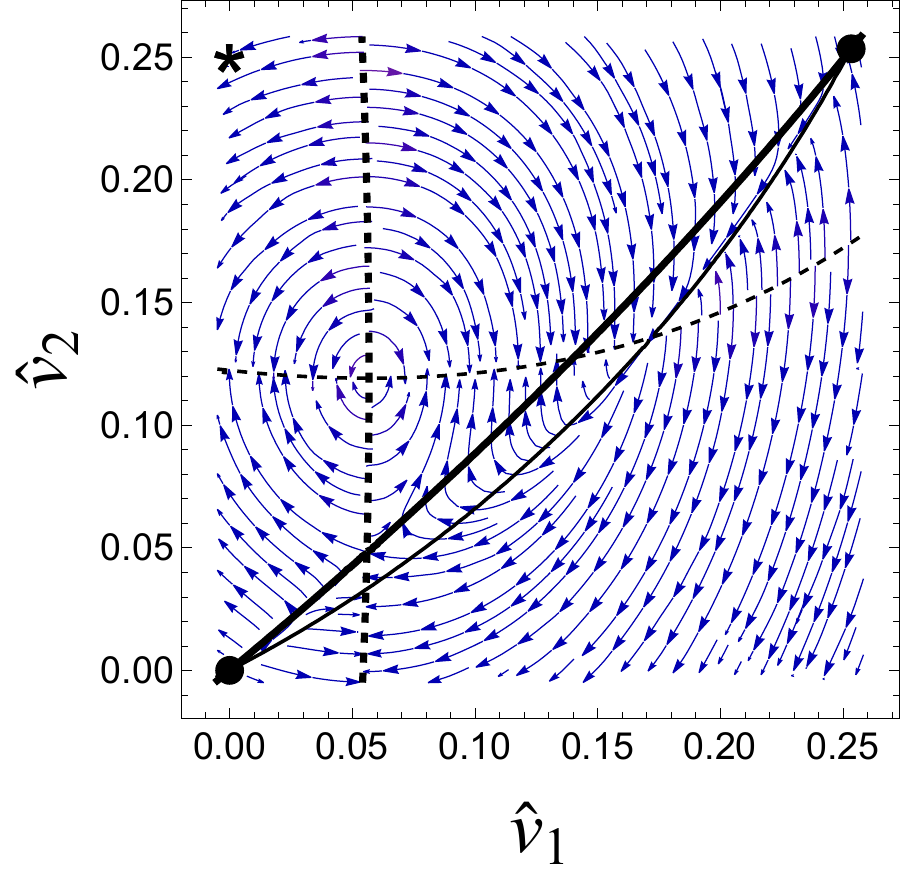} %
		\includegraphics[width=0.45\linewidth]{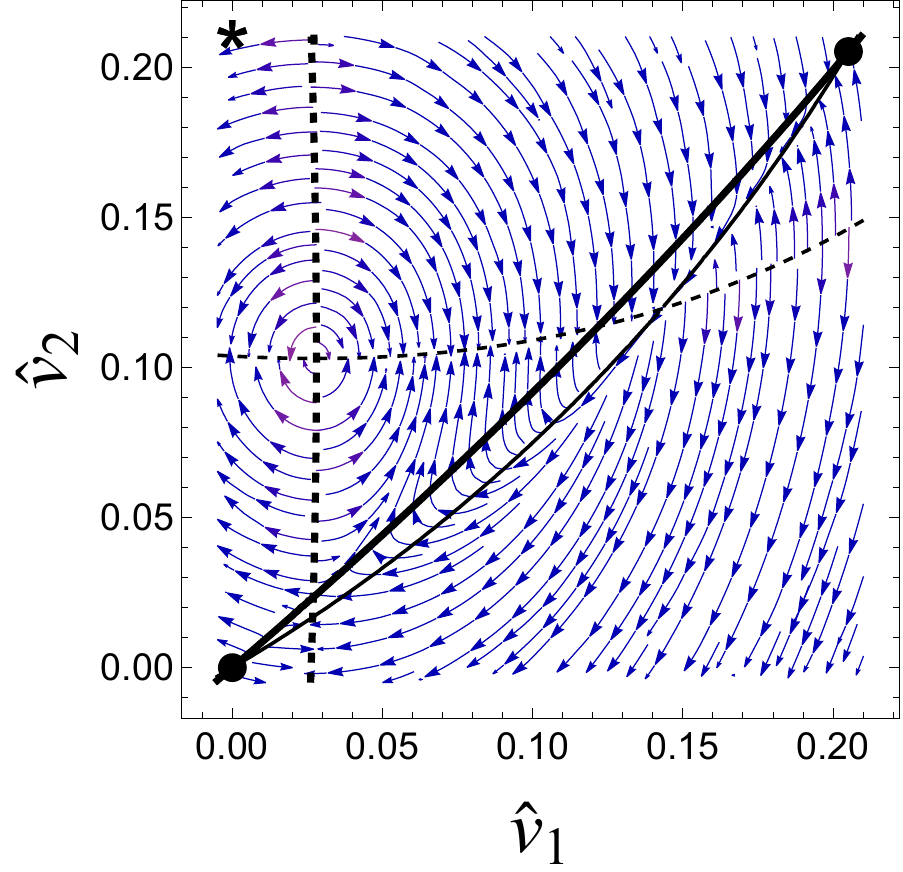} %
	\end{center}
	\caption{Direction fields in the case of $\gamma_1 = 7/6$, $\gamma_2 = 5/3$, $\mu = 0.55$ and $c_0 = 0.57$ for $M_0 = 1.075$ (above left), $1.12$ (below), $1.15$ (above right). 
		The curves of $M_0 = M_{10}$ (thick dashed curve), $M_0 = M_{20}$ (thin dashed curve), $(d v_1)/(d \varphi) = 0$ (thick solid curve) and $(d v_2)/(d \varphi) = 0$ (thick solid curves) are also shown. 
		The black marks represent the unperturbed and unperturbed states and the star is the state just after a sub-shock $s_1^{A}$. }
	\label{fig:dynamical_mu055_c057}
\end{figure}

In this section, we summarize a method to classify the shock-structure solutions numerically from the viewpoint of the dynamical analysis of the ODE system \eqref{eq:shockstructure_general}. 
In the present case, the ODE system \eqref{eq:shockstructure_general} is written in the following explicit form with the independent variables $\mathbf{u}=(\rho_1, \rho_2, v_1, v_2, T_1, T_2)$: 
\begin{widetext}
\begin{equation}\label{finaleODE}
\begin{split}
&\frac{d}{d \varphi}\left( - s \rho + \rho v \right)= 0,\\
&\frac{d}{d \varphi}\left( - s \rho v+ \rho v^2 + p + \Pi - \sigma
\right) = 0,\\
&\frac{d}{d \varphi}\left\{ - s \left(\frac{1}{2}\rho v^2 + \rho \varepsilon\right)
+\left(\frac{1}{2}\rho v^2 + \rho \varepsilon +  p + \Pi -\sigma
\right)v + q \right\} = 0,\\
&\frac{d}{d \varphi}\left(  - s \rho_1 + \rho_1  v_1 \right) 
= 0, \\
&\frac{d}{d \varphi}\left( - s \rho_1 v_1 + \rho_1  v^2_1 + p_1  \right) = \hat{m}_1, \\
&\frac{d}{d \varphi}\left\{ - s \left(\rho_1  v^2_1 + 2 \rho_1  \varepsilon_1 \right) +
\left( \rho v^2_1 + 2\rho_1  \varepsilon_1  + 2 p_1  \right) v_1  \right\} 
=  2(\hat{e}_1 +  \hat{m}_1 v),
\end{split}
\end{equation}
\end{widetext}
where \eqref{campiG}, \eqref{eq:production_interaction} and \eqref{eq:main_field} hold. 
We can immediately integrate \eqref{finaleODE}$_{1-4}$ and obtain the expressions of $\rho_1$, $\rho_2$, $T_1$ and $T_2$ in terms of $v_1$ and $v_2$. 
By inserting the obtained expressions into \eqref{finaleODE}$_{1-4}$, we have 
\begin{equation*}
\begin{split}
&\frac{d v_1}{d \varphi} = f_1(v_1, v_2), \\
&\frac{d v_2}{d \varphi} = f_2(v_1, v_2).   
\end{split}
\end{equation*}
Although the functional forms of $f_1$ and $f_2$ are too complicated to write down here explicitly, we can draw the direction fields for the corresponding shock-structure solution numerically. 

Figures \ref{fig:dynamical_mu055_c057} shows the direction fields in the case of $\gamma_1 = 7/6$, $\gamma_2 = 5/3$, $\mu = 0.55$ and $c_0 = 0.57$. 
The Mach numbers are adopted $M_0 = 1.075, 1.12$, and $1.15$, respectively. 
In this figure, we show the state just after the sub-shock $s_1^{A}$ and the curves of $M_0 = M_{10}$, $M_0 = M_{20}$, $(d v_1)/(d \varphi) = 0$ and $(d v_2)/(d \varphi) = 0$. 
The shock structure solution connecting two equilibrium state pass the meeting point between the curves of $M_0 = M_{20}$ and of $(d v_2)/(d \varphi) = 0$ for $M_0 = 1.075$. 
Therefore the solution becomes continuous as we see in Figure \ref{fig:c057_M0-1_075}. 
In the case of $M_0 = 1.15$, the Mach number is larger than the maximum characteristic velocity, and the solution should be connected with the state just after the sub-shock $s_1^{A}$. 
However, as we see from Figure \ref{fig:c057_M0-1_075}, the solution does not across the meeting point between the curves of $M_0 = M_{20}$ and of $(d v_2)/(d \varphi) = 0$. 
This fact implies that the sub-shock should emerge. 
In such a way, we can interpret the behavior of the shock-structure solution without solving the Riemann problem and identify the critical Mach number as $M_0 \simeq 1.12$ by searching a threshold value such that the solution passes the meeting point between the curves of $M_0 = M_{20}$ and of $(d v_2)/(d \varphi) = 0$.

Figures \ref{fig:dynamical_mu055_c057} shows the direction fields in the case of $\gamma_1 = 7/6$, $\gamma_2 = 5/3$, $\mu = 0.81$ and $c_0 = 0.08$ for $M_0 = 1.08, 1.13$, and $1.2$. 
As all $M_0 = 1.08, 1.13, 1.2$ are larger than $M_{20}$, a sub-shock for constituent $2$ emerges. 
In a similar way, we can identify the boundary is $M_0 \simeq 1.13$ also in Case B. 

\begin{figure}
\begin{center}
\includegraphics[width=0.45\linewidth]{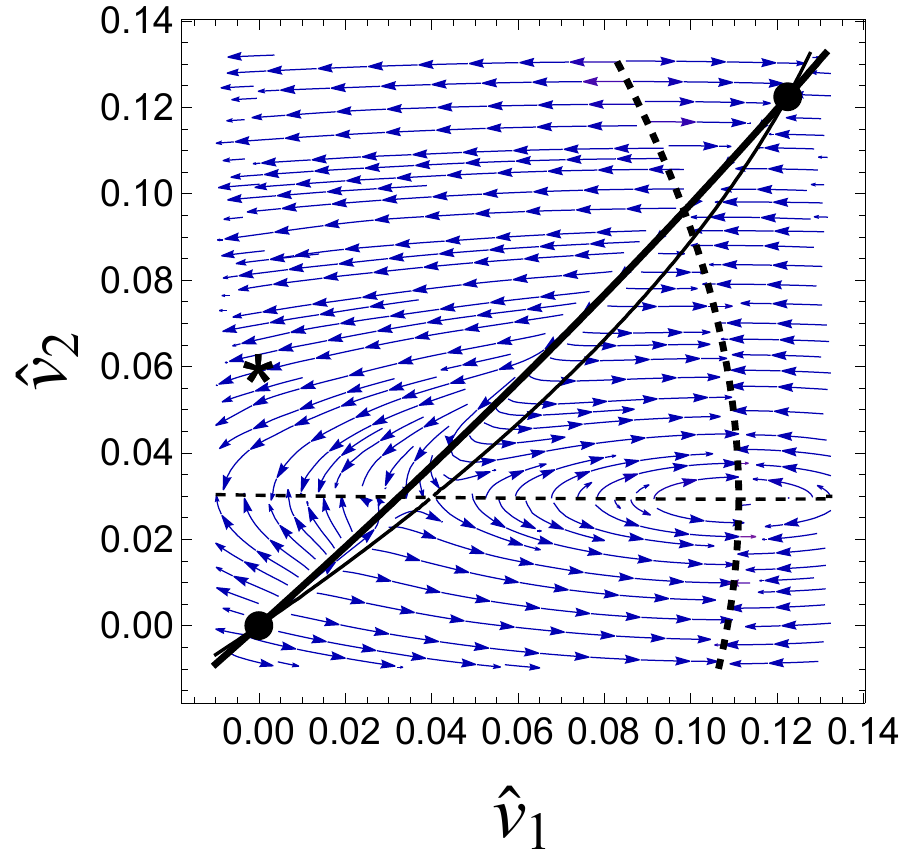} %
\includegraphics[width=0.45\linewidth]{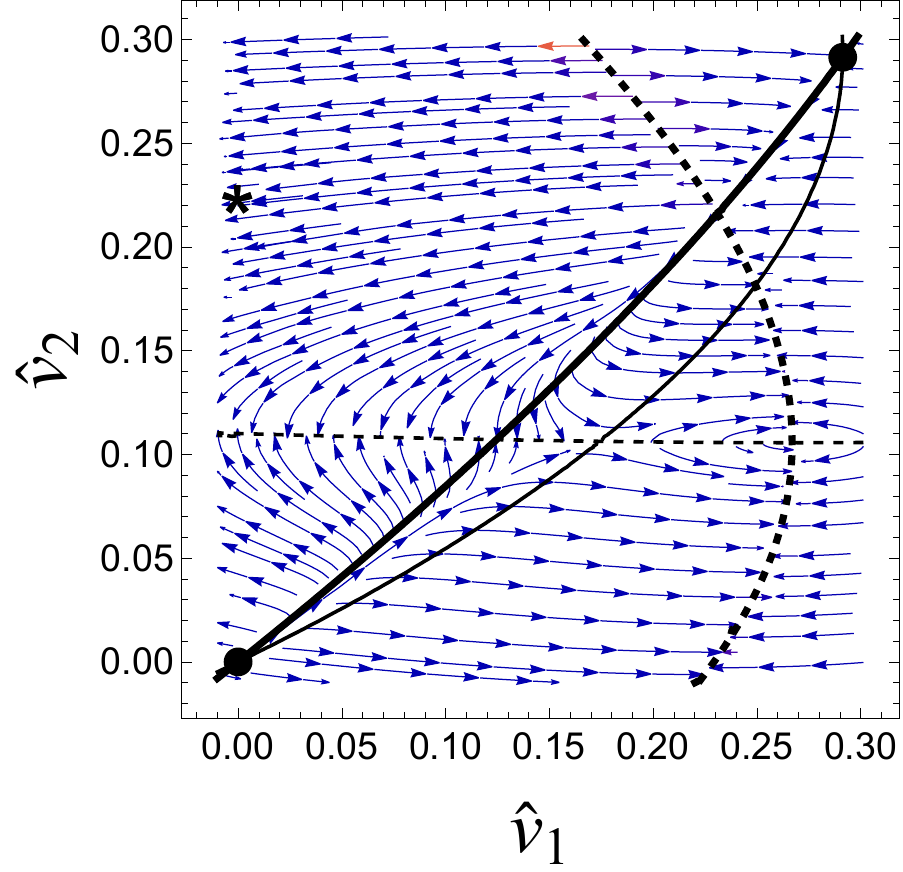} %
\includegraphics[width=0.45\linewidth]{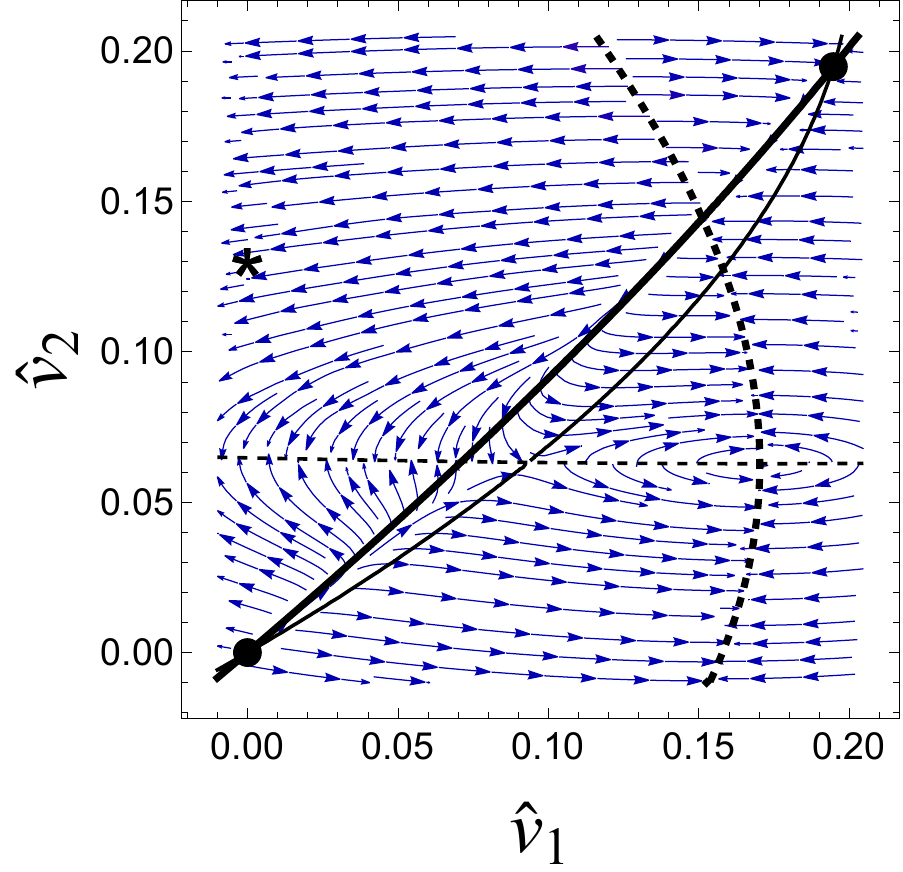} %
\end{center}
\caption{Direction fields in the case of $\gamma_1 = 7/6$, $\gamma_2 = 5/3$, $\mu = 0.81$ and $c_0 = 0.08$ for $M_0 = 1.08$ (above left), $1.13$ (below), $1.2$ (above right). 
The curves of $M_0 = M_{10}$ (thick dashed curve), $M_0 = M_{20}$ (thin dashed curve), $(d v_1)/(d \varphi) = 0$ (thick solid curve) and $(d v_2)/(d \varphi) = 0$ (thick solid curves) are also shown. 
The black marks represent the unperturbed and unperturbed states and the star is the state just after a sub-shock $s_2^{A}$.}
\label{fig:dynamical_mu081_c008}
\end{figure}

\end{document}